\documentclass[mixedtoc]{msu-thesis}
\usepackage[T1]{fontenc}
\input{glyphtounicode}
\usepackage{datatool}
\usepackage[super]{nth}
\usepackage[table]{xcolor}
\usepackage[binary-units]{siunitx}
\AtBeginDocument{\DeclareSIUnit{\kWh}{kWh}}
\AtBeginDocument{\DeclareSIUnit{\MWh}{MWh}}
\usepackage{fnpct}
\setfnpct{after-dot-space=-0.16em,after-comma-space=-0.16em}
\makepagenote
\usepackage{xparse}
\usepackage{bm}
\usepackage{pifont}
\usepackage{afterpage}
\usepackage{enumitem} %
\usepackage{tikz}
\usetikzlibrary{tikzmark, calc}

\usepackage{graphicx} %

\usepackage{tabularx} %
\usepackage{multirow} %
\usepackage[section]{placeins} %
\usepackage{caption} %
\makeatletter
\AtBeginDocument{%
  \expandafter\renewcommand\expandafter\subsection\expandafter
    {\expandafter\@fb@subsecFB\subsection}%
  \newcommand\@fb@subsecFB{\FloatBarrier
    \gdef\@fb@afterHHook{\@fb@topbarrier \gdef\@fb@afterHHook{}}}%
  \g@addto@macro\@afterheading{\@fb@afterHHook}%
  \gdef\@fb@afterHHook{}%
}
\makeatother
\makeatletter
\AtBeginDocument{%
  \expandafter\renewcommand\expandafter\subsubsection\expandafter
    {\expandafter\@fb@subsubsecFB\subsubsection}%
  \newcommand\@fb@subsubsecFB{\FloatBarrier
    \gdef\@fb@afterHHook{\@fb@topbarrier \gdef\@fb@afterHHook{}}}%
  \g@addto@macro\@afterheading{\@fb@afterHHook}%
  \gdef\@fb@afterHHook{}%
}
\makeatother

\usepackage{amssymb} %
\usepackage{amsmath}
\usepackage{mleftright}

\renewcommand{\left}{\mleft}
\renewcommand{\right}{\mright}

\usepackage{amsthm} %
\usepackage{thmtools} %

\makeatletter
\patchcmd\thmt@mklistcmd
  {\thmt@thmname}
  {\check@optarg{\thmt@thmname}}
  {}{}
\patchcmd\thmt@mklistcmd
  {\thmt@thmname\ifx}
  {\check@optarg{\thmt@thmname}\ifx}
  {}{}
\protected\def\check@optarg#1{%
  \@ifnextchar\thmtformatoptarg\@secondoftwo{#1}%
}
\makeatother
\newtheorem*{definition}{Definition}

\PassOptionsToPackage{hyphens}{url}
\usepackage{url}
\usepackage{hyperref} %
\usepackage{needspace}
\usepackage{footnotehyper} %
\usepackage{bookmark} %
\bookmarksetup{numbered, depth=3}
\makesavenoteenv{tabular} %
\makesavenoteenv{table} %
\makesavenoteenv{figure} %
\makesavenoteenv{minipage} %
\usepackage{xurl}
\usepackage[hyperref, backend=biber, minnames=2, sorting=none, style=numeric-comp, backref]{biblatex} %
\usepackage{etoolbox}
\usepackage{tabstackengine}

\makeatletter
\DeclareFieldFormat{eprint:arxiv}{%
  arXiv\addcolon\space
  \ifhyperref
    {\href{https://arxiv.org/\abx@arxivpath/#1}{%
       \nolinkurl{#1}%
       \iffieldundef{eprintclass}
         {}
         {\addspace\mkbibbrackets{\thefield{eprintclass}}}}}
    {\nolinkurl{#1}%
     \iffieldundef{eprintclass}
       {}
       {\addspace\mkbibbrackets{\thefield{eprintclass}}}}}
\makeatother

\stackMath
\makeatletter
\newcommand\sqmatrix[2][c]{%
  \fixTABwidth{T}%
  \setbox0=\hbox{$\tabbedCenterstack{#2}$}%
  \setstackgap{L}{\dimexpr\maxTABwd+\tabbed@gap}%
  \tabbedCenterstack[#1]{#2}%
}
\makeatother

\makeatletter
\renewcommand{\@pnumwidth}{2.5em} %
\setpnumwidth{2.5em}              %
\makeatother

\appto{\bibfont}{\sloppy}
\AtEveryBibitem{\filbreak}

\hypersetup{colorlinks=false,pdfborder={0 0 0}}

\makeatletter
\AtBeginDocument{
  \hypersetup{
    pdftitle = {\@title},
    pdfauthor = {\@author},
    pdfsubject = {PhD Dissertation},
    pdfkeywords = {PhD Dissertation, Thesis, Michigan State University, Nuclear Physics, Many-Body Physics, Emulation, Parametric Matrix Models},
  }
}
\makeatother

\usepackage{acro} %
\acsetup{single=1,single-style=long,make-links=true}

\newcommand{\hiddenacp}[1]{\begin{NoHyper}\acp{#1}\end{NoHyper}}

\makeatletter
\newcommand{\colorstrut}{%
  \rule[-\dimexpr\dp\@arstrutbox+\aboverulesep\relax]%
       {0pt}%
       {\dimexpr\ht\@arstrutbox+\dp\@arstrutbox+\aboverulesep+\belowrulesep\relax}%
}
\makeatother

\let\comment\undefined
\newcommand{\comment}[1]{}

\newcommand{\bigO}[1]{\mathcal{O}\left(#1\right)}

\let\oldunderline\underline
\renewcommand{\underline}[1]{%
  \relax\ifmmode
    {\mkern 1.5mu \oldunderline{\mkern -1.5mu \relax #1\mskip -1.5mu}\mskip 1.5mu}%
  \else
    \oldunderline{#1}%
  \fi
}

\newcommand{\ul}[1]{\underline{#1}}
\newcommand{\abs}[1]{\left|#1\right|}
\newcommand{\norm}[1]{\left\Vert#1\right\Vert}
\newcommand{\diag}[1]{\mathrm{diag}\left(#1\right)}

\let\P\undefined
\newcommand{\P}{{\vphantom{P}^{\downharpoonright\!}P}}
\newcommand{\Pdown}{\P}
\newcommand{\Pinv}{{P^{\upharpoonleft}}}
\newcommand{\Pup}{\Pinv}
\newcommand{\Pdag}{{P^{\dagger}}}
\newcommand{\V}{\mathcal{V}}

\newcommand{\mathset}[1]{\ensuremath{\left\{#1\right\}}}
\newcommand{\mathlist}[1]{\ensuremath{\left[#1\right]}}

\newcommand{\Cbb}{\mathbb{C}}
\newcommand{\Rbb}{\mathbb{R}}

\newcommand{\Hbb}{\mathbb{H}}

\newcommand{\bra}[1]{\ensuremath{\left\langle #1 \right|}}
\newcommand{\ket}[1]{\ensuremath{\left| #1 \right\rangle}}
\newcommand{\braket}[2]{\ensuremath{\left\langle #1 \middle| #2 \right\rangle}}
\newcommand{\ketbra}[2]{\ensuremath{\left| #1 \right\rangle\left\langle #2 \right|}}
\newcommand{\sandwich}[3]{\ensuremath{\left\langle #1 \middle| #2 \middle| #3 \right\rangle}}
\newcommand{\expect}[1]{\ensuremath{\left\langle #1 \right\rangle}}

\newcommand{\pypackage}[1]{\textsc{#1}}

\DeclareDocumentCommand{\Loss}{d()}{%
    \mathcal{L}%
    \IfNoValueF{#1}{\left(#1\right)}%
}

\DeclareMathOperator*{\argmax}{arg\,max}
\DeclareMathOperator*{\argmin}{arg\,min}
\DeclareMathOperator*{\mean}{mean}
\DeclareMathOperator*{\std}{std}

\DeclareMathOperator*{\mad}{MAD}
\DeclareMathOperator*{\median}{median}
\DeclareMathOperator*{\clipfunc}{clip}

\newcommand{\bin}[1]{\ensuremath{#1}_2}
\newcommand{\dec}[1]{\ensuremath{#1}_{10}}
\newcommand{\bitstr}[1]{\ensuremath{\left(#1\right)}_2}

\newcommand{\subfigref}[2]{\hyperref[#1]{\ref*{#1}(#2)}}

\newcommand{\emulatorsummary}[9]{%
    \def\tempa{#1}%
    \def\tempb{#2}%
    \def\tempc{#3}%
    \def\tempd{#4}%
    \def\tempe{#5}%
    \def\tempf{#6}%
    \def\tempg{#7}%
    \def\temph{#8}%
    \def\tempi{#9}%
    \emulatorsummarycontinued
    }
\newcommand{\emulatorsummarycontinued}[3]{%
    {%
        \renewcommand{\arraystretch}{1.5}
        \setlength{\tabcolsep}{8pt}
        \begin{tabularx}{\linewidth}{p{3.5cm}X}
            \toprule
            \multicolumn{2}{c}{\begin{minipage}{\dimexpr\linewidth - 3.5cm - 4\tabcolsep - 4\arrayrulewidth\relax}%
                \vspace{0.5em}%
                \centering\textbf{Emulator Summary for \tempa}%
                \vspace{0.5em}%
            \end{minipage}} \\
            \midrule
            \textbf{Mathematically\newline Parametric} & \tempb \\
            \midrule
            \textbf{Numerically\newline Parametric} & \tempc \\
            \midrule
            \textbf{Nonlinear-Capable} & \tempd \\
            \midrule
            \textbf{Computationally\newline Efficient} & \tempe \\
            \midrule
            \textbf{Data Efficient} & \tempf \\
            \midrule
            \textbf{Hyperparameter\newline Efficient} & \tempg \\
            \midrule
            \textbf{Systematically\newline Improvable} & \temph \\
            \midrule
            \textbf{Interpretable} & \tempi \\
            \midrule
            \textbf{Physically\newline Consistent} & #1 \\
            \midrule
            \textbf{Intrusiveness} & #2 \\
            \midrule
            \textbf{Quantified\newline Uncertainties} & #3 \\
            \bottomrule
        \end{tabularx}
    }
}

\DeclareRobustCommand{\IfIsInteger}[3]{\DTLifint{#1}{#2}{#3}}

\newcommand{\ord}[1]{%
    \IfIsInteger{#1}{%
        \nth{#1}%
    }{%
        $#1^{\mathrm{th}}$%
    }%
}
\newcommand{\nlo}[1]{%
    \IfIsInteger{#1}{%
        \ifcase#1\relax
            LO%
        \or NLO%
        \else
            N$^{#1}$LO%
        \fi
    }{%
        N$^{#1}$LO%
    }%
}
\newcommand{\cmark}{\textcolor{green}{\ding{51}}}%
\newcommand{\xmark}{\textcolor{red}{\ding{55}}}%
\newcommand{\bsim}{\textcolor{blue}{\ding{93}}}%
\newcommand{\sumI}{\textbf{I}}%
\newcommand{\sumN}{\textbf{N}}%
\newcommand{\sumH}{\textbf{H}}%
\newcommand{\sumE}{\textcolor{blue}{\textbf{E}}}%
\newcommand{\allemsummary}{%
            \toprule
            \textbf{Method} {\hfill\begin{tabular}{|cl|}\hline\multicolumn{2}{|c|}{\textbf{Legend}}\\\hline \cmark & Yes\\\xmark & No\\ \bsim & Partial or conditional\\\sumI & Intrusive \\ \sumN & Non-intrusive\\ \sumH & Hybrid\\ \textbf{A} & Adaptive\\ \sumE & Error bounds or estimates\\\hline\end{tabular}\hfill} & %
                \rotatebox[origin=c]{270}{\hyperref[sec:parametricity]{\textbf{Mathematically Parametric}}} & %
            \rotatebox[origin=c]{270}{\hyperref[sec:parametricity]{\textbf{Numerically Parametric}}} & %
            \rotatebox[origin=c]{270}{\hyperref[sec:linearity]{\textbf{Nonlinear-Capable}}} & %
            \rotatebox[origin=c]{270}{\hyperref[sec:comp_efficiency]{\textbf{Computationally Efficient}}} & %
            \rotatebox[origin=c]{270}{\hyperref[sec:data_efficiency]{\textbf{Data Efficient}}} & %
            \rotatebox[origin=c]{270}{\hyperref[sec:hyper_efficiency]{\textbf{Hyperparameter Efficient}}} & %
            \rotatebox[origin=c]{270}{\hyperref[sec:sys_improvability]{\textbf{Systematically Improvable}}} & %
            \rotatebox[origin=c]{270}{\hyperref[sec:interpretability]{\textbf{Interpretable}}} & %
            \rotatebox[origin=c]{270}{\hyperref[sec:phys_consistency]{\textbf{Physically Consistent}}} & %
            \rotatebox[origin=c]{270}{\hyperref[sec:intrusiveness]{\textbf{Intrusiveness}}} & %
            \rotatebox[origin=c]{270}{\hyperref[sec:quantifiable_uncertainties]{\textbf{Quantified Uncertainties}}}\\
            \midrule
            \hyperref[sec:erbm]{\textbf{Galerkin}} & & & & & & & & & & &\\
            \hspace{2em}\hyperref[sec:pod]{\textbf{\Acl{pod}}} & \cmark & \xmark & \bsim & \cmark & \cmark & \cmark & \cmark & \cmark & \cmark & \sumI & \sumE\\
            \hspace{2em}\hyperref[sec:dmd]{\textbf{\Acl{dmd}}} & \bsim & \xmark & \bsim & \cmark & \cmark & \cmark & \bsim & \cmark & \xmark & \sumN & \xmark\\
                \hspace{2em}\hyperref[sec:apdr]{\textbf{Method in Section~\ref{sec:apdr}}} & \xmark & \xmark & \xmark & \cmark & \cmark & \cmark & \cmark & \cmark & \cmark & \sumI & \sumE\\
            \hspace{2em}\hyperref[sec:ec]{\textbf{\Acl{ec}}} & \cmark & \xmark & \xmark & \cmark & \cmark & \cmark & \cmark & \cmark & \cmark & \sumI & \sumE\\
            \hyperref[sec:gpr]{\textbf{\Acl{gpr}}} & \cmark & \xmark & \cmark & \xmark & \xmark & \bsim & \xmark & \bsim & \xmark & \sumN & \cmark\\
            \hyperref[sec:pinn]{\textbf{\Aclp{pinn}}} & \cmark & \cmark & \cmark & \bsim & \bsim & \xmark & \xmark & \xmark & \bsim & \sumH & \xmark\\
}

\title{Parametric Matrix Models \\\protect for Emulation in \\\protect Nuclear and Many-Body Physics}
\author{Patrick Cook}
\dualmajor{Physics}{Computational Mathematics, Science, and Engineering} %
\date{2026}

\DeclareAcronym{ad}{%
  short=AD,
  long=automatic differentiation,
}
\DeclareAcronym{psd}{%
  short=PSD,
  long=positive semidefinite,
}
\DeclareAcronym{pmm}{%
  short=PMM,
  long=parametric matrix model,
}
\DeclareAcronym{lmg}{%
    short=LMG,
    long=Lipkin--Meshkov--Glick,
    }
\DeclareAcronym{svd}{%
  short=SVD,
  long=singular value decomposition,
}
\DeclareAcronym{pca}{%
  short=PCA,
  long=principal component analysis,
}
\DeclareAcronym{msu}{%
  short=MSU,
  long=Michigan State University,
}
\DeclareAcronym{frib}{%
  short=FRIB,
  long=Facility for Rare Isotope Beams,
}
\DeclareAcronym{dmd}{%
    short=DMD,
    long=dynamic mode decomposition,
}
\DeclareAcronym{sindy}{%
    short=SINDy,
    long=sparse identification of nonlinear dynamics,
}
\DeclareAcronym{pod}{%
  short=POD,
  long=proper orthogonal decomposition,
}
\DeclareAcronym{deim}{%
  short=DEIM,
  long=discrete empirical interpolation method,
}
\DeclareAcronym{nmf}{%
  short=NMF,
  long=non-negative matrix factorization,
}
\DeclareAcronym{ann}{%
  short=ANN,
  long=artificial neural network,
}
\DeclareAcronym{cnn}{%
  short=CNN,
  long=convolutional neural network,
}
\DeclareAcronym{ae}{%
  short=AE,
  long=autoencoder,
}
\DeclareAcronym{mlp}{%
  short=MLP,
  long=multi-layer perceptron,
}
\DeclareAcronym{ml}{%
  short=ML,
  long=machine learning,
}
\DeclareAcronym{nn}{%
  short=NN,
  long=neural network,
}
\DeclareAcronym{ai}{%
  short=AI,
  long=artificial intelligence,
}
\DeclareAcronym{gpu}{%
  short=GPU,
  long=graphics processing unit,
}
\DeclareAcronym{cc}{%
  short=CC,
  long=coupled cluster,
}
\DeclareAcronym{imsrg}{%
  short=IMSRG,
  long=in-medium similarity renormalization group,
}
\DeclareAcronym{eos}{%
  short=EOS,
  long=equation of state,
}
\DeclareAcronym{eft}{%
  short=EFT,
  long=effective field theory,
}
\DeclareAcronym{ceft}{%
  short=$\chi$EFT,
  long=chiral effective field theory,
}
\DeclareAcronym{mad}{%
  short=MAD,
  long=median absolute deviation,
}
\DeclareAcronym{pnm}{%
  short=PNM,
  long=pure neutron matter,
}
\DeclareAcronym{snm}{%
  short=SNM,
  long=symmetric nuclear matter,
}
\DeclareAcronym{emn}{%
  short=EMN,
  long=Entem--Machleidt--Nosyk,
}
\DeclareAcronym{dft}{%
  short=DFT,
  long=density functional theory,
  short-plural=s,
  long-plural-form=density functional theories,
}
\DeclareAcronym{mbpt}{%
  short=MBPT,
  long=many-body perturbation theory,
  short-plural=s,
  long-plural-form=many-body perturbation theories,
}
\DeclareAcronym{nqs}{%
  short=NQS,
  long=neural quantum state,
}
\DeclareAcronym{ec}{%
  short=EC,
  long=eigenvector continuation,
}
\DeclareAcronym{uq}{%
  short=UQ,
  long=uncertainty quantification,
}
\DeclareAcronym{lec}{%
  short=LEC,
  long=low-energy constant,
}
\DeclareAcronym{gp}{%
  short=GP,
  long=Gaussian process,
  short-plural=s,
  long-plural=es,
}
\DeclareAcronym{gpr}{%
  short=GPR,
  long=Gaussian process regression,
}
\DeclareAcronym{llm}{%
    short=LLM,
    long=large language model,
    tag=hidden,
}
\DeclareAcronym{tddft}{%
    short=TDDFT,
    long=time-dependent density functional theory,
}
\DeclareAcronym{hf}{%
    short=HF,
    long=Hartree--Fock,
}
\DeclareAcronym{hfb}{%
    short=HFB,
    long=Hartree--Fock--Bogoliubov,
}
\DeclareAcronym{wmse}{%
    short=wMSE,
    long=weighted mean squared error,
}
\DeclareAcronym{wvar}{%
    short=wVAR,
    long=variance of the weighted absolute error,
}
\DeclareAcronym{erbm}{%
    short=eRBM,
    long=explicit reduced basis method,
}
\DeclareAcronym{irbm}{%
    short=iRBM,
    long=implicit reduced basis method,
}
\DeclareAcronym{rbm}{%
    short=RBM,
    long=reduced basis method,
}
\DeclareAcronym{pinn}{%
    short=PINN,
    long=physics-informed neural network,
}
\DeclareAcronym{apdr}{%
    short=APDR,
    long=\textit{a priori} dimensionality reduction
}

\begin{document}

\frontmatter

\pdfbookmark[0]{Title Page}{titlepage}
\maketitlepage

\pdfbookmark[0]{Abstract}{abstract}
\begin{abstract}
Progress in nuclear and many-body physics today is predicated on the ability to solve large-scale, strongly correlated quantum many-body problems. As the theoretical models become more sophisticated, they also become more computationally complex. Simultaneously, quantifying uncertainty in model predictions and fitting free parameters to experimental observations requires repeated evaluation of these expensive models. The desire for progress in the theoretical understanding of such systems will always outpace the rate at which computational resources become available, and so improvements in the efficiency of our high-fidelity methods alone are insufficient. Instead, surrogate models---known as emulators---provide the means of accomplishing these goals.

This thesis provides an introduction into the current state of emulation in nuclear and many-body physics. The motivations, goals, and origins of currently popular emulation methods are discussed along with selected examples. We see how many methods are closely mathematically related and how trade-offs are made to optimize specific properties or applications.

The central work in this thesis is the method of \acfp{pmm}, an emulation and general \acl{ml} framework which combines aspects of traditional \acl{rbm} with modern parametric \acl{ml}. \Acp{pmm} are able to retain as much or as little physical information about the underlying system as desired, yielding not only excellent performance but also nearly unparalleled adaptability, interpretability, and trustworthiness as an emulation method.

A formal mathematical framework for \acp{pmm} is developed and accompanied by practical step-by-step procedures for the application of the method. Extensions and current research into applying the method to a variety of problems are presented, culminating in the to-date most comprehensive \ac{pmm}-emulator which is applied to the problem of emulating \acl{imsrg} calculations of nuclear matter.

As part of this thesis, the open-source \pypackage{pyPMM} package was developed~\cite{pyPMM2026}. This package enables any researcher to construct, train, share, and deploy \ac{pmm}-based emulators with modular, extendable, and \ac{gpu}-optimized code. All \ac{pmm} examples in this thesis were created using this package.

\end{abstract}

\clearpage
\setcounter{tocdepth}{1}

\SingleSpacing
\pdfbookmark[0]{Table of Contents}{toc}
\tableofcontents* %

\clearpage
\printacronyms[heading=chapter, name=List of Abbreviations, exclude={hidden}, preamble={\vspace*{1em}}]
\mainmatter

\chapter{Introduction}\label{ch:intro}
\acresetall %

\section{Foreword}
The fields of nuclear theory, many-body theory, computational physics, \ac{ai}, and \ac{ml} are all rapidly evolving. This dissertation captures only a partial snapshot of these fields and how this work forced itself into them. Many of the statements made in this dissertation---especially in regard to \ac{ai} and \ac{ml}---will likely be outdated at the time of reading\footnote{I'm sure they'll either age like wine or milk, nothing in between.}. Moreover, much of the work itself is still evolving. The reader is encouraged to look up the latest research regarding the topics discussed herein.

It is unfortunate that deliberately deceptive and unquantifiably harmful modern trends in the commercialization of a very specific subfield of \ac{ai} and \ac{ml} have brought such a negative connotation to the entire field. Almost nothing in this thesis is relevant to the application of \hiddenacp{llm} and generative \ac{ai}. The extent to which this work is related to \hiddenacp{llm} begins and ends at the fact that both are in the subfield of \ac{ml}. The research, engineering, algorithms, mathematics, and science behind \hiddenacp{llm} are all exceedingly interesting and impressive and deserve to be appreciated in a vacuum separate from their marketed products.

\subsection*{Declaration on the use of Generative AI Tools}
Despite the use of em-dashes\footnote{A punctuation that some today associate with \ac{ai}-generated text.}---a punctuation I have always made extensive use of---absolutely no text or material content in this thesis was created with the use of generative \ac{ai} tools. Spelling, grammar, and other typographical errors in drafts of this thesis were detected in part with generative \ac{ai} tools. Some \LaTeX{} code for the formatting of this document was created with assistance from generative \ac{ai} tools.

\section{Organization}
This dissertation is organized as follows: Chapter~\ref{ch:intro} is this. Chapter~\ref{ch:emu} introduces emulation and establishes a list of properties, goals, and considerations for emulators in computational physics and concludes with a summary of commonly used methods with selected examples. Chapter~\ref{ch:pmm} introduces \acp{pmm}, a new class of emulators and machine learning models that use trainable versions of the known or supposed governing equations of a system to learn from data. We discuss the origins of \acp{pmm}, formalize the mathematical framework for the method, detail the practical aspects of constructing and training \acp{pmm}, and walk through varied example applications. Two uncertainty quantification methods for emulation are discussed in Chapter~\ref{ch:uq}, with particular attention to conformal predictions with \acp{pmm}. Finally, Chapter~\ref{ch:conclusion} summarizes the work and outlines future directions.

\section{Reading Guide}
This thesis is written for readers with strong backgrounds in quantum mechanics, linear algebra, and numerical methods, passing familiarity with \ac{ml}, and little to no background in emulation or \acp{pmm}. A linear reading should provide a comprehensive understanding of the theory and practice of \acp{pmm}. The minimum required reading for just the theory of \acp{pmm} is Sections~\ref{sec:erbm},~\ref{sec:pod},~and~\ref{sec:pmm_irbm}. Those interested in using \acp{pmm} should additionally read Sections~\ref{sec:training},~\ref{sec:parameterization},~and~\ref{sec:pitfalls}. For the latest advances in \acp{pmm} at time of writing, see Sections~\ref{sec:smoothing},~\ref{sec:pmm_reg},~\ref{sec:dft},~\ref{sec:tddft},~and~\ref{sec:pmm_pdd}.

\chapter{Emulation in Nuclear and Many-Body Physics}\label{ch:emu}
\acresetall
Our current best models in nuclear physics---both \textit{ab initio} and phenomenological---have free parameters that must be fit to experimental data. These parameters often come in the form of \acp{lec} arising from the truncation of effective field theories, but also appear in phenomenological models such as energy density functionals~\cite{Epelbaum09,Hebeler11,Furnstahl20,Marino21}. The process of fitting these parameters, and just as importantly quantifying the uncertainties in these parameters, requires searching high-dimensional parameter spaces. Each trial point in parameter space requires a full model evaluation. It is not uncommon for a single model evaluation to take hours or days on high-performance computing clusters~\cite{He24,Morris15,Yoshida23,Miyagi22,Schunck11,Ryssens16,Bender20,Schunck16,Schunck22}. With hundreds of thousands to millions of trial evaluations required to adequately sample the parameter space, this process quickly becomes intractable.

One may be tempted to replace the full---also called \textit{high-fidelity} or \textit{full-order}---models with relatively simple and fast regression models such as \acp*{mlp} or \acp*{gp} fit directly to available experimental data. After all, the high-fidelity models are also fit to experimental data. There are many issues with this approach. Depending on your persuasion, you may focus first on the philosophical issues regarding what we are actually learning about the universe from such models or more on the practical issues in training such models to sufficient accuracy. Our goal is ultimately to understand the physics which leads to the properties of nuclei, and even an \ac{ann} that perfectly reproduces experimental data cannot provide this understanding\footnote{Unless it turns out that the universe is just a big neural network.}. Such a perfect \ac{ann} is likely unobtainable anyway, due to the limited amount of experimental data available and the complexity of nuclear systems. Consider an overly optimistic scenario where we have access to experimental data of the energies, decay modes, quadrupole deformation, and any ten other observables imaginable for several low-lying states of all known nuclei. A quick Fermi estimate shows that this would be about $10^5$ data points. As a comparison, one of the most widely used and now-solved datasets is the ImageNet Large Scale Visual Recognition Challenge which contains more than $10^6$ full-color images~\cite{ILSVRC15}. Assuming that just $0.1\%$ of the raw values of that dataset are actual useful information, it would contain about $10^8$ data points. This is three orders of magnitude larger than our exceedingly optimistic estimate for all the nuclear data we could ever hope to have. Even if we had enough data, the complexity of nuclear systems---strongly coupled quantum many-body systems---is substantially greater than any computer vision task. All this is to say that our high-fidelity models play irreplaceable roles both in the theoretical understanding of nuclear systems and in the practical prediction of nuclear properties.

High-fidelity models are irreplaceable but simultaneously too expensive to fit directly to data. The solution to this dilemma comes in the form of \textit{emulators} or \textit{reduced-order models}. The core idea is to use the high-fidelity model to generate a relatively large amount of synthetic data, then use that synthetic data to train a fast surrogate model that approximates the high-fidelity model. Traditionally, if this surrogate model was a low-dimensional (usually projected) approximation of the high-fidelity model it would be called a reduced-order model, and conversely if the surrogate model was a trained ``black-box'' machine learning model it would be known as an emulator. Recent advancements have blurred the line between reduced-order models and emulators in the traditional sense, and as such we will refer to all surrogate models collectively as emulators and classify them via a series of other properties~\cite{Melendez2022}.

While slow, generating data with the high-fidelity model is still far cheaper than obtaining experimental data. Once trained, the emulator can be used in place of the high-fidelity model for expensive tasks such as parameter fitting and uncertainty quantification. If accurate and trustworthy enough, the emulator may even replace the high-fidelity model for most evaluations as is the case in traditional model order reduction and the concept of digital twins. Beyond raw accuracy, there are several important properties that a good emulator should have which are discussed in the following sections.

\section{Parametricity}\label{sec:parametricity}

Physical models are \textit{parametric}. That is, the behavior, dynamics, or physics of the system depend on some parameters. For instance, the mass of a body, the interaction strength between bodies, or even the number of particles are all parameters of a system. These parameters may be physical, like the preceding examples, or numerical, such as the resolution and tolerance of the numerical representation. One might describe the parametric equations governing such a system as actually defining a family of systems, where each system corresponds to a specific realization of those parameters. For example, consider the family of quadratic polynomials $f(x; b, c) = x^2 + bx + c$. A specific quadratic in this family is determined by the parameters $b$ and $c$.

There are several distinct definitions for ``parametricity'' and ``parametric'' in the context of machine learning, physics, and emulation. The most common definition from machine learning describes whether or not a machine learning model's size depends on the amount of data provided to it. We will distinguish between the mathematical or physical parametricity and the machine learning or numerical parametricity as follows.

\begin{definition}[Mathematically Parametric Emulator]
    An emulator which is able to model an entire family of systems as a function of some parameters is a mathematically parametric emulator. The opposite is a mathematically non-parametric emulator.
\end{definition}

\begin{definition}[Numerically Parametric Emulator]
    An emulator or machine learning model with a size that is independent of the amount of data available to it is called a numerically parametric emulator. The opposite is a numerically non-parametric emulator. Equivalently, a numerically parametric emulator is one whose parameters are not explicitly determined from data.
\end{definition}

Numerical parametricity is a particularly useful property in a machine learning model since it often allows for progressive training. A numerically parametric model trained on a set of data can later update its numerical parameters when new data become available---without needing to consult the original data. This means that as new training examples become available, one can adapt the model with comparatively small changes rather than needing to completely refit the model.

There is an additional overloading of the term ``parameter'' in machine learning. Namely, the trainable parameters of a machine learning model are often referred to simply as ``parameters''. It is likely for this reason that specific models will use more specialized terms for trainable parameters, like ``weights'' and ``biases'' in the case of \acp{ann}. Most of the time however, it is clear from the context of the problem which parameters are the physical or mathematical ones and which are the trainable ones.

\section{Linearity}\label{sec:linearity}

Linear systems are by far the most well-studied types of problems. Most of our mathematical and numerical techniques are maximally effective only for linear systems. Many real-world problems of interest are also linear---or at least well-approximated by linear problems. Modern quantum mechanics, many-body theory, and \textit{ab initio} nuclear theory make heavy use of linear operators. It is perhaps unsurprising then that emulators themselves, especially more traditional methods, are often linear. There are advantages to linear emulators, stemming from the aforementioned effectiveness of tools for linear systems. However, nuclear and many-body physics contain plenty of nonlinear processes. Furthermore as the fields have improved both understanding and measurement, they have increasingly introduced beyond-linear theories.

It is possible to approximate a nonlinear process by a linear one. This is known as \textit{linearization}, and several emulator methods rely on it. However, in order to properly capture the nonlinear effects of a given model, it is best if the emulator can preserve the exact form of the nonlinearity from the high-fidelity model. At minimum, we want the emulator to be able to capture some nonlinear effects.

\section{Computational Efficiency}\label{sec:comp_efficiency}

The justification for emulation is computational efficiency. While words like ``fast'' and ``cheap'' are often used to describe emulators, computational efficiency comes in many forms and is evaluated at different stages of the emulation process. There are scenarios in which a slower algorithm that requires less memory is preferable to a faster one that requires more memory. The cost of an algorithm is often measured in terms of \textit{complexity}, specifically time and space complexity. Complexity measures how the resources required by an algorithm scale with certain parameters, often the size of the input data or the size of the algorithm itself. Complexity theory and analysis is a deep and rich field of study~\cite{sipser13}, but for our purposes we will use only a loose interpretation of complexity through asymptotic notation, or more commonly \textit{big-O notation}.

Suppose an algorithm $\mathcal{F}$ requires $f(n)$ steps\footnote{The precise definition of ``steps'' here is usually related to Turing machines. However, in scientific computing we often take a ``step'' to mean a single floating point operation, since these can be done in constant time.} to complete, where $n$ is some parameter regarding the size of the computation such as the size of the input. We say that the time complexity of $\mathcal{F}$ is $\bigO{g(n)}$ if there exist $c, n_0$ such that $f(n) \leq c g(n)$ for all $n\geq n_0$. That is to say that past a certain size the runtime of $\mathcal{F}$ does not grow faster than some constant multiple of $g(n)$. Big-O notation is particularly useful since the determination of the exact number of steps required by an algorithm is often difficult, and the leading order behavior is often sufficient to compare algorithms\footnote{I am glossing over the issue of computability and halting. We will assume that all algorithms here terminate unless otherwise stated.}. Table~\ref{tab:complexity} lists the time complexities of some common operations and algorithms used in practice in scientific computing as an example and reference.

Similar to time complexity, space complexity measures how the memory requirements of an algorithm scale with certain size parameters. This can include the size of the inputs (total space complexity) or only the size of any intermediate and output data structures (auxiliary space complexity). Auxiliary space complexity is often more relevant in practice, and is the convention used here. Table~\ref{tab:complexity} also lists the space complexities of the same common operations and algorithms.

\begin{figure}
  \centering
    \begin{tabularx}{\textwidth}{lXXXX}
    \toprule
        Algorithm & Input\newline Size & Output\newline Size & Time\newline Complexity & Space\newline Complexity\\
    \midrule
        Floating Point $+$/$\times$ & $1$ & $1$ & $\bigO{1}$ & $\bigO{1}$ \\
        Vector Addition/Scaling & $n,\,n\text{ or }1$ & $n$ & $\bigO{n}$ & $\bigO{n}$ \\
        Matrix-Vector Multiplication & $(n,n),\,n$ & $n$ & $\bigO{n^2}$ & $\bigO{n}$ \\
        Matrix-Matrix Multiplication~\cite{Strassen1969} & $(n,n),\,(n,n)$ & $(n,n)$ & $\bigO{n^{\log_2 7}}$ & $\bigO{n^2}$ \\
        Full Eigendecomposition & $(n,n)$ & $n$, $(n,n)$ & $\bigO{n^3}$ & $\bigO{n^2}$ \\
        Partial Eigendecomposition & $(n,n)$ & $k$, $(n,k)$ & $\bigO{kn^2}$ & $\bigO{kn}$\\
    \bottomrule
  \end{tabularx}
    \captionof{table}{In-practice complexities of common algorithms.}
  \label{tab:complexity}
\end{figure}

\begin{figure}
    \centering
    \includegraphics[height=0.2\textheight]{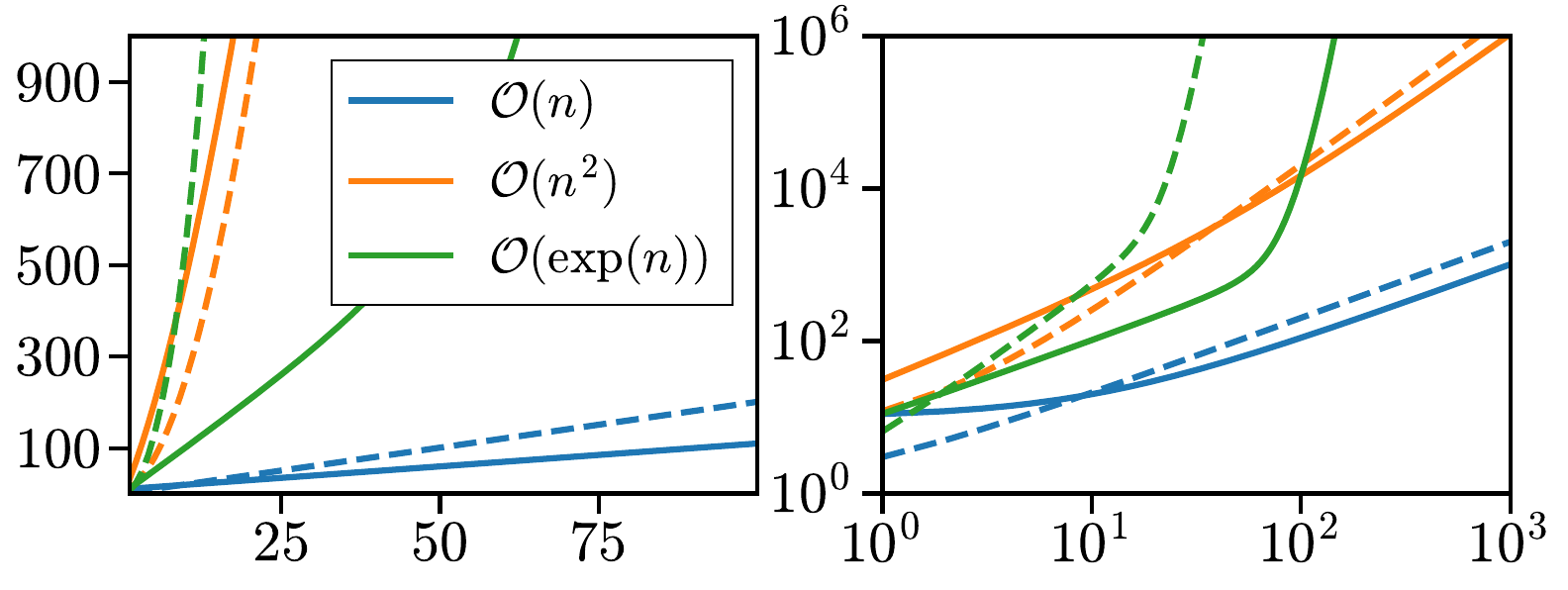}
    \caption[Examples of various functions with different big-O complexities.]{Examples of various functions with different big-O complexities. Note that since big-O notation is an asymptotic measure, constant factors and sub-leading-order terms mean that functions in the same big-O class may behave very differently for small $n$, and some functions in a ``worse'' big-O class may be smaller than functions in a ``better'' big-O class for small $n$.}
    \label{fig:complexity}
\end{figure}

In broader computational contexts, we usually say an algorithm is ``efficient'' if it has polynomial time and space complexity, i.e. $\bigO{n^k}$ for some constant $k$. However, in the context of emulation we are instead interested in outperforming the high-fidelity model---often substantially. So while the complexity of an emulator itself is often still of interest, the constant factors hidden by big-O notation may become important since the size of emulators are often significantly smaller than their high-fidelity counterparts. For example, a high-fidelity model may be efficient in the sense that it requires $\bigO{N}$ time and $\bigO{N^2}$ space, but an algorithmically inefficient emulator that operates in $\bigO{\exp(n)}$ time and $\bigO{n^3}$ space may still be preferable if $n$ is always much smaller than $N$.

\subsection{Online and Offline Costs}

A complete definition of computational efficiency in the context of an emulator requires consideration of both ``online'' and ``offline'' costs. The online cost of an emulator encompasses the resources required to use the deployed emulator. That is, once the emulator is constructed, trained, fit, or otherwise prepared and sent to an end-user, the online cost is the cost of using the emulator for inference. In the context of this work, the offline cost of an emulator encompasses the resources required to prepare it for deployment, excluding the cost of acquiring any training data. The offline cost includes the cost of constructing the emulator, training it, and any other necessary steps to prepare it for deployment. Note that some sources will include the collection of training data---and therefore the cost of evaluating the high-fidelity model the requisite number of times---in the offline costs. Including data collection in the offline costs emphasizes the importance of minimizing the amount of data required, which we will do separately in the following section. The offline cost is a one-time cost, while the online cost is incurred every time the emulator is used. It can then be said that the offline cost is amortized over the course of the online phase. Some examples of online and offline costs include:
\begin{itemize}
    \item \textbf{Online}
        \begin{itemize}
            \item Storing the emulator
            \item Performing inference with the emulator, including any necessary preprocessing of the input data
            \item Postprocessing the output of the emulator, including any necessary rescaling or dimensionality expansion
        \end{itemize}

    \item \textbf{Offline}
        \begin{itemize}
            \item Storing the training data
            \item Rescaling and preprocessing the training data, including dimensionality reduction techniques such as \ac{pca}
            \item Training via gradient descent or other optimization algorithms
            \item Pruning and other post-training optimizations
        \end{itemize}
\end{itemize}
This gives us the tools to define computational efficiency in the context of emulation.

\begin{definition}[Computationally Efficient Emulator]
    An emulator of size $n$ for a high-fidelity model of size $N$ is computationally efficient if $n\ll N$, its offline costs are at most $\bigO{N^2n^k}$, and its online costs are at most $\bigO{n^k}$ for some constant $k$, unless the output of the emulator is of size $\bigO{N}$, in which case allowance is made only for the costs of projecting the final reduced solution back to the full space and so the online costs must be at most $\bigO{Nn + n^k}$ for some constant $k$. That is, a computationally efficient emulator must be much smaller than its high-fidelity counterpart, scale at most quadratically with the size of the high-fidelity model and polynomially with its own size during the offline phase, and---unless the output size scales with the high-fidelity model---scale independently of the high-fidelity model and at most polynomially with its own size during the online phase.
\end{definition}

\section{Data Efficiency}\label{sec:data_efficiency}

A very similar concept to computational efficiency is data efficiency. However, while the number of floating point operations at our disposal is at our discretion to an extent (if we need more, we can just run our code for longer), this is often not the case for data. Especially in the context of nuclear physics emulation, data is at a premium. It is often the case that acquiring more data is exceedingly expensive monetarily or temporally, if possible at all.

\begin{definition}[Data Efficient Emulator]
    An emulator is data efficient if the error of the emulator converges with the number of training examples, $m$, at a rate $\bigO{m^{-k}}$ for some constant $k>1$ when not limited by emulator size. That is, a data efficient emulator can be made arbitrarily accurate by increasing the amount of training data at a power-law rate. More efficient emulators have larger values of $k$.
\end{definition}

While this definition is quite similar to that of computational efficiency, particular effort is focused on improving or optimizing the structure of an emulator in order to improve its rate of convergence in this context. Even the constant factors hidden by big-O notation may be the deciding factor as to whether or not a particular emulator is successful.

Recent interest in \hiddenacp{llm} highlighted the importance of data efficiency and the exact values of the constant prefactor and exponent of the convergence rate. Such attention was given to this that it became known as ``neural scaling laws'' in the study of \hiddenacp{llm}~\cite{Kaplan2020}. Scaling laws like these are not limited to machine learning, emulation, or even physics as they appear in a seemingly universal assortment of systems that exhibit phase transitions~\cite{De25,Nishimori2011}. Broadly, they are known as critical phenomena or universal phenomena.

\section{Hyperparameter Efficiency}\label{sec:hyper_efficiency}

The final type of efficiency is the efficiency of the emulator's usage of \textit{hyperparameters}---parameters that are not determined by fitting or training the emulator. Common hyperparameters include the emulator size and the choice of loss function. Hyperparameter tuning can be exceedingly difficult and expensive while also introducing opportunities for human bias from bad intuition about what values might be best. Hyperparameters are an unavoidable part of machine learning, but we generally prefer emulators with fewer possible hyperparameter values to choose from. Additionally, emulators with clear procedures and guidance on choosing the optimal hyperparameters (or at least significantly narrowing down the search space) are generally more desirable.

\section{Systematic Improvability}\label{sec:sys_improvability}

Rounding out the scaling laws is the notion of systematic improvability. In general machine learning, the idea of ``universal approximators'' most closely resembles this concept. While universal approximation theorems for \acp{ann} provide only a proof of existence, in emulation we seek not only the existence but also a description of how the emulator could reach convergence. This is possible because of the existence and knowledge of the high-fidelity model.

When considering systematic improvability, we ignore the issues of data availability and emulator fitting. Instead, focus is placed squarely on whether or not such an emulator could be sufficiently expressive that it would be equivalent to the high-fidelity model given the right parameters.

\begin{definition}[Systematically Improvable Emulator]
    An emulator of size $n$ for a high-fidelity model of size $N$ is systematically improvable if for $n\to \bigO{N}$ it converges to the high-fidelity model. That is, a systematically improvable emulator can be made arbitrarily accurate by increasing its size---up to a constant multiple of the high-fidelity model size.
\end{definition}

\section{Interpretability}\label{sec:interpretability}

Interpretability is somewhat subjective, but has nevertheless become a sought-after property in the \ac{ml} community. An interpretable emulator, or machine learning model in general, is one whose constituent parts and parameters can be ascribed meaning in the context of the system being modeled. This meaning must be beyond the direct structure of the machine learning model itself, and identifying such properties often requires an expert in both the system being modeled and the machine learning model. For example, the layers of a feedforward \ac{ann} fit to a dataset of nuclear binding energies have little to no inherent meaning in the context of nuclear physics: the parameters of the model do not correspond to physical constants or encode physical processes. On the other hand, a convolutional \ac{ann} fit to a dataset of images often learns several convolutional filters in the first layer that correspond to edge detectors, which have a clear meaning in the context of image processing.

Interpretability is desirable for several philosophical and practical reasons. In scientific contexts, our goals center around understanding the universe and the laws that govern it. A model that is not interpretable may be able to fit some data without providing any insight into the underlying processes. In this context, we've simply replaced one unknown with another, possibly more complicated, unknown. Moreover, while other avenues like \ac{uq} can guide the trustworthiness of a model, an interpretable model can be more easily interrogated and have its failure modes identified and understood. Perhaps the most practical concern is adoption. People of all backgrounds are less likely to trust, adopt, and invest in a model that they cannot understand.

\section{Physical Consistency}\label{sec:phys_consistency}

More specific to physics emulation, we want our emulators to be physically consistent. One might think that if an emulator is accurate enough, it will automatically respect the underlying conservation and symmetry laws of the system. This is seldom the case however, as the same data may be generated from many different underlying processes with different conservation and symmetry properties.

It is often impossible to guarantee exact adherence to physical laws with an emulator. Instead, we want our emulators to have quantifiable bounds on the degree to which they violate said laws and for these bounds to systematically improve with the size of the emulator and the amount of training data. The exact form and scaling of these bounds will depend on the specifics of the emulator and system being modeled, but the key point is that a physically consistent emulator should have a well-defined procedure for quantifying these bounds. Furthermore, the emulator should be able to be modified to accommodate additional physical constraints if desired.

\section{Intrusiveness}\label{sec:intrusiveness}

Emulators have been broadly grouped into two classes: \textit{intrusive} (or \textit{model-driven}) and \textit{non-intrusive} (or \textit{data-driven})~\cite{Duguet2024,Somasundaram2025}. The fact that each class has two commonly used names reflects the convergent history from multiple fields independently researching methods which were only later realized to be equivalent.

\begin{definition}[Intrusive or Model-Driven Emulator]
    An emulator which uses complete knowledge of the associated high-fidelity model is an intrusive or model-driven emulator.
\end{definition}
\begin{definition}[Non-Intrusive or Data-Driven Emulator]
    An emulator which makes no assumptions about the associated high-fidelity model is a non-intrusive or data-driven emulator.
\end{definition}

Intrusiveness describes the structure, or form, of the emulator as well as the data required to fit or train it. Emulators with a static form that does not change when the form of the high-fidelity model changes (i.e.\ when applying the same emulator to a completely distinct problem) are non-intrusive. For example, \acp{ann} are nearly maximally non-intrusive as their structure always mimics biological neural networks regardless of the data they aim to fit. Although the specific layers and activation functions of the \ac{ann} may be changed when applying the method to different problems, those choices are guided purely by the properties of the data and not of the high-fidelity model.

Recent research developments have introduced and increasingly focused on \textit{hybrid} models, which use limited knowledge or assumptions about the high-fidelity model to combine aspects of both data- and model-driven emulators~\cite{Chen2021}. Perhaps the most well-known hybrid emulators are \acp{pinn}, which combine a data-driven \ac{ann} with a ``physics-informed'' loss function which penalizes violations of equalities required by the high-fidelity model.

\begin{definition}[Hybrid Emulator]
    In the context of the intrusiveness of an emulator, one which uses limited or selected information about the high-fidelity model and combines aspects of both data-driven and model-driven emulators is called a hybrid emulator.
\end{definition}

All of the three classes presented so far describe emulators which dictate what information (or lack thereof) they require. One might wonder if a class of emulators exists that, rather than specifying precise information and data requirements, instead can be adapted to any amount and type of information. In other words, is there a class of emulators that are always applicable (like data-driven emulators) and can make use of varying amounts of information provided about the high-fidelity model, all the way from limited constraints (like hybrid emulators) to complete access to the high-fidelity model (like intrusive emulators)? Whether or not this class is empty will be left as an open question for now, but we will label such a class as \textit{adaptive} emulators.

\begin{definition}[Adaptive Emulator]
    In the context of the intrusiveness of an emulator, one which does not have any requirements as to the type or extent of high-fidelity model information necessary to apply the emulator while at the same time being able to effectively use any high-fidelity model information is called an adaptive emulator.
\end{definition}

\begin{figure}
    \centering
    \includegraphics[width=0.5\textwidth]{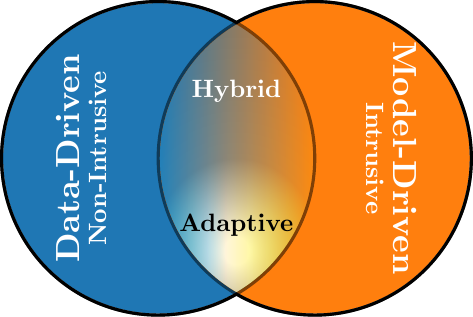}
    \caption[Illustration of various classes of emulation techniques.]{\label{fig:intrusiveness}Illustration of various classes of emulation techniques. The left circle contains data-driven (non-intrusive) methods, the right circle encompasses model-driven (intrusive) methods, the overlapping region includes hybrid methods with varying degrees of partially data-driven and partially model-driven techniques, and finally a subset of this middle region---highlighted at the bottom---shows adaptive methods which are not fixed to a single point in the diagram but instead have the ability to adapt to be as data-driven or model-driven as required. Inspired by a similar figure in Ref.~\cite{Duguet2024}.}
\end{figure}

Figure~\ref{fig:intrusiveness} summarizes the relations between the four described classes of emulators, emphasizing the mixing of intrusive and non-intrusive concepts in hybrid emulators and the close relationship between hybrid and adaptive emulators.

Within the context of a given problem, an emulator which makes use of as much information as possible is generally the best one. Remember that the motivation for emulation is the relative lack of data, so any emulator that ignores or forgoes information is at an additional disadvantage. This is why so much focus has been placed on hybrid emulators, as full knowledge of the high-fidelity model is often impossible or impractical but access to limited knowledge or assumptions about it is practically universal.

\section{Quantifiable Uncertainties}\label{sec:quantifiable_uncertainties}

The ability of any machine learning method to communicate its uncertainty is crucial. This once again touches on the idea of trustworthiness. Beyond this, emulators trained on physical data should account for the uncertainty in the data itself and be able to propagate this uncertainty through to the predictions. A phrase often repeated in machine learning and \ac{uq} circles is some variation of ``a bad model isn't necessarily so if it tells you it's bad.'' This strikes at the key point: a successful emulator may make mistakes sometimes, but it should be able to give some indication that it is doing so.

It can be difficult to find a single quantitative definition of uncertainty. Many different entirely distinct things can be reported as uncertainties, such as standard deviations, confidence intervals, and relative errors. The best definition of uncertainty comes from the field of metrology, which treats all measurements as coming from some probability distribution around the true value and the uncertainty as some metric for the dispersion of this distribution. With this definition, it is clear that even unusual things like the full width half maximum (FWHM) or the median absolute deviation (MAD) can be the basis of an uncertainty measure.

With such an open-ended definition for uncertainty, it may seem impossible to identify what makes some more useful or more precise than others. Generally, it is better if fewer unproven assumptions about the data and model are made. This is where the ubiquitous standard deviation may break down, as it relies on the assumption that the data and the model predictions come from normal distributions. Uncertainty measurements that correspond directly to a notion of ``confidence'' are popular for good reason. Again calling back to interpretability and trustworthiness, an uncertainty measure that communicates ``$X\%$ of the time the value will be within $Y$ of the prediction'' is clear and understandable. This is so understandable that it is used for weather forecasts presented to the public, e.g. ``$20\%$ chance of rain on Saturday.''

\begin{figure}
    \centering
    \includegraphics[width=\textwidth]{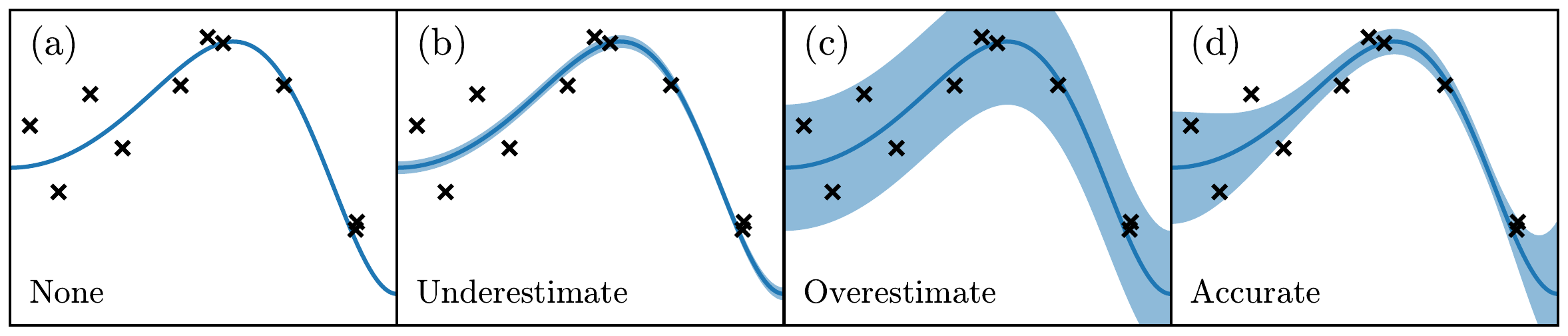}
    \caption[Illustration of models with varying amounts of \ac{uq}.]{\label{fig:uq_ex}Illustration of varying amounts of \ac{uq} for a model which is inaccurate on the left half of the data and reasonably accurate on the right half of the data. (a) A model with no \ac{uq}. (b) A model with underestimated (or overly optimistic) \ac{uq}. (c) A model with overestimated (or overly pessimistic) \ac{uq}. (d) A model with accurate (or well-calibrated) \ac{uq} that adapts to show the model's different degrees of confidence over the domain.}
\end{figure}

Altogether, we desire in our emulators the ability---either inherent or as an additional step---to accurately convey the uncertainty associated with any and all predictions. This uncertainty must be clearly presented and should be easily understandable. Figure~\ref{fig:uq_ex} shows an illustration of different levels of \ac{uq} for the same model and data.

\section{Existing Methods}

This section provides a brief overview of some existing methods for emulation in nuclear and many-body physics. Rather than a complete corpus of all methods, only those which are the most widely used (with one exception in Section~\ref{sec:apdr}) are included. Each method has its own strengths and weaknesses, and certainly all are useful in specific contexts. However, the topic of this thesis necessarily draws focus to the downsides of each of these methods so that in Chapter~\ref{ch:pmm} they may be addressed by the titular method.

\subsection{Explicit Reduced Basis Methods (Galerkin's Method)}\label{sec:erbm}

Referred to in literature simply as \acp{rbm} (the cause for the ``explicit'' modifier will be made clear in Chapter~\ref{ch:pmm}), nearly all \acp{erbm} can trace their origins back to what became known in the West as Galerkin's Method or The Galerkin Method~\cite{Galerkin1915,Duncan1937}. The basic statement of Galerkin's Method is that one can project a high-dimensional (potentially infinite-dimensional) linear problem onto a lower-dimensional subspace and solve the resulting projected problem~\cite{NASA1967,Langtangen2019}. In familiar matrix-vector notation, the original problem is that given a vector space $\V$, a linear operator $A: \V\mapsto \V$ and vector $b\in \V$, find $x\in \V$ such that
\begin{equation}\label{eq:linear}
    Ax = b.
\end{equation}
Galerkin realized that one could choose a set of $n$ linearly independent vectors $\mathset{u_i}$ in $\V$ representing a basis of a subspace $\V_n\subseteq \V$ and project onto it while retaining the linear structure. We call $\V_n$ the \textit{reduced space} and $\left\{u_i\right\}$ the \textit{reduced basis}. Let $\tilde{x}\in \V$ be the projection of $x$ into this subspace, given by
\begin{equation}\label{eq:xtilde}
    \tilde{x} = \sum_i c_i u_i
\end{equation}
where $c_i$ are to-be-determined scalar coefficients. The residual of $\tilde{x}$ is
\begin{equation}
    \varepsilon = b - A\tilde{x} = Ax - A\tilde{x} = A\left(x-\tilde{x}\right).
\end{equation}
Finally, we impose the Galerkin orthogonality condition on $\tilde{x}$ which states that this residual must be orthogonal to $\V_n$ (i.e.\ $v_i^\dagger \varepsilon = 0\, \forall v\in\V_n$)\footnote{Alternatively in Petrov-Galerkin methods, a separate subspace is introduced that is orthogonal to the residual, $\mathcal{W}_n$~\cite{Antoulas2009,Rowley2017}. This subspace is sometimes referred to as the test space, while $\V_n$ is referred to as the trial space.}. This implies
\begin{equation}\label{eq:galerkin}
    \sum_i c_i u_j^\dagger Au_i = u_j^\dagger b,
\end{equation}
which is an equation with only as many unknowns, $c_i$, as vectors in our chosen subspace. These relatively few unknowns can be used via Eq.~\eqref{eq:xtilde} to construct an approximate solution to the original problem. Of course, $A$ need not be a matrix and $b$ need not be a vector. Everything here applies identically for any well-posed linear operator acting on abstract objects in a vector space with a consistent inner product. A linear operator $A$ is well-posed if for all $x,y\in \V$ there exist positive constants $\alpha(A)$ and $\beta(A)$ such that
\begin{equation}
    \begin{aligned}
        \abs{x^\dagger A y} &\leq \alpha(A) \Vert x \Vert \Vert y \Vert\\
        \abs{x^\dagger A x} &\geq \beta(A) \Vert x \Vert^2.
    \end{aligned}
\end{equation}
If $A$ is a Hermitian matrix, then $\alpha(A)$ is the absolute value of the eigenvalue with largest magnitude and $\beta(A)$ is the absolute value of the eigenvalue with smallest magnitude. Crucially for many of the original applications of this method, derivatives are well-posed linear operators over functions and functions live in a vector space with a consistent inner product. Equation~\eqref{eq:linear} may represent a differential equation of a continuous function just as much as it may represent a matrix inversion problem.

In the context of this thesis, there is a much more suggestive form of Eq.~\eqref{eq:galerkin}. One can rewrite the equation in vector form as
\begin{equation}\label{eq:galerkinvector}
    \ul{A} \ul{x} = \ul{b}
\end{equation}
where $\ul{x} \equiv \left[c_1, c_2, \ldots \right] \in \mathbb{C}^n$ is the vector of to-be-determined coefficients\footnote{The notation of underlining to denote the Galerkin-projected objects is not arbitrary, but instead inspired by the notation used in Ref.~\cite{NASA1967} where an overline was used to denote the projected objects. To avoid confusion with the common overline notation used to denote complex conjugation, we use an underline instead.}, $\ul{A}\in\Cbb^{n\times n}$ is given by $\ul{A}_{ij} = u_i^\dagger A u_j$, and similarly $\ul{b}\in\Cbb^n$ is given by $\ul{b}_j = u_j^\dagger b$. Regardless of what sort of mathematical objects $A$, $x$, and $b$ were originally, their low-dimensional counterparts $\ul{A}$, $\ul{x}$, and $\ul{b}$ are a matrix and two vectors. This is guaranteed simply by the fact that the inner product defined in the original space must return a scalar value and the elements of these low-dimensional representations are defined purely by those inner products. Equation~\eqref{eq:galerkinvector} is often referred to as the \ac{rbm} of the original problem in Eq.~\eqref{eq:linear}.

Now consider parameterizing the projection $\mathbb{P}: \V\mapsto \V_n$ by two operators $Q$ and $R$,
\begin{equation}
    \mathbb{P} \equiv Q\left(R^\dagger Q\right)^{-1} R^\dagger,
\end{equation}
such that
\begin{equation}
    \mathbb{P}x = Q\left(R^\dagger Q\right)^{-1} R^\dagger x = \tilde{x},
\end{equation}
and
\begin{equation}
    R^\dagger x = \ul{x}.
\end{equation}
Finally, it is useful to identify and define the \textit{projector}\footnote{In broader mathematical contexts, a projector is something that---among other properties---is idempotent (${PP = P}$). This is not the case here as the product $\P\P$ is not even defined. In general machine learning this object is often referred to as a ``linear encoder''.} $\P:\V\mapsto\Cbb^n\equiv R^\dagger$ and its ``inverse''\footnote{Likewise, $\Pinv$ is not a true inverse since $\Pinv \P\neq I$ (but $\P\Pinv=I$ is always true by definition). As the subspace approaches the full space ($\V_n\rightarrow \V$) it behaves more and more like an inverse ($\Pinv \P\rightarrow I$). In general machine learning this object is often referred to as the ``linear decoder'' associated with $\P$.} $\Pinv:\Cbb^n\mapsto \V_n\equiv Q\left(R^\dagger Q\right)^{-1}$. This notation emphasizes the role of these operators: $\P$ projects objects ``down'' into the reduced space and $\Pinv$ projects back ``up'' into the original space. $\P$ may be said as ``$P$-down'' and $\Pinv$ may be said as ``$P$-up''. With these definitions we are able to write Eqs.~\eqref{eq:galerkin}~and~\eqref{eq:galerkinvector} as
\begin{equation}\label{eq:galerkinprojector}
    \begin{aligned}
        \P A \Pinv \P x &= \P b\\
        \text{or, }&\\
        \ul{A}\ul{x} &= \ul{b}\\
        \text{where }&\\
        \ul{A}&\equiv \P A \Pinv\\
        \ul{x}&\equiv \P x\\
        \ul{b}&\equiv \P b\\
        \text{and }&\\
        x&\approx\tilde{x}=\Pinv\ul{x}
    \end{aligned}
\end{equation}
For Galerkin methods, $R$ and $Q$ are equal\footnote{Petrov-Galerkin methods have $R\neq Q$~\cite{Antoulas2009,Rowley2017}.}. In the case that the $\mathset{u_i}$ are vectors and the inner product is the usual vector dot product, then both $R$ and $Q$ are the matrices whose columns are the $\mathset{u_i}$. This makes $\P$ the matrix whose rows are $\mathset{u_i^\dagger}$ and $\Pinv$ the matrix whose columns are $\mathset{u_i}$ normalized by their inner products, ${\Pinv = R\left(R^\dagger R\right)^{-1} = \P^\dagger \left(\P \P^\dagger\right)^{-1}}$.

Owing to the relative simplicity of Galerkin's Method, one can derive error bounds on the error between the approximate solution $\tilde{x}$ and the true solution $x$
\begin{equation}
    \left\Vert x-\tilde{x}\right\Vert \leq \kappa(A) \inf_{v\in \V_n}\left\Vert x - v\right\Vert
\end{equation}
for positive constant $\kappa(A)=\alpha(A)/\beta(A)$ which is the condition number of $A$~\cite{Cea1964}. That is, the approximate solution from Galerkin's Method is at least as good as any vector in the space spanned by the chosen $\left\{u_i\right\}$, up to a constant. Additionally, this proves that Galerkin's Method is systematically improvable, as successively increasing the dimensionality of $\V_n$ by adding new $u_i$ that have at least some component which is linearly independent of all previous $\left\{u_i\right\}$, thereby driving $\V_n\rightarrow \V$, the error of the approximation goes to $0$, reaching $0$ when $\V_n = \V$. The rate of this convergence depends on both the specific $A$ and chosen $\left\{u_i\right\}$.

We can additionally quantify the computational complexity of Galerkin's Method. We denote the dimensionality of the original space $\V$ by $N$ and assume that $Ax$ and $x^\dagger y$ can be computed in $\bigO{N^2}$ and $\bigO{N}$ time respectively, for all $x,y\in \V$, regardless of what type of mathematical objects $x$, $y$, and $A$ represent. Then the offline cost of Galerkin's Method is comprised only of the $\bigO{N^2n}$ and $\bigO{N^2}$ time of computing $\ul{A}$ and $\ul{b}$ respectively. A precise complexity of the online phase would depend on the method used to solve the linear problem, but even naive approaches would be at most $\bigO{n^3}$. The online phase would also include transforming $\ul{x}$ to $\tilde{x}$, which requires $\bigO{Nn}$ time. The vectors $\left\{u_i\right\}$ would require $\bigO{Nn}$ space both in the offline and online phases. So, assuming $n\ll N$, the offline costs are $\bigO{N^2n}$ time and $\bigO{Nn}$ space and the online costs are $\bigO{Nn}$ time and $\bigO{Nn}$ space.

Galerkin's Method allows us to take a high- or infinite-dimensional linear problem and approximate it by a smaller linear problem with a dimensionality of our choosing. Importantly, this applies equally to \textit{parametric} linear problems---those where the problem itself is parameterized by some list of parameters, $\mu\equiv\mathlist{\mu_1,\,\mu_2,\,\ldots}$.  The most general form of such a problem is
\begin{equation}\label{eq:parametriclinear}
    A(\mu)x(\mu) = b(\mu).
\end{equation}
Consider the simplest case where $\mu$ is a scalar and $A$ is an affine combination of two linear operators $A_0$ and $A_1$,
\begin{equation}
    A(\mu)x(\mu) = \left(A_0 + \mu A_1\right) x(\mu) = b.
\end{equation}
The application of Eq.~\eqref{eq:galerkinprojector} is unchanged from before,
\begin{equation}
    \begin{aligned}
        \P A(\mu) \Pinv \P x(\mu) &= \P b\\
        \iff \left(\P A_0 \Pinv + \mu \P A_1 \Pinv\right) \P x(\mu) &= \P b\\
        \iff \left(\ul{A_0} + \mu \ul{A_1}\right) \ul{x(\mu)} &= \ul{b}.
    \end{aligned}
\end{equation}
This generalizes to significantly more complex parameterizations where $\mu$ can represent an entire array of parameters as well as parameterizations which are nonlinear in $\mu$, so long as $A(\mu)$ is a linear operator for all $\mu$ and $A(\mu)$ and $b(\mu)$ are expressible linearly in terms of any functions of $\mu$.
\begin{equation}\label{eq:galerkinparametric}
    \begin{aligned}
        \text{For }&\\
        A(\mu)x(\mu)&=b(\mu)\\
        \text{with }&\\
        A(\mu) &\equiv \sum_i f_i(\mu)A_i\\
        b(\mu) &\equiv \sum_j g_j(\mu)b_j\\
        \text{Galerkin's Method yields }&\\
        \ul{A(\mu)}\ul{x(\mu)} &= \ul{b(\mu)}\\
        \iff \left(\sum_i f_i(\mu)\ul{A_i}\right)\ul{x(\mu)} &= \sum_j g_j(\mu)\ul{b_j}\\
        \text{where }&\\
        \ul{A_i} &\equiv \P A_i \Pinv\\
        \ul{b_j} &\equiv \P b_j\\
        \text{and }&\\
        x(\mu)&\approx\tilde{x}(\mu)=\Pinv\ul{x(\mu)}
    \end{aligned}
\end{equation}
Note that no assumption on the parametric dependence of $x$ on $\mu$ was made. Although the problem itself is a linear one for fixed $\mu$, the dependence of the problem may be highly nonlinear in $\mu$, and thus the solution $x(\mu)$ may be arbitrarily nonlinear in $\mu$. The distinction between the linearity of the problem and the linearity of the parameterization is an important one that will appear several times throughout this thesis.

One can see that Galerkin's Method does not change the form or structure of the original problem. This allows for the transformed equation to be interpretable in much the same way as the original problem. It also allows for secondary operators or equations, such as those corresponding to conservation or symmetry laws, to be projected in the same way using the same projector in order for the degree to which they are conserved to be analyzed just as the approximate solution $\tilde{x}$ was.

Table~\ref{tab:galerkin} contains a brief summary of each of the relevant emulator properties as they relate to Galerkin's Method.

What Galerkin's Method stops short of is providing guidance on how to find or choose the basis vectors that define the projection, how to proceed when the linear operator, $A$, or inhomogeneous part of the problem, $b$, are not fully known, how the method generalizes to nonlinear problems, if at all, or how to apply the method when the problem setup is not exactly like that in Eq.~\eqref{eq:linear}. The methods discussed herein can fundamentally be equated to some formulation of Galerkin's Method but are distinguished by their approaches to each of these unanswered questions.

\begin{figure}
    \centering
    \emulatorsummary{Galerkin's Method}
    {%
        Yes, so long as the form of the parametric dependence is known.
    }
    {%
        No. Reduced objects have dimensions corresponding to the subspace $\V_n$.
    }
    {%
        Strictly speaking in the original description, no. Extensions to \acs{pod}-Galerkin and \ac{dmd} exist (see Sections~\ref{sec:pod}~and~\ref{sec:dmd}) for nonlinear systems.
    }
    {%
        Dependent on the cost of the high-fidelity linear operator and inner product. If those time complexities are $\bigO{A(N)}$ and $\bigO{\text{dot}(N)}$ respectively and have space complexity $\bigO{N}$ (typical of linear operators and inner products), then the method has $\bigO{A(N)\text{dot}(N)n}$ time and $\bigO{N + n^2}$ space offline complexity and $\bigO{n^3}$ time and $\bigO{n}$ space online complexity when using standard iterative solvers. An additional $\bigO{Nn}$ online time cost is incurred when projecting the Galerkin solution back to the original space $\V$.
    }
    {%
        Heavily dependent on the choice of the subspace $\V_n$.
    }
    {%
        The only potential hyperparameter is the process one chooses to form $\V_n$ from available data.
    }
    {%
        Yes. Convergence rate heavily dependent on the specific choice of successive $\V_n$.
    }
    {%
        Yes. The projected system has a form identical to the original.
    }
    {%
        Tied directly to the accuracy of the reduced equations, which depends heavily on $\V_n$.
    }
    {%
        Maximally intrusive. Requires computing $\P A \Pinv$ and $\P b$, which requires full knowledge of the high-fidelity model.
    }
    {%
        Not inherent. Can be added on top of the method. Error bounds and knowledge of the high-fidelity model in conjunction with $\V_n$ may greatly improve the accuracy of uncertainty quantification.
    }
    \captionof{table}{Summary of properties for emulators based on Galerkin's Method.}
    \label{tab:galerkin}
\end{figure}

\subsubsection{Proper Orthogonal Decomposition}\label{sec:pod}

Originally introduced in the field of computational fluid dynamics, \ac{pod} guides the selection of the reduced space $\V_n$ which is then used via Galerkin's Method to create a reduced-order model~\cite{Kahlbacher2007, Berkooz1993}. This method is often referred to as the \acs{pod}-Galerkin method. The context for \ac{pod} is a vector-valued parametric linear problem of the form in Eq.~\eqref{eq:parametriclinear}---often, the operator is constant and the only parameter is time---with some vector data that are solution \textit{snapshots}, that is, a set of solutions for some set of parameter realizations. Let $\mu^{(i)}$ denote a specific realization for the parameters of the problem, then $x\left(\mu^{(i)}\right)$ which satisfies Eq.~\eqref{eq:parametriclinear} with $\mu=\mu^{(i)}$ is the snapshot of the system for parameters $\mu^{(i)}$. Then the data available to the \ac{pod} method are the $m$ snapshots $\mathset{x\left(\mu^{(1)}\right), x\left(\mu^{(2)}\right), \ldots, x\left(\mu^{(m)}\right)}$ and associated parameter realizations $\mathset{\mu^{(1)}, \mu^{(2)}, \ldots, \mu^{(m)}}$. In the simplest implementation, one forms the $N\times m$ matrix $X$ whose columns are the snapshots and computes its \ac{svd}
\begin{equation}\label{eq:podsvd}
    X = U\Sigma V^\dagger.
\end{equation}
By convention the singular values and singular vectors are ordered by decreasing singular value, so $\Sigma=\text{diag}(\sigma_1, \sigma_2, \ldots)$ with $\sigma_1\geq \sigma_2\geq\cdots$. The matrix whose columns are the first $n$ columns of $U$---often denoted as $U[:, {}\mathbin{:}n]$, $U[:, 1\mathbin{:}n]$, or $U(:, 1\mathbin{:}n)$---is then taken to be $\Pinv$. That is, the $n$ orthonormal left singular vectors of $X$ corresponding to the $n$ largest singular values of $X$ are the $\mathset{u_i}$ that form a reduced basis via Galerkin's Method. Since the columns of $\Pinv$ are orthonormal, $\left(\Pinv\right)^\dagger\Pinv = I_n$ and so $\P = \left(\Pinv\right)^\dagger$. For notational simplicity, it is common to use the notation $P\equiv \Pinv$ and work in terms of $P$ and $P^\dagger$. Computing $P$ can be nontrivial or even intractable via the exact \ac{svd} if the size or number of snapshots is too great. Several approximate or iterative methods exist to find $P$ in such cases. \Ac{pod} has close ties to \ac*{pca}; the columns of $P$ are exactly the principal components of the snapshots. This allows the number of components $n$ to be automatically determined from a specified desired cumulative explained variance---that is, the proportion of the variance in the snapshots that is retained under the projection $P\Pdag$. In fact, the Eckart-Young-Mirsky theorem proves that the best possible rank-$n$ approximation to the matrix of snapshots is the one given by \ac{pca}, $X_n = P\Pdag X$~\cite{Eckart1936,Mirsky1960}.

Orthogonal projections have several properties that make them particularly attractive for the basis of an \ac{rbm}. Perhaps the most important is the preservation of inner products. For any two vectors $u$ and $v$, we have $u^\dagger v \approx \tilde{u}^\dagger \tilde{v} = \left(P \ul{u}\right)^\dagger P \ul{v} = \ul{u}^\dagger \left(\Pdag P\right) \ul{v} = \ul{u}^\dagger \ul{v}$. What makes this property so important is that it renders projecting back to the full space unnecessary if the quantity of interest is a scalar which is the result of only inner products---all because $\tilde{u}^\dagger\tilde{v}=\ul{u}^\dagger\ul{v}$. This includes eigenvalues, singular values, $\ell_2$ norms, expectation values, and much more. For operators, Hermiticity (or symmetry in the case of real-valued operators and projectors) is exactly preserved purely from the semi-unitarity ($\Pdag P = I$) of the projector. For Hermitian operators, the Poincar\'e separation theorem\footnote{Also referred to as the Cauchy interlacing theorem due to the result of the special case where $n=N-1$.} guarantees that if $\lambda_i(A)$ is the \ord{i} eigenvalue of $A\in\mathbb{H}(N)$ and $\lambda_j\left(\ul{A}\right)$ is the \ord{j} eigenvalue of $\ul{A}\in\mathbb{H}(n)$ with both sets of eigenvalues in ascending order ($\lambda_1 \leq \lambda_2 \leq \cdots$) then
\begin{equation}\label{eq:pst}
    \lambda_j(A)\leq \lambda_j\left(\ul{A}\right)\leq \lambda_{N-n+j}(A)
\end{equation}
for $j=1,\,2,\,\ldots,\,n$~\cite{Rao1973,Rao1979,Bellman1997}. This result can be generalized to the singular values of any matrix\footnote{It can also be generalized to transformations of the form $\Pdown A \Pup$ where $\Pdown$ and $\Pup$ are both semi-unitary, possibly with differing dimensions.}
\begin{equation}\label{eq:pst_singular}
    \sigma_j(A) \geq \sigma_j\left(\ul{A}\right) \geq \sigma_{2(N -n) +j}(A)
\end{equation}
where the singular values are in the standard descending order $\sigma_1\geq\sigma_2\geq\cdots\geq 0$ and $j=1,\,2,\,\ldots,\,n$~\cite{Rao1979,Queiro1987}. Equations~\eqref{eq:pst}~and~\eqref{eq:pst_singular} guarantee too many exact and approximate conservations between the full and reduced models to exhaustively list here. Some of the most important properties are listed below.
\begin{itemize}
    \item The condition number, $\kappa(\cdot)$, of $\ul{A}$ is at most the condition number of $A$---that is, $1\leq\kappa\left(\ul{A}\right)\leq\kappa(A)$. This guarantees that the reduced model will be either numerically easier or no harder than the original.
    \item The definiteness of $A$ and $\ul{A}$ are identical---unless $A$ is indefinite in which case $\ul{A}$ may have any definiteness.
    \item For any unitarily invariant norm $\norm{\cdot}$, the norm of $\ul{A}$ is bounded by that of $A$. That is, $\norm{A} \geq \norm{\ul{A}}$. This includes the spectral norm\footnote{Also known as the induced $\ell_2$-norm, $\norm{\cdot}_2$. For Hermitian matrices this is equal to the maximum singular value.}, Frobenius norm, nuclear norm\footnote{Also called the trace norm.}, and all Schatten $p$-norms~\cite{Schatten2012}.
\end{itemize}

The \ac{pod}-Galerkin method can be naively applied to nonlinear systems in the following way. Consider a general parametric nonlinear problem
\begin{equation}
    A(\mu) x(\mu) + F\left(x(\mu); \mu\right) = b(\mu),
\end{equation}
where $F\left(~\cdot~;\mu\right):\V\rightarrow\V$ is the nonlinear part. Applying \ac{pod}-Galerkin to this problem yields the partially reduced model
\begin{equation}\label{eq:podnonlin}
    \begin{aligned}
        \Pdag A(\mu) P \Pdag x(\mu) + \Pdag F\left(P \Pdag x(\mu); \mu\right) &= \Pdag b(\mu)\\
        \iff \ul{A(\mu)}\ul{x(\mu)} + \Pdag F\left(P \ul{x(\mu)}; \mu\right) &= \ul{b(\mu)}.
    \end{aligned}
\end{equation}
This is ``partially'' reduced because the evaluation of the nonlinear term cannot directly be reduced without more specific knowledge about it. Evaluating this term requires projecting back to the full space, evaluating the full-dimensional nonlinearity, and then projecting back down---a process which alone is at least $\bigO{C_F(N)+Nn^2}$, where $C_F(N)$ is the complexity of evaluating $F$ on a size-$N$ input. In practice, even the simplest nonlinearities have complexities no better than $\bigO{N}$. Combined with the cost of solving Eq.~\eqref{eq:podnonlin} with an iterative solver, this means that an emulator formed in this fashion would have an online time complexity exceeding $\bigO{C_F(N)n + Nn^3}$. When applied this way, a \ac{pod}-Galerkin emulator for a nonlinear problem can actually be just as, if not more, expensive than the full high-fidelity model~\cite{Chaturantabut2009,Lu2020,Kramer2019}. The usual workaround for this problem is an approximation of $F\left(~\cdot~;\mu\right)$ on top of the existing reduced-basis approximation. This additional approximation is referred to as a ``hyper-reduction'' and the most popular technique is the \ac{deim}~\cite{Chaturantabut2009}.

\Ac{deim} assumes that the nonlinear term is evaluated \textit{elementwise}. That is, if $x$ is the vector $\left[x_1, x_2, \ldots, x_N\right]$ then $F(x) = \left[F(x_1), F(x_2), \ldots, F(x_N)\right]$. While seemingly a limiting assumption, many useful and realistic nonlinear terms behave this way. However, \ac{deim} also assumes that $x$ and $F(x)$ are component-wise smooth---which can be a very limiting assumption. The method avoids scaling with the size of the full space, $N$, by approximating $F$ by utilizing not just snapshots of $x(\mu)$ but also of $F(x(\mu);\mu)$ (which usually incur no additional cost as they are required for the high-fidelity model to produce the snapshots of $x$). During the online phase, $x$ is ``downsampled'' by selecting a fixed set of only $m\ll N$ indices to evaluate $F$ at and \ac{deim} then attempts to linearly interpolate the rest of the indices based on the set of example nonlinear snapshots. This is where the component-wise smoothness assumption becomes necessary. Written out explicitly, denote the $N\times m$ matrix whose columns are the standard one-hot basis vectors which encode the chosen $m$ indices by
\begin{equation}
    B = \begin{bmatrix}
        \vline & \vline & &\vline\\
        e_{d_1} & e_{d_2} & \cdots & e_{d_m}\\
        \vline & \vline & &\vline
    \end{bmatrix},
\end{equation}
where $d_i$ are the chosen $m$ indices, and $e_j$ is the \ord{j} standard basis vector. $B$ selects the encoded indices of $x(\mu)$ from $\ul{x(\mu)}$ simply by
\begin{equation}
    B^\dagger P \ul{x(\mu)}.
\end{equation}
Note that $B^\dagger P$ can be precomputed in the offline phase by taking the $d_i$ columns of $P$ and is only $m\times n$. The elementwise nonlinearity is then evaluated over the $m$ components of this vector and interpolated using the basis spanned by the $m$ nonlinear snapshots. Finally, the result is projected back down to the original size-$n$ \ac{pod} basis. Denote the $m \times N$ Galerkin projector for the nonlinear snapshots---that is the matrix whose columns are the nonlinear snapshots---by $Q$, then \ac{deim} approximates the nonlinear term in Eq.~\eqref{eq:podnonlin} by
\begin{equation}\label{eq:deim}
    \Pdag F\left(P \ul{x(\mu)};\mu\right) \approx \tikzmarknode{pdag}{\Pdag} \tikzmarknode{qbq}{\underbrace{Q\left(B^\dagger Q\right)^{-1}}} F\left(\tikzmarknode{b}{B^\dagger} \tikzmarknode{p}{P} \ul{x(\mu)};\mu\right).
\end{equation}
\begin{tikzpicture}[remember picture, overlay]
    \draw ([yshift=-2pt]pic cs:pdag) -- ++(0, -5em) -| ++(1em, 0) node[anchor=west] {\textit{Project back down to \ac{pod}-Galerkin basis}};
    \draw ([yshift=-1.5em]pic cs:qbq) -- ++(0, -2.4em) -| ++(1em, 0) node[anchor=west] {\textit{Empirical interpolation}};
    \draw ([yshift=-2pt]pic cs:b) -- ++(0, -2.5em) -| ++(1em, 0) node[anchor=west] {\textit{Select $m$ indices}};
    \draw ([yshift=-2pt]pic cs:p) -- ++(0, -1.5em) -| ++(1em, 0) node[anchor=west] {\textit{Project to full space}};

\end{tikzpicture}
\vspace{3em}

\noindent Again note that $\Pdag Q\left(B^\dagger Q\right)^{-1}$ is an $n\times m$ matrix which can be precomputed during the offline phase, just like $B^\dagger P$. Utilizing these precomputations, the cost of evaluating Eq.~\eqref{eq:deim} is only $\bigO{C_F(m) + nm}$, now completely independent of the size of the high-fidelity model. Thus from the extra assumptions that $F$ is an elementwise nonlinearity and $x$ and $F(x)$ are component-wise smooth and the use of additional data that is usually already available, \ac{deim} is able to efficiently extend the \ac{pod}-Galerkin method to a wide array of nonlinear problems. Further details and analysis of \ac{deim} can be found in Ref.~\cite{Chaturantabut2009}.

\ac{deim} is not without its downsides. For strongly nonlinear problems---those where the contribution of the nonlinear terms is comparable or potentially even greater than that of the linear terms---it can be the case that the required number of nonlinear snapshots, $m$, approaches the size of the high-fidelity model, $N$~\cite{Huang2018,Kramer2019}. Additionally, introducing a second independent approximation scheme on top of \ac{pod}-Galerkin complicates convergence, stability, and improvability analyses.

To handle nonlinear terms without additional approximations, the concept of lifting transformations was proposed in Ref.~\cite{Gu2011}. Lifting transformations introduce auxiliary variables to transform the original form of the problem to one which is more amenable to \ac{pod}-Galerkin. This is particularly powerful in emulators for systems of nonlinear differential equations~\cite{Gu2011,Kramer2019}. First, we note that polynomial nonlinearities can be handled directly by the standard \ac{pod}-Galerkin method. Consider the quartic function
\begin{equation}
    F(x) = Ax + Bx^2 + Cx^3 + Dx^4
\end{equation}
where $A, B, C, D$ are linear operators and $x^k\equiv x\odot\stackrel{k}{\cdots}\odot x$ denotes the elementwise \ord{k} power of $x$. We can directly apply \ac{pod}-Galerkin to this nonlinear function as in Eq.~\eqref{eq:podnonlin},
\begin{equation}\label{eq:quarticpod}
    \begin{aligned}
        \Pdag F(x) & \approx \Pdag F\left(P\ul{x}\right)\\
        &= \Pdag A P \ul{x}
        \begin{aligned}[t]
            &+ \Pdag B \left[\left(P\ul{x}\right)\odot\left(P\ul{x}\right)\right] \\&+ \Pdag C \left[\left(P\ul{x}\right)\odot\left(P\ul{x}\right)\odot\left(P\ul{x}\right)\right] \\
            &+ \Pdag D \left[\left(P\ul{x}\right)\odot\left(P\ul{x}\right)\odot\left(P\ul{x}\right)\odot\left(P\ul{x}\right)\right]
        \end{aligned}\\
        \iff \left(\Pdag F(x)\right)_i &\approx \ul{A}_{i\mu}\ul{x}_\mu
        \begin{aligned}[t]
            &+ \underbrace{\left[\Pdag_{i\nu}B_{\nu\mu}P_{\mu\sigma}P_{\mu\omega}\right]}_{\ul{B}}\underbrace{\ul{x}_\sigma\ul{x}_\omega}_{\ul{x}\otimes\ul{x}}\\
            &+ \underbrace{\left[\Pdag_{i\nu}C_{\nu\mu}P_{\mu\sigma}P_{\mu\omega}P_{\mu\alpha}\right]}_{\ul{C}}\underbrace{\ul{x}_\sigma\ul{x}_\omega\ul{x}_\alpha}_{\ul{x}\otimes\ul{x}\otimes\ul{x}}\\
            &+ \underbrace{\left[\Pdag_{i\nu}D_{\nu\mu}P_{\mu\sigma}P_{\mu\omega}P_{\mu\alpha}P_{\mu\beta}\right]}_{\ul{D}}\underbrace{\ul{x}_\sigma\ul{x}_\omega\ul{x}_\alpha\ul{x}_\beta}_{\ul{x}\otimes\ul{x}\otimes\ul{x}\otimes\ul{x}}
        \end{aligned}\\
        \iff \Pdag F(x) &\approx \ul{A}\ul{x} + \ul{B}\left(\ul{x}\otimes\ul{x}\right) + \ul{C}\left(\ul{x}\otimes\ul{x}\otimes\ul{x}\right) + \ul{D}\left(\ul{x}\otimes\ul{x}\otimes\ul{x}\otimes\ul{x}\right),
    \end{aligned}
\end{equation}
where $\otimes$ denotes the Kronecker product and the third step is written in Einstein summation notation for clarity, with all indices denoted by Greek letters being summed over. $\ul{B}$, $\ul{C}$, $\ul{D}$ are rank $3,4,5$ tensors respectively and are contracted over the Greek indices with the corresponding rank $2,3,4$ $\left(\ul{x}\otimes\cdots\otimes\ul{x}\right)$. It is not uncommon to reshape the tensors to $n\times n^k$ matrices and the $\left(\ul{x}\otimes\cdots\otimes\ul{x}\right)$ to $n^k$ vectors, where $k=2,\,3,\,4$, so that all operations in the last line of Eq.~\eqref{eq:quarticpod} are matrix-vector multiplications. All the operators, $\ul{A}$, $\ul{B}$, $\ul{C}$, and $\ul{D}$ can be precomputed during the offline phase at costs of $\bigO{N^2 n^2},\,\bigO{N^2n^3},\,\bigO{N^2n^4},\,\text{and}$ $\bigO{N^2n^5}$ respectively and the online cost of evaluating this reduced form of the nonlinearity is $\bigO{n^5}$. In principle, any polynomial nonlinearity can be handled this way, but in practice it is often more beneficial to apply lifting transformations to limit the order of the polynomial.

To illustrate how auxiliary variables and lifting transformations can make an arbitrary nonlinearity into a polynomial one, consider the following toy problem,
\begin{equation}
    \frac{dx}{dt} = Ax + B\sin(x).
\end{equation}
Using the auxiliary variables $y=\sin(x)$ and $z=\cos(x)$, we are able to lift the problem from a nonlinear one of size $N$ to a quadratic one of size $3N$,
\begin{equation}\label{eq:pod_ex_lifted}
    \left\{\begin{aligned}
        \frac{dx}{dt} &= Ax + By\\
        \frac{dy}{dt} &= \cos(x)\odot\frac{dx}{dt} = z\odot\left(Ax + By\right)\\
        \frac{dz}{dt} &= -\sin(x)\odot\frac{dx}{dt} = -y\odot\left(Ax + By\right)
    \end{aligned}\right.
\end{equation}
which in matrix form is
\begin{equation}\label{eq:pod_ex_lifted_matrix}
    \begin{aligned}
        \frac{d}{dt}\underbrace{\begin{bmatrix}
            x\\y\\z
        \end{bmatrix}}_{v} &=
        \underbrace{\begin{bmatrix}
            A & B & 0\\
            0 & 0 & 0\\
            0 & 0 & 0
        \end{bmatrix}}_{\mathcal{A}}
        \begin{bmatrix}
            x\\y\\z
        \end{bmatrix} +\left(
        \underbrace{\begin{bmatrix}
            0 & 0 & 0\\
            0 & 0 & 1\\
            0 & -1 & 0
        \end{bmatrix}}_{\mathcal{B}}
        \begin{bmatrix}
            x\\y\\z
        \end{bmatrix}\right)\odot\left(
        \underbrace{\begin{bmatrix}
            0 & 0 & 0\\
            A & B & 0\\
            A & B & 0
        \end{bmatrix}}_{\mathcal{C}}
        \begin{bmatrix}
            x\\y\\z
        \end{bmatrix}\right)\\
        \iff \frac{dv}{dt} &= \mathcal{A}v + \left(\mathcal{B} v\right)\odot\left(\mathcal{C} v\right),
    \end{aligned}
\end{equation}
which can then be reduced via \ac{pod}-Galerkin for polynomial nonlinearities in a few different ways. The first is to construct the projector from stacked snapshots of the original and auxiliary variables $v=\left[x\,y\,z\right]^\mathsf{T}$ and project Eq.~\eqref{eq:pod_ex_lifted_matrix} as
\begin{equation}
    \begin{aligned}
        \frac{d\ul{v}}{dt} &= \ul{\mathcal{A}} \ul{v} + \Pdag\left(\left[P\ul{\mathcal{B}}\ul{v}\right]\odot\left[P\ul{\mathcal{C}}\ul{v}\right]\right)\\
        \iff \left(\frac{d\ul{v}}{dt}\right)_i &= \ul{\mathcal{A}}_{i\mu}\ul{v}_\mu + \Pdag_{i\mu}P_{\mu\nu}\ul{\mathcal{B}}_{\nu\omega}\ul{v}_\omega P_{\mu\sigma}\ul{\mathcal{C}}_{\sigma\lambda}\ul{v}_\lambda\\
         &= \ul{\mathcal{A}}_{i\mu}\ul{v}_\mu + \underbrace{\Pdag_{i\mu}P_{\mu\nu}P_{\mu\sigma}}_{\ul{G}^{(2)}}\underbrace{\ul{\mathcal{B}}_{\nu\omega}\ul{v}_\omega \ul{\mathcal{C}}_{\sigma\lambda}\ul{v}_\lambda}_{\left(\ul{\mathcal{B}}\ul{v}\right)\otimes\left(\ul{\mathcal{C}}\ul{v}\right)}\\
         \iff \frac{d\ul{v}}{dt} &= \ul{\mathcal{A}}\ul{v} + \ul{G}^{(2)}\left[\left(\ul{\mathcal{B}}\ul{v}\right)\otimes\left(\ul{\mathcal{C}}\ul{v}\right)\right].
    \end{aligned}
\end{equation}
The second way treats snapshots of $x$, $y$, and $z$ on equal footing, considering them essentially as independent snapshots of the same variable, and constructing the \ac{pod} projector from this data and reducing Eq.~\eqref{eq:pod_ex_lifted} directly
\begin{equation}
    \begin{aligned}
        &\left\{\begin{aligned}
            \frac{d\ul{x}}{dt} &= \ul{A}\ul{x} + \ul{B}\ul{y}\\
            \frac{d\ul{y}}{dt} &= \Pdag\left(\left[P\ul{z}\right]\odot\left[P\left(\ul{A}\ul{x} + \ul{B}\ul{y}\right)\right]\right)\\
            \frac{d\ul{z}}{dt} &= -\Pdag\left(\left[P\ul{y}\right]\odot\left[P\left(\ul{A}\ul{x} + \ul{B}\ul{y}\right)\right]\right)
        \end{aligned}\right.\\
        \iff&\left\{\begin{aligned}
            \frac{d\ul{x}}{dt} &= \ul{A}\ul{x} + \ul{B}\ul{y}\\
            \left(\frac{d\ul{y}}{dt}\right)_i &= \underbrace{\Pdag_{i\mu}P_{\mu\nu}P_{\mu\omega}}_{\ul{G}^{(2)}}\underbrace{\ul{z}_\nu\left(\ul{A}\ul{x} + \ul{B}\ul{y}\right)_\omega}_{\ul{z}\otimes\left(\ul{A}\ul{x} + \ul{B}\ul{y}\right)}\\
            \left(\frac{d\ul{z}}{dt}\right)_i &= -\underbrace{\Pdag_{i\mu}P_{\mu\nu}P_{\mu\omega}}_{\ul{G}^{(2)}}\underbrace{\ul{y}_\nu\left(\ul{A}\ul{x} + \ul{B}\ul{y}\right)_\omega}_{\ul{y}\otimes\left(\ul{A}\ul{x} + \ul{B}\ul{y}\right)}
        \end{aligned}\right.\\
        \iff&\left\{\begin{aligned}
            \frac{d\ul{x}}{dt} &= \ul{A}\ul{x} + \ul{B}\ul{y}\\
            \frac{d\ul{y}}{dt} &= \ul{G}^{(2)}\left[\ul{z}\otimes\left(\ul{A}\ul{x} + \ul{B}\ul{y}\right)\right]\\
            \frac{d\ul{z}}{dt} &= -\ul{G}^{(2)}\left[\ul{y}\otimes\left(\ul{A}\ul{x} + \ul{B}\ul{y}\right)\right].
        \end{aligned}\right.
    \end{aligned}
\end{equation}
There is a third option involving multiple projectors, one for each of the original and auxiliary variables (i.e.\ $P_x,\,P_y,\,P_z$). For an example of this technique see the Appendix of Ref.~\cite{Jammooa2025}. Which of the three methods to use comes down to problem specifics and the choices of auxiliary variables. Often, multiple options are tried and the one with the best performance on some held-out data is selected.

This example is somewhat contrived, and it may seem unlikely that such a small number of simple variable substitutions could lift an arbitrary problem to polynomial form. Nevertheless, it is known that the number of auxiliary variables necessary to transform any nonlinear problem to a polynomially nonlinear one is proportional to the number of elementary nonlinear functions in the problem~\cite{Gu2011}. Thus the size of the lifted problem is $\bigO{n_f N}$ where $n_f$ is the number of elementary nonlinear functions. This linear factor is manageable by the \ac{pod}-Galerkin reduction that follows.

In the case that the mathematical parametricity of the high-fidelity model is unknown\footnote{That is, $A(\mu)$ in Eq.~\eqref{eq:parametriclinear} cannot be expanded and thus $\ul{A}(\mu)$ at new values of $\mu$ requires the computation of $A(\mu)$ in the full space before the Galerkin projection.}, several extensions to \ac{pod} exist to either efficiently approximate the high-fidelity problem by a more amenable one or efficiently interpolate either the emulated prediction, reduced basis itself, or the reduced operators using data-driven interpolation techniques~\cite{Lieu2004,Lieu2007,Amsallem2007,Weickum2006,Galletti2004, Grepl2005,Ghattas2008,Ly2001,Degroote2010}.

\begin{figure}
    \centering
    \includegraphics[width=0.85\textwidth]{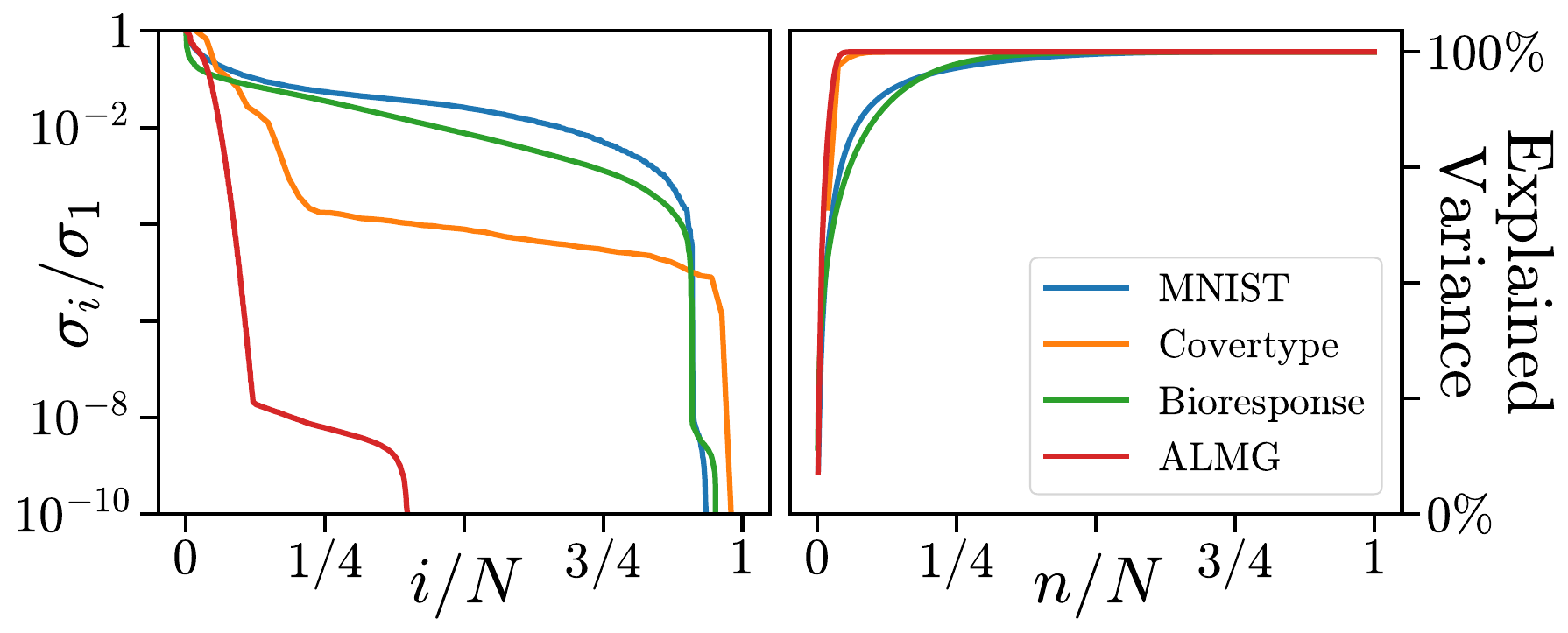}
    \caption[Singular values and cumulative explained variance for several real-world datasets.]{\label{fig:svd_decay}%
    Singular values and cumulative explained variance for four real-world datasets: images of handwritten digits (MNIST)~\cite{Lecun1998}, cartographic data from the US Geological Survey and US Forest Service (Covertype)~\cite{Blackard1998}, molecular descriptors (Bioresponse)~\cite{Hamner2012}, and low-lying energy eigenstates of the anharmonic \ac*{lmg} model for various coupling strengths (ALMG)~\cite{Gamito2022,Cook25}. The left plot shows the normalized singular value as a function of the normalized index $i/N$ where $N$ is the dimensionality of the data (the number of features). The right plot shows the cumulative explained variance from the first $n$ principal components as a function of $n/N$.}
\end{figure}

Error and convergence analysis of \ac{pod}-Galerkin comes down to the convergence of \ac{pca}. We first derive the error between $X$ (the $N\times m$ matrix of training snapshots from Eq.~\eqref{eq:podsvd}) and the associated rank-$n$ \ac{pca} approximation $X_n = P \Pdag X$. By the definition of the $2$-norm, it is straightforward to verify that the error in this approximation is exactly $\norm{X - P \Pdag X}_2^2 = \sigma_{n+1}^2$. Any vector $v\in\V$, can be written as a sum of the part that is in the column space of $X$ and the part perpendicular to that space: $v = v_X + v_\perp$ where $v_X = X w$ for some $w$. The error from the \ac{pca} projection will serve as a rough upper bound for the error of the \ac{pod}-Galerkin method,
\begin{equation}
    \label{eq:podexperr}
    \begin{aligned}
        \norm{v - P P^\dagger v}_2^2 &= \norm{Xw - P \Pdag Xw}_2^2 + \norm{v_\perp - P \Pdag v_\perp}_2^2\\
        &\leq \norm{X - P\Pdag X}_2^2\norm{w}_2^2 + \norm{v_\perp}_2^2\\
        &= \sigma_{n+1}^2 \norm{w}_2^2 + \norm{v_\perp}_2^2.
    \end{aligned}
\end{equation}
Thus for vectors that are almost exactly normalized linear combinations of the snapshots (pure linear interpolation), in which case $\norm{w}_2^2\approx 1$ and $\norm{v_\perp}_2^2\approx 0$, the error is approximately bounded by $\sigma_{n+1}^2$. However, if we assume that the snapshots were drawn from some distribution and this test vector $v$ is drawn from the same distribution, then we have an expected relative error given by
\begin{equation}\label{eq:podexprelerr}
    \mathbb{E}\left[\frac{\norm{v - P \Pdag v}_2^2}{\norm{v}_2^2}\right] = \frac{\sum_{i=n+1}^N \sigma_i^2}{\sum_{i=1}^N \sigma_i^2},
\end{equation}
which is exactly $1$ minus the cumulative explained variance of the \ac{pca} projector~\cite{Holmes2023}. This is a reasonable assumption that holds well in nearly all practical applications---one essentially just ensures that the snapshots are representative of the space of interest. Another in-practice truth is that the singular values of $X$ decay exponentially or as a power-law. This is not true of an arbitrary $X$, and many pathological examples can be made, such as random matrices, where the singular values do not decay at all. However, it is overwhelmingly true in practice that either
\begin{equation}\label{eq:typicalsingular}
    \sigma_k \sim \sigma_1 \alpha^{-\beta (k-1)}\quad\text{or}\quad \sigma_k\sim\sigma_1 k^{-\gamma},
\end{equation}
for some constants $\alpha$, $\beta$, or $\gamma$. This fact is why \ac{pod}-Galerkin emulators can be so accurate with $n\ll N$. Figure~\ref{fig:svd_decay} shows the singular values and cumulative explained variance as a function of $n/N$ for various real-world datasets spanning applications in image recognition, forestry, biochemistry, and quantum mechanics. Beyond increasing the number of retained principal components or snapshots, the accuracy of the method can also be improved by the addition of data-driven correction terms fit to additional data~\cite{Couplet2005}.

\subsubsection*{Example}

Consider the problem of diffusion in one dimension, which is governed by the partial differential equation
\begin{equation}
    \begin{aligned}
        \frac{\partial u}{\partial t} &= \alpha \frac{\partial^2 u}{\partial x^2},\\
        u(x, t=0) &= u_0(x),
    \end{aligned}
\end{equation}
for $0<x<L$, $t>0$, and diffusivity parameter $\alpha>0$. A typical context is that where the boundaries are held fixed, which imposes Dirichlet boundary conditions,
\begin{equation}
    \left.\frac{\partial u}{\partial t}\right|_{x=0} = \left.\frac{\partial u}{\partial t}\right|_{x=L} = 0
\end{equation}
Typical numerical methods will discretize the system over the spatial dimension, rendering the problem a system of coupled ordinary differential equations in time. Choosing a uniform grid of $N$ parts over $x$ makes $u$ a size-$N$ vector and the spatial derivative an $N\times N$ matrix representing the finite difference approximation of the derivative with spacing $L/(N-1)$.
\begin{equation}
    \begin{aligned}
        \frac{\partial u}{\partial t} &=\alpha%
        \underbrace{\left(\frac{N-1}{L}\right)^2\begin{bmatrix}
            0 &  &  & & & & \\
            1 & -2 & 1 & & & & \\
              &  1 &-2 &1& & & \\
              &    &   &\ddots& & & \\
            & &  & 1 &-2 &1& \\
            & & & & 1 & -2 & 1\\
            & & & &  &  & 0
        \end{bmatrix}}_{D}%
        \underbrace{\begin{bmatrix}
            u_0\\u_1\\\vdots\\u_{N-1}
        \end{bmatrix}}_{u}\\
        \iff \frac{\partial u}{\partial t} &= \alpha D u,
    \end{aligned}
\end{equation}
where we have used the Dirichlet boundary condition at the boundaries, the central finite difference everywhere else, and the elements of $u$ relate to the original continuous problem by $u_k(t) \equiv u\left(x=\frac{kL}{N-1},t\right),\,k=0,\,1,\,\ldots,\,N-1$. We can now solve for $u_k(t)$ for any $N$, $t$, and $u_0(x)$ via numerical integration.

This process so far is the typical numerical treatment of partial differential equations. The form of the problem is now amenable to \ac{pod}-Galerkin parameterized by $\alpha$. Suppose our goal is to find the evolution of a Gaussian $u_0(x)$ for some $\alpha_\text{test}$, but due to stability constraints with numerical integration we are only able to run the high-fidelity simulation for $\alpha \leq \alpha_\text{test}/2$. Choosing $N=\num{1000}$ for a sufficiently high-resolution high-fidelity solution, we run just five full simulations using a matrix-free implementation\footnote{See Appendix~\ref{app:matrixfree}.} at uniformly spaced sample values of $\alpha$ between $0.1$ and $\alpha_\text{test}/2$ for $N_t$ timesteps from $0$ to our desired final time $T$. From these high-fidelity simulations we now have $5 N_t$ snapshots. As is typical in time-dependent \ac{pod}-Galerkin, we temporally downsample by taking every $f_t=100$ snapshot to lessen the cost of computing the \ac{svd} of the matrix of snapshots\footnote{For systems that are smooth in time and simulations with suitably small timesteps, both of which are often true in practice, the snapshots $u\left(x,t_i\right)$ and $u\left(x, t_{i+1}\right)$ contain functionally identical information. This justifies temporal downsampling before \ac{svd} without appreciable information loss.}. We thus form the $5N_t/f_t \times N$ snapshot matrix $X$ and compute the leading $n=10$ principal components using a standard \ac{svd} algorithm. These $n$ vectors form the columns of the \ac{pod}-Galerkin projector $P$. Applying the projector to our discretized equations, we have the reduced $n$-dimensional system of coupled differential equations
\begin{equation}
    \begin{aligned}
        \frac{\partial}{\partial t} \ul{u} &= \alpha \ul{D} \ul{u}\\
        \ul{u}(x, t=0) &= \ul{u_0}
    \end{aligned}
\end{equation}
with $\ul{D} = P^\dagger D P$ and $\ul{u_0} = P^\dagger u_0$. The approximate \ac{pod} solution to the original problem is $\tilde{u}=P\ul{u}\approx u$. This is a system just $1\%$ the size of the original, and thus can be quickly numerically integrated with a much smaller timestep if necessary\footnote{Recall, however, that the condition number of the reduced model will be at most the condition number of the original. Even if this were not the case, the dimensionality reduction allows us to use more expensive methods or resolutions while maintaining a fast and efficient emulator.}. So to approximate $u\left(x, t; \alpha_\text{test}\right)$, we integrate our reduced system with $\alpha=\alpha_\text{test}$ and obtain $\tilde{u}\left(x, t; \alpha_\text{test}\right) = P \ul{u}\left(x, t; \alpha_\text{test}\right)$.

\begin{figure}
    \centering
    \includegraphics[width=\textwidth]{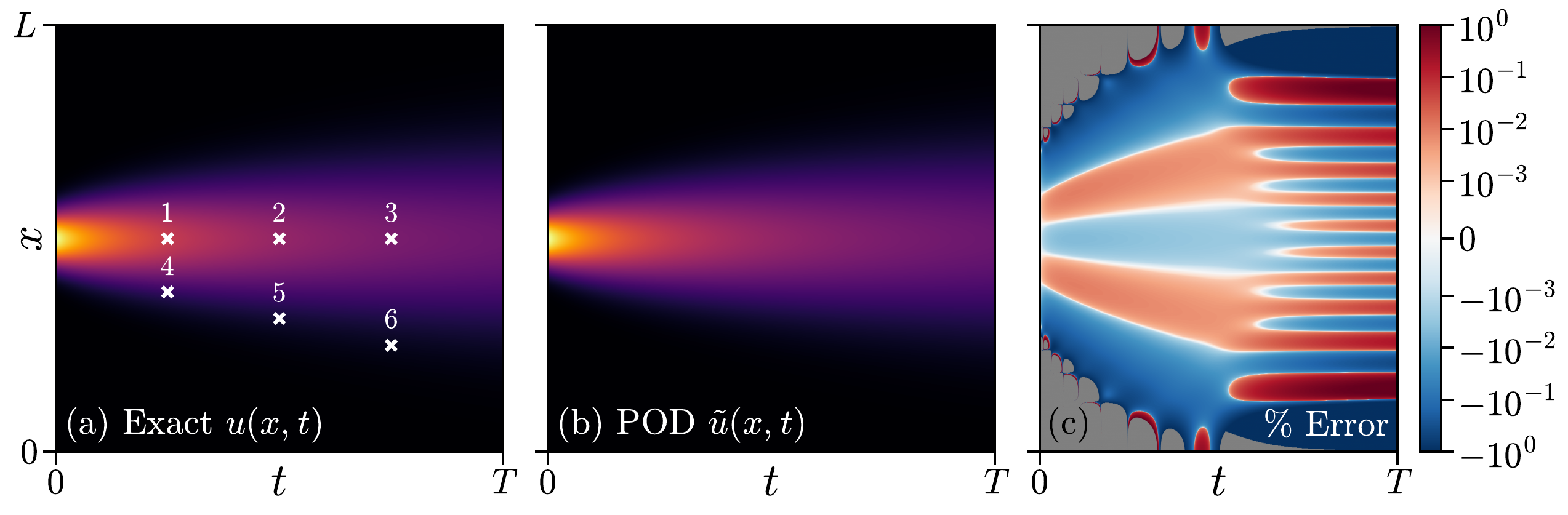}
    \includegraphics[width=0.7\textwidth]{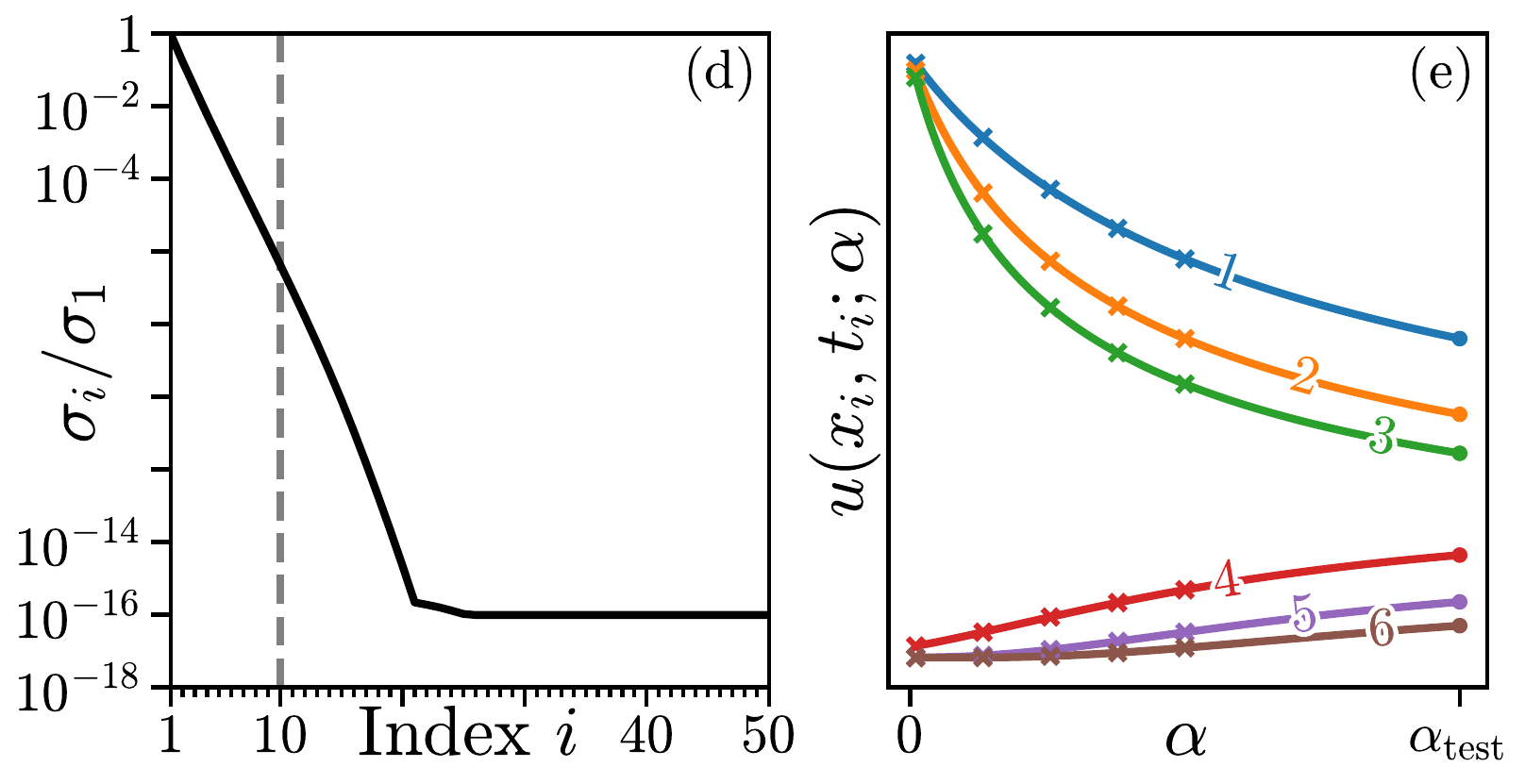}
    \caption[Results of applying the \ac{pod}-Galerkin method to the 1D diffusion differential equation.]{\label{fig:podex} Results of applying the \ac{pod}-Galerkin method to the problem of 1D diffusion. (a) The exact solution to the 1D diffusion equation for diffusivity $\alpha_\text{test}$ that was not used when training the \ac{pod}-Galerkin emulator. The six points marked in (a) are the $(x_i, t_i)$ that are sampled as a function of $\alpha$ in (e). The \ac{pod}-Galerkin approximation to the system for the same diffusivity $\alpha_\text{test}$ is shown in (b). (c) The percent error between (b) the \ac{pod}-Galerkin solution and (a) the true solution at $\alpha_\text{test}$---errors above $1\%$ where the exact value was very small are shown in grey since a relative error metric is not meaningful for such points. The first $\num{50}$ singular values of the matrix of snapshots, $X$, used for training the emulator are shown as a function of their index $i$ in (d), normalized by the largest singular value $\sigma_1$. The vertical dashed line in (d) denotes the index of the smallest singular value associated with any singular vector used to construct the \ac{pod} projector; that is, the vertical line denotes the emulator dimension, $n$. Finally, (e) shows values of the approximate solution as a function of $\alpha$ at the six points indicated in (a). Crosses in (e) denote the values of the snapshots used to train the emulator at the same locations, showing only five unique values of $\alpha$ were used to compute the projector. The circles in (e) indicate the values of the exact solution at $\alpha_\text{test}$, from (a), at the corresponding locations, showing the accuracy of the \ac{pod}-Galerkin emulator.
    }
\end{figure}

Figure~\subfigref{fig:podex}{a} shows the results of the high-fidelity $u(x,t)$ solution at $\alpha_\text{test}$ using a smaller step size (twice the number of timesteps) such that the simulation converged. The diffusion of the Gaussian initial state over time is clearly visible. The process described here is used to find the \ac{pod} solution at $\alpha_\text{test}$, which is shown in Fig.~\subfigref{fig:podex}{b} and exhibits strong qualitative agreement. Quantitative errors are shown in Fig.~\subfigref{fig:podex}{c} as the elementwise percent error
\begin{equation}
    \text{\% Error} = \frac{\tilde{u}\left(x, t; \alpha_\text{test}\right) - u\left(x, t; \alpha_\text{test}\right)}{\abs{u\left(x, t; \alpha_\text{test}\right)}} \cdot 100\%.
\end{equation}
Figure~\subfigref{fig:podex}{c} demonstrates that despite a reduction of $99\%$ in the size of the problem, the \ac{pod} approximate solution is within $1\%$ of the exact solution\footnote{Ignoring regions where $\abs{u\left(x, t;\alpha_\text{test}\right)}$ is so small as to cause the percent error to diverge, in which case the absolute difference is a better metric, but these regions are not of interest in this problem.}. Further comparisons as a function of $\alpha$ are shown in Fig.~\subfigref{fig:podex}{e}, where the values of $u(x, t; \alpha)$ and $\tilde{u}(x, t; \alpha)$ are shown as a function of $\alpha$ at six $(x, t)$ points indicated in Fig.~\subfigref{fig:podex}{a}. This figure shows the accuracy of the \ac{pod} approximation on the training data as well as the smoothness of the solution as a function of $\alpha$. Finally, the decay of the singular values of the data matrix is shown in Fig.~\subfigref{fig:podex}{d}. The rapid exponential decay of the singular values is not specific to this example and is what allows for subspaces with $n\ll N$ to accurately approximate the high-fidelity model. Recall from Eq.~\eqref{eq:podexprelerr} that the expected relative error in the approximation\footnote{Over the norm of the vector, not elementwise as is shown in Fig.~\subfigref{fig:podex}{c}.} is the ratio of the area under the squared curve for $i>n$ to the area under the entire squared curve in Fig.~\subfigref{fig:podex}{d}.

\begin{figure}
    \centering
    \begin{tabularx}{\textwidth}{Xcr}
        \toprule
        Quantity & Symbol & Value\\
        \midrule
        High-fidelity spatial discretization & $N$ & $\num{1000}$\\
        Domain size & $L$ & $\num{10}$\\
        Final time & $T$ & $\num{10}$\\
        Diffusivities used for snapshots & $\alpha_\text{train}$ & $0.1,\,1.325,\,2.55,\,3.775,\,5$\\
        Timesteps from $t=0$ to $T$ for snapshot generation & $N_t$ & $\num{10000}$\\
        Target or testing diffusivity & $\alpha_\text{test}$ & $10$\\
        Timesteps used for exact simulation of test diffusivity & & $\num{20000}$\\
        Initial state & $u_0(x)$ & $\exp\left\{-\left(\frac{16}{L}\right)^2\left(x - \frac{L}{2}\right)^2\right\}$\\
        \midrule
        Temporal downsampling factor & $f_t$ & $\num{100}$\\
        \ac{pod}-Galerkin emulator dimension & $n$ & $\num{10}$\\
        Timesteps for each emulator simulation & & $\num{10000}$\\
        \midrule
        Numerical integration method & & Explicit (forward) Euler\\
        \bottomrule
    \end{tabularx}
    \captionof{table}{\label{tbl:podex}Exact values of all quantities used in the 1D diffusion equation \ac{pod}-Galerkin example.}
\end{figure}

\begin{figure}
    \centering
    \emulatorsummary{Proper Orthogonal Decomposition (\ac{pod}-Galerkin)}
    {%
        Yes, so long as the form of the parametric dependence is known.
    }
    {%
        No. Even with the choice of the number of retained principal vectors, $n$, the addition of new snapshots changes all principal vectors. Extensions that introduce some numerically parametric elements, such as those in Refs.~\cite{Ly2001,Lieu2004,Lieu2007,Degroote2010,Couplet2005}, must still contend with the dominating non-numerically parametric \ac{pod} projection.
    }
    {%
        Can be naively applied to nonlinear systems inefficiently. Nonlinear terms can be efficiently approximated via \ac{deim}~\cite{Chaturantabut2009} or treated exactly via lifting transformations~\cite{Kramer2019}.
    }
    {%
        Broadly, yes. Nonlinearities must be treated carefully to preserve efficiency.
    }
    {%
        Yes. Size of subspace determines the retained explained variance of the snapshots, which in practice converges to $1$ either exponentially or polynomially.
    }
    {%
        Only inherent hyperparameter is the size of the subspace, or equivalently the cumulative explained variance of the snapshots.
    }
    {%
        Yes. Convergence rate in practice is either exponential or polynomial.
    }
    {%
        Yes. The projected system has a form identical to the original.
    }
    {%
        Tied directly to the accuracy of the reduced equations. Important properties like normalization and inner products are approximately conserved.
    }
    {%
        Maximally intrusive. Extensions with data-driven components are less intrusive, but still require complete descriptions of the high-fidelity model at selected parameters~\cite{Ly2001,Lieu2004,Lieu2007,Degroote2010}.
    }
    {%
        Not inherent, but can be added on top of the method. Error bounds and error estimates available.
    }
    \captionof{table}[Summary of properties for the \ac{pod}-Galerkin method.]{Summary of properties for the \ac{pod}-Galerkin method. Many properties are inherited from Galerkin's Method.}
    \label{tab:pod}
\end{figure}

\subsubsection{Dynamic Mode Decomposition}\label{sec:dmd}

Another method originally introduced in the field of computational fluid dynamics, \ac{dmd} in its original form is a purely data-driven technique for sequential data~\cite{Schmid2010,Tu2013}. That is, the data is of the form $\mathset{z_1,\,z_2,\,\ldots,\,z_n}$ where the true underlying process is assumed to transform $z_i$ to $z_{i+1}$. This was generalized to ``Exact'' \ac{dmd}, where instead of a set sequence, the data are in the familiar input $\mathset{x_1,\,\ldots,\,x_n}$ and output $\mathset{y_1,\,\ldots,\,y_n}$ sets and the underlying process is assumed to connect them via $x_i\rightarrow y_i$~\cite{Tu2013}. Being a purely data-driven method, \ac{dmd} requires (and uses) no knowledge of the underlying problem or high-fidelity model. Instead, it imposes an assumption on the underlying process, namely that it is purely linear,
\begin{equation}
    Ax_i = y_i,
\end{equation}
for some unknown $A$. This assumption is equal parts simple and powerful as many systems---even nonlinear ones---can be well-approximated in this way. The goal of \ac{dmd} will be to find, or at least approximate, $A$. In practice, one usually obtains and returns only a limited number of ``dynamic modes'' or \ac{dmd} modes and \ac{dmd} eigenvalues, which are the leading few eigenvectors and eigenvalues of this operator. These dynamic modes become the basis of a Galerkin subspace when \ac{dmd} is used for dimensionality reduction.

There are many methods used in practice to compute $A$, and the one used often depends on the dimensionality of the data, the amount of data, and desired accuracy versus speed of the approximation of $A$. Fundamentally, these methods all aim to approximate the exact solution which is
\begin{equation}
    A = YX^+
\end{equation}
where $X$ is the matrix whose columns are $x_i$, $Y$ is the matrix whose columns are $y_i$, and $X^+$ denotes the Moore-Penrose pseudoinverse of $X$. This is simply the least-squares solution, which minimizes $\norm{AX-Y}_F$~\cite{Tu2013}.

Generalizing \ac{dmd} to parametric problems is nontrivial and a properly parametric version was only introduced relatively recently~\cite{Huhn2023, Andreuzzi2023}. As \ac{dmd} has no knowledge of the underlying equations, the parametric dependence is unknown and so parametric \ac{dmd} utilizes data-driven interpolations of the \ac{dmd} modes at known parameters to approximate the modes at new parameters\footnote{This is similar to some of the extensions to \ac{pod}-Galerkin mentioned in Section~\ref{sec:pod}.}.

\ac{dmd} can be generalized beyond linear systems, particularly in the case of dynamical systems. Through deliberate feature engineering and careful regularization, the method known as \ac{sindy} aims to discover the form of any nonlinearities from data alone~\cite{Brunton2016,Brunton2022}. \Ac{sindy} first constructs a ``library'' of candidate nonlinearities of the input data. For a single snapshot this is something like
\begin{equation}
    \Theta\left(x\right) =
    \begin{bmatrix}
        1 \\ x \\ \left(x\otimes x\right) \\ \left(x\otimes x \otimes x\right) \\ \vdots \\ \sin\left(x\right) \\ \vdots
    \end{bmatrix}
\end{equation}
where $x\otimes x$ represents quadratic polynomials of the elements of the snapshot, i.e.\ the vector containing all quadratic combinations of the elements of $x$. The specific candidate nonlinearities are hyperparameters and can vary from problem to problem. We then suppose that the problem is linear in these new features
\begin{equation}
    y_i = \Xi^\mathsf{T} \Theta\left(x_i\right),
\end{equation}
for some unknown matrix $\Xi^\mathsf{T}$. The size of $\Theta(x)$ can far exceed that of $x$ depending on the type and amount of candidate nonlinearities, so \ac{sindy} makes the assumption that only a very small number of the candidate nonlinearities will contribute for each element of $y_i$. This corresponds to the rows of $\Xi$ being very sparse---consisting of mostly $0$. Finding $\Xi$ is now just a relatively simple matter of solving the sparse regression problem
\begin{equation}
    Y^\mathsf{T} = \Theta\left(X^\mathsf{T}\right)\Xi
\end{equation}
for $\Xi$. While this method does not reduce the dimensionality of the problem directly, it can in principle be combined with \acp{ann} such as \acp{ae} to achieve effective reductions or the resulting linear system may be treated with standard \ac{dmd}. However, it appears in literature that extensions to \ac{sindy} are limited to $\bigO{10}$-dimensional systems---though this is an area of active research~\cite{Champion2019,Zolman2025,Viknesh2026}.

An alternative nonlinear extension to \ac{dmd}, known variously as Hankel \ac{dmd} or Time-Delay \ac{dmd}\footnote{Hankel \ac{dmd} and Time-Delay \ac{dmd} are not strictly the same method, but are exceedingly closely related.}---and nearly mathematically identical to the Eigensystem Realization Algorithm\footnote{The Eigensystem Realization Algorithm is itself closely related to the matrix pencil method and Prony analysis~\cite{Almunif2020}.}~\cite{Juang1985,Almunif2020}---leverages Takens' Theorem to identify the nonlinear dynamics by describing the data as a linear model not just from state $i$ to state $i+1$ but from state $i$ and the entire history---or a window of the history---of states to state $i+1$~\cite{Arbabi2017,Takens1981}. In contrast to \ac{sindy}, Hankel \ac{dmd} is computationally efficient as it only increases the costs by up to a factor of the number of snapshots.

\section*{Example}

\begin{figure}
    \centering
    \includegraphics[width=\textwidth]{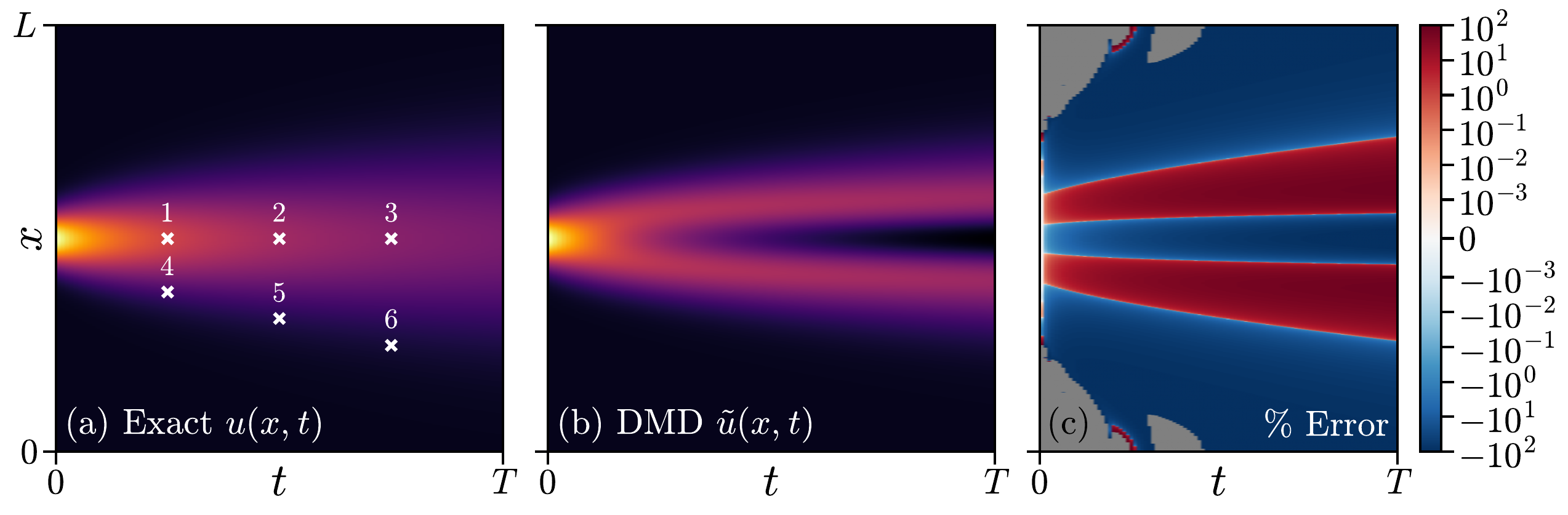}
    \includegraphics[width=0.35\textwidth]{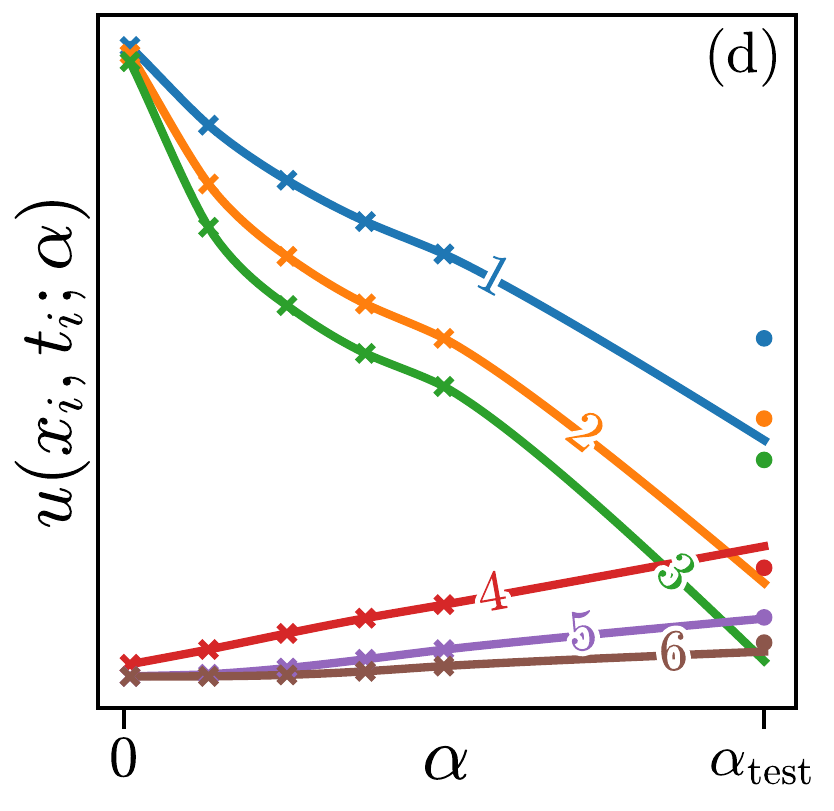}
    \caption[Results of applying the \ac{dmd} method to the 1D diffusion differential equation.]{\label{fig:dmdex} Results of applying parametric \ac{dmd} to the problem of 1D diffusion. (a) The exact solution to the 1D diffusion equation for diffusivity $\alpha_\text{test}$ that was not used when training the \ac{dmd} emulator. The six points marked in (a) are the $(x_i, t_i)$ that are sampled as a function of $\alpha$ in (d). The \ac{dmd} approximation to the system for the same diffusivity $\alpha_\text{test}$ is shown in (b). (c) The percent error between (b) the \ac{dmd} solution and (a) the true solution at $\alpha_\text{test}$---errors above $100\%$ where the exact value was very small are shown in grey since a relative error metric is not meaningful for such points. Note that this limit is larger than in Fig.~\ref{fig:podex}. Finally, (d) shows values of the approximate solution as a function of $\alpha$ at the six points indicated in (a). Crosses in (d) denote the values of the snapshots used to train the emulator at the same locations, showing only five unique values of $\alpha$ were used to fit the emulator. The circles in (d) indicate the values of the exact solution at $\alpha_\text{test}$, from (a), at the corresponding locations, showing that the emulator struggles with extrapolating in $\alpha$.
    }
\end{figure}

We consider the same example as in the previous section (Section~\ref{sec:pod}), that of diffusion in one dimension. This is a linear system and thus should be well-approximated by \ac{dmd}. All parameters of the high-fidelity simulations and training data are unchanged, including temporal downsampling. We create the emulator using the implementation of parametric \ac{dmd} provided in the \pypackage{PyDMD} package~\cite{Ichinaga2024,Andreuzzi2023} with a reduced subspace of size $n=10$ to match the \ac{pod}-Galerkin example. Significantly more preprocessing and hyperparameter tuning is required to achieve a reasonable emulator result compared to the previous example. We utilize an \ac{svd} preprocessor (\ac{pca}) to reduce both the dimensionality and the condition number of the training data. Additionally, there are many choices for the interpolation method used to achieve the parametric dependence. The one chosen---radial basis function interpolation with the standard multiquadratic kernel---represented the most accurate for both interpolation and extrapolation in $\alpha$. It is likely that more careful hyperparameter tuning, preprocessing, and the use of constrained \ac{dmd} methods could improve the accuracy of the emulator. Figure~\ref{fig:dmdex} shows the results of the \ac{dmd} emulator for the same test diffusivity $\alpha_\text{test}$ as in the previous example. The \ac{dmd} emulator interpolates in $\alpha$ well, especially given the fact that it has no knowledge of the underlying system since it is a fully data-driven method. However, the emulator struggles with extrapolation, especially at late times, as shown in Fig.~\subfigref{fig:dmdex}{d}. This is a known limitation of \ac{dmd} and is not specific to this example.

\begin{figure}
    \centering
    \emulatorsummary{Dynamic Mode Decomposition}
    {%
        Recent extensions for parametric problems exist~\cite{Huhn2023, Andreuzzi2023}.
    }
    {%
        No. Similar to \ac{pod}-Galerkin, the addition of new snapshots changes all existing \ac{dmd} modes. Some extensions, like parametric \ac{dmd} introduce numerically parametric elements (in order to introduce mathematical parametricity) but are still dominated by the non-numerically parametric components~\cite{Huhn2023, Andreuzzi2023}.
    }
    {%
        Exact \ac{dmd} is limited to linear systems, or systems that can be linearized well. Extensions like Hankel \ac{dmd} and \ac{sindy} can handle nonlinear systems.
    }
    {%
        Yes.
    }
    {%
        Yes, \ac{dmd} modes and \ac{dmd} eigenvalues behave similarly to the principal values and principal components.
    }
    {%
        Only inherent hyperparameter is the number of \ac{dmd} modes kept.
    }
    {%
        Yes for linear systems. Only in the infinite-data limit for nonlinear systems. In practice only for linear systems.
    }
    {%
        Yes. The projected system is a simple linear map.
    }
    {%
        Not necessarily. The projected system is a simple linear map, which may not preserve physical properties of the original system, especially if the underlying snapshots are produced by a nonlinear map.
    }
    {%
        Purely data-driven.
    }
    {%
        Not inherent, but can be added on top of the method.
    }
    \captionof{table}{Summary of properties for \ac{dmd}.}
    \label{tab:dmd}
\end{figure}

\subsubsection[Dimensionality Reduction for Quantum Dynamics]{Tractable \textit{a priori} Dimensionality Reduction for Quantum Dynamics}\label{sec:apdr}
{%
\newcommand{\oneover}[1]{\frac{1}{#1}}
\newcommand{\abssq}[1]{\left|#1\right|^2}
\newcommand{\normsq}[1]{\left\|#1\right\|^2}
\newcommand{\nxn}[1]{\ensuremath{#1 \times #1}}
\newcommand{\Sbb}{\ensuremath{\mathbb{S}}}
\newcommand{\Dbb}{\ensuremath{\mathbb{D}}}
\newcommand{\Psinaught}{\ensuremath{\ket{\Psi_0}}}
\newcommand{\Psit}{\ensuremath{\ket{\Psi(t)}}}
\newcommand{\psit}{\ensuremath{\ket{\psi(t)}}}
\newcommand{\ktpsit}{\ensuremath{\ket{\widetilde{\psi}(t)}}}
\newcommand{\btpsit}{\ensuremath{\bra{\widetilde{\psi}(t)}}}
\newcommand{\tpsit}{\ensuremath{\widetilde{\psi}(t)}}
\newcommand{\tO}{\ensuremath{\widetilde{O}}}
\newcommand{\phit}{\ensuremath{\ket{\phi(t)}}}
\newcommand{\Tr}[1]{\ensuremath{\text{Tr}\left[#1\right]}}

This section makes an exception to the stated goal of covering methods which are widely used and contains the relevant parts of the method described in Ref.~\cite{Cook2025apdr}. A common task in nuclear and many-body theory is the time evolution of a known arbitrary initial state $\Psinaught$ under a Hamiltonian $H$ of dimension $N\times N$, as well as the expectation value of any set of observables as a function of time.
\begin{equation}
    \left\{
    \begin{aligned}
        \Psinaught\\
        H\\
        \left\{O_i\right\}
    \end{aligned}\right\}
    \qquad\longrightarrow\qquad
    \left\{
    \begin{aligned}
        \Psit\\
        \left\{\sandwich{\Psi(t)}{O_i}{\Psi(t)}\right\}
    \end{aligned}\right\}
\end{equation}
There are two traditional approaches to this problem. The first is the full eigenvalue decomposition (diagonalization) of the Hamiltonian,
\begin{equation}
    H\ket{E_i} = E_i\ket{E_i},
\end{equation}
followed by writing the initial state as a linear combination of the energy eigenstates
\begin{equation}
    \label{eq:apdr_linearcomb}
    \Psinaught = \sum\limits_{i=1}^N c_i\ket{E_i} = \sum\limits_{i=1}^N \braket{E_i}{\Psi_0}\ket{E_i}.
\end{equation}
Finally, time propagation is accomplished by independently evolving each energy eigenstate,
\begin{equation}
    \Psit = \sum\limits_{i=1}^N\braket{E_i}{\Psi_0}e^{-iE_it}\ket{E_i}.
\end{equation}
The time complexity of this approach is about $\bigO{N^3}$ in practice owing to the full dense eigenvalue decomposition.

The second way is the direct computation and application of the time evolution operator
\begin{equation}
    \Psit = \exp\left\{-i H t\right\}\Psinaught.
\end{equation}
This approach is $\bigO{N^3}$ in practice as well, but the complexity is much more dependent on the specific structure (or lack thereof) of the Hamiltonian due to the difficulties in numerical matrix exponentiation~\cite{Moler03}.

All previously discussed Galerkin-like methods in this chapter have operated in the context that some snapshots were provided without any choice in how those snapshots were collected. By contrast, for a desired number of snapshots $n$ we now seek to find a way to collect only the best possible snapshots for the problem of Hamiltonian dynamics. To accomplish this, we seek to find the $n$ vectors such that $\Psit$ projected onto these vectors is best conserved.

Consider Eq.~\eqref{eq:apdr_linearcomb} with a reordering of the coefficients such that ${\abs{c_i} \geq \abs{c_{i+1}} \geq 0}$, meaning the energy eigenvalues are no longer in the traditional ascending order. Define the approximate wavefunction to order $n\ll N$,
\begin{equation}
    \label{eq:apdr_approx}
    \begin{aligned}
        \ket{\psi} &= \frac{1}{\mathcal{N}}\sum\limits_{i=1}^n c_i\ket{E_i},\\
        \mathcal{N}^2 &\equiv \sum\limits_{i=1}^n \abs{c_i}^2,\,\mathcal{N}\in\Rbb^+.
    \end{aligned}
\end{equation}
Here I show that this approximation minimizes the error in the $\ell^2$ norm of the dynamics for a given number of contributing basis states $n$, given that the error is time-independent. Consider the most general form of such an \ord{n} order approximation,
\begin{equation}
    \begin{aligned}
        \ket{\phi(t)} &= \frac{1}{\mathcal{M}(t)}\sum\limits_{q=1}^n \ket{a_q}\braket{a_q}{\Psi(t)},\\
        \mathcal{M}(t)^2 &\equiv \sum\limits_{q=1}^n \abs{\braket{a_q}{\Psi(t)}}^2,\,\mathcal{M}(t)\in\Rbb^+,
    \end{aligned}
\end{equation}
where $\mathset{\ket{a_q}}$ is an arbitrary set of $n$ orthonormal basis vectors. This can equivalently be written as ${\ket{\phi(t)} = PP^\dagger\ket{\Psi(t)}/\mathcal{M}(t)}$ where ${P\in\mathbb{U}(N,n)}$ is the \ac{pod}-Galerkin-like projector formed by taking $\mathset{\ket{a_q}}$ as the columns. Note that $P^\dagger P = I_n$. The $\ell^2$ error of this approximation is
\begin{equation}
    \label{eq:apdr_err}
    \begin{aligned}
        \norm{\Psit - \phit}_2^2 &= 2 - \frac{2}{\mathcal{M}(t)}\Re\left\{\sandwich{\Psi(t)}{PP^\dagger}{\Psi(t)}\right\}\\
        &= 2\left(1-\mathcal{M}(t)\right).
    \end{aligned}
\end{equation}
For the error to be independent of time we must have
\begin{equation}
    \begin{aligned}
        \mathcal{M}(t)^2 &= \mathcal{M}^2\\
                        &= \sandwich{\Psi(t)}{PP^\dagger}{\Psi(t)}\\
                        &= \sum\limits_{q=1;\,a,b=1}^{n;\,N}c_a^\ast c_b \braket{E_a}{a_q}\braket{a_q}{E_b}e^{-i\left(E_b - E_a\right)t}\\
                        &= \sum\limits_{q=1;\,a=1}^{n;\,N}\abs{c_a}^2\abs{\braket{E_a}{a_q}}^2.
    \end{aligned}
\end{equation}
That is, all the time-dependent terms in $\mathcal{M}(t)$ must cancel. Since $\mathcal{M}\leq 1$, the error is minimized when $\mathcal{M}$, or equivalently $\mathcal{M}^2$, is maximized. So we must solve
\begin{equation}
    \argmax\limits_{\mathset{\ket{a_q}}}\left\{\mathcal{M}^2\right\}.
\end{equation}
This is equivalent to solving the Rayleigh quotient problem
\begin{equation}
    \argmax\limits_{\mathset{\ket{a_q}}}\left\{\sum_{q=1}^n\sandwich{a_q}{\Xi}{a_q}\right\},
\end{equation}
where ${\Xi\equiv\sum\limits_{a=1}^N\abs{c_a}^2\ketbra{E_a}{E_a}}$ is the matrix with eigenvalues $\mathset{\abs{c_a}^2}$ and eigenvectors $\mathset{\ket{E_a}}$. The solution to this problem is the $n$ eigenvectors of $\Xi$ with the largest eigenvalues, and so the error is minimized when ${\mathset{\ket{a_q}} = \mathset{\ket{E_1}, \ket{E_2},\ldots,\ket{E_n}}}$ up to arbitrary phase factors.
This proves that Eq.~\eqref{eq:apdr_approx} is the approximation that minimizes the time-independent approximation error. The resulting error is
\begin{equation}
    \label{eq:apdr_minerr}
    \norm{\Psit - \psit}_2^2 = 2\left(1-\mathcal{N}\right).
\end{equation}
To derive an error bound for any observable, $O$, let $\expect{O}_{\theta} \equiv \sandwich{\theta(t)}{O}{\theta(t)}$, $\rho_\theta \equiv \ketbra{\theta}{\theta}$, note that $\braket{\Psi(t)}{\psi(t)}=\mathcal{N}$, and recall that $\expect{O}_\theta = \Tr{O\rho_\theta}$. Then we can write
\begin{equation}
    \label{eq:apdr_obserr}
    \begin{aligned}
        \abs{\expect{O}_\Psi - \expect{O}_\psi} &= \abs{\Tr{O\left(\rho_\Psi - \rho_\psi\right)}}\\
        &\leq \norm{O}_2\norm{\rho_\Psi - \rho_\psi}_\ast\\
        &\leq 2\sqrt{1-\abs{\braket{\Psi(t)}{\psi(t)}}^2}\norm{O}_2\\
        &= 2\sqrt{1-\mathcal{N}^2}\norm{O}_2,
    \end{aligned}
\end{equation}
where $\norm{\cdot}_\ast$ is the nuclear norm\footnote{Also known as the trace norm or the Schatten $1$-norm and sometimes denoted $\norm{\cdot}_1$.}, $\norm{O}_2$ is the matrix $\ell^2$ norm of $O$---which for Hermitian operators is equal to the maximum absolute eigenvalue---and the second inequality is a standard result for the trace distance of two pure states~\cite{Wilde2016}.

So we have shown that the $n$ eigenstates of the Hamiltonian with maximum overlap with the target state are the best possible snapshots and derived error bounds for the Galerkin emulator based on those snapshots. However, it may seem that the only way to obtain these snapshots is by a full diagonalization of the Hamiltonian, rendering this a purely academic exercise. Indeed, most existing implementations of numerical eigensolvers either compute the complete eigendecomposition or a subset of eigenpairs selected by eigenvalue. For example, the ubiquitous ARPACK library provides subroutines \texttt{SSEUPD} and \texttt{DSEUPD} for the implicitly restarted Arnoldi iteration and allows for targeting a subset of eigenpairs based on eigenvalue~\cite{Lehoucq1998}. Fortunately, a lesser-used algorithm for the generalized eigenvalue decomposition, the Jacobi-Davidson algorithm, can be modified to target eigenpairs by any property of either the eigenvalue or eigenstate~\cite{SVo00, Geus02, Tackett02}. This modification allows the targeting of eigenpairs by $\abssq{\braket{\Psi_0}{E_q}} = \abssq{c_q}$ to compute only the $n$ eigenpairs with maximum overlap with the target state without the need to over-compute extra eigenpairs. The algorithm can be trivially modified for the case where multiple target states are of interest.

The computation of the $n$ eigenstates of interest requires $\bigO{nN^2}$ time and $\bigO{N^2}$ space for dense representations of the Hamiltonian. In practice however, physical operators are nearly universally local and can be implemented in a matrix-free way such that the cost of matrix-vector multiplication is only $\bigO{N}$ time and space. Leveraging such a matrix-free implementation of the Hamiltonian reduces the cost of computing the $n$ eigenstates to just $\bigO{nN}$ time and $\bigO{N}$ space. Since $n\ll N$ and in practice can be fixed to a small value, the time complexity is roughly $\bigO{N}$. This is the same time and space complexity of just directly writing down the time-evolved state.

It is fortuitous that the optimal reduced basis states are energy eigenstates, as this renders the Galerkin-projected Hamiltonian diagonal and therefore the time evolution trivial in the reduced space,
\begin{equation}
    \Pdag H P = \ul{H} = \text{diag}\left(E_1, E_2, \ldots, E_n\right).
\end{equation}
Which makes the time evolution in the reduced basis just $\bigO{n}$. Recall that due to the reordering of the energy eigenstates, $E_1$ is not necessarily the ground state and it is not necessarily the case that the $E_q$ are ordered. Any other operator(s) of interest can be projected down to the reduced space, allowing the dynamics of any expectation value to be calculated in just $\bigO{n^2}$ time.

Including the cost of the Galerkin projection of the initial state and the observable(s) of interest, as well as the time evolution in the reduced space, the entire method requires $\bigO{Nn^2}$ time and $\bigO{N}$ space for matrix-free implementations of the Hamiltonian and observable(s).

\subsubsection*{Example}

Consider the Hamiltonian detailed in Appendix~\ref{app:spin}, a $1$D spin chain of $L$ spins with couplings $J_{ij}$ and transverse field $B$,
\begin{equation}\label{eq:longrange_ex}
    H = -\frac{1}{\mathcal{J}} \sum_{i<j}^L J_{ij} \left(\gamma_x \sigma_i^x \sigma_j^x + \gamma_y \sigma_i^y \sigma_j^y\right) - B \sum_i^L \sigma_{i}^z,
\end{equation}
where $\gamma_{x/y}$ are the relative strengths of the $x$ and $y$ couplings, $\sigma_i^v$ is the Pauli spin operator acting on the \ord{i} site in the $v$ axis, and $\mathcal{J}$ is the Kac normalization factor
\begin{equation}
    \mathcal{J} \equiv \frac{1}{L-1}\sum_{i\neq j}^L J_{ij}.
\end{equation}
For this example we will take the couplings to be a power-law decay with scale $J$ and exponent $p$,
\begin{equation}
    J_{ij} = \frac{J}{\abs{i-j}^p}.
\end{equation}
We choose $J=\gamma_x=B=1$, $\gamma_y=0$, the initial state to be the fully polarized-up state
\begin{equation}\label{eq:apdr_psi0}
    \Psinaught \equiv \ket{\uparrow \uparrow \stackrel{L}{\cdots} \uparrow},
\end{equation}
and the observable of interest to be
\begin{equation}
    S_x^2/L \equiv \frac{1}{L} \sum_{i,\,j}^L \sigma_i^x\sigma_j^x.
\end{equation}
We will consider two values for the decay exponent, a long range decay $p=0.1$ and a shorter-range decay $p=0.9$. The significance of this Hamiltonian, these specific values, and why they make for an interesting problem are discussed in Appendix~\ref{app:spin} and Refs.~\cite{De25,Cook2025apdr}.

To visualize how the Jacobi-Davidson method and the optimal approximate state work, we consider the case of $L=12$ spins. The dimension of the Hilbert space, and therefore the Hamiltonian, is $N=2^L =\num{4096}$. We choose $n=8$ for the size of the Galerkin subspace. Figure~\ref{fig:apdr_spectrum} shows the squared magnitude of the components of the exact and optimal approximate initial states in the energy basis of the Hamiltonian with $p=0.1$ and $p=0.9$. That is, Fig.~\ref{fig:apdr_spectrum} shows $\abssq{c_i}$ for the $c_i$ in Eq.~\eqref{eq:apdr_linearcomb} for both values of the decay exponent. It is immediately clear that the eigenstates with maximum overlap with the initial state are not simply the lowest lying states, nor are they grouped by their associated energies. The $n=8$ energy eigenstates with maximum overlap with Eq.~\eqref{eq:apdr_psi0}, which make up the optimal approximate state $\ket{\psi_0}$, are denoted with crosses and span more than a fifth (more than $\num{800}$ states) of the spectrum. This means that simply approximating the initial state with the lowest lying eigenstates would require more than $\num{100}$ times as many basis states to achieve roughly the same accuracy. A strategy of randomly sampling eigenstates would not work either, as the proportion of eigenstates with an initial state overlap greater than just $10^{-6}$ is less than $2\%$ for both Hamiltonians. Note that the magnitude of the components decay exponentially, with the \ord{8} largest component five orders of magnitude smaller than the first for $p=0.1$ and three orders of magnitude smaller for $p=0.9$. This is generally the case in practice and is what allows for $n\ll N$ to still produce an accurate prediction.

\begin{figure}
    \centering
    \includegraphics[width=0.75\textwidth]{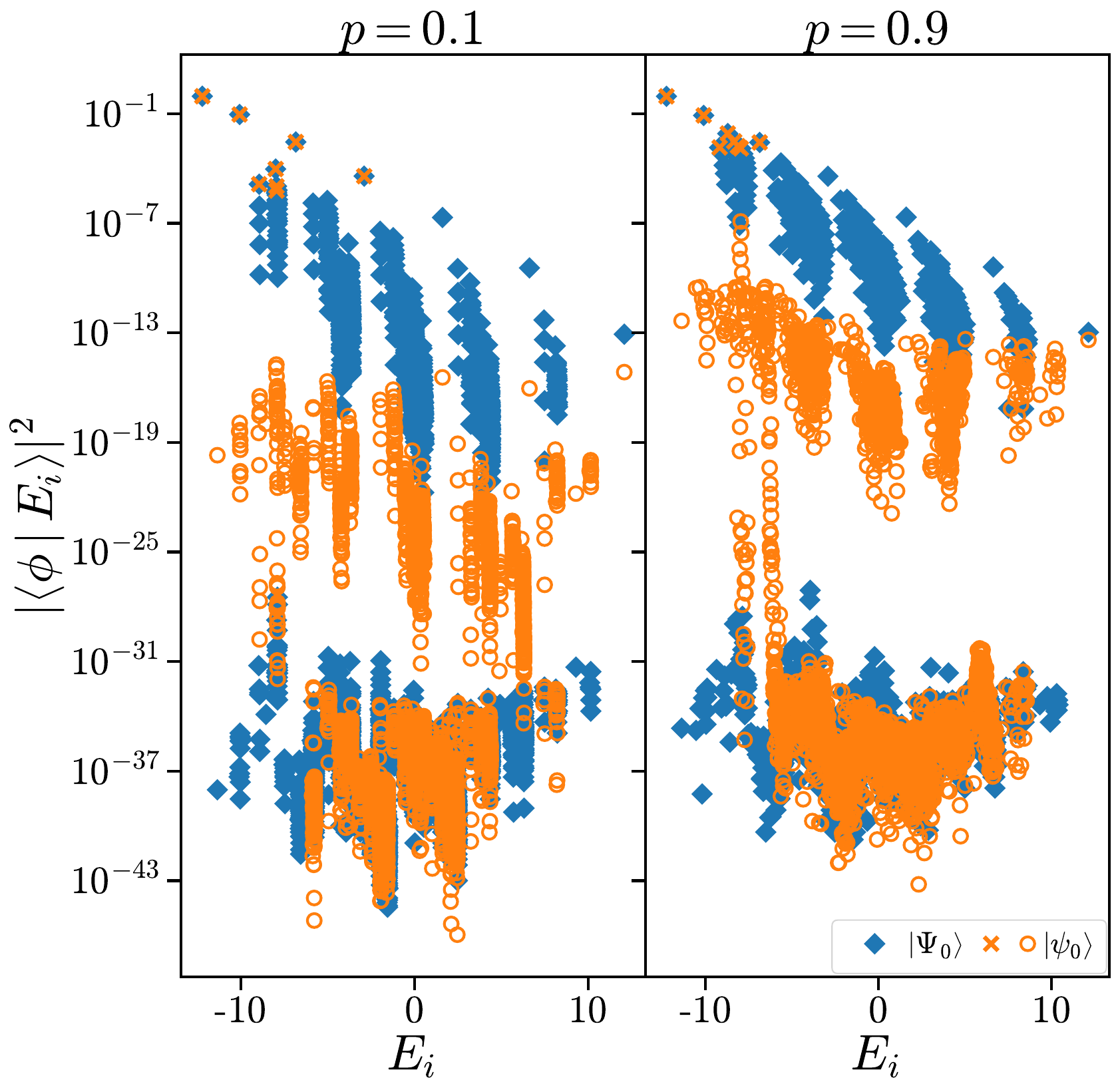}
    \caption[Spectrum of optimally approximated quantum state for time dynamics.]{\label{fig:apdr_spectrum}Squared magnitude of the components of the $L=12$ initial state in Eq.~\eqref{eq:apdr_psi0} in the energy basis of the Hamiltonian with $p=0.1$ (left) and $p=0.9$ (right) power-law couplings and $\mathlist{L, J, \gamma_x, \gamma_y, B}=\mathlist{12,1,1,0,1}$. The exact state ($\ket{\phi}\equiv\Psinaught$) is shown as blue diamonds. The optimal approximate state ($\ket{\phi}\equiv\ket{\psi_0}$) is shown as orange crosses and circles, with crosses denoting the automatically targeted states with maximum overlap. Nonzero solver tolerances result in nonzero overlap with the non-targeted states.}
\end{figure}

The efficacy of the method is demonstrated with a larger $L=14$ system, corresponding to $N=\num{16384}$, with the same $n=8$ number of Galerkin basis states. The exact evolution of the state under the two Hamiltonians is computed and the expectation value of $S_x^2/L$ is found at each time $t$. Then for each Hamiltonian, the optimal projector $P$ is found and used to project the initial state, Hamiltonian, and $S_x^2/L$ to the reduced space where time evolution is performed. Once the state and operators are projected to the reduced space, no computations need to be performed in the full space to carry out the time evolution. Figure~\ref{fig:apdr_dynamics} shows these dynamics along with the relative error in the observable. The relative error is on the order of $0.01\%$ for $p=0.1$ and $1\%$ for $p=0.9$ while the size of the Galerkin subspace is less than $0.05\%$ that of the full space. Finally, Fig.~\ref{fig:apdr_error} shows the numerically computed and analytically predicted error (or error bound) in the time evolution of the approximate state, validating Eqs.~\eqref{eq:apdr_minerr}~and~\eqref{eq:apdr_obserr}.

\begin{figure}
    \centering
    \begin{minipage}[t]{0.49\textwidth}
        \centering
        \includegraphics[width=\textwidth]{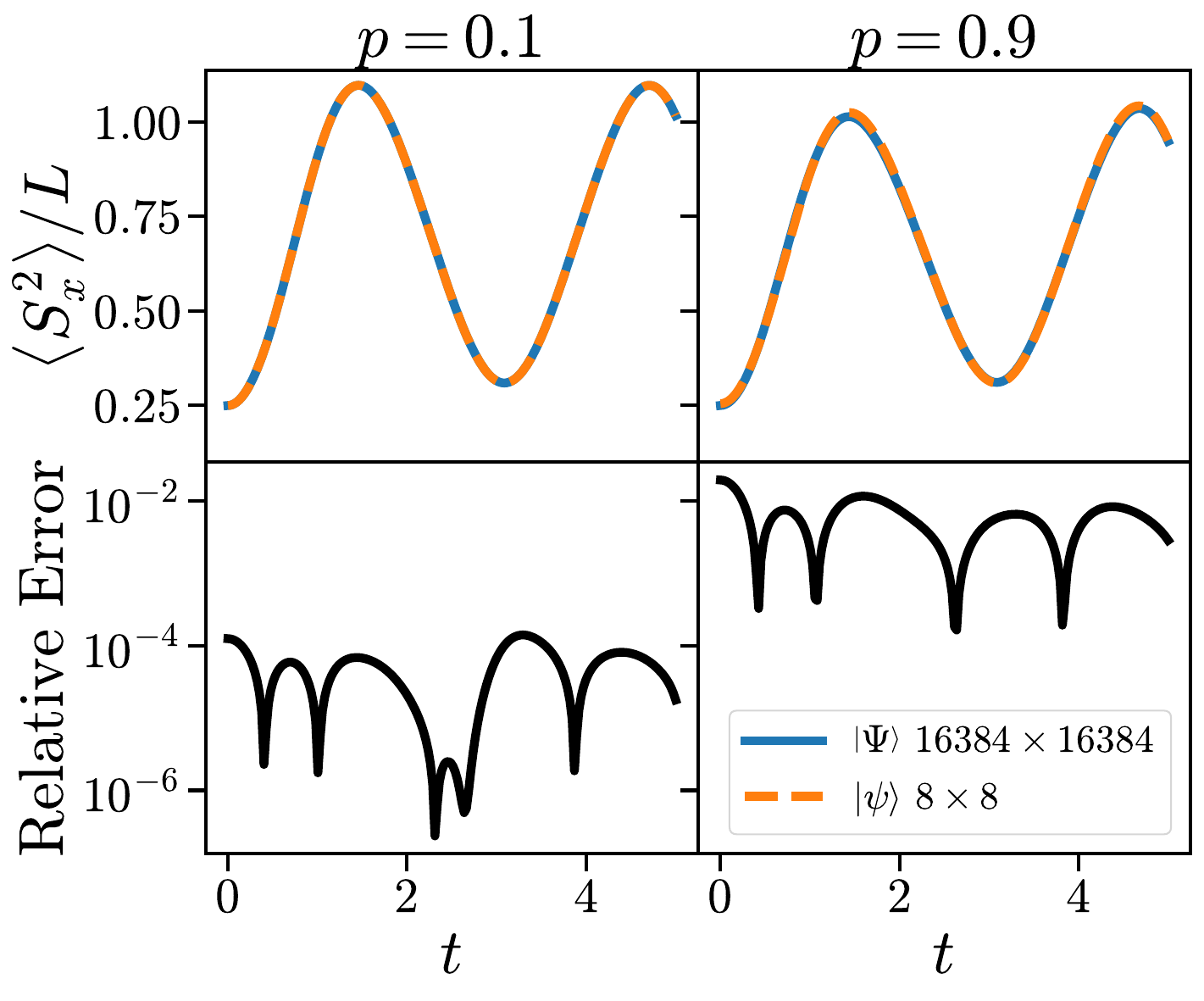}
        \caption[Dynamics of optimally approximated quantum state.]{\label{fig:apdr_dynamics}Exact (solid blue) and approximate (dashed orange) dynamics of $S_x^2/L$ for the Hamiltonian with $p=0.1$ (left) and $p=0.9$ (right). Relative errors in the approximations are shown in the lower plots.}
    \end{minipage}\hfill%
    \begin{minipage}[t]{0.49\textwidth}
        \centering
        \includegraphics[width=\textwidth]{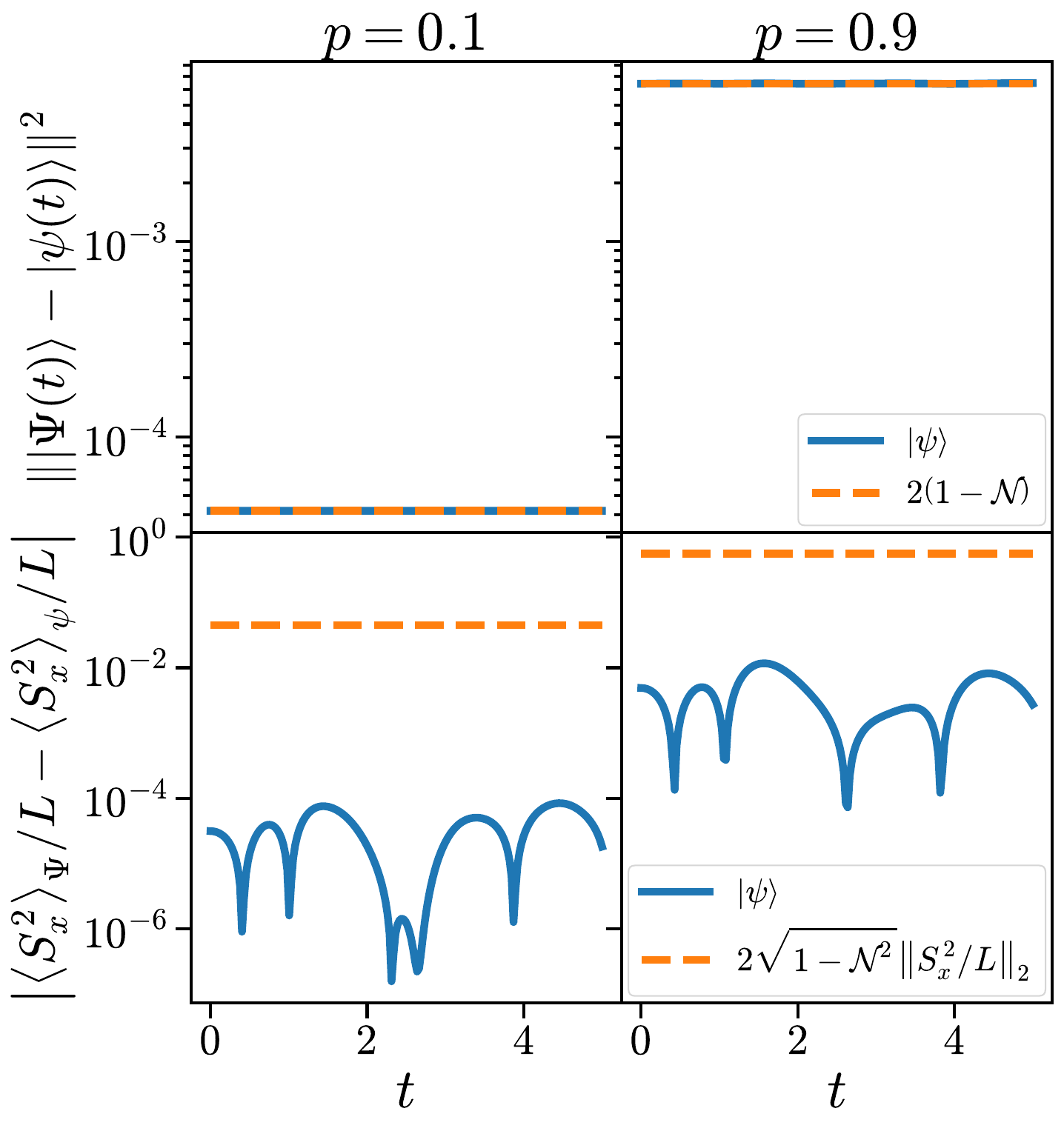}
        \caption[Error in the dynamics of optimally approximated quantum state.]{\label{fig:apdr_error}Numerically computed (solid blue) and analytically predicted (dashed orange) error for the dynamics of the state (top) and observable (bottom) for the Hamiltonian with $p=0.1$ (left) and $p=0.9$ (right).}
    \end{minipage}
\end{figure}

\begin{figure}
    \centering
    \emulatorsummary{Tractable \textit{a priori} Dimensionality Reduction for Quantum Dynamics}
    {%
        No.
    }
    {%
        No. Follows the same reasons as \ac{pod}-Galerkin.
    }
    {%
        No. Limited to dynamics of Hermitian systems with known initial states.
    }
    {%
        Yes.
    }
    {%
        Yes, maximally, as the method dictates what data are chosen.
    }
    {%
        Only inherent hyperparameter is the Galerkin subspace dimension.
    }
    {%
        Yes, converges absolutely to the exact time dynamics as $n\rightarrow N$.
    }
    {%
        Yes, the projected system is exactly a truncated form of the original system.
    }
    {%
        Yes, retained eigenvalues are equivalent, normalization is preserved, and energy is conserved.
    }
    {%
        Maximally intrusive. Requires partial diagonalization of the exact Hamiltonian.
    }
    {%
        \Ac{uq} not applicable, but error estimates and bounds are available and practically useful.
    }
    \captionof{table}[Summary of properties for the tractable \textit{a priori} dimensionality reduction for quantum dynamics method.]{Summary of properties for the method presented in this section.}
    \label{tab:apdr}
\end{figure}

} %
\subsubsection{Eigenvector Continuation}\label{sec:ec}
Within the field of \textit{ab initio} nuclear theory, \ac{ec} has perhaps been the most popular Galerkin-type emulator since its introduction~\cite{Frame2018,Sarkar2021,Konig2020,Demol2020,Furnstahl2020,Drischler2021,Bai2021,Sarkar2020,Franzke2022,Duguet2024,Sarkar2022,Frame2019}. While developed independently of the Galerkin method, it can nevertheless be viewed as a Galerkin projection of the standard parametric Hermitian eigenproblem
\begin{equation}\label{eq:ec_eigen}
    H(\mu)\Psi_k(\mu) = E_k(\mu)\Psi_k(\mu)
\end{equation}
where $H(\mu)$ is some parametric Hamiltonian with parameters $\mu$ and $\Psi_k(\mu)$ is the energy eigenstate associated with the energy eigenvalue $E_k(\mu)$~\cite{Bonilla2022}. \Ac{ec} utilizes snapshots, or ``training vectors'', of the eigenstates of interest at various parameter realizations. Originally this consisted of only the ground state, but later extensions showed that excited states could be included without modification~\cite{Frame2018,Franzke2022}. The \ac{ec} method proceeds as follows. Let $H(\mu)\in\Hbb(N)$ be the parametric Hamiltonian of interest and $v_j\left(\mu^{(i)}\right)\in\Cbb^N$ with $\norm{v_j\left(\mu^{(i)}\right)}=1$ be the snapshot eigenvector for parameter realization $\mu^{(i)}$ associated with the \ord{j} eigenvalue. For $n$ snapshots, we form the Galerkin projectors $\P$ and $\Pinv$ as before in Eq.~\eqref{eq:galerkinprojector}, where $\P$ is the matrix whose rows are $\mathset{v_j\left(\mu^{(i)}\right)^\dagger}$ and $\Pinv = \P^\dagger\left(\P\P^\dagger\right)^{-1}$. Following the conventional \ac{ec} notation, let $T_V$ be the matrix whose columns are the training vectors, so $\P = T_V^\dagger$, and ${\mathcal{N}=T_V^\dagger T_V=\P \P^\dagger}$ be the \textit{norm matrix}. The elements of the norm matrix are simply all the inner products of the training vectors. The relation to the Galerkin projectors from Eq.~\eqref{eq:galerkinprojector} is then $\P=T_V^\dagger$ and $\Pinv = T_V \mathcal{N}^{-1}$. Projecting Eq.~\eqref{eq:ec_eigen} onto the Galerkin subspace with these projectors yields
\begin{equation}\label{eq:ec}
    \begin{aligned}
        \underbrace{\left(\P H(\mu) \Pinv\right)}_{\ul{H(\mu)}}\underbrace{\left(\P \Psi_k(\mu)\right)}_{\ul{\Psi_k(\mu)}} &= E_k(\mu)\underbrace{\left(\P \Psi_k(\mu) \right)}_{\ul{\Psi_k(\mu)}}\\
        \iff \Bigl(\underbrace{T_V^\dagger H(\mu) T_V}_{\tilde{H}(\mu)} \mathcal{N}^{-1}\Bigr)\left(T_V^\dagger \Psi_k(\mu)\right) &= E_k(\mu) \left(T_V^\dagger \Psi_k(\mu)\right)\\
        \iff \tilde{H}(\mu) \Bigl(\underbrace{\mathcal{N}^{-1} T^\dagger_V \Psi_k(\mu)}_{\psi_k(\mu)}\Bigr) &= E_k(\mu) \overbrace{\mathcal{N} \Bigl( \mathcal{N}^{-1}}^{I_n} T_V^\dagger \Psi_k(\mu)\Bigr)\\
        \iff \tilde{H}(\mu) \psi_k(\mu) &= E_k(\mu)\mathcal{N}\psi_k(\mu).
    \end{aligned}
\end{equation}
Notice that this is simply the generalized eigenvalue problem\footnote{Also notice that this would have been the standard eigenvalue problem if the method used the \ac{pod}-Galerkin projector and stopped after the first line of Eq.~\eqref{eq:ec}. In practice, the difficulty of the generalized eigenvalue problem leads to implementations of \ac{ec} orthonormalizing the matrix of training vectors, essentially recovering the \ac{pod}-Galerkin method for the eigenvalue problem~\cite{Duguet2024}.} for $n\times n$ matrices $\tilde{H}(\mu)$ and $\mathcal{N}$. One can reconstruct the full-space solution with $\Psi_k(\mu)\approx\Psi^{\text{EC}}_k(\mu) = T_V \psi_k(\mu)$.

There are many recent extensions, applications, and analyses of \ac{ec} throughout the literature~\cite{Frame2018,Sarkar2021,Konig2020,Demol2020,Furnstahl2020,Drischler2021,Bai2021,Sarkar2020,Franzke2022,Duguet2024,Sarkar2022,Frame2019,Yoshida2022,Yapa2022,Yapa2023,Wesolowski2021,Xilin2022,Djarv2022}. Perhaps the interpretation that led to its rapid adoption is that of its close ties to perturbation theory and its success either as a faster-converging alternative to perturbation theory or in drastically improving the convergence rate of perturbation theory calculations~\cite{Sarkar2021,Demol2020}. This same interpretation---that of a series expansion of the eigenstate---gives rise to rigorous bounds on the convergence rate of \ac{ec}~\cite{Duguet2024}. See Ref.~\cite{Duguet2024} for an in-depth discussion and derivation of the error bounds as well as an excellent review of \ac{ec}. The impressive result is that \ac{ec} converges exponentially with the number of snapshots. For $d$ parameters, i.e.\ $\mu = \left[\mu_1,\,\mu_2,\,\ldots,\,\mu_d\right]$, the norm of the residual vector between the true and \ac{ec}-approximate eigenstate is
\begin{equation}
    \norm{\Psi_k(\mu) - \Psi_k^{\text{EC}}(\mu)}_2 \sim \alpha^{n^{1/d}},
\end{equation}
for some $0<\alpha<1$, where $n$ is the number of snapshots. This bound is valid for large $n$ and converges faster than any polynomial for sufficiently large $n$. Since $n$ is both the amount of data and the size of an \ac{ec} emulator, this shows that \ac{ec} is both computationally and data efficient as well as systematically improvable.

\subsubsection*{Example}

\begin{figure}
    \centering
    \includegraphics[width=0.8\textwidth]{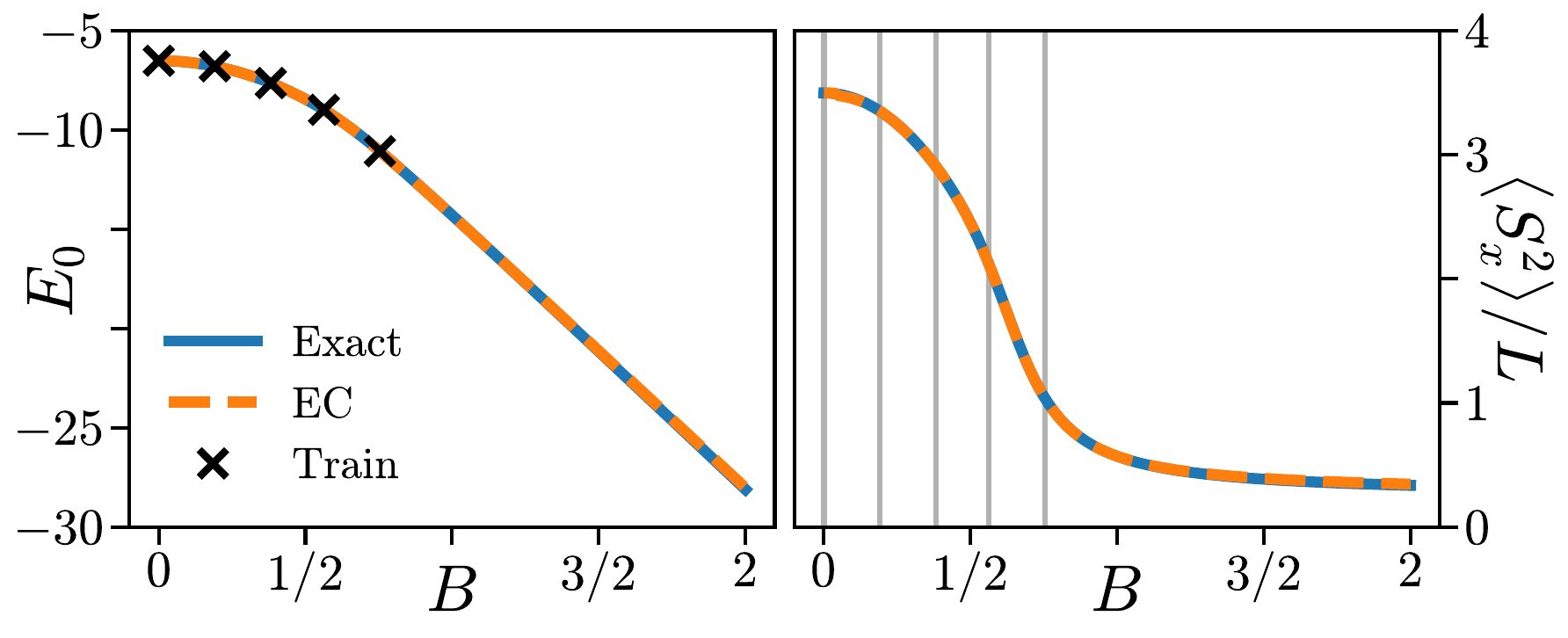}
    \caption[Results of applying \ac{ec} to a parametric many-body quantum system.]{\label{fig:ec_ex}Exact and \ac{ec}-predicted results for the ground state energy (left) and expectation value of $S_x^2/L$ in the ground state (right) for the Hamiltonian detailed in Appendix~\ref{app:spin} as a function of $B$. All other Hamiltonian parameters are fixed $\left\{L,\,\gamma_x,\,\gamma_y\right\}=\{14,\,0,\,1\}$ with a power-law coupling defined by $J=0$ and $p=0.9$. The locations of the training eigenvectors are shown as crosses (left) and vertical lines (right).}
\end{figure}

Consider the same Hamiltonian from the example in Section~\ref{sec:apdr} shown in Eq.~\eqref{eq:longrange_ex} and detailed in Appendix~\ref{app:spin}. As in the previous example, we choose $L=14$. We suppose that we want to find the ground state and the expectation value of the observable $S_x^2/L$ in the ground state as a function of $B$ while all other Hamiltonian parameters remain fixed ($\left\{J,\,p,\,\gamma_x,\,\gamma_y\right\}=\left\{1,\,0.9,\,1,\,0\right\}$). This allows us to frame the problem as an affine family of Hamiltonians
\begin{equation}
    H(B) = H_0 + B H_1
\end{equation}
where
\begin{equation}
    \begin{aligned}
        H_0 &\equiv -\frac{1}{\mathcal{J}} \sum_{i<j}^L J_{ij} \left(\gamma_x \sigma_i^x \sigma_j^x + \gamma_y \sigma_i^y \sigma_j^y\right),\\
        H_1 &\equiv - \sum_i^L \sigma_{i}^z.
    \end{aligned}
\end{equation}
We sample $n=5$ values of $B$ between $0$ and $0.75$ to solve the exact eigenproblem at and use the resulting eigenstates to form the $N\times n$ matrix of snapshots,
\begin{equation}
    T_V =%
    \begin{bmatrix}
    \vline & \vline & \vline & \vline & \vline \\
        \Psi_0\left(B_1\right) & \Psi_0\left(B_2\right) & \Psi_0\left(B_3\right) & \Psi_0\left(B_4\right) & \Psi_0\left(B_5\right)\\
        \vline & \vline & \vline & \vline & \vline
    \end{bmatrix}
\end{equation}
and orthonormalize the columns via the QR decomposition or other orthogonalization algorithm. This orthonormalization ensures that $\mathcal{N}=I$, greatly simplifying the projected problem. To make a prediction for the ground state energy using the \ac{ec} emulator for this system at a new target $B_t$, we solve the projected eigenproblem
\begin{equation}
    \begin{aligned}
        \left(T_V^\dagger H\left(B_t\right) T_V\right)\left(T_V^\dagger \Psi_0\left(B_t\right)\right) &= E^{\text{EC}}_0\left(T_V^\dagger \Psi_0\left(B_t\right)\right)\\
        \iff \left(T_V^\dagger H_0 T_V + B_t T_V^\dagger H_1 T_V\right)\left(T_V^\dagger \Psi_0\left(B_t\right)\right) &= E^{\text{EC}}_0\left(T_V^\dagger \Psi_0\left(B_t\right)\right)\\
        \iff \left(\tilde{H}_0 + B_t\tilde{H}_1\right) \psi_0\left(B_t\right) &= E_0^{\text{EC}} \psi_0\left(B_t\right).
    \end{aligned}
\end{equation}
Leveraging knowledge of the structure of $H(B)$ to project each of its constituent operators ($H_0$ and $H_1$) during the offline phase greatly decreases the online cost of the \ac{ec} emulator since the construction and subsequent projection of $H(B)$ can be exceedingly expensive and scales with $N$. Similarly, instead of reconstructing the $N$-dimensional \ac{ec} approximation of the eigenstate in order to make a prediction for the expectation value of $S_x^2/L$, it is much more efficient to simply project $S_x^2/L$ during the offline phase,
\begin{equation}
    \tilde{S}_x^2/L = T_V^\dagger S_x^2 T_V / L,
\end{equation}
and use the reduced-space eigenvector to compute expectation values with this reduced operator. It is straightforward to show that this is equivalent to the more expensive alternative,
{%
    \newcommand{\PsiB}{\ensuremath{\Psi_0(B)}}%
    \newcommand{\psib}{\ensuremath{\psi_0(B)}}%
    \newcommand{\PsiECB}{\ensuremath{\Psi_0^{\text{EC}}(B)}}%
    \begin{equation}
        \begin{aligned}
            \sandwich{\PsiB}{S_x^2}{\PsiB} &\approx\sandwich{\PsiECB}{S_x^2}{\PsiECB} \\&= \sandwich{T_V\psib}{S_x^2}{T_V\psib} \\&= \sandwich{\psib}{T_V^\dagger S_x^2 T_V}{\psib} \\&= \sandwich{\psib}{\tilde{S}_x^2}{\psib}.
        \end{aligned}
    \end{equation}%
}%
This process results in an \ac{ec} emulator for the $N=2^L=16384$-dimensional Hamiltonian which for any target value $B_t$ requires only the solution of a $5\times 5$ standard eigenproblem and the expectation value with a $5\times 5$ operator. The results for this example are shown in Fig.~\ref{fig:ec_ex}, demonstrating the predictive power of \ac{ec} for parametric Hamiltonian problems such as this one.

\begin{figure}
    \centering
    \emulatorsummary{Eigenvector Continuation}
    {%
        Yes.
    }
    {%
        No. Size of the emulator is the number of snapshots, $n$.
    }
    {%
        No. Limited to standard eigenvalue problems.
    }
    {%
        Only if the form of $H(\mu)$ is known and the parameter dependence is separable, allowing for the projection(s) to be done during the offline phase.
    }
    {%
        Yes. Exponential convergence.
    }
    {%
        Only inherent hyperparameter is the number of snapshots.
    }
    {%
        Yes.
    }
    {%
        Yes. The projected system is a generalized eigenvalue problem as opposed to the original standard eigenvalue problem, but the interpretation of all operators are largely unchanged. The projected system can be a standard eigenproblem of identical form to the original if the matrix of snapshots is orthonormalized first.
    }
    {%
        Yes, eigenvalues represented in the snapshots are equivalent, normalization is preserved, and energy is conserved.
    }
    {%
        Maximally intrusive. Requires evaluation of the high-fidelity $H(\mu)$ at all parameter realizations of interest. If the form of $H(\mu)$ is known and the parameter dependence separable, much of this cost can be moved to the offline phase.
    }
    {%
        \Ac{uq} not applicable, but error bounds are available and practically useful. Can be combined with \ac{uq} methods~\cite{Djarv2022}.
    }
    \captionof{table}{Summary of properties for \ac{ec}.}
    \label{tab:ec}
\end{figure}

\subsection{Gaussian Processes}\label{sec:gpr}
In data-driven machine learning and emulation, the usual paradigm is to prescribe a functional form for the model with some unknown parameters---and then fit those parameters to data. There can be several downsides to this typical approach, of which \ac{gp}---or more specifically in the context of emulation, \ac{gpr}---addresses two: interpretability and trustworthiness or \ac{uq}. \acp{gp} instead frame the problem not as determining a single function but instead as determining a probability distribution of functions themselves. A single \ac{gp} represents infinitely many functions, called realizations, each with some associated probability. Importantly, \ac{gpr} sidesteps explicitly enumerating these functions and instead works only in correlations between the data samples---a method ubiquitous in machine learning known as the \textit{kernel trick}~\cite{Rasmussen2005,Hofmann2008}. The point-wise prediction from \ac{gpr} often takes the form of a mean value---the value of the \textit{mean function} of the \ac{gp}---and confidence intervals which describe the distribution over the pointwise prediction. For a comprehensive introductory reference on \acp{gp} and \ac{gpr}, see Refs.~\cite{Barber2012,Beckers2021}.

\ac{gpr} has seen enormous success in varied applications in nuclear physics emulation and model \ac{uq}~\cite{Drischler2020,Semposki2025,Millican2024,Millican2025,Semposki2022}. It can be applied as the emulator directly, as part of an emulation framework, or as a \ac{uq} model fit to the underlying emulator. Despite its numerical non-parametricity, an excellent and unique use-case for \ac{gpr} in scientific contexts is in informing where in feature space data or snapshots should be collected~\cite{Boehnlein2022}. By guiding either experimental measurements or high-fidelity models to regions where an associated \ac{gpr}-based emulator is most uncertain, one can effectively maximize the informativeness of each additional sample. In applications where every data point is expensive, maximizing the usefulness of each sample is critical.

In Ref.~\cite{Kanagawa2018}, the authors show that the convergence rate of the mean function as the number of training samples, $m$, approaches infinity is $\bigO{m^{-k}}$ with $1/2 < k < 1$. The exact value of $k$ is determined by the dimensionality of the data, among other things, and generally decreases as the dimensionality increases. This can limit the effectiveness of \ac{gpr} in high-dimensional regression.

\begin{figure}
    \centering
    \emulatorsummary{Gaussian Process Regression}
    {%
        Yes.
    }
    {%
        No.
    }
    {%
        Yes.
    }
    {%
        No, though factorized and approximate methods exist which improve efficiency~\cite{Belyaev2014,Fradi2025}.
    }
    {%
        No. Converges as $\bigO{m^{-k}}$ with $1/2 < k < 1$.
    }
    {%
        Determined by the choice of kernel function. Usually only one or two relevant hyperparameters that each have clear interpretable meaning.
    }
    {%
        No. Given infinite data, \ac{gpr} will converge to the true function, but there is no guarantee of convergence with finite data.
    }
    {%
        Interpretable only in ``data-space,'' as predictions are purely combinations of training examples. No guaranteed interpretability in the context of the original problem or high-fidelity model.
    }
    {%
        No.
    }
    {%
        Purely data-driven.
    }
    {%
        All predictions are inherently realizations from probability distributions. The output of \ac{gpr} can be the probability distribution itself, leading to built-in uncertainty estimates.
    }
    \captionof{table}{Summary of properties for \ac{gpr}-based emulators.}
    \label{tab:gpr}
\end{figure}

\subsection{Artificial Neural Networks}\label{sec:pinn}

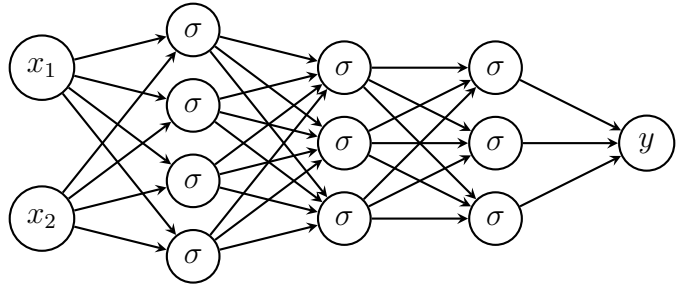
\begin{figure}
    \centering
    \begin{tikzpicture}
        \node[draw, circle, thick] (x1) at (0, 1) {$x_1$};
        \node[draw, circle, thick] (x2) at (0, -1) {$x_2$};

        \foreach \i in {1, 2, 3, 4} {%
            \node[draw, circle, thick] (h\i) at (2, 2.5 - \i) {$\sigma$};
        }

        \foreach \i in {1, 2, 3} {%
            \node[draw, circle, thick] (g\i) at (4, 2 - \i) {$\sigma$};
        }

        \foreach \i in {1, 2, 3} {%
            \node[draw, circle, thick] (e\i) at (6, 2 - \i) {$\sigma$};
        }

        \node[draw, circle, thick] (y) at (8, 0) {$y$};

        \foreach \i in {1, 2} {%
            \foreach \j in {1, 2, 3, 4} {%
                \draw[-stealth, thick] (x\i) -- (h\j);
            }
        }

        \foreach \i in {1, 2, 3, 4} {%
            \foreach \j in {1, 2, 3} {%
                \draw[-stealth, thick] (h\i) -- (g\j);
            }
        }
        \foreach \i in {1, 2, 3} {%
            \foreach \j in {1, 2, 3} {%
                \draw[-stealth, thick] (g\i) -- (e\j);
            }
        }

        \foreach \i in {1, 2, 3} {%
            \draw[-stealth, thick] (e\i) -- (y);
        }
    \end{tikzpicture}
    \caption[Schematic illustration of an \ac{mlp}.]{\label{fig:mlp}Schematic illustration of a \acl{mlp} with two inputs $(x_1, x_2)$, three hidden layers, and one output $y$. The hidden layers consist of four, three, and three neurons, respectively. Each edge represents a scalar multiplication by a \textit{weight}. When multiple edges converge on a single neuron, the sum of the weighted inputs is passed through a nonlinear \textit{activation function} $\sigma$, and then a scalar \textit{bias} is added. For example, the output of the top node in the first hidden layer would be $\sigma(w_{11}x_1 + w_{12}x_2) + b_1$, where $w_{11}$ and $w_{12}$ are the weights of the edges connecting $x_1$ and $x_2$ to the node, and $b_1$ is the bias.}
\end{figure}

\Acfp{ann} are a now-ubiquitous class of machine learning models that need little introduction. See Refs.~\cite{Geron2022,Zhang2023} for an introduction to \acp{ann}. Perhaps the simplest useful and most common \ac{ann} is the \ac{mlp}, a type of feedforward neural network where each layer is fully connected to the next. Figure~\ref{fig:mlp} shows a schematic of an \ac{mlp} with two inputs, three hidden layers, and one output. \Acp{mlp} can be used for nonlinear regression and classification, and therefore can be directly applied to emulation tasks. However, a more sophisticated approach which aims to incorporate the underlying physics has gained significant traction relatively recently. This method is called the \acf{pinn} and combines both the versatility and variety of \acp{ann} with available physics knowledge in order to produce models and emulators that are more accurate than just directly fitting the \ac{ann} as a black-box regression model.

For a review of \ac{pinn} methods see Ref.~\cite{Karniadakis2021}, and for a practical introduction see Ref.~\cite{Baty2024}. The core principle of \acp{pinn} is the modification of the typical data-driven loss function with the addition of the residual of the underlying physics equations. This makes \acp{pinn} a hybrid emulation technique---being part data-driven and part intrusive. Uninformed \acp{ann} are fit directly to data, such as snapshots of the system as in Galerkin's Method, such that the difference between the \ac{ann}'s predictions and the data is minimized. In contrast, \acp{pinn} are fit such that the sum of this ``data loss'' and the residuals of the physics equations is minimized. For example, consider a \ac{pinn} for a 2D parametric partial differential equation of the form
\begin{equation}\label{eq:pde_pinn}
    \frac{\partial^2}{\partial x^2}u(x,y; c) + c \frac{\partial}{\partial y}u(x,y; c) = b(x,y),
\end{equation}
for some function $b(x,y)$. We represent $u(x,y; c)$ as an \ac{ann} with three inputs $x$, $y$, and $c$, and one output $u(x,y;c)$. Given $m_\text{data}$ snapshots of solutions of Eq.~\eqref{eq:pde_pinn} for various values of $x$, $y$, and $c$, which we denote by $\mathset{(x_i, y_i, c_i, u_i)}$ with $i=1,2,\ldots,m$, the ``data loss'' would be
\begin{equation}
    \mathcal{L}_\text{data} = \frac{1}{m_\text{data}}\sum_i\abs{u(x_i, y_i, c_i) - u_i}^2,
\end{equation}
which is known as the mean squared error\footnote{Many different loss functions are equally valid here.}. The ``physics loss'' is given by the average squared residual of Eq.~\eqref{eq:pde_pinn} over some set of $m_\text{physics}$ points $(x_j, y_j, c_j)$ which need not be the same points as in the snapshots,
\begin{equation}\label{eq:pde_residual}
    \mathcal{L}_\text{physics} = \frac{1}{m_\text{physics}}\sum_j\abs{\frac{\partial^2}{\partial x^2}u(x_j,y_j; c_j) + c_j \frac{\partial}{\partial y}u(x_j,y_j; c_j) - b(x_j,y_j)}^2,
\end{equation}
where the derivatives of the output of the \ac{ann}, $\frac{\partial^2}{\partial x^2}u(x,y;c)$ and $\frac{\partial}{\partial y}u(x,y;c)$, are computed via \ac{ad} (see Section~\ref{sec:ad}). The \ac{ann} is then fit, or trained, to minimize a weighted sum of these two losses,
\begin{equation}
    u = \argmin_u\left[w_\text{data} \mathcal{L}_\text{data} + w_\text{physics}\mathcal{L}_\text{physics}\right].
\end{equation}
Setting $w_\text{physics}=0$ is equivalent to the traditional uninformed \ac{ann} approach.

This illustrates only the core idea of \acp{pinn}. Additional terms penalizing violations of desired physical properties can be devised and added to the loss function to improve the accuracy of the trained model. Significant research effort has been put in to devising specific \ac{ann} architectures that have certain symmetries or other properties that each physical system should have.

Both the offline and online efficiency of \acp{pinn} depends heavily on the specific \ac{ann} architecture. The offline cost is also strongly influenced by the cost of evaluating the residual of the underlying equations. Compared to \acp{rbm}, some sacrifices are made, particularly regarding data efficiency, hyperparameter efficiency, and interpretability. Highly over-parameterized \acp{ann}---the usual regime---can require significant amounts of data to train. Additionally, the performance of a \ac{pinn} is highly dependent on a suitable set of hyperparameters and different target problems may require very different hyperparameters even for the same \ac{ann} architecture~\cite{Hao2024}.

However, more recent works have indicated that many \ac{pinn} results are overoptimistic in both accuracy and computational advantage~\cite{Grossmann2024,McGreivy2024}. Traditional numerical methods---not emulators---still outperform \acp{pinn} in many cases~\cite{Grossmann2024}.

\begin{figure}
    \centering
    \emulatorsummary{Physics-Informed Neural Networks}
    {%
        Yes.
    }
    {%
        Yes.
    }
    {%
        Yes.
    }
    {%
        Highly dependent on the specific \ac{ann} architecture and evaluation of the residual of the informing equations.
    }
    {%
        Usually only with sophisticated \ac{ann} architectures tailored to the target problem.
    }
    {%
        No. A vast continuum of \ac{ann} architectures exist to choose from. Even within a single \ac{ann} architecture there can be several hyperparameters.
    }
    {%
        No. In theory, the universal approximation theorem guarantees that a large enough \ac{ann} will converge absolutely to the data. However, converging to the data does not guarantee that the \ac{ann} will converge to the high-fidelity model.
    }
    {%
        No. In general, the \ac{ann} architecture has little to do with the true physical system.
    }
    {%
        Approximately physically consistent, to the degree that the physics loss represents all relevant physical properties and is sufficiently minimized.
    }
    {%
        Hybrid.
    }
    {%
        Not inherent, but can be added on top of the method.
    }
    \captionof{table}{Summary of properties for \ac{pinn}-based emulators.}
    \label{tab:pinn}
\end{figure}

\section{Summary}
    This chapter introduced the concept of emulators, established the definitions for various properties of emulators, and presented several commonly used methods. Many of these methods can either trace their origins to, or be reinterpreted as, variations on Galerkin's Method---which here we call the \acf{erbm} (Section~\ref{sec:erbm}). It is still useful to differentiate these methods in name from Galerkin's Method, not only for conciseness\footnote{For instance, it is much simpler to say ``\acf{pod}'' instead of ``Galerkin's Method with snapshots chosen as the first $n$ principal components of the full matrix of snapshots'' while conveying the same information.} but also for analyses which are only valid under the assumptions of specific methods\footnote{Such as the exponential convergence of \acf{ec} which is not true in general for Galerkin's Method.}. \Acf{gpr} and \Acfp{pinn} are outliers as they cannot be cast as Galerkin-like methods and instead are formulated from probability distributions and \acfp{ann} respectively (Sections~\ref{sec:gpr}~and~\ref{sec:pinn}). Each emulator has its own strengths, weaknesses, use-cases, variations, and ongoing research. Table~\ref{tbl:emusummary} provides a summary of the properties of each emulator as discussed in this chapter.

\begin{figure}
    \centering
    \setlength\tabcolsep{5pt}
    \begin{tabularx}{\linewidth}{Xccccccccccc}
        \allemsummary
        \bottomrule
    \end{tabularx}
    \captionof{table}[Summary of the properties of emulator methods covered in Chapter~\ref{ch:emu}.]{\label{tbl:emusummary}Summary of the properties of emulator methods discussed in this chapter. There are a few noteworthy trends: the methods that can be equated to some variation of Galerkin's Method often share many properties. Of the methods here, only the \ac{pinn} is numerically parametric---an often highly desirable property---though this is at the expense of many other properties.}%
\end{figure}

\chapter{Parametric Matrix Models}\label{ch:pmm}
\acresetall

\Acp{pmm} were first introduced in Ref.~\cite{Cook25} by me, Dr.\ Danny Jammooa, Dr.\ Morten Hjorth-Jensen, Dr.\ Daniel D.\ Lee, and Dr.\ Dean Lee. Originally developed simply as the next-generation \ac{ec}-like emulator, the scope of \acp{pmm} reach far beyond standard eigenproblem emulation to nonlinear dynamics and general machine learning.

\section{Parametric Matrix Models as \textit{Implicit} Reduced Basis Methods}\label{sec:pmm_irbm}

In the previous chapter, Galerkin-like methods were referred to as \acp{erbm}, despite literature referring to them simply as \acp{rbm}. In each of the Galerkin-like methods, both the \textit{reduced space} $\V_n$ and the \textit{reduced basis} $\left\{u_i\right\}$ were explicitly specified or chosen. However, note that the form of the Galerkin-projected problem in Eq.~\eqref{eq:galerkinprojector} is entirely independent of both the reduced space and reduced basis. Beyond just the form, there are cases where the predictions themselves are independent of the reduced basis or reduced space.

To illustrate this, suppose only the eigenvalues from a standard Hermitian eigenproblem are to be emulated. Using a \ac{pod}-Galerkin approach, following the same notation from Chapter~\ref{ch:emu} where $P$ is the \ac{pod} projector and $\ul{X} = \Pdag X P$ and $\ul{x} = \Pdag x$ denote reduced operators and vectors respectively, we have
\begin{equation}
    H\Psi_k = E_k\Psi_k\quad\implies\quad\left(\Pdag H P\right)\left(\Pdag \Psi_k\right) = E_k \left(\Pdag \Psi_k\right)\quad\implies\quad \ul{H}\ul{\Psi}_k = \ul{E}_k \ul{\Psi}_k.
\end{equation}
Recall the most important property of the \ac{pod}-Galerkin method: scalars, inner products, and norms are approximately preserved. This property guarantees that the projected problem is still a standard eigenproblem---that is, the $\ul{\Psi}_k$ are orthonormal---and $\ul{E}_k\approx E_k$. Now, consider the two sets of all the transformations of the original problem and of the projected problem that leave $\ul{E}_k$ unchanged. The former is the set of all unitary transformations on the space spanned by the projection\footnote{This is not the same as all unitary transformations on the original space. One can construct a unitary transformation on the original space that rotates elements such that they are orthogonal to $P$ and thus the eigenvalue of the projected problem would not be conserved.} induced by $P$, and the latter is all unitary transformations on the reduced space. Denote a particular pair of these transformations by $U$ and $\ul{U}$, then we know
{%
    \newcommand{\chatul}[1]{\check{\ul{#1}}}%
    \begin{equation}
        \begin{aligned}
            H\Psi_k &= E_k\Psi_k\\
            \iff U H U^\dagger U \Psi_k &= E_k U \Psi_k\\
            \implies \underbrace{\left(\Pdag U H U^\dagger P\right)}_{\hat{\ul{H}}}\underbrace{\left(\Pdag U \Psi_k\right)}_{\hat{\ul{\Psi}}_k} &= \ul{E}_k \left(\Pdag U\Psi_k\right)\\
            \iff \hat{\ul{H}} \hat{\ul{\Psi}}_k &= \ul{E}_k \hat{\ul{\Psi}}_k\\
            \iff \underbrace{\ul{U} \hat{\ul{H}} \ul{U}^\dagger}_{\chatul{H}} \underbrace{\ul{U} \hat{\ul{\Psi}}_k}_{\chatul{\Psi}_k} &= \ul{E}_k \ul{U} \hat{\ul{\Psi}}_k\\
            \iff \chatul{H}\chatul{\Psi}_k &= \ul{E}_k \chatul{\Psi}_k
        \end{aligned}
    \end{equation}%
}%
for the same $E_k\approx \ul{E}_k$ as before. That is, there are as many reduced models of the original problem with identical form that produce identical results as there are possible combinations of $U$ and $\ul{U}$ (infinitely many). Instead of applying these transformations to the reduced model, we can apply them to the \ac{pod} projector, $\hat{P} = U^\dagger P\ul{U}$, thus showing that for this problem there are infinitely many reduced bases---all corresponding to orthonormal bases of the same reduced space---that yield an emulator with identical predictions. Thus there exist many more emulators with identical forms and predictions than \ac{pod}-Galerkin, or any Galerkin method---any \textit{explicit} reduced basis method---can produce. Although illustrated with a particular example, this is true in general for any system with \textit{symmetries} that leave the target values unchanged.

We have thus far neglected any parametric dependence of the system. Despite these transformations leaving the predictions of the emulator invariant at specific parameter realizations, the predictions may be significantly affected at other values of the parameters. Consider expanding upon the previous example by introducing the simplest affine parameter dependence,
\begin{equation}
    H(\mu) = H_0 + \mu H_1.
\end{equation}
It is again true that unitary transformations of $H(\mu)$ leave the eigenvalues $E_k(\mu)$ unchanged for all $\mu$. However, any mixed unitary transformations of $H_0$ and $H_1$ trivially leave $E_k(\mu=0)$ invariant while potentially altering $E_k(\mu\neq 0)$. Once more, this is not a specific quality of this example but true in general for any parametric system whose constituent components exhibit independent symmetries.

So we now see that any Galerkin emulator is actually just one element from a vast family of emulators with identical forms and equivalent predictions at some countable set of parameter realizations---usually the parameter realizations associated with training data. This observation leads us to two fundamental questions:
\begin{enumerate}[label=Q.\arabic*]
    \item\label{itm:exist} Do there exist emulators in this family that are more useful than the Galerkin ones?
\end{enumerate}
And if so,
\begin{enumerate}[resume*]
    \item\label{itm:method} Can we describe an emulation method that is able to take advantage of these non-Galerkin emulators?
\end{enumerate}

There are two complementary ways to answer {}\ref{itm:exist} in the affirmative. The first is the recognition that there are many real-world problems where the information required to construct any \ac{erbm} is unavailable either due to practical or fundamental reasons---despite the known existence of the reduced model. Multitudes of many-body methods for computing eigenenergies do not compute the full associated eigenstates---which would become the snapshots an \ac{erbm} uses---often because they are impractically large\footnote{Just one eigenvector from a $9\times9\times9$ lattice chiral effective field theory calculation of ${}^6\text{Li}$ would require more than $\si{1000~\tera\byte}$ to store.}. However, the complete state vector of a system---especially a quantum one---is rarely the quantity of interest. Instead, relatively few scalar quantities like energies, expectation values, and transition amplitudes are the more common goal. Even problems which require the full ``microstate'' solution are often solved without direct access to the full space operators and instead use ensemble or Monte Carlo methods, such is the case in fluid dynamics with the lattice Boltzmann method~\cite{Chen1998}. In either of these cases, the form of the full model is still known, and given unlimited computational power we usually could attain the information necessary to construct an \ac{erbm}, so we know that reduced-space models exist. Having a method that allows us to shortcut directly to the reduced model without the need to explicitly construct a basis and project the full model onto it would allow \ac{rbm} emulation in these ubiquitous contexts.

The second approach to answering {}\ref{itm:exist} is to show there exists at least one case where the \ac{erbm} approach fails to find a suitable reduced model despite the existence of one. Following the example laid out in Ref.~\cite[Methods]{Cook25}, we again consider a standard parametric Hermitian eigenproblem. Consider a system of $S$ non-interacting spin-$1/2$ particles with the parametric Hamiltonian
\begin{equation}\label{eq:noninteracting}
    H(c) = \frac{1}{2S}\sum_i^S\left(\sigma_i^z + c \sigma_i^x\right) = \frac{1}{2S}\left(Z + cX\right),
\end{equation}
where $c$ is the real-valued parameter, $\sigma_i^u$ is the standard $u$ Pauli matrix for spin $i$, and $Z,\,X$ are the total $z$-spin and total $x$-spin operators respectively. We seek an emulator for the ground state energy of this system as a function of $c$ for a given system size $S$ based on data from $n=5$ samples for $c<0$. Constructing the $5\times 5$ \ac{ec}---or equivalently \ac{pod}---emulator $\ul{H}(c) = \frac{1}{2S}\left(\ul{Z} + c\ul{X}\right)$ for this problem shows significant deviations in the prediction from the true energy for $c>0$ when $S$ becomes large, as shown in Fig.~\ref{fig:implicit}. However, there are infinitely many solutions of the same $5\times 5$ form as this \ac{ec} emulator which reproduce the true ground state energies exactly. One such family of emulators is given by the ground state energy of the block-diagonal parametric matrix
\begin{equation}\label{eq:noninteractingemulator}
    M(c) =\frac{1}{2}\left(%
    \begin{bmatrix}
        \sigma^z & 0 \\
        0 & h
    \end{bmatrix} + c%
    \begin{bmatrix}
        \sigma^x & 0\\
        0 & 0
    \end{bmatrix}\right),
\end{equation}
where $\sigma^z$ and $\sigma^x$ are the standard $2\times 2$ $z$ and $x$ Pauli matrices and $h$ is any $3\times 3$ Hermitian matrix with all eigenvalues strictly greater than $-1$. The predictions from this emulator are shown in Fig.~\ref{fig:implicit} as well and match the true energies exactly.

{%
    \newcommand{\da}{\downarrow}
    \newcommand{\ua}{\uparrow}
    \begin{figure}
        \centering
        \includegraphics[width=0.6\textwidth]{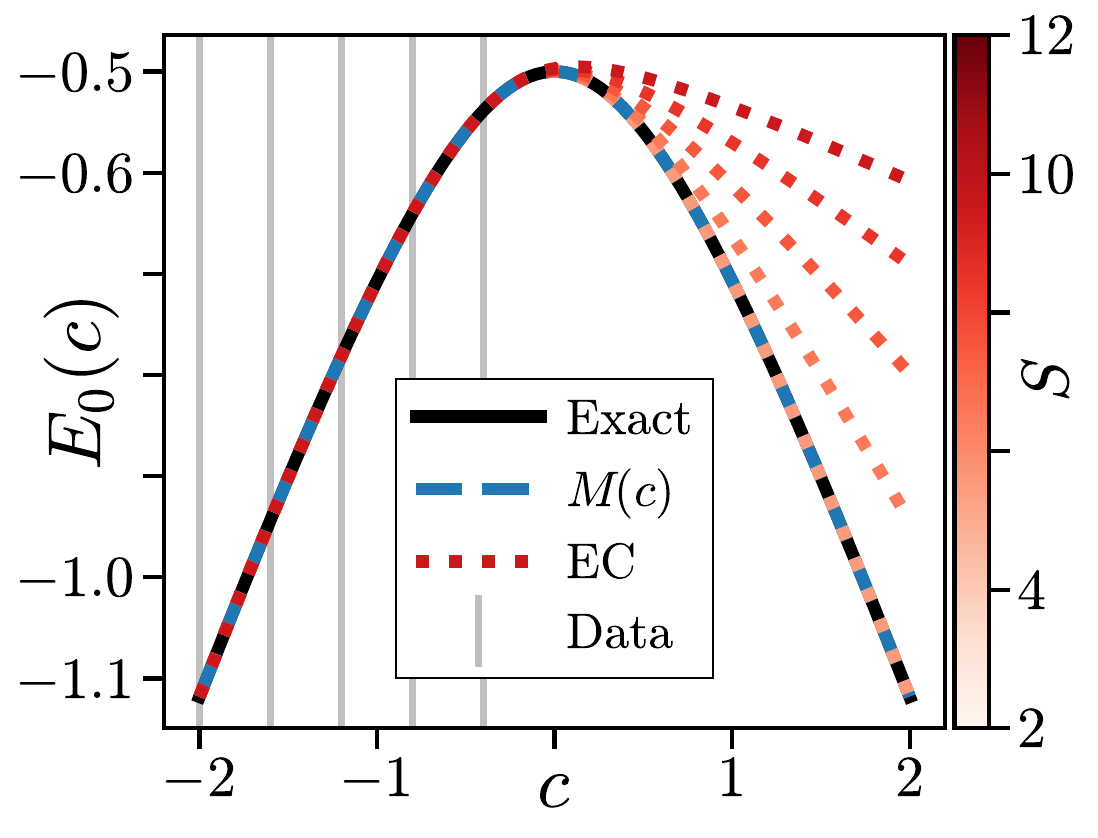}
        \caption[Failure case of \ac{erbm} approaches illustrated by \ac{ec}-based emulation of a non-interacting spin system.]{\label{fig:implicit}Ground state energy as a function of $c$ for the parametric Hamiltonian in Eq.~\eqref{eq:noninteracting} (solid black) as well as the predictions from an \ac{erbm} (specifically \ac{ec}, dotted red) and from one realization from the family of emulators described by Eq.~\eqref{eq:noninteractingemulator} (dashed blue) for different system sizes $S$. The location of training data for the emulators is indicated by vertical gray lines. Based on a similar figure from Ref.~\cite{Cook25} with permission\footnote{In Ref.~\cite{Cook25} the total $z$-spin basis (basis vectors $\ket{Z}\in\left\{\ket{-S/2},\,\ket{-S/2+1},\,\ldots,\,\ket{S/2-1},\,\ket{S/2}\right\}$) was used for evaluations of the Hamiltonian in Eq.~\eqref{eq:noninteracting} which allowed for system sizes as large as $S=256$ while only requiring exact eigenvectors of size $N=257$. By contrast, here we use the individual $z$-spin basis (basis vectors $\ket{\sigma^z_1\sigma^z_2\cdots\sigma^z_S}\in\mathset{\ket{\da\da\cdots\da\da},\,\ket{\da\da\cdots\da\ua},\,\ldots,\,\ket{\ua\ua\cdots\ua\da},\,\ket{\ua\ua\cdots\ua\ua}}$) which require exact eigenvectors of size $N=2^S$. Thus, Fig.~\ref{fig:implicit} represents a dimensionality reduction from at most $N=2^{12}=4096$ to $n=5$ while the equivalent figure in Ref.~\cite{Cook25} represents only a reduction from at most $N=S+1=257$ to $n=5$. The conclusions are unchanged in either case.}\footnote{I am the author of the original figure in Ref.~\cite{Cook25} and gave permission.}.}
    \end{figure}%
}

The answer to {}\ref{itm:method} is the central topic of this thesis, \acfp*{pmm}, which can be viewed as a class of \acp{irbm}. Instead of specifying a subspace $\V_n$ and projection operators $\Pup$ and $\Pdown$, a \ac{pmm} emulator specifies only the absolute minimum to describe the form of the reduced model: the size of the reduced space $n$ and properties of the projection operators (e.g.\ orthogonality). This mathematical form is then concretized by solving an optimization problem to fit the reduced model to available data---known as \textit{training} the model. In this sense the basis of the resulting reduced model is \textit{implicit}: the information for the reduced basis is contained within and inseparable from the operators of the reduced model. Following the previous example, we saw that a reduced model of the ground state energy of Eq.~\eqref{eq:noninteracting} for an $n=5$ reduced space arising from an orthogonal projector had the form of the ground state energy of
\begin{equation}\label{eq:noninteracting_pmm}
    \ul{H}(c) = \frac{1}{2S}\left(\ul{Z}+c\ul{X}\right).
\end{equation}
In the PMM approach, it is not the case that $\ul{A} = \Pdag A P$ since $P$ is not known and may vary between constituent components of the reduced model. Instead, the reduced operators are the solution to the minimization of some loss function relating the predictions of the emulator to available data. For this example,
\begin{equation}\label{eq:exmin}
        \ul{Z},\ul{X} = \argmin_{\ul{Z},\,\ul{X}\in\mathbb{S}(n)}\mathcal{L}\bigl(\mathset{E_0\left(c_i\right)},\mathset{\ul{E}_0\left(c_i\right)}\bigr)
\end{equation}
where $\ul{E}_0(c_i)$ is the ground state energy of Eq.~\eqref{eq:noninteracting_pmm} for $c=c_i$, $\mathset{E_0(c_i)}$ is the set of data to fit the reduced model to, $\mathset{\ul{E}_0(c_i)}$ are the predictions from the reduced model at the same values of $c$ as the data, $\mathcal{L}$ is some \textit{loss function} which attains its minimum when $\mathset{E_0(c_i)}=\mathset{\ul{E}_0(c_i)}$, and $\mathbb{S}(n)$ is the set of all $n\times n$ symmetric matrices. In this example, we know $\ul{Z}$ and $\ul{X}$ are symmetric because we specified that they are the result of an orthogonal projection on $Z$ and $X$ which are also symmetric. There is no unique solution to Eq.~\eqref{eq:exmin}, and by construction the less-desirable \ac{erbm} is a solution. Guiding the minimization to more robust solutions is an exercise in machine learning solved through various techniques such as regularization and validation which will be discussed later.

In the broadest possible terms, the steps of using the \ac{pmm} method to form an emulator are as follows:
\begin{enumerate}[label=(\arabic*.), ref=\arabic*]
\setcounter{enumi}{-1}
\item\label{itm:pmm0} Provide the \textit{form} of the full model\footnote{It is also possible to prescribe or suppose a form for the underlying model if it is not wholly known.} that produces target values as well as a set of training examples of those target values.
\end{enumerate}%
\begin{enumerate}[resume*, label=\arabic*., ref=\arabic*]
    \item\label{itm:pmm1} Select the desired reduced-space dimension $n$ and properties of an unknown (implicit) projection from the full space to a reduced space of this size.
    \item\label{itm:pmm2} Follow the process of deriving the form of the \ac{erbm} that would result from projecting the full model with the chosen projection. This includes identifying any known properties (such as matrix symmetry and vector normalization) of the objects in the reduced model.
    \item\label{itm:pmm3} Replace all unknown projected objects with unknown numerically parameterized ones that respect the identified known properties.
    \item\label{itm:pmm4} Identify a suitable \textit{loss function} which encodes the dissimilarity between predictions from the parameterized reduced model and the training examples\footnote{The problem statement here is in the context of \textit{supervised learning}---that is, the context in which the target predictions for the training data are known. However, all discussions in this chapter generalize to \textit{unsupervised} contexts---those where the exact target output values are not known and instead it is often properties of the distribution of output values like clustering that are desired---only by changing to a suitable unsupervised loss function.}.
    \item\label{itm:pmm5} Find the location of the minimum of the loss function over all possible values of the parameterized objects.
    \item\label{itm:pmm6} The values of the parameterized objects at the minimum are the \textit{trained} objects and the reduced model with these parameters is the trained \ac{pmm} emulator.
\end{enumerate}
A projection with many of the properties of an orthogonal projection is nearly universally chosen in step~\ref{itm:pmm1} due to the numerous properties that are preserved under such transformations---see Section~\ref{sec:pod} in Chapter~\ref{ch:emu}. Unless stated otherwise, in this thesis we will assume all \acp{pmm} choose a projection that exactly preserves Hermiticity and approximately preserves some eigenvalues (if applicable) and inner products\footnote{There is research that suggests oblique (non-orthogonal) projections can perform just as well or better than orthogonal ones for certain problems~\cite{Padovan2024}. However, orthogonal-like projections are still currently preferred in practice due to the numerous conserved properties of operators under orthogonal projections as well as the added difficulty in training an oblique \ac{pmm}.}---which is similar to but not necessarily exactly an orthogonal projection. As a consequence of the \ac{pmm} following the same form as the associated \ac{erbm}, all the parameterized objects in the reduced model will be scalars, vectors, matrices, or tensors\footnote{Of course, these are all tensors: order-$0$, order-$1$, order-$2$, and above. Equivalently, it can be said that the reduced objects are always multidimensional arrays.}. Just like in many other machine learning tasks, the loss function chosen in step~\ref{itm:pmm4} may include additional penalties; however, this is usually unnecessary for \acp{pmm}. The potential drawbacks of the \ac{pmm} method lie in steps~\ref{itm:pmm2},~\ref{itm:pmm3},~and~\ref{itm:pmm5}. Compared to purely data-driven methods and \acp{erbm}, \acp{pmm} require work on the part of the implementer to carefully derive the form for any given problem. Laying out a \ac{pmm} for a new problem requires intimate understanding of both the problem itself and the theory behind \acp{rbm}. Through Chapter~\ref{ch:emu} and the numerous examples in this chapter, the reader should be well-prepared and find this potential drawback a non-issue. The optimization problem in step~\ref{itm:pmm5} is exceedingly nontrivial for all but the smallest and simplest \acp{pmm}. Fortunately, decades of research in training ever-larger \acp{ann} have produced both the knowledge and the software necessary to obtain satisfactory approximate minima in a controlled and efficient manner.

\section{Numerical Parametricity and Reduced Basis Methods in the Literature}

As stated in Chapter~\ref{ch:emu}, numerical parametricity is a very desirable property for any machine learning---and therefore emulation---method. Simultaneously, the mathematical rigor, interpretability, and convergence guarantees of \acp{erbm} are perhaps even more important for physics-emulation applications. It is no surprise then that the \ac{pmm} method is not the first to mix numerically parametric elements with \acp{erbm}\footnote{See discussions in Sections~\ref{sec:pod}~and~\ref{sec:dmd} and Refs.~\cite{Lieu2004,Lieu2007,Amsallem2007,Weickum2006,Galletti2004, Grepl2005,Ghattas2008,Ly2001,Degroote2010,Couplet2005,Huhn2023,Andreuzzi2023} as well as the rest of this section.}. However, existing hybrid \ac{erbm}-based methods either extend a non-numerically parametric \ac{erbm} core~\cite{Lieu2004,Lieu2007,Amsallem2007,Weickum2006,Galletti2004, Grepl2005,Ghattas2008,Ly2001,Degroote2010,Couplet2005,Huhn2023,Andreuzzi2023,Brunton2016,Burohman2023}, rely on specific ans\"atze for the form of the full or reduced equations\footnote{References~\cite{Huhn2023,Andreuzzi2023} and all \ac{dmd}-based methods assume a linear map from the current state or history of states to the new state, potentially with nonlinear feature engineering often chosen by some heuristics. Reference~\cite{Bradde2022}, as well as many emulation methods in the field of electronic circuit design, assume a rational function form known as the Parameterized Sanathanan-Koerner model~\cite{Sanathanan1963}. Reference~\cite{Padovan2024} assumes a polynomial reduced-order model, although this is practical limitation as opposed to a fundamental one. Reference~\cite{Burohman2023} assumes a very general form using only linear operations on the inputs and some hidden state variables.}~\cite{Huhn2023,Andreuzzi2023,Bradde2022,Burohman2023,Padovan2024}, learn the explicit projection operators from data~\cite{Padovan2024}, or utilize nonlinear transformations which break the connection between the reduced model and the high-fidelity model~\cite{Brunton2016,Karniadakis2021,Sharma2022}. Conceptually, the Petrov-Galerkin-based methods described in Refs.~\cite{Burohman2023,Padovan2024} are the most closely related methods in the literature to \acp{pmm}.

In Ref.~\cite{Burohman2023}, the authors describe a fully data-driven emulation method for systems with noisy snapshots. By formulating the emulator as a Petrov-Galerkin projection of an infinitely large set of models that are capable of explaining the observed data---described by a quadratic matrix inequality formed from the noisy data alone---the method provides an entire set of emulators with error bounds and stability guarantees. The concept of representing infinitely many possible emulators is closely related to the \ac{irbm} interpretation of \acp{pmm}. However, in Ref.~\cite{Burohman2023} this arises strictly from some assumed noise in the snapshots as opposed to inherent symmetries in the model or representation of the data. Additionally, as a fully data-driven method, it does not preserve the structure of any high-fidelity model---this is stated as a possible direction for future research of the method.

In Ref.~\cite{Padovan2024}, the authors describe a Petrov-Galerkin-based emulation method by which the projectors and the reduced-space operators are learned via the gradient-based minimization of a loss function. The form of the reduced-space model is assumed to be polynomial, allowing for closed-form gradients of the loss function. Learning the reduced-space objects is also a central aspect of the \ac{pmm} method. However, the \ac{pmm} method does not assume a particular reduced-space form, instead deriving the form from the known or supposed form of the high-fidelity model. This maintains the connections between the emulator and any known properties of the high-fidelity model. Additionally, the \ac{pmm} method does not learn the projectors themselves, instead only learning the reduced-space objects which are implicitly related to the projectors by any imposed constraints on the reduced objects. Especially for problems where the emulated quantities are not full snapshots of the high-fidelity model, this is significantly more efficient and practical than learning the projectors themselves, which scale with the size of the full space.

The \ac{pmm} method is the first to describe a universal framework for emulation that is fully numerically parametric while also preserving all known structure of the high-fidelity model. The method is also the first \textit{adaptive} emulation method as discussed in Section~\ref{sec:pmm_intrusiveness}.

\section{Variations}

The typical \ac{pmm} application, as described in Section~\ref{sec:pmm_irbm}, is one where scalar values resulting from a process with a known form is emulated. However, \acp{pmm} which produce vector outputs like the associated \ac{erbm} as well as fully and partially data-driven \acp{pmm} are possible and useful. These variations are not mutually exclusive; a vector-valued \ac{pmm} can also be partially data-driven. There are more variations than discussed in this section.

\subsection{Vector-Valued Parametric Matrix Models}\label{sec:vec_pmm}

In order for a \ac{pmm} to produce an output vector in the full space, there is little choice but to make use of the \ac{erbm} projector\footnote{Or have the dimensionality of the \ac{pmm} be the same as the full space, defeating the point of emulation in nearly all cases.} to take predictions from the reduced space back to the full space. Recall from the statement of Galerkin's Method in Eq.~\eqref{eq:galerkinparametric} that the reduced (linear) problem $\ul{A(\mu)}\ul{x(\mu)}=\ul{b(\mu)}$ is solved independently of the full space and only afterwards projected back to the full space via $x(\mu)\approx \Pup \ul{x(\mu)}$. For a vector-valued \ac{pmm}, the operators and various objects in the reduced space can still be learned directly from data as opposed to constructed like in the \ac{erbm} approach, and this learning can happen entirely in the reduced space. But some of the expressiveness granted by the implicit reduced basis approach is lost. Still only the form of the full model is necessary, which does grant these \acp{pmm} more flexibility compared to their associated \acp{erbm}. The exact modifications necessary to the steps outlined in Section~\ref{sec:pmm_irbm} are as follows. Two additional steps between steps~\ref{itm:pmm1}~and~\ref{itm:pmm2} are added which state,
\begin{enumerate}[label=+1.\arabic*., ref=1.\arabic*]
    \item\label{itm:vpmm1.1} Compute the chosen projectors $\Pup$ and $\Pdown$ (almost always the \ac{pod} projectors $\Pup=P$ and $\Pdown=\Pdag$) for the chosen reduced-space dimension $n$ using the available training snapshots.
    \item\label{itm:vpmm1.2} Project the training snapshots to the reduced space using the computed projector $\Pdown$.
\end{enumerate}
Then the loss function in step~\ref{itm:pmm4} is modified to encode the difference between the projected training snapshots from step~\ref{itm:vpmm1.2} and the \ac{pmm} predictions still in the reduced space ($\ul{x(\mu)}$ following the linear example discussed before). Finally, after step~\ref{itm:pmm6} we make it clear that the \ac{pmm} predictions must be projected back to the full space,
\begin{enumerate}[label=+6.\arabic*.,ref=6.\arabic*]
    \item\label{itm:vpmm:6.1} Denoting the predictions of the \ac{pmm} in the reduced space as $\ul{x}$, the predictions in the full space are $\Pup \ul{x}$.
\end{enumerate}

\subsection{Fully and Partially Data-Driven Parametric Matrix Models}\label{sec:var_data_driven}

There are many applications where even just the form of the full space model is not wholly known. On one end of the spectrum, problems like purely data-driven regression have no information about the equations of the underlying system---if one even exists. Instead, machine learning methods for these problems must assume suitably expressive---or \textit{universal}---forms that can be tuned to fit almost any data. This is what techniques like least-squared regression or multi-layer perceptrons do by supposing a model of linear combinations of the data in the former and a feedforward \ac{ann} in the latter. \acp{pmm} for general non-sequential data-driven regression and classification can take many forms, but one proven option is that which is detailed in Ref.~\cite{Cook25} where the process from inputs to outputs is described by a set of observables in a quantum system whose Hamiltonian is parameterized by the inputs. This is discussed in detail in Section~\ref{sec:pmm_reg}. Near the opposite end of the spectrum, there are countless problems where most of the equations are known but a single or few important terms are not. In this case, missing terms can be replaced by data-driven machine learning methods while the known terms remain intact. This technique of a partially data-driven \ac{pmm} is discussed in Section~\ref{sec:pmm_pdd}.

\section{Properties}

For now we will put aside the task of training and consider the properties of trained \acp{pmm}. By construction, a \ac{pmm} inherits almost all of the properties of the associated \ac{erbm}. This is why the choice of an implicit orthogonal projection is so common; as discussed in Section~\ref{sec:pod} orthogonal projections preserve many properties and operations. However, a key advantage of \acp{pmm} is that they can strategically depart from the associated \ac{erbm} construction to avoid downsides or deficiencies. We have already seen this to some extent, but will see more when discussing nonlinear \acp{pmm} in this section. The central takeaway is that \acp{pmm} can be constructed to inherit all of the advantages of the associated \ac{erbm} with little or none of the downsides.

To avoid unnecessary repetition, only properties of \acp{pmm} which differ or warrant clarification compared to the associated \ac{erbm} are included in this section.

\subsection{Parametricity}
\acp{pmm} inherit mathematical parametricity from the associated \ac{erbm}. Just as discussed in Section~\ref{sec:erbm}, the parametric dependence on any control parameters can be arbitrarily nonlinear and complicated without affecting the complexity of the emulator. On the other hand, \acp{pmm} are numerically parametric by design. No method in Chapter~\ref{ch:emu}, besides the \ac{pinn}, was numerically parametric. In emulation and broader machine learning, numerical parametricity is a highly desirable property both for computational efficiency and model flexibility.

The dimensionality of a \ac{pmm} is a completely free hyperparameter independent of the data, and the introduction of additional data consistent with the original does not necessitate re-training the parameters of the \ac{pmm}---similar to \acp{ann}. For a general order-$d$ linear operator\footnote{For example, a matrix is an order-$2$ linear operator.} acting on elements of an $N$-dimensional vector space, the number of parameters for the \ac{pmm} representation of the operator is $n^{d}$ where $n$ is the chosen dimensionality of the \ac{pmm}. This is the same for \acp{erbm}, but now the chosen $n$ and emulator parameters are independent of the data or snapshots.
\subsection{Linearity}\label{sec:pmm_linearity}
Following the treatment of nonlinear operations in Section~\ref{sec:pod}, \acp{pmm} can also efficiently handle polynomially nonlinear equations directly and arbitrary elementwise nonlinearities via lifting transformations.

Additionally, a more efficient approach with close ties to the original eigenvalue emulator interpretation of \acp{pmm} exists. Strictly speaking, this approach would be available to the \ac{pod}-Galerkin \ac{erbm} as well, but it does not appear in literature\footnote{To the best of my knowledge at the time of writing.}\footnote{I include it in this chapter rather than the previous since it was developed in the context of \acp{pmm} and has only been verified with \acp{pmm}.}. This alternative approach relies on the canonical matrix-valued version of the nonlinearity to transform the full-space problem before constructing the reduced basis form. Consider the problem of computing
\begin{equation}\label{eq:vec_nonlin}
    F(v) = y(v) \odot v
\end{equation}
in the full space. The only requirement of $y$ is that it is elementwise and analytic; it does not need to be elementwise smooth like in the \ac{deim}. This form of nonlinearity---a vector-valued nonlinear term multiplied elementwise with the vector itself---is ubiquitous throughout scientific computing. In nuclear and quantum physics, several high-profile nonlinear systems exhibit this form including the exchange operator in the Hartree--Fock method, the mean-field term in the energy functional in density functional theory, and the mean-field term in the Gross--Pitaevskii equation~\cite{Lietz2017,Flynn2025,Foot2005}. The first step is to replace the elementwise product in Eq.~\eqref{eq:vec_nonlin} with a different nonlinearity, the $\diag{\cdot}$ operator,
\begin{equation}\label{eq:diag_vec_nonlin}
    F(v) = y(v) \odot v = \diag{y(v)}v,
\end{equation}
where $\diag{x}$ is a diagonal matrix whose elements are the elements of $x$. The product between $\diag{y(v)}$ and $v$ is the standard inner product. We now introduce the canonical matrix-valued version of $y$. For any elementwise and analytic $y$, we define $Y$ to be the matrix-valued version as
\begin{equation}
    \label{eq:canon_mat_f}
    Y(A) \equiv V_A \diag{y\left(\lambda_A\right)} V_A^{-1} \quad \text{for} \quad A = V_A \diag{\lambda_A} V_A^{-1},
\end{equation}
where $\lambda_A$ is the vector of eigenvalues of $A$. $Y(A)$ is only defined in this way for diagonalizable $A$ with eigenvalues that $y$ is analytic over. Note that by definition $Y(A)$ commutes with similarity transformations of $A$---that is, $Y\left(QAQ^{-1}\right) = QY(A)Q^{-1}$. This is not a new concept; the matrix exponential, logarithm, power, inverse, trigonometric functions, and many more are often expressed and implemented this way. Making explicit use of this allows us to factor the nonlinear function $y$ out of the $\diag{\cdot}$ operation in Eq.~\eqref{eq:diag_vec_nonlin},
\begin{equation}
    \label{eq:mat_nonlin}
    F(v) = \diag{y(v)}v = Y\left(\diag{v}\right)v.
\end{equation}
While these transformations may initially seem to have made the problem more complicated, we can now derive the form of the reduced model assuming an orthogonal projector, $P$. We follow the naive handling\footnote{Since the nonlinear term is operator-valued (matrix-valued) here instead of vector-valued as in Section~\ref{sec:pod}, we have additional right multiplications on both the argument and the result: $Y(X) \rightarrow \Pdag Y(P \ul{X} \Pdag) P$.} of the nonlinearity from Section~\ref{sec:pod} and project any remaining objects down to the reduced space. Denote the reduced-space representation of $\diag{v}$ by $\ul{D_v}$. Then,
\begin{equation}
    \label{eq:F_nonlin_reduced}
    \begin{aligned}
        Y\left(\diag{v}\right) \rightarrow \Pdag Y\left(P \ul{D_v} \Pdag \right) P &\approx \Pdag P Y\left(\ul{D_v}\right) \Pdag P\\
        &= Y\left(\ul{D_v}\right),
    \end{aligned}
\end{equation}
where the approximation is valid from the fact that $Y$ commutes with similarity transforms, which translates to an approximate commutation with the compression. We find $\ul{D_v}$ similarly\footnote{Here, we only have an additional right multiplication by $P$ on the result of the nonlinear operation $\diag{\cdot}$ since its argument is a vector.},
\begin{equation}
    \label{eq:diag_reduced}
    \begin{aligned}
        \diag{v} \rightarrow \ul{D_v} &= \Pdag \diag{P \ul{v}} P\\
        \iff \ul{D_v}_{ij} &= \underbrace{\Pdag_{i\nu}P_{\nu\mu}P_{\nu j}}_{\ul{G}^{(2)}}\ul{v}_\mu\\
        \iff \ul{D_v}&=\ul{G}^{(2)} \ul{v},
    \end{aligned}
\end{equation}
where $\ul{G}^{(2)}$ is the same order-$3$ tensor of the same name from Section~\ref{sec:pod} which enabled nonlinearities of the form $a \odot b$---the difference here is that only one index is contracted with the reduced vector $\ul{v}$ instead of two. Combining Eqs.~\eqref{eq:F_nonlin_reduced}~and~\eqref{eq:diag_reduced} to form the complete reduced nonlinearity which multiplies $\ul{v}$ we have
\begin{equation}
    F(v) = y(v)\odot v \rightarrow Y\left(\ul{G}^{(2)}\ul{v}\right)\ul{v},
\end{equation}
which only requires operations in the reduced space. Note that despite the fact that we have made no assumptions on the content of $y$ other than that it is elementwise and analytic over all relevant $v$, the largest tensor is the order-$3$ tensor $\ul{G}^{(2)}$. Previously in Section~\ref{sec:pod} we saw that $\ul{G}^{(2)}$ was the reduced form of the elementwise product nonlinear operation and higher-order tensors were required for higher powers. Now we see that we can instead use $y(x) = x^p$ for any $p\in\Cbb$ and treat any exponential nonlinearity with only this single order-$3$ tensor. Moreover, nonlinear operations that may have required numerous lifting transformations can instead be treated with the same computational overhead as the simplest polynomial nonlinearity. Importantly, the nonlinear operation itself has been preserved. This method requires no linearization, additional data, or hyper-reduction and converges to the exact nonlinearity as the dimension of the reduced space approaches the dimension of the full space.

\subsection{Computational Efficiency}

The complexity of making a prediction with a \ac{pmm} is the same as the cost of the associated \ac{erbm}. Previously in Section~\ref{sec:pod} we saw that predictions from \ac{pod}-Galerkin emulators scaled polynomially in the size of the emulator so long as all nonlinearities were handled efficiently.

Prediction complexity only reflects the online costs of the model. Since \acp{pmm} do not necessarily construct the projectors, the only offline costs are those of fitting the unknown operators. Just as for the method of \acp{pinn} in Section~\ref{sec:pinn}, if we assume an iterative fitting scheme then the offline time complexity of training the model is merely the online time complexity times the number of training examples, and the space complexity is unchanged from the online phase. The desired accuracy sets the prefactor of this scaling. Therefore both the online and offline complexities are independent of the full space dimension and scale polynomially with the reduced-space dimension.

As will be seen later this chapter, \acp{pmm} that target vector-valued outputs the size of the full space will make use of the associated \ac{erbm} projector and thus for \ac{pod}-Galerkin-based \acp{pmm} will require an additional $\bigO{N^2 n}$ offline and $\bigO{Nn}$ online time complexity---where $N$ is the dimension of the full space and $n$ is the dimension of the reduced space. Importantly, many loss functions used during training can be computed directly in the reduced space. This bypasses the additional $\bigO{Nn}$ prediction cost during training.

In the case of purely data-driven \acp{pmm}, the forms presented in Ref.~\cite{Cook25} are shown to have online costs which scale polynomially in $n$.

\subsection{Data Efficiency}

The convergence rate of a \ac{pmm} is highly dependent on the underpinning \ac{erbm}. By construction, the resulting \ac{pmm}---if trained properly---will perform at least as well as the associated \ac{erbm}. For general \ac{pod}-Galerkin-based \acp{pmm}, following the discussion surrounding Eqs.~\eqref{eq:podexperr}~and~\eqref{eq:podexprelerr}, approximate error bounds depend on the singular values of the matrix of training vectors. As emphasized before, there may not be any training vectors when using a \ac{pmm} and therefore seemingly no concept of singular values of the training data. However, just as the projectors are guaranteed to exist even when they are not obtainable, all scalar ``snapshots'' have some unknown (possibly unknowable) associated vector which together constitute this hidden matrix of snapshot vectors. The associated singular values, while unknown, come from this data. If this data is typical real-world data, then just as discussed previously around Eq.~\eqref{eq:typicalsingular} the singular values will approximately decay either polynomially or exponentially. This allows us to say that the approximate upper bound on the relative error for scalar values is either $\alpha^{-n}$ or $n^{-\beta}$ for a \ac{pmm} of dimension $n$ and some positive constants $\alpha$ and $\beta$. In either case, this shows that the typical upper bound on the error is $\bigO{n^{-\beta}}$. At first glance, this appears to be independent of the number of ``snapshots'' and to depend only on the size of the \ac{pmm}. This is due to the simplification we made in ignoring the training process. As expected by intuition, larger \acp{pmm} (larger values of $n$) require more data to accurately fit the unknown parameters. A simple upper bound for the error in terms of the number of training examples is found by taking the minimum of the \ac{pmm} dimension $n$ and the number of training examples $m$, since the assumed underlying \ac{erbm} for $n>m$ must have a rank-deficient projector. Thus we can state the final approximate error scaling for \ac{pod}-Galerkin-based \acp{pmm} to be $\bigO{\mathrm{min}\left(m,n\right)^{-\beta}}$ for some problem-dependent $\beta>0$.

\subsection{Hyperparameter Efficiency}

The only hyperparameter inherent in the \ac{pmm} method is the size of the reduced space $n$. Purely data-driven \ac{pmm} forms in Ref.~\cite{Cook25} introduce just one additional hyperparameter with a very restricted set of possible values. Improvements on these data-driven forms, discussed in Section~\ref{sec:pmm_reg} introduce one more hyperparameter.

Partially data-driven \acp{pmm} can have many more hyperparameters due to the freedom in choosing universal functions. Whether or not the hyperparameters introduced are few and interpretable is problem- and architecture-specific.

\subsection{Intrusiveness}\label{sec:pmm_intrusiveness}

As alluded to in Chapter~\ref{ch:emu}, \acp{pmm} do not fit cleanly into one of the three existing categories of emulator intrusiveness. Without leaving the \ac{pmm} framework, intrusive, data-driven, and hybrid emulators can be created. If the training process leverages the same information and techniques as the underlying \ac{erbm}, then the \ac{pmm} is functionally identical to the maximally intrusive \ac{erbm}. Conversely, if absolutely no information of the full system is known or used and instead a universal \ac{pmm} form is substituted, then the resulting emulator will be a fully data-driven \ac{pmm}. The spectrum of hybrid \ac{pmm} emulators follows from relaxing the training process and potentially substituting parts of the \ac{pmm} emulator for data-driven components.

This qualifies \acp{pmm} as hybrid emulators. What sets \acp{pmm} apart from existing hybrid emulators is the idea that the ``hybrid-ness'' of the \ac{pmm} method is not fixed and instead the extent to which a particular \ac{pmm} is intrusive or data-driven is determined by the context of the problem and available data. Existing hybrid emulation techniques, like \acp{pinn}, have limits to how intrusive or data-driven they can be. For example, a \ac{pinn} can be purely data-driven---in which case it is equivalent to a feedforward \ac{ann} regression model. But a \ac{pinn} cannot be fully intrusive, as by design it will always represent the reduced-space system as an \ac{ann}.

It is for this reason we introduce a fourth class of emulators called \textit{adaptive} emulators, defined previously in Section~\ref{sec:intrusiveness}. Emulators in this class adapt to the types of data available and are able to use all relevant data to inform the reduced-space model. Adaptive emulation methods represent large frameworks that are always applicable\footnote{Note that this is not the same as saying they are always effective or the best method.} regardless of the specific problem statement.

\subsection{Quantifiable Uncertainties}

The useful error bounds and estimates of the underlying \ac{erbm} are still applicable for \acp{pmm}. However, one important distinction is that---without more careful parameterization to explicitly restrict the possible values for the reduced operators---the condition number of the reduced operators are not necessarily smaller than their full-space counterparts. This is a consequence of choosing a projector with only some of the properties of an orthogonal projector and a central example of the freedom and importance in choosing a ``good'' parameterization for the reduced objects.

In theory, uncertainty estimates can be obtained directly from the Hessian of the loss function~\cite{Pumplin2001}. This arises through a quadratic expansion of the loss around the solution to the minimization problem. However, this is almost universally prohibitively expensive due to the potentially large number of fit parameters.

Compared to traditional hybrid emulation methods that utilize \acp{ann}, \acp{pmm} are particularly amenable to uncertainty quantification via conformal prediction---covered in Section~\ref{sec:conformal}.

\section{Training}\label{sec:training}

The process of determining the values of the numerically parameterized objects in a \ac{pmm}, or any machine learning model, in order to minimize some loss function---step~\ref{itm:pmm5} in Section~\ref{sec:pmm_irbm}---is called \textit{optimizing}\footnote{``Optimizing'' in this context is distinct from and should not be confused with the optimization of the architecture itself or the implementation (code).}, \textit{fitting}, or \textit{training}. Since there are as many different loss functions and \acp{pmm} as there are \acp{rbm} and as many \acp{rbm} as possible problems and data, we seek an optimization method that is highly versatile and minimally intrusive\footnote{Not to be confused with emulator intrusiveness.}. To accommodate diverse \acp{pmm} from just a few trainable parameters through to millions of trainable parameters\footnote{\acp{pmm} with trainable parameter counts beyond several million would also be trainable by these methods, but no work to date has done this.} and datasets of equally variable size, we adopt the deep learning approach of gradient-based optimization, usually called \textit{gradient descent}.

\begin{figure}
    \centering
    \includegraphics[width=0.6\linewidth]{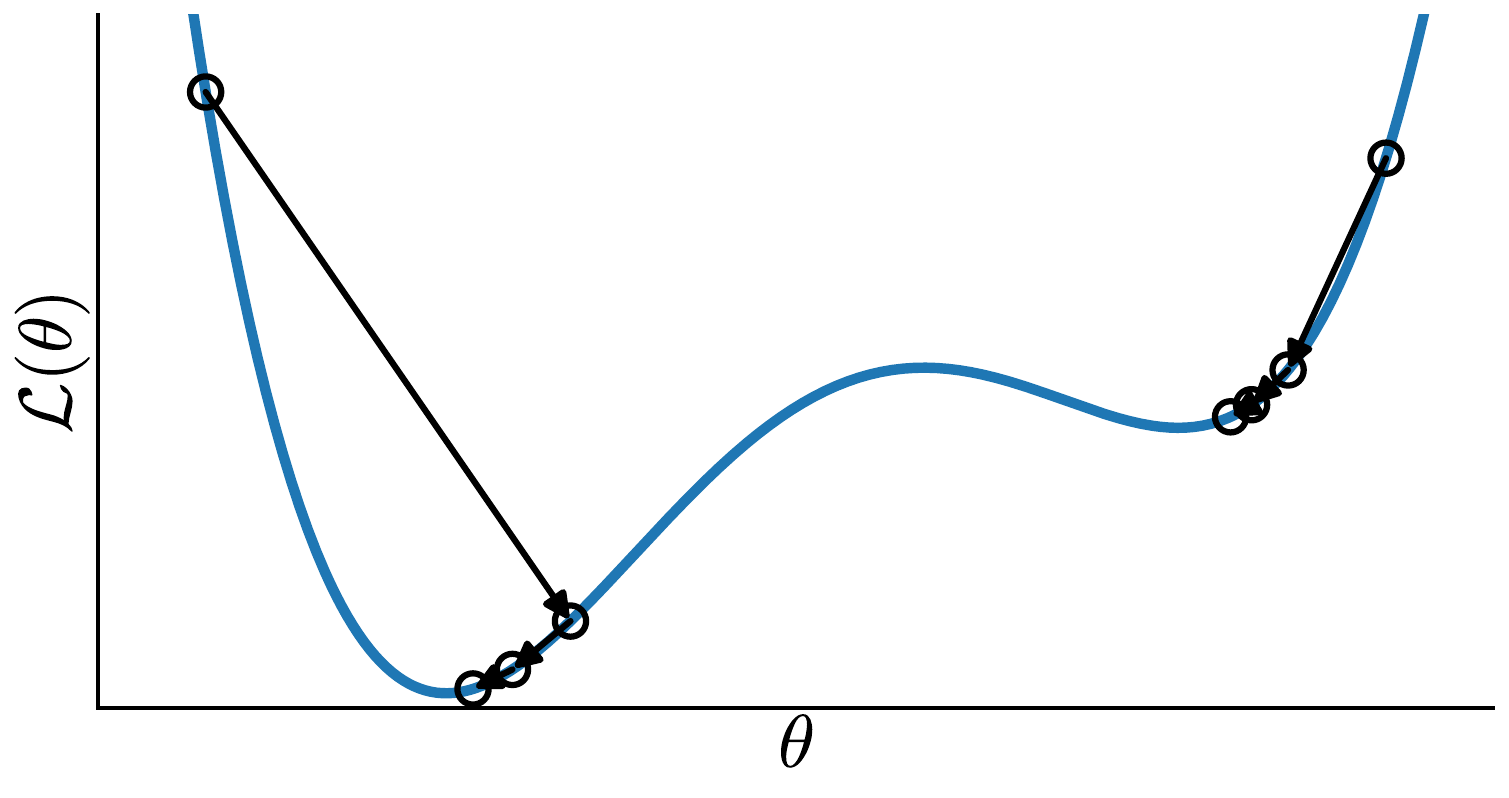}
    \caption[Illustrative example of the steepest descent method.]{Illustrative example of three steps of the steepest descent method in 1D. Two initial points are shown. The initial point on the left converges to the global minimum while the initial point on the right converges to a local minimum.}
    \label{fig:steepestdescent}
\end{figure}

Consider the loss function for a particular model as a function of the trainable parameters, $\Loss(\theta)$. The loss function must be bounded from below---it must have a minimum value. Gradient descent methods are iterative methods that update the parameters based on the gradient of the loss function, $\frac{\partial\Loss}{\partial\theta}$, and potentially higher-order derivatives. The loss function is a real-valued function of which we want to find the location of the minimum. The simplest gradient descent method is the method of steepest descent\footnote{This method was originally just called ``gradient descent,'' but this term now refers to a family of methods. The original gradient descent method is sometimes called ``vanilla gradient descent'' or ``method of steepest descent'' as used here.}, which proposes the iterative update
\begin{equation}\label{eq:gd}
    \theta \leftarrow \theta - \eta \left.\frac{\partial \Loss}{\partial \theta}\right|_{\theta}
\end{equation}
where $\eta>0$ is the \textit{learning rate}, the step size for each iteration. Figure~\ref{fig:steepestdescent} shows an illustrative example for this update in 1D. The figure additionally illustrates a fundamental challenge of gradient descent methods: local minima and saddle points. At these critical points, the gradient of the loss function is zero $(\frac{\partial\Loss}{\partial\theta}=0)$ but the function has yet to achieve its global minimum. When optimizing any model, we are interested in the global minimum---the best possible model given the model architecture and loss function. Higher-order derivatives of the loss function can help overcome this challenge, but are significantly more expensive and quickly become computationally impractical. It is for this reason that first-order methods---those which only use the first derivative of the loss function---will often include momentum-like terms or adaptive learning rates to encourage the iterations to pass through or by any critical points that are not the global minimum.

A local minimum with a loss value close to the global minimum is also an acceptable result from the training process, since it would yield a model with nearly identical performance. Luckily, as the number of trainable parameters increases, the likelihood that any critical point is a ``bad'' local minimum---that is, one with a loss value significantly larger than the global minimum value---is vanishingly small~\cite{Dauphin2014}. Instead, almost all critical points are either saddle points or nearly equal to the global minimum. A sketch of the proof for this starts with the identification that, at a critical point, any one direction in parameter space has a non-zero, non-unit, probability $0<p<1$ to have positive curvature (and therefore would be a local minimum along this direction only). This probability is around $0.5$ initially and approaches $1$ as the loss function nears its global minimum value---independent of the location $\theta$. Away from the global minimum value where $p$ is not yet $1$, the probability that a particular critical point is a local minimum is $p^N$ with $N$ being the number of trainable parameters---which approaches $0$ as $N$ increases. Near the global minimum value however, $p\approx 1$, and nearly every critical point is a local minimum. This is acceptable however, since their values are very near the global minimum value.

To support \acp{pmm} with complex parameters, we treat the real and imaginary components of each parameter as independent real variables. Let the vector of complex parameters be decomposed as $\theta = a + ib$ with $a$ and $b$ purely real vectors. We can then update $a$ and $b$ independently using Eq.~\eqref{eq:gd},
\begin{equation}\label{eq:gdii}
    \begin{aligned}
        a &\leftarrow a - \eta \left.\frac{\partial \Loss}{\partial a}\right|_{a},\\
        b &\leftarrow b - \eta \left.\frac{\partial \Loss}{\partial b}\right|_{b}.
    \end{aligned}
\end{equation}
We now switch to using Wirtinger derivatives for notational compactness, but this will also be practically impactful shortly. In Wirtinger calculus, a complex variable $z$ and its complex conjugate $z^\ast$ are treated as independent variables. Suppose $z=x+iy$ with $x$ and $y$ real numbers, then the Wirtinger derivatives are defined as~\cite{Wirtinger1927}
\begin{equation}\label{eq:wirtinger}
    \begin{aligned}
        \frac{\partial}{\partial z} &\equiv \frac{1}{2}\left(\frac{\partial}{\partial x} - i \frac{\partial}{\partial y}\right),\\
        \frac{\partial}{\partial z^\ast} &\equiv \frac{1}{2}\left(\frac{\partial}{\partial x} + i \frac{\partial}{\partial y}\right).
    \end{aligned}
\end{equation}
This readily generalizes to vector derivatives with complex vectors. Using these operators, we can simplify Eq.~\eqref{eq:gdii} to
\begin{equation}\label{eq:gdi}
    \theta \leftarrow \theta - \eta\left.\frac{\partial \Loss}{\partial \theta^\ast}\right|_{\theta},
\end{equation}
where the factor of $1/2$ in the definition of the Wirtinger derivative has been absorbed into the learning rate $\eta$~\cite{Boeddeker2019}. Note that the derivative is now with respect to the conjugate of the trainable parameters, $\theta^\ast$. Since the loss function is a purely real-valued function\footnote{Since complex numbers cannot be ordered, the concept of ``minimizing'' a complex-valued loss function is undefined.}, we have
\begin{equation}
    \left(\frac{\partial \Loss}{\partial \theta}\right)^{\!\ast} = \frac{\partial \Loss^\ast}{\partial \theta^\ast}=\frac{\partial \Loss}{\partial \theta^\ast}.
\end{equation}
And thus the complex-valued update can be equivalently written as
\begin{equation}
    \theta \leftarrow \theta - \eta\left(\left.\frac{\partial \Loss}{\partial \theta}\right|_{\theta}\right)^{\!\ast},
\end{equation}
where the derivative is the Wirtinger derivative~\cite{Hirose2013}. Beyond just a notational convenience, major libraries including \pypackage{JAX}, \pypackage{PyTorch}, and \pypackage{TensorFlow} all implement automatic differentiation of real functions\footnote{And complex-valued holomorphic functions, but the details of the implementations here are only in the context of real-valued functions.} of complex variables via Wirtinger derivatives~\cite{JAX,PyTorch,TensorFlow}\footnote{It is worth noting some inconsistencies between the three libraries mentioned here at the time of writing: \pypackage{JAX} omits the factor of $1/2$ from the definitions of the derivatives in Eq.~\eqref{eq:wirtinger} and the provided \texttt{grad} method implements $\partial/\partial z$, while the equivalent methods in \pypackage{PyTorch} and \pypackage{TensorFlow} include the $1/2$ factor but implement the conjugate gradient $\partial/\partial z^\ast$.}.

Significantly more advanced and successful gradient descent methods have been developed, and the training of machine learning models remains a vast and evolving topic. For an excellent introduction and overview of gradient descent techniques, see Refs.~\cite{Goodfellow2016,Zhang2023}\footnote{Specifically, Chapters 4 and 8 in Ref.~\cite{Goodfellow2016} and Chapter 12 in Ref.~\cite{Zhang2023}.}. For the training of \acp{pmm} we utilize a modification of the Adam\footnote{From ``\underline{\textbf{ada}}ptive \underline{\textbf{m}}oments.''} method of mini-batch stochastic gradient descent~\cite{Kingma2017}---modified to support complex-valued parameters. Adam uses two auxiliary vectors, $m$ and $v$, of the same shape as the vector of trainable parameters to track a running average of the gradients and second moment of the gradients respectively for each trainable parameter. These running averages additionally ``forget'' the history exponentially---with rates set by two independent hyperparameters, $\beta_1$ and $\beta_2$. This allows Adam to adjust the learning rate independently for each parameter based on the trajectory of the training process. ``Mini-batch stochastic'' gradient descent refers to using a subset of the training data (a mini-batch) in the loss function at each update\footnote{Pure stochastic gradient descent uses just a single training example per update.}. In practice, training is divided into \textit{epochs}. During each epoch, the whole of the training data is randomly shuffled and divided into these mini-batches, and the model parameters are updated using the update rule for each mini-batch sequentially until the model has seen all of the training samples. Denoting the epoch number by $t$ (starting at epoch $1$) and taking the initial values of $m$ and $v$ to be the zero vector, the update rule for Adam with complex parameters is
\begin{equation}
    \begin{alignedat}{3}
        g&\equiv\left(\left.\frac{\partial \Loss}{\partial \theta}\right|_{\theta}\right)^\ast&\quad&\text{(conjugate of the gradient of the loss)}\\
        m &\leftarrow \beta_1 m + \left(1 - \beta_1\right) g&\quad&\text{(update gradient running average)}\\
        v &\leftarrow \beta_2 v + \left(1 - \beta_2\right) \abs{g}^2&\quad&\text{(update second moment running average)}\\
        \hat{m} &\equiv \frac{m}{1-\beta_1^t}&\quad&\text{(bias correction)}\\
        \hat{v} &\equiv \frac{v}{1-\beta_2^t}&\quad&\text{(bias correction)}\\
        \theta &\leftarrow \theta - \eta \frac{\hat{m}}{\sqrt{\hat{v}} + \varepsilon}&\quad&\text{(update parameters)}
    \end{alignedat}
\end{equation}
where $\abs{g}^2$ denotes the elementwise absolute value squared ($\left(\abs{g}^2\right)_i \equiv \abs{g_i}^2 = g_i^\ast g_i$), the two bias correction steps account for the bias introduced by initializing the running averages to zero, $\eta$ is the learning rate, and $0<\varepsilon\ll 1$ is a small positive value to prevent division by zero. Despite the introduction of three new hyperparameters ($\beta_1$, $\beta_2$, and $\varepsilon$), it is rarely the case that these need to be changed from their commonly\footnote{These values are common default values across many different libraries and publications, not just for \acp{pmm}. The default value of $\varepsilon$ is slightly less consistent across literature, but present only for numerical stability and otherwise does not significantly change the behavior of the algorithm.} used values of $\beta_1 = 0.9$, $\beta_2=0.999$, $\varepsilon=10^{-8}$. This leaves only the learning rate, $\eta$, as the important hyperparameter for the training algorithm. Common values for the learning rate range from $10^{-2}$ to $10^{-4}$, but values outside this range are not unreasonable. Additionally, the learning rate may be epoch-dependent, $\eta(t)$, which is known as \textit{learning rate scheduling} and is outside the scope of this work~\cite{Zhang2023}.

\subsection{Automatic Differentiation}\label{sec:ad}

The update rules in gradient descent algorithms require the derivative of the loss function with respect to the trainable parameters of the model, which by the chain rule requires the gradient of the model itself. Luckily, instead of requiring that each implementation of a model comes with an implementation of its gradient, we can use \ac[format/long=\itshape]{ad} to automatically create this gradient implementation from the model's implementation. Another vast topic far beyond the scope of this work, \ac{ad} has been a central tool in machine learning for several decades~\cite{Rumelhart1986}. Here, we cover only a sketch of the core principle of \ac{ad}. See Ref.~\cite{Baydin2018} for a review of \ac{ad} and Ref.~\cite{Zhang2023} for a practical introduction to \ac{ad}.

\Ac{ad} encapsulates a large family of techniques and is distinct from both symbolic differentiation and numerical differentiation. In contrast, \ac{ad} is significantly more computationally efficient than symbolic differentiation and more accurate than numerical differentiation---in fact, it is exact to numerical precision. Perhaps most importantly for the use in gradient descent, it is computationally inexpensive even for functions with a large number of independent variables. In the broadest possible terms, \ac{ad} exploits the chain rule of differentiation and the fact that all operations can be rewritten as function compositions. Regardless of the complexity of any function, numerically it must be written as a sequence of basic operations. So long as the derivatives of these basic operations are known, then the gradient of the entire function can be found via the chain rule. As an illustrative example, consider the function
\begin{equation}
    f(x,y) = x\sin(y).
\end{equation}
This is actually a composition of the multiplication function, $m(u,v) = uv$, with the sine function,
\begin{equation}
    f(x,y) = m(x, \sin(y)).
\end{equation}
Now that $f$ is written purely as compositions of basic operations with known derivatives, we can apply the chain rule during the evaluation,
\begin{equation}
    \begin{alignedat}{3}
        u &= x&\qquad \partial u &= \partial x\\
        v &= \sin(y)&\qquad \partial v &= \cos(y)\partial y\\
        m(u,v)&=uv&\qquad \partial m(u,v) &= u\partial v + (\partial u) v\\
        \implies f(x,y)&=x\sin(y)&\qquad \partial f(x,y) &= x\cos(y)\partial y + (\partial x)\sin(y).
    \end{alignedat}
\end{equation}
The same basic principle allows libraries like \pypackage{JAX}, \pypackage{PyTorch}, and \pypackage{TensorFlow} to automatically compute the gradients of loss functions for significantly more complex models than in this example~\cite{JAX,PyTorch,TensorFlow}.

\subsubsection{Vanishing and Exploding Gradients}

There are many difficulties in training deep and recurrent \acp{ann} and perhaps the most well-known are the problems of the vanishing or exploding gradient~\cite{Zhang2023,Bengio1994,Pascanu2013}. These occur due to the long chain of multiplications in the derivative of the loss function which arise from the chain rule. If the gradients at each step are routinely less than $1$, then the gradient approaches $0$ exponentially, yielding no or very slow training progress. If the gradients at each step are routinely greater than $1$, then the gradient diverges exponentially, yielding unstable training, overflows, and other numerical errors. These are problems which worsen with larger and deeper \acp{ann}. However, most \acp{pmm} do not involve significant levels of nested function evaluations---since most physical problems do not. Even operations that are implemented via iterative methods often have non-iterative gradient implementations or implementations that otherwise avoid long chain rule applications. For instance, the \ord{k} eigenvalue, $E_k$, of a Hermitian matrix, $H$, is computed iteratively---possibly with many iterations---which would lead to vanishing gradients if \ac{ad} was performed through the iterative algorithm itself. Instead, \ac{ad} treats the eigenvalue decomposition as a basic operation and implements its gradients directly. The Hellmann--Feynman theorem---first published in~\cite[Eq. (11)]{Guttinger1932}---provides the gradient as
\begin{equation}
    \frac{\partial E_k}{\partial z} = v_k^\dagger\left(\frac{\partial H}{\partial z}\right)v_k,
\end{equation}
where $z$ is some independent variable that $H$ (and therefore $E_k$) depends on and $v_k$ is the eigenvector of $H$ associated with the \ord{k} eigenvalue of $H$, $E_k$. Expressions for the gradients for the eigenvectors themselves also exist~\cite{Boeddeker2019}. These kinds of gradient implementations, which are already present in major \ac{ad} libraries~\cite{JAX,TensorFlow,PyTorch}, and the relatively shallow ``depth'' of most \acp{pmm} solve the problem of vanishing and exploding gradients when training \acp{pmm} by avoiding the long chain of gradient multiplications.

\subsection{Parameter Initialization}

In the field of deep learning, significant effort is spent on initializing the trainable parameters of models in such a way that the subsequent iterative training algorithm performs as well as possible~\cite{Goodfellow2016}. There are many different heuristics in this regard with varying levels of success and formalization. At the moment, none of this has been brought to \acp{pmm}. It is unclear what knowledge from \ac{ann} parameter initialization is applicable to \acp{pmm} in general. However, two trivial heuristics have proven successful.
\begin{itemize}
    \item The magnitudes of the initial values for trainable parameters should be small. For a complex parameter $z=a+ib$, typical initialization scales for $a$ and $b$ are $10^{-1}$ or $10^{-2}$.
    \item The parameter initialization must break any model symmetries. That is, if two independent parameters contribute equally to the model output, they must be initialized with distinct values. Otherwise, the training process will update these parameters identically and not break the symmetry automatically.
\end{itemize}
Currently, \ac{pmm} initialization is usually done by drawing the real and imaginary components of all trainable parameters from independent and identical distributions---either a zero-mean normal distribution or a uniform distribution centered around zero.

\subsection{Data Preprocessing}\label{sec:preprocess}

To vastly improve both the numerical precision, stability, and convergence of the training algorithm, both the features (inputs) and targets (expected outputs) of the training data are often rescaled. This is common practice in general machine learning~\cite{Zhang2023,Geron2022}. However, when using \acp{pmm}---or any \ac{rbm}---to emulate physical systems, extra care must be taken to ensure that the link between the original equations of the high-fidelity model and the reduced equations of the emulator is not broken. That link is of course the linear projection which is implicit for \acp{pmm} but otherwise must remain linear. In practice, this means that any preprocessing applied to projected objects (e.g.\ vectors and operators) is restricted to linear rescaling. Due to scalar invariance under the assumed orthogonal projection for \acp{pmm} as discussed previously, both input and output scalars can be transformed in any consistent way without the linear restriction.

Common affine rescaling transformations include \textit{standardization}, \textit{min-max scaling}, and \textit{robust scaling}. Standardization, or standard scaling, is a preprocessing method that shifts and scales the data so that each feature has mean $0$ and standard deviation $1$. For a set of training data $T$, the standard-scaled data would be~\cite{Hvitfeldt2024,Geron2022}
\begin{equation}
    T_\text{standard} \equiv \left\{\frac{t - \mean(T)}{\std(T)}~\text{for}~t\in T\right\}.
\end{equation}
Min-max scaling transforms the data so that the minimum and maximum of the rescaled data correspond to some chosen $a$ and $b$ with $a<b$,
\begin{equation}
    T_\text{min-max} \equiv \left\{a + \frac{t - \min(T)}{\max(T)-\min(T)}\left(b-a\right)~\text{for}~t\in T\right\}.
\end{equation}
Popular choices for $[a,b]$ are $[0,1]$ and $[-1, 1]$~\cite{Hvitfeldt2024,Geron2022}. Finally, robust scaling transforms the data such that the median of the transformed data is $0$ and the interquartile range (the distance between the \ord{75} and \ord{25} percentiles of the data) is $1$. This transformation is designed to be less sensitive to outliers.
\begin{equation}
    T_\text{robust} \equiv \left\{\frac{t - P_{50}(T)}{P_{75}(T) - P_{25}(T)}~\text{for}~t\in T\right\},
\end{equation}
where $P_{p}(T)$ denotes the \ord{p} percentile of $T$~\cite{Hvitfeldt2024}. While these three transformations represent some of the most common for continuous variables, there are in principle an unlimited number of different useful preprocessing methods depending on the specific data and context. See Ref.~\cite{Hvitfeldt2024} for a large compendium of different data preprocessing methods. In practice, the training of \acp{pmm} has performed best with min-max scaling on the range $[-1,1]$ for both inputs and outputs---but this is far from a rule and just like parameter initialization warrants further investigation.

When rescaling the targets (output values) in the training data, the model will be trained to produce predictions in this rescaled regime. This necessitates inverse scaling to produce predictions that correspond to the original unscaled data. The three transformations described here are trivially invertible. However, a vital point which segues nicely into the next section is that the parameters of the transformation---the values which arose from statistics on the training data such as $\mean(T)$, $\max(T)$, and $P_{50}(T)$---are always determined only by the original training data, even when rescaling predictions from the model for previously unseen data.

\subsection{Generalization, Calibration, and Evaluation}\label{sec:gen_cal_eval}

We have thus far only discussed the training set, $T$. However, if a model is trained on all of the available data and performs perfectly on it, then how can we know it will generalize well to new, unseen data? A model that performs perfectly on the training data but poorly on new data is called an \textit{overfit} model and is highly undesirable. Instead, we want to ensure that our emulators generalize well to unseen data. Similarly, how can we transform pointwise predictions to probabilistic ones, or ensure that already probabilistic predictions match the true probabilities? Finally, how can we accurately and honestly evaluate the performance of a trained model? An answer to all of these questions is data splitting---the process of dividing the whole of the available data into \underline{\textbf{disjoint}} subsets. In general, there are four subsets that data can be split into.
\begin{enumerate}
    \Needspace{5\baselineskip}
    \item Testing\footnote{Also referred to as ``test'', ``evaluation'', ``unseen'', or ``holdout'' data.} Data, $E$
        \begin{itemize}
            \item The subset of data which the model's performance is evaluated with to obtain unbiased performance metrics
            \item Ideally, is representative of data the model is expected to encounter when deployed
            \item Must be handled with the most care to ensure no information from any other subsets contaminates it or vice versa, and for that reason should be set aside first before any further splitting, preprocessing, training, or other work
        \end{itemize}
    \Needspace{5\baselineskip}
    \item Training Data, $T$
        \begin{itemize}
            \item The only subset that can be used to determine any constants involved in data preprocessing and to fit the model's trainable parameters
            \item Usually the largest subset
        \end{itemize}
    \Needspace{5\baselineskip}
    \item Validation Data, $V$
        \begin{itemize}
            \item The subset used to evaluate the performance of the model during training and to make decisions about the training process such as when to stop training, which hyperparameters to use, or even which model architecture to use
            \item Functions similarly to the test set, but only used for evaluating models during training and model selection and not for final performance metrics
            \item Should be representative of the test set, and therefore of the data the model will encounter when deployed
        \end{itemize}
    \Needspace{5\baselineskip}
    \item Calibration Data, $C$
        \begin{itemize}
            \item The subset used to calculate any stratified bias corrections or calibrate the model's probabilistic predictions, for example via conformal prediction~\cite{Vovk2005,Angelopoulos2023}
            \item Least common, often entirely absent when only pointwise predictions are required
            \item Should be statistically similar to the test set, and therefore to the data the model will encounter when deployed
        \end{itemize}
\end{enumerate}

We have already seen how the training set, $T$, is used. The validation set, $V$, is used as a proxy for the testing set, $E$, when designing and training an emulator. The relative performance on the validation set is used to select a specific architecture or hyperparameters as well as for stopping criteria during training. On the latter point, iterative training algorithms will track the loss on the validation set\footnote{It is not necessarily the case that the loss function here is the same as the one being minimized by the training process. This is particularly useful when the desired loss metric is not differentiable. In this case, a proxy loss function which is differentiable is minimized during training and the original non-differentiable loss function is tracked with the validation data. Many classification tasks follow this exact scheme.}, called the validation loss, and stop iterating and return the parameters with the lowest validation loss when the validation loss fails to improve for a set number of iterations. By evaluating the model on a dataset disjoint from the training data and ideally representative of the test data, this process drastically improves the generalization performance of the trained model. The validation set is also used for more manual failure-case exploration, which assists in debugging and model improvement. The calibration set is usually only present when the model's predictions are probabilistic, when a pointwise-predictive model is to be calibrated to produce probabilistic outputs, or for stratified bias correction. Sometimes in machine learning, the validation set is used for these tasks, but this borders on data leakage and can quickly lead to overfitting and overly optimistic model evaluations. More advanced schemes like cross-validation---which are popular and powerful in data-scarce contexts---will treat the training, validation, and calibration datasets as a single large dataset from which to train multiple models (usually the same model multiple times) with different choices for the specific training, validation, and calibration sets from this larger set~\cite{Zhang2023,Geron2022}. The best performing model in this ensemble of models can often generalize much better than using a fixed split. Alternatively, the models can be combined using averaging or another statistical aggregation to produce a single model with strong generalization performance.

A complementary technique for improving generalization performance is \textit{regularization}~\cite{Goodfellow2016}. The simplest and most common form of regularization involves constraining the trainable parameters of the model during training---either by hard algorithmic constraints like using the same parameter for multiple parts of the model or by penalty terms in the loss function like adding the $\ell^2$ norm of the vector of trainable parameters to the loss function. Currently, these techniques have not shown to improve the performance of any \acp{pmm}. When approaching \acp{pmm} as just machine learning models this may seem surprising, but when taking the \ac{irbm} approach this becomes almost obvious. A form of regularization for \acp{pmm} comes in the form of effective parameterizations that purposely conserve specific properties of the projected objects and operators, rather than of the vector of all trainable parameters. This is discussed further in Section~\ref{sec:parameterization}.

The test set is used only once and only after all data processing, model fitting, and model calibration steps are done and locked in. Adjusting the model or data pipeline in any way based on results from the test set constitutes data leakage and will result in overly optimistic performance metrics at best~\cite{Geron2022,Zhang2023}. This last point is worth restating: at no point should any amount of information from or about the test data play any role in the design, fitting, calibration, or selection of any model.

\section{Effective Parameterization}\label{sec:parameterization}
A vital consideration when building the reduced equations for a \ac{pmm} which goes beyond simply following the \ac{erbm} procedure is choosing how the reduced objects will be parameterized by the trainable parameters. For instance, consider a single square operator $A$ in a full space of dimension $N$ and its size-$n$ reduced-space representation, the $n\times n$ matrix $\ul{A}$. There are two extremes for the parameterization of $\ul{A}$, the first is to treat every element as a completely independent trainable complex value---yielding the maximum flexibility for $\ul{A}$ to adjust during the training process in order to accommodate the training data. The opposite extreme, which is not always available, is to determine all of the elements of $\ul{A}$ directly from the \ac{erbm} projection of $A$, that is $\ul{A} = \Pdown A \Pup$---leaving no trainable parameters whatsoever but conforming exactly to the \ac{erbm} prescription. Note that the first parameterization discards all properties of both $A$ and the chosen implicit projector while the second preserves all properties of $A$ that the corresponding \ac{erbm} would. Striking a balance between these two extremes is perhaps the only part of the \ac{pmm} method where model ``design'' exists beyond simple hyperparameter tuning. It is here we must take care to preserve---or add---any important properties of the high-fidelity model without over-constraining our parameterizations and sacrificing the flexibility of our emulator. Intimate knowledge and familiarity with the high-fidelity model is necessary here, as is knowing how to impose constraints in an exact and computationally efficient way.

Consider a trivial non-parametric example: the \ac{pmm} for the lowest lying eigenvalue of a Hermitian matrix, $H$. Suppose we have chosen to base the \ac{pmm} for this on the \ac{pod} \ac{erbm}. We start by enumerating some of the properties of $H$ that would be preserved under that projection: Hermiticity and definiteness as well as the consequences of the Poincar\'e separation theorem (see Section~\ref{sec:pod}). There are many more, but we restrict our effort to properties that would meaningfully change the predictive ability of the emulator. Consider the consequences of \textit{not} incorporating each property: Would an emulator for a Hermitian problem that has the ability to produce complex eigenvalues be useful or sensible? Often not, and for this example we will say no, so we should incorporate Hermiticity into our parameterization. Suppose we do not know the definiteness of the high-fidelity $H$ nor the value of any of the eigenvalues beyond the lowest. In that case we have no reason or ability to impose the other two listed constraints. With the properties---or in this example property---that we want to impose, we want to parameterize $\ul{H}$ such that they are exactly adhered to. One way to do this is simply labeling the $n(n-1)/2$ complex values of the strictly upper triangular portion of $\ul{H}$ along with the $n$ real values of the diagonal of $\ul{H}$ as the trainable parameters. Identifying the trainable parameter vector as $\theta=\mathlist{\theta_1, \theta_2, \ldots}$ with $\theta_i\in\Rbb$ for $i \leq n$ and $\theta_i\in\Cbb$ for $i>n$, we could then construct $\ul{H}$ from these parameters as
\begin{equation}
    \ul{H}\equiv \begin{bmatrix}
        \theta_1          & \theta_{n+1} & \theta_{2n}   & \theta_{3n-2} & \cdots \\
        \theta^\ast_{n+1} & \theta_2     & \theta_{n+2} & \theta_{2n+1} & \ddots\\
        \theta^\ast_{2n}   & \theta^\ast_{n+2} & \theta_3     & \theta_{n+3}   & \ddots\\
        \theta^\ast_{3n-2} & \theta^\ast_{2n+1} & \theta^\ast_{n+3} & \theta_4     & \ddots\\
        \vdots             & \ddots        & \ddots       & \ddots         & \ddots
    \end{bmatrix},
\end{equation}
which is guaranteed to be Hermitian for any values of the trainable parameters. This method requires the minimum number of trainable real values, but manually constructing $\ul{H}$ element-by-element is cumbersome\footnote{This method especially complicates debugging and portability.} and can be computationally inefficient. An alternative way that is less parameter efficient but significantly simpler and often more performant is to represent $\ul{H}$ as the result of some matrix-valued function of an unconstrained trainable complex matrix $\Theta\in\Cbb^{n\times n}$. This function should be a surjective function from the space of all complex $n\times n$ matrices to the space of all $n\times n$ Hermitian matrices---that is, it should be defined for any $n\times n$ complex matrix input and be able to produce as output any $n\times n$ Hermitian matrix. For effective training with gradient-descent-type methods, it must also be differentiable and well-behaved. This function is simply $h(X)\equiv \left(X+X^\dagger\right)/2$. Then for $\Theta\in\Cbb^{n\times n}$ being the trainable matrix, we construct $\ul{H}$ as
\begin{equation}
    \ul{H} \equiv h(\Theta) \equiv \frac{\Theta + \Theta^\dagger}{2}.
\end{equation}
Although this requires about twice as many trainable real values as the elementwise parameterization, the number of independent real values is exactly the same and thus the dimensionality of the parameter space which the training algorithm must explore is unchanged. Constructing the constrained reduced-space objects from closed-form expressions of unconstrained objects is often simpler, more understandable, and more performant. This example illustrates the core principles of effective parameterizations, but is by no means comprehensive. The following sections cover some of the most common constraints and ways to parameterize objects subject to them.

\subsection{Hermiticity and Symmetry}

As discussed in the example, an effective way to parameterize Hermitian or symmetric operators is to represent them by suitably shaped unconstrained trainable matrices and then ``Hermitianize'' or ``symmetrize'' them. For a Hermitian matrix $\ul{H}\in\Hbb(n)$, we parameterize it by the unconstrained complex matrix $\Theta\in\Cbb^{n \times n}$ as
\begin{equation}
    \ul{H} \equiv h(\Theta) \equiv  \frac{\Theta + \Theta^\dagger}{2},\qquad\Theta\in\Cbb^{n\times n}.
\end{equation}
Similarly for a symmetric matrix $\ul{S}$, we parameterize it by the unconstrained real matrix $\Theta\in\Rbb^{n\times n}$ as
\begin{equation}
    \ul{S} \equiv s(\Theta) \equiv \frac{\Theta + \Theta^\mathsf{T}}{2},\qquad\Theta\in\Rbb^{n\times n}.
\end{equation}
Note that both of these parameterizations are idempotent---that is, $h(h(\Theta))=h(\Theta)=\ul{H}$ and $s(s(\Theta))=s(\Theta)=\ul{S}$. Anti-Hermitian and antisymmetric matrices are treated similarly by simply subtracting the Hermitian conjugate or transpose instead of adding.

This concept generalizes to symmetries in arbitrary indices of higher-order tensors, and can be composed with additional Hermitianization or symmetrization expressions to handle multiple symmetries simultaneously. For example, consider an order-$4$ tensor $\ul{T}_{ijkl}$ which should be symmetric in the first two indices, symmetric in the last two indices, and Hermitian in $ij\leftrightarrow kl$. That is,
\begin{equation}
    \begin{alignedat}{3}
        \ul{T}_{ijkl} &= \ul{T}_{jikl} &\qquad&(\text{symmetric in}~i\leftrightarrow j),\\
        \ul{T}_{ijkl} &= \ul{T}_{ijlk} &\qquad&(\text{symmetric in}~k\leftrightarrow l),\\
        \ul{T}_{ijkl} &= \ul{T}^\ast_{klij} &\qquad&(\text{Hermitian in}~ij\leftrightarrow kl).
    \end{alignedat}
\end{equation}
We can enforce these symmetries by applying them sequentially to an unconstrained trainable order-$4$ tensor,
\begin{equation}
    \begin{alignedat}{3}
        \Theta&\in\Cbb^{n\times n\times n\times n}&\qquad&(\text{unconstrained trainable tensor}),\\
        \ul{T}^{(1)}_{ijkl}&\equiv \frac{\Theta_{ijkl}+\Theta_{jikl}}{2}&\qquad&(\text{symmetric in}~i\leftrightarrow j),\\
        \ul{T}^{(2)}_{ijkl} &\equiv \frac{\ul{T}^{(1)}_{ijkl} + \ul{T}^{(1)}_{ijlk}}{2} &\qquad&(\text{symmetric in}~k\leftrightarrow l),\\
        \ul{T}_{ijkl} &\equiv \frac{\ul{T}^{(2)}_{ijkl} + \ul{T}^{(2)\ast}_{klij}}{2} &\qquad&(\text{Hermitian in}~ij\leftrightarrow kl).
    \end{alignedat}
\end{equation}

\subsection{Unitarity and Matrix Orthogonality}\label{sec:param_unitary}

One may be tempted to parameterize unitary or orthogonal matrices by the result of some Gram-Schmidt process or QR decomposition on an unconstrained matrix. However, the resulting matrix from this approach can have abrupt changes for relatively small changes in the unconstrained matrix---in other words, this approach would not be well-behaved in the way necessary for gradient descent. In fact, the group of all orthogonal matrices is actually two disconnected sets (the set with determinant $+1$ and the set with determinant $-1$) and thus there is no well-behaved expression mapping unconstrained matrices to the entire set of orthogonal matrices. The solution to this problem is to leverage the knowledge of the high-fidelity model to restrict the matrix to one of the two types of orthogonal matrix first. Any real orthogonal matrix $\ul{R}$ with $\det\left(\ul{R}\right)=+1$ can be parameterized as
\begin{equation}\label{eq:param_rotations}
    \ul{R} \equiv \exp\left(A\right),\qquad A\equiv \frac{\Theta - \Theta^\mathsf{T}}{2},
\end{equation}
where $\exp(\cdot)$ is the matrix exponential and $\Theta\in\Rbb^{n\times n}$ is the unconstrained trainable matrix as before. Similarly, a real orthogonal matrix $\ul{O}$ with $\det\left(\ul{O}\right)=-1$ can be parameterized as
\begin{equation}
    \ul{O}\equiv\diag{\mathlist{-1, 1, 1, \ldots, 1}}\exp\left(A\right),\qquad A\equiv\frac{\Theta - \Theta^\mathsf{T}}{2},
\end{equation}
where $\diag{\mathlist{-1, 1, 1, \ldots, 1}}$ is the diagonal matrix with all $+1$ on the diagonal except for the first element, which is $-1$.

In contrast, unitary matrices are a continuous group that can be parameterized from an unconstrained complex matrix $\Theta\in\Cbb^{n\times n}$ by
\begin{equation}
    \ul{U}\equiv\exp\left(iH\right),\qquad H\equiv\frac{\Theta + \Theta^\dagger}{2},
\end{equation}
where $i$ is the imaginary unit\footnote{Of course, $iH$ could be replaced by the parameterization of an anti-Hermitian matrix. However, this form is more familiar to those with a background in quantum mechanics.}.

For semi-unitary and semi-orthogonal matrices, one simply takes the leading few columns or rows of the respective parameterized unitary or orthogonal matrix\footnote{Choosing $\ul{R}$ from Eq.~\eqref{eq:param_rotations} as the base orthogonal matrix is sufficient to parameterize all possible semi-orthogonal matrices.}. This is often denoted as $X(:,1\mathbin{:}k)$ for the first $k$ columns of $X$ and $X(1\mathbin{:}k,:)$ for the first $k$ rows of $X$.

The parameterizations described here have been used to apply orthogonality constrains to the weight matrices of \acp{ann} without any appreciable computational overhead~\cite{LezcanoCasado2019}.

\subsection{Vector Orthogonality and Normalization}

Parameterizing vectors with a fixed norm can be done straightforwardly by directly normalizing an unconstrained vector $\theta\in\Cbb^{n}$,
\begin{equation}
    \ul{v} \equiv \mathcal{N} \frac{\theta}{\norm{\theta}},
\end{equation}
where $\mathcal{N}$ is the desired value for the norm of the vector and $\norm{\cdot}$ is any well-defined vector norm.

When parameterizing a set of orthonormal vectors, one follows the parameterization for parameterizing semi-unitary (for complex vectors) or semi-orthogonal (for real vectors) matrices in the previous section, taking the columns (or rows) of the parameterized semi-unitary/semi-orthogonal matrix as the parameterized orthonormal vectors. For non-normalized orthogonal vectors, one introduces an additional trainable real-valued vector for the norms of the vectors and uses these elements to rescale the parameterized orthonormal vectors.

\subsection{Eigenvalue and Singular Value Constraints}
The most general and straightforward way to impose constraints directly on the eigenvalues or singular values of an operator is to first parameterize the operator following its other constraints, such as Hermiticity\footnote{Note that \ac{ad} for non-Hermitian eigenvalue decomposition is not implemented in all \ac{ad} libraries. At the time of writing, support was only recently added to \pypackage{JAX} in version v0.10.1.}, while ignoring the eigenvalue and singular value constraints. Then by computing the corresponding decomposition of the operator, constraints can be applied directly to the eigenvalues or singular values before reconstructing the now fully constrained matrix. For example, a Hermitian, positive semi-definite matrix $\ul{H}^{(+)}$ would be parameterized from an unconstrained complex matrix $\Theta\in\Cbb^{n\times n}$ by
\begin{equation}
    \begin{alignedat}{3}
        H &\equiv \frac{\Theta + \Theta^\dagger}{2}&\qquad&\text{(Hermitian matrix)},\\
        H &= V\diag{\lambda_1, \lambda_2, \ldots} V^H&\qquad&\text{(eigenvalue decomposition)},\\
        \ul{H}^{(+)} &\equiv V \diag{\lambda_1^2, \lambda_2^2, \ldots} V^H&\qquad&\text{(positive semi-definite matrix)}.
    \end{alignedat}
\end{equation}
This parameterization of $\ul{H}^{(+)}$ has the same eigenvectors as $H$, but with non-negative eigenvalues\footnote{For positive definite matrices (strictly positive eigenvalues), one would simply add the smallest positive machine-representable number to each of the squared eigenvalues. For single-precision floats this is $2^{-149}$ and for double-precision floats this is $2^{-1074}$.}. Note that an equivalent, but more performant, parameterization of a Hermitian positive semi-definite matrix would be
\begin{equation}
    \begin{alignedat}{3}
        H &\equiv \frac{\Theta + \Theta^\dagger}{2}&\qquad&\text{(Hermitian matrix)},\\
        \ul{H}^{(+)} &\equiv H H^\dagger&\qquad&\text{(positive semi-definite matrix)}.
    \end{alignedat}
\end{equation}

\begin{figure}
    \centering
    \includegraphics[height=0.25\textheight]{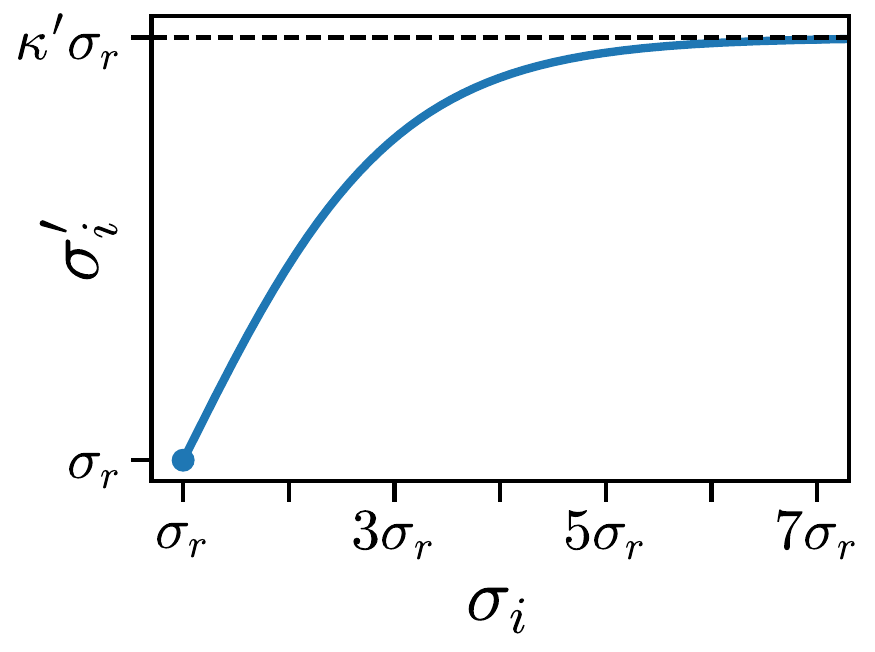}
    \caption[Illustration of the logistic function applied to the problem of constraining the condition number of a parameterized operator.]{\label{fig:condition_param}Illustration of the logistic function in Eq.~\eqref{eq:cond_logistic} applied to the problem of constraining the condition number of a parameterized operator. The horizontal axis shows the unconstrained value of an arbitrary singular value $\sigma_i$ while the vertical axis shows the resulting singular value $\sigma^\prime_i$. The ticks on both axes show the scale relative to the smallest nonzero singular value, $\sigma_r$. This function ensures that the new largest singular value, $\sigma_1^\prime$ over the new smallest nonzero singular value, which is unchanged from the original $\sigma_r$, is less than or equal to some desired maximum condition number $\kappa^\prime$.}
\end{figure}

A powerful and important singular value constraint is limiting the condition number of an operator, the ratio of the maximum and minimum singular values $\kappa = \sigma_\text{max}/\sigma_\text{min}$. As discussed in Section~\ref{sec:pod}, the Poincar\'e separation theorem guarantees that \ac{pod}-projected operators must have a condition number no greater than that of their full-space counterparts. Imposing an inequality constraint is less straightforward than the previous constraints, and a common tool for this---and a staple in machine learning---is the \textit{logistic} or \textit{sigmoid} function defined by
\begin{equation}
    s(x)\equiv\frac{1}{1+e^{-x}}.
\end{equation}
Defined on the domain $x\in(-\infty, \infty)$, this function is monotonically increasing and smoothly varies from $0$ to $1$, asymptotically approaching each value at the respective infinity. We can use this to limit the ratio of minimum and maximum values of any list of numbers, such as the singular values of an operator. Consider the parameterized operator $X$ which we wish to constrain such that the constrained operator, $\ul{X}$, satisfies $\kappa\left(\ul{X}\right)\leq \kappa^\prime$ for some chosen $\kappa^\prime$. We start by taking the singular value decomposition of $X$,
\begin{equation}
    X = U\diag{\sigma_1, \sigma_2, \ldots, \sigma_r, 0, \ldots, 0}V^\dagger
\end{equation}
where $\sigma_1\geq \sigma_2\geq\cdots\sigma_r> 0$ are the singular values of $X$ and $r$ is the rank of $X$. The condition number of $X$ is $\sigma_1/\sigma_r$. Notice that normalizing the singular values by the smallest (nonzero) singular value, $\sigma_r$, does not change the condition number and ensures that all the singular values are greater than or equal to $1$. By applying a shifted and rescaled logistic function to each of the normalized singular values then un-normalizing them, the minimum singular value remains unchanged while large singular values are smoothly reduced in order to keep the condition number less than or equal to $\kappa^\prime$,
\begin{equation}\label{eq:cond_logistic}
    \sigma^\prime_i \equiv \sigma_r\left[2(\kappa^\prime - 1)s\left(\frac{\sigma_i}{\sigma_r} - 1\right)-\kappa^\prime+2\right].
\end{equation}
This is shown in Fig.~\ref{fig:condition_param}. These rescaled singular values can then be taken to be the singular values of $\ul{X}$ with the same left and right singular vectors as $X$,
\begin{equation}
    \ul{X}\equiv U\diag{\sigma_1^\prime,\sigma_2^\prime,\ldots,\sigma_r^\prime,0,\ldots,0}V^\dagger.
\end{equation}

\subsection{Rank}

One way to enforce a particular rank for a parameterized operator would be to follow the method of constraining singular values and zero out all singular values beyond the desired rank. However, there is a far more parameter and computationally efficient method based on rank-$1$ matrices.

For a general rank-$r$ operator of shape $n\times m$, $\ul{X}$, we parameterize it by two sets of $r$ unconstrained trainable complex (or real) vectors $\theta_i\in\Cbb^{n}$ and $\phi_i\in\Cbb^m$ for $i=1,2,\ldots,r$,
\begin{equation}
    \ul{X}\equiv\sum_i^r \theta_i \phi_i^\dagger.
\end{equation}
That is, the sum of the outer products of each pair of $\theta_i,\phi_i$. For an $n\times n$ Hermitian (or symmetric) rank-$r$ operator, $\ul{Y}$, we need only a single set of $r$ unconstrained trainable complex (or real) vectors $\theta_i\in\Cbb^n$ and $r$ unconstrained trainable real values $\varphi\in\Rbb^r$,
\begin{equation}
    \ul{Y}\equiv\sum_i^r \varphi_i \theta_i \theta_i^\dagger.
\end{equation}
Anti-Hermitian rank-deficient operators are handled similarly, where each term in the sum is multiplied by $i$. Antisymmetric rank-deficient operators are a special case, since all antisymmetric real matrices must have even rank. Thus, rank-deficient real antisymmetric matrices, $\ul{Z}$, must be parameterized as the sum of rank-$2$ matrices,
\begin{equation}
    \ul{Z} \equiv\sum_i^{r/2} \theta_i\phi_i^\mathsf{T} - \phi_i\theta_i^\mathsf{T},\qquad \text{for}~r~\text{even}.
\end{equation}

\subsection{Tensor Hypernetworks}\label{sec:hypernetwork_param}

While not a common object in high-fidelity models directly, tensor hypernetworks\footnote{Tensor hypernetworks are expressions of tensor contractions where any contracted index appears more than twice. For an example using standard Einstein summation notation, $A_{ij}B_{jk}C_{klm}D_{m}$ is a standard tensor network since each index appears at most twice, but $A_{ij}B_{jk}C_{jl}D_{j}$ and $A_{jj}B_{jj}C_{jj}D_j$ are tensor hypernetworks since the index $j$ appears three or more times.}  appear in the reduced-space representation of nearly all high-fidelity models with nonlinear terms. See Sections~\ref{sec:pod}~and~\ref{sec:pmm_linearity} for discussions of nonlinear terms \acp{rbm} and \acp{pmm} respectively. The resulting tensor for nonlinear operations can be prohibitively large---the smallest and simplest, $\ul{G}^{(2)}$, which encodes the elementwise product of two vectors in the full space, is an order-$3$ tensor which would require $n^3$ trainable parameters if parameterized naively. Instead however, we can use the expression for these tensors which arises from the \ac{erbm} formulation of the problem. For $\ul{G}^{(2)}$ this is
\begin{equation}
    \ul{G}^{(2)}_{ijk} \equiv \left(\Pdag\right)_{i\mu} P_{\mu j} P_{\mu k},
\end{equation}
using Einstein summation notation, where $P\in\mathbb{U}\left(N,n\right)$ is the $N\times n$ unitary \ac{pod} projector from the $N$-dimensional full-space to the $n$-dimensional reduced space. We can approximate this tensor in the reduced space by a method similar to ``trimming'' in tensor networks~\cite{Bridgeman2017}. We replace each $P$ in the expression with an unconstrained trainable $m\times n$ complex (or real for purely real-valued data) matrix $\Theta\in\Cbb^{m\times n}$ where $n\leq m \ll N$ and take the approximation of $\ul{G}^{(2)}$ to be
\begin{equation}
    \ul{G}^{(2)}_{ijk}\approx \left(\Theta^\dagger\right)_{i\mu}\Theta_{\mu j}\Theta_{\mu k},
\end{equation}
which requires only $nm$ trainable complex parameters. Higher-order tensors for nonlinearities in the reduced space follow similar forms and can be efficiently approximated in the same manner. See Section~\ref{sec:dft} for a worked example.

\section{Potential Pitfalls}\label{sec:pitfalls}

There are several mistakes one can make when parameterizing objects with constraints or interpreting the post-training values of parameterized objects. The most common is to ascribe some meaning to the individual elements of the reduced-space objects. A usual manifestation of this is assuming that a diagonal operator in the full space will also be a diagonal operator in the reduced space, which is not true in general.
\setstacktabbedgap{1ex}%
\begin{equation}
    \renewcommand{\arraystretch}{1.5}
    \begin{array}{%
    >{\displaystyle}r @{\;} >{\displaystyle}l @{\qquad}
        >{\displaystyle}c @{\qquad}
        >{\displaystyle}r @{\;} >{\displaystyle}l}
        D &= \left[\sqmatrix{%
            d_1  & & \\
            &  \ddots & \\
            &  & d_N}\right]_{N\times N} &\scalebox{2}{$\nrightarrow$} & \ul{D}&=\left[\sqmatrix{%
                \ul{d}_1  & & \\
            &  \ddots & \\
            &  & \ul{d}_n}\right]_{n\times n}
    \end{array}
\end{equation}
Instead, these kinds of operators should be parameterized as general symmetric operators---potentially with definiteness constraints.

The core issue is over-constraining parameterized objects. Unless the constraints represent real physical properties or other properties guaranteed by the full-space model, parameterizations of objects in the reduced space should be as general as possible. An example of over-constraining the parameterized objects which has appeared many times comes in the form of breaking the unitary freedom of eigenvalue problems. Recall from Section~\ref{sec:pmm_irbm} that \acp{pmm} and \acp{irbm} rely on symmetries in both the full and reduced spaces for their flexibility. In a parametric affine eigenproblem, $H(c) = H_0 + \sum_{i>0} c_i H_i$, it is true that the eigenvalues are invariant under any unitary transformation and thus one could choose to always work in the basis where $H_0$ is diagonal. However, restricting $H_0$ or its reduced-space representation, $\ul{H}_0$, to be diagonal unnecessarily constrains not only $\ul{H}_0$, but all the $\ul{H}_i$, and can severely hinder training performance. Closely related is the mistake of replacing complex Hermitian operators with real symmetric ones. This most often appears in data-driven \acp{pmm}, discussed in Sections~\ref{sec:var_data_driven}~and~\ref{sec:pmm_reg}. The consequences of these unnecessary constraints often become more detrimental for more complex problems and emulators, potentially preventing successful emulation altogether.

One may be tempted to impose constraints on the reduced-space objects by adding penalty terms to the loss function, as is common practice in machine learning. Except for in very limited applications\footnote{Such as penalizing, but not constraining, the condition number of an operator.}, this is not recommended for physical constraints or properties which usually can be more efficiently and accurately imposed via the methods discussed in Section~\ref{sec:parameterization}.

When constructing an emulator\footnote{This pitfall applies to all \acp{rbm}, not just the \ac{pmm} method.} for vector-valued predictions, the projector must represent a suitable subspace for all target objects. For example, if the emulator is to produce predictions for some vector-valued functions $f(\cdot)$ and $g(\cdot)$, the subspace that the projector represents must approximately span snapshots of both $f(\cdot)$ and $g(\cdot)$. In this example, if the subspace only approximately spanned snapshots of $f(\cdot)$ then there would be no guarantee or expectation that the emulator would be able to produce reasonable predictions for $g(\cdot)$. Similarly, if the projector came from snapshots of an entirely distinct third function $h(\cdot)$, then in general the emulator would not produce accurate predictions for either $f(\cdot)$ or $g(\cdot)$. This is due to the nature of the linear transformation that the projector represents: the predictions from the emulator can be only linear combinations of the snapshots used to construct the projector. See Section~\ref{sec:tddft} for an example of the consequences of this and how \acp{pmm} can adapt when the ideal snapshots are not available.

Finally, one of the most challenging parts of effective \ac{pmm} emulation is deriving the form of the associated \ac{erbm} when nonlinear terms are present in the high-fidelity model. Whether following the \ac{pod} method of handling polynomially nonlinear terms (see Section~\ref{sec:pod}) or the matrix-valued-function method described in Section~\ref{sec:pmm_linearity}, it can help to replace the nonlinear operations with a black-box function $F$---which operates only on objects in the full space and returns an object in the full space---and perform as much of the \ac{erbm} derivation first before substituting the known form for $F$ and switching to Einstein summation notation. For example, there are three common nonlinear function \textit{signatures}: vector-input vector-output, vector-input matrix-output, and matrix-input matrix-output. Denoting these by $F^{(\text{vv})}$, $F^{(\text{vM})}$, and $F^{(\text{MM})}$ respectively and the full- and reduced-space vectors and matrices as $v$, $\ul{v}$, $M$, and $\ul{M}$, the \ac{erbm} derivation for each of the three nonlinearities would proceed as
{%
    \newcommand{\Fvv}[1]{\ensuremath{F^{(\text{vv})}\left(#1\right)}}%
    \newcommand{\FvM}[1]{\ensuremath{F^{(\text{vM})}\left(#1\right)}}%
    \newcommand{\FMM}[1]{\ensuremath{F^{(\text{MM})}\left(#1\right)}}%
\begin{equation}
    \begin{aligned}
        \Fvv{v} &\rightarrow \Pdag \Fvv{P\ul{v}},\\ \FvM{v} &\rightarrow \Pdag \FvM{P\ul{v}} P, \\ \FMM{M} &\rightarrow \Pdag \FMM{P \ul{M} \Pdag} P.
    \end{aligned}
\end{equation}%
}%
As we have seen, the form of the \ac{pmm} is not merely a guess or ansatz based on the high-fidelity model form, nor is it just the same form as the high-fidelity equations. It is a concrete result of the careful derivation of an associated \ac{erbm}.

\section{Affine Eigenproblems}\label{sec:affine}

The original context in which \acp{pmm} were developed, affine eigenproblems are problems which seek the eigenvalues and some sesquilinear forms\footnote{The sesquilinear form represented by a matrix $A$ on two complex vectors $u$ and $v$ is $v^\dagger A u$~\cite{Lang2005}. This is sometimes confused with bilinear forms, which are represented as $v^\mathsf{T} A u$---note the lack of complex conjugation on $v$. We can see the distinction between bilinear forms and sesquilinear forms if we take $f(u,v)\equiv v^\dagger A u$ and $a$ to be some complex scalar, then $f(au,v)=af(u,v)$ and $f(u,av)=a^{\ast} f(u,v)$. That is, a sesquilinear form $f$ is linear in the first argument and anti-linear in the second---compared to a bilinear form which is linear in both arguments. The word ``sesquilinear'' means one and one half times linear, just as ``bilinear'' means twice linear. For completeness, Hermitian forms are represented as $u^\dagger A u$ and quadratic forms by $u^\mathsf{T} A u$.} of the associated eigenvectors of a (usually Hermitian) matrix which is affine in some control parameters. That is, for control parameters $c \equiv \left[c_1,\, c_2,\,\ldots\right]$ the matrix is of the form
\begin{equation}\label{eq:pmm_affine}
    H(c) = H_0 + \sum_{i>0} c_i H_i
\end{equation}
and the quantities of interest are a subset of the eigenvalues $\mathset{E_j}$ (usually a few of the lowest lying) and the values of some set of particular sesquilinear forms $v_a^\dagger O_k v_b$ where $v_j$ is the \ord{j} eigenvector of $H\left(c\right)$ and $O_k$ are matrices.

The \ac{pmm} for these problems has a form identical to that of the original problem. The emulated eigenvalues are the eigenvalues of the reduced-space parametric matrix
\begin{equation}
    \ul{H(c)} = \ul{H_0} + \sum_{i>0} c_i \ul{H_i}
\end{equation}
and the emulated sesquilinear forms are the values of the sesquilinear forms of the eigenvectors of $\ul{H(c)}$ with the reduced-space matrices $\ul{O_k}$.

\subsection*{Example}
\begin{figure}
    \centering
    \includegraphics[width=0.8\textwidth]{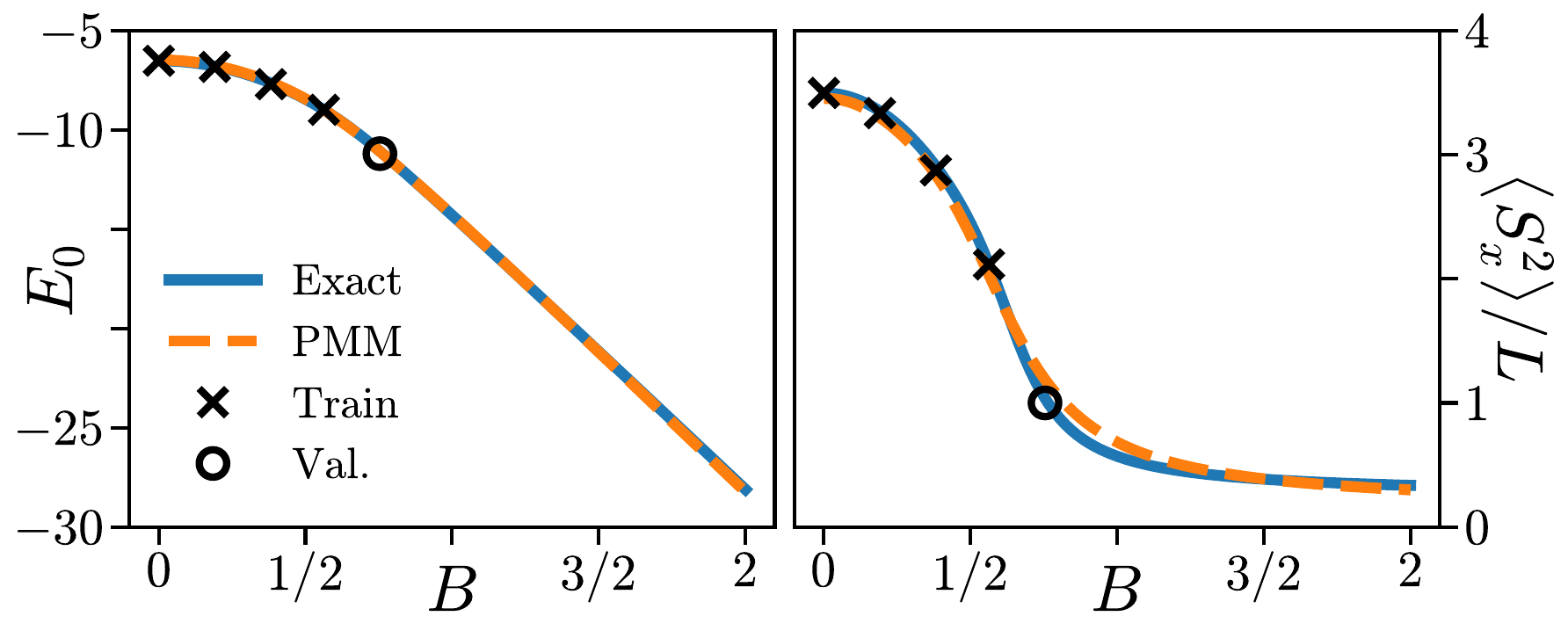}
    \caption[Results of applying a \ac{pmm} to a parametric many-body quantum system.]{\label{fig:pmm_affine_ex}Exact and \ac{pmm}-predicted results for the ground state energy (left) and expectation value of $S_x^2/L$ in the ground state (right) for the Hamiltonian detailed in Appendix~\ref{app:spin} as a function of $B$. All other Hamiltonian parameters are fixed $\left\{L,\,\gamma_x,\,\gamma_y\right\}=\{14,\,0,\,1\}$ with a power-law coupling defined by $J=0$ and $p=0.9$. The locations of the training and validation values are shown as crosses and open circles respectively.}
\end{figure}

Consider the same example problem as in Section~\ref{sec:ec}: a Hamiltonian describing a 1D spin chain of $L=14$ spins with couplings $J_{ij}$ in the influence of a transverse field with strength $B$,
\begin{equation}\label{eq:longrange_ex_pmm}
    H = -\frac{1}{\mathcal{J}} \sum_{i<j}^L J_{ij} \left(\gamma_x \sigma_i^x \sigma_j^x + \gamma_y \sigma_i^y \sigma_j^y\right) - B \sum_i^L \sigma_{i}^z,
\end{equation}
where $\gamma_{x/y}$ are the relative strengths of the $x$ and $y$ couplings, $\sigma_i^u$ is the Pauli spin operator acting on the \ord{i} spin in the $u$ axis, and $\mathcal{J}$ is the Kac normalization factor
\begin{equation}
    \mathcal{J}\equiv \frac{1}{L-1}\sum_{i\neq j}^L J_{ij}.
\end{equation}
This Hamiltonian is detailed further in Appendix~\ref{app:spin}. We are interested in emulating the ground state eigenvalue as a function of $B$ as well as the expectation value of $S_x^2/L$ in the ground state, where
\begin{equation}
    S_x^2/L\equiv\frac{1}{L}\sum_{i,j}^L \sigma_i^x\sigma_j^x.
\end{equation}
For this example we take the couplings to be a power-law decay with scale $J=1$ and exponent $p=0.9$, so $J_{ij} = J/\abs{i-j}^p$, with $\gamma_x=1$ and $\gamma_y=0$. We now follow the steps to emulate this problem with a \ac{pmm} as described in Section~\ref{sec:pmm_irbm}. For step~\ref{itm:pmm0}, we note that the form of the problem in terms of the varying parameter $B$ is simply
\begin{equation}
    \begin{aligned}
        H(B) &= H_0 + B H_1,\\
        H(B)\Psi_0(B) &= E_0(B)\Psi_0(B),\\
        \left\langle S_x^2/L\right\rangle(B) &= \frac{1}{L}\Psi_0(B)^\dagger S_x^2 \Psi_0(B),
    \end{aligned}
\end{equation}
and we suppose we have a known set of $B$ values $\mathset{B_i}$ and data for the ground state of energy of Eq.~\eqref{eq:longrange_ex_pmm} at those $B$ values, $\mathset{E_0\left(B_i\right)}$, as well as data for the expectation value of $S_x^2/L$ in the ground state for the same set of $B$ values, $\mathset{\left\langle S_x^2/L\right\rangle\left(B_i\right)}$. Note that we do not require access to any of the operators in the full problem. For step~\ref{itm:pmm1} we choose a reduced-space size of $n=5$ and an orthogonal implicit projection. In step~\ref{itm:pmm2} we note that the projected problem will be
\begin{equation}
    \begin{aligned}
        \ul{H(B)} &= \ul{H_0} + B \ul{H_1},\\
        \ul{H(B)}\ul{\Psi_0(B)} &= \ul{E_0(B)}\ul{\Psi_0(B)},\\
        \ul{\left\langle S_x^2/L\right\rangle(B)} &= \ul{\Psi_0(B)}^\dagger \ul{\left(S_x^2/L\right)} \ul{\Psi_0(B)},
    \end{aligned}
\end{equation}
where we have absorbed the factor of $1/L$ into the operator for the expectation value. Due to the properties of orthogonal projections discussed before, we know $E_0(B)\approx \ul{E_0(B)}$, $\left\langle S_x^2/L\right\rangle(B)\approx\ul{\left\langle S_x^2/L\right\rangle(B)}$, and that $\ul{H_0}$ and $\ul{H_1}$ must be Hermitian and $\ul{S_x^2/L}$ must be positive semi-definite since $H_0$ and $H_1$, are Hermitian and $S_x^2/L$ is positive semi-definite. We parameterize these reduced-space operators in step~\ref{itm:pmm3} by representing them as unknown general square complex matrices $X,\,Y,\,Z\in\mathbb{C}^{n\times n}$ and enforcing Hermiticity and positive semi-definiteness via
\begin{equation}
    \begin{aligned}
        \ul{H_0} &= \left(X + X^\dagger\right)/2\\
        \ul{H_1} &= \left(Y + Y^\dagger\right)/2\\
        \ul{S_x^2/L} &= Z^\dagger Z,
    \end{aligned}
\end{equation}
which allows us to perform unconstrained optimization over $X$, $Y$, and $Z$. We identify the mean squared error between the training data and the \ac{pmm} predictions as the loss function for step~\ref{itm:pmm4},
\begin{equation}
    \mathcal{L}(X,Y,Z) = \frac{1}{m}\sum_i^m \left[\left(E_0\left(B_i\right) - \ul{E_0\left(B_i\right)}\right)^2 + \left(\left\langle S_x^2/L\right\rangle\left(B_i\right) - \ul{\left\langle S_x^2/L\right\rangle\left(B_i\right)}\right)^2\right],
\end{equation}
where $m$ is the number of training examples. We use complex-valued gradient descent to find a suitable approximation to the location of the minimum of this loss function in step~\ref{itm:pmm5}. We can then use the optimized values for $X,\,Y,$ and $Z$ to compute the reduced-space operators and make predictions with this trained \ac{pmm}. Figure~\ref{fig:pmm_affine_ex} shows the results of this \ac{pmm}, demonstrating strong interpolation and extrapolation performance. The predictions are less accurate than the \ac{ec} emulator for this same problem (shown in Section~\ref{sec:ec}), but in contrast the \ac{pmm} emulator requires significantly less information, about $0.01\%$ as much as for \ac{ec}: it required only $15$ real values ($\mathset{B_i, E_0(B_i), \left\langle S_x^2/L\right\rangle(B_i)}$) compared to $2\times 2^{15} + 5\times 2^{14} + 5=\num{147461}$ complex values ($\mathset{H_0, H_1, V_T, B_i}$).

\subsection{Smoothing via Level Repulsion}\label{sec:smoothing}
A particularly useful demonstration of the \ac{pmm} method's ability to intentionally deviate from the \ac{erbm} involves the re-parameterization of the bias matrix ($\ul{H_0}$) in an affine Hermitian eigenproblem to encourage smooth eigenvalue and eigenvector functions. Depending on the problem context, this re-parameterization can instead be interpreted as encoding more information from the full-dimensional model rather than a deviation from it. As a function of the control parameters $c$, consider the eigenvalues $\lambda_k(c)$ and arbitrary Hermitian forms\footnote{This smoothing method equally applies to quadratic ($v_k(c)^\mathsf{T} X v_k(c)$), sesquilinear ($v_k(c)^\dagger X v_{k^\prime}(c)$), and bilinear ($v_k(c)^\mathsf{T} X v_{k^\prime}(c)$) forms as well.} of the associated eigenvectors $x_k\equiv v_k(c)^\dagger X v_k(c)$ of Eq.~\eqref{eq:pmm_affine} where $X$ is any arbitrary Hermitian matrix of compatible shape. As a function of $c$, $\lambda_k(c)$ for a fixed $k$ will exhibit sharp behavior (large second derivative magnitude) when the eigenvalues of $H(c)$ are degenerate or nearly degenerate. This behavior is potentially more pronounced for any Hermitian forms of the associated eigenvector. These are known as ``avoided crossings'' and have great significance in quantum mechanics~\cite{Landau1991}. The sharpness of these avoided crossings are determined by the coupling of the associated eigenvector to adjacent levels. Denoting the location of the avoided crossing by $c=c^\prime$, then the magnitude of
{%
    \newcommand{\cp}{\ensuremath{\left(c^\prime\right)}}
    \begin{equation}\label{eq:avoided_crossing_coupling}
        v_k\cp^\dagger H\cp v_{k^\prime}\cp \quad\text{and}\quad v_{k^\prime}\cp^\dagger H\cp v_k\cp
    \end{equation}%
}%
determines the sharpness of the avoided crossing, where $k^\prime = k\pm 1$, with smaller magnitudes indicating sharper behavior. When $c$ represents only a single control parameter, then only when the magnitude of both parts of Eq.~\eqref{eq:avoided_crossing_coupling} for a fixed $k^\prime$ is $0$ can the eigenvalue be degenerate---that is $\lambda_k\left(c^\prime\right)=\lambda_{k^\prime}\left(c^\prime\right)$. In quantum mechanics, this corresponds to the respective eigenstates having different symmetries.

There are various reasons why one might want to deliberately encourage more smooth behavior near an avoided crossing in a \ac{pmm} for the affine eigenproblem. From a machine learning perspective, encouraging smoother output is akin to regularization and can help prevent overfitting. The straightforward and physically motivated way to do this is with the commutator between each of the $H_i$. A true non-avoided crossing (the sharpest case) occurs only when a pair of $H_i$ commute, and the sharpness of the avoided crossing decreases as the magnitude\footnote{Here we use ``magnitude'' of an operator to mean any consistent norm.} of the commutator increases. To promote smoothness then, we want to control the magnitude of all the commutators between pairs of $H_i$. For a given pair of operators $H_k$ and $H_l$ with $k\neq l$ then, we add
\begin{equation}
    i\left[H_k, H_l\right] = i\left(H_kH_l - H_lH_k\right)
\end{equation}
to the affine Hermitian form of $H(c)$. The imaginary unit appears since the commutator of two Hermitian operators is anti-Hermitian. We can scale this term by a tunable hyperparameter $s$ to give us finer control over the smoothing, where larger values of $s$ correspond with more aggressive encouragement of smooth eigenpair behavior. This term can be interpreted as increasing the ``level repulsion'' between the eigenvalue levels $k$ and $l$ by deliberately introducing a gap (energy difference). We can efficiently compute this term for all pairs of $H_i$ by exploiting the linearity of the commutator, giving a revised form for the \ac{pmm},
\begin{equation}\label{eq:affine_smoothed}
    \begin{aligned}
        \ul{H(c)} &= \ul{H_0} + sC + \sum_{i>0} c_i \ul{H_i}\\
        C&\equiv i\sum_{k>0}\left[\ul{H_k},\sum_{l<k} \ul{H_l}\right].
    \end{aligned}
\end{equation}
Some important remarks about this revised \ac{pmm} form,
\begin{itemize}
    \item The additional smoothing term is independent of the control parameters $c$, and thus can be thought of purely as a modification to the constant $\ul{H_0}$ matrix.
    \item All operations are still linear and hence the form of the \ac{pmm} and the supposed full-space model are still equivalent.
    \item Regardless of the value of $s$, this term does not guarantee smoothness. If any pair of $\ul{H_k}$ and $\ul{H_l}$ commute, then the associated contribution to this term will be $0$. This does not represent a shortcoming but instead a desirable property of this approach. We have not restricted the space of eigenproblems the \ac{pmm} can represent. Instead, we have lessened the proportion of those representable eigenproblems which exhibit non-avoided crossings or sharp avoided crossings. This term's effects are felt during the training phase of the emulator, as it makes it much more difficult for the smooth training process to converge to a solution with sharp behavior. It is still possible and will happen if the specific training data cannot be fit any other way.
\end{itemize}

\begin{figure}
    \centering
    \includegraphics[width=0.8\textwidth]{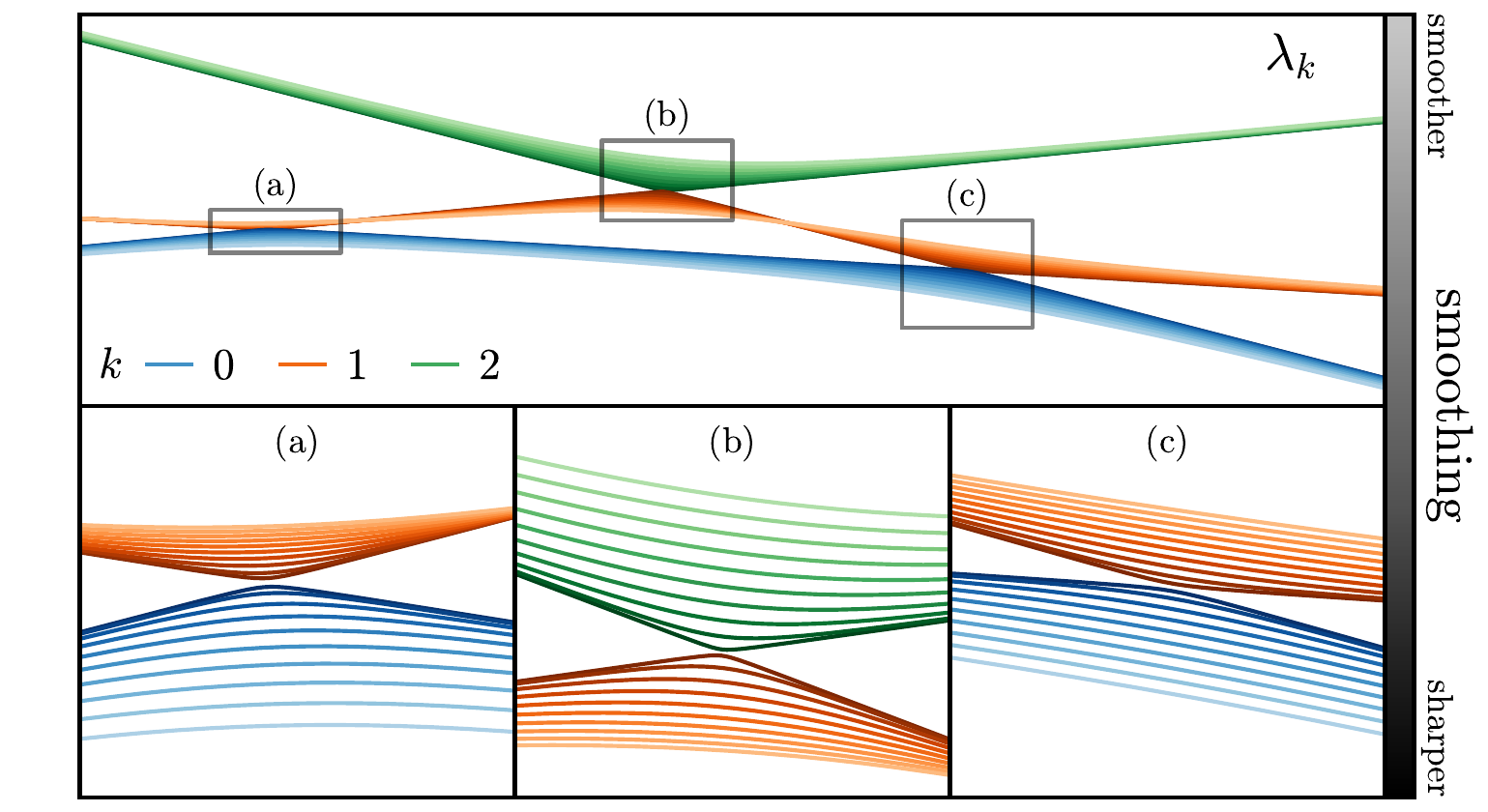}
    \caption[Demonstration of smoothing the eigenvalues of a parametric affine Hermitian eigenproblem.]{\label{fig:pmm_smoothing}Demonstration of the smoothing term in Eq.~\eqref{eq:affine_smoothed} on the eigenvalues of a parametric affine Hermitian eigenproblem. The top panel ($\lambda_k$) shows the three eigenvalues as a function of the control parameter for values of the smoothing parameter starting from no smoothing (darkest) to strong smoothing (lightest). The insets provide a zoomed-in view of each avoided crossing.}
\end{figure}

\begin{figure}
    \centering
    \includegraphics[width=0.8\textwidth]{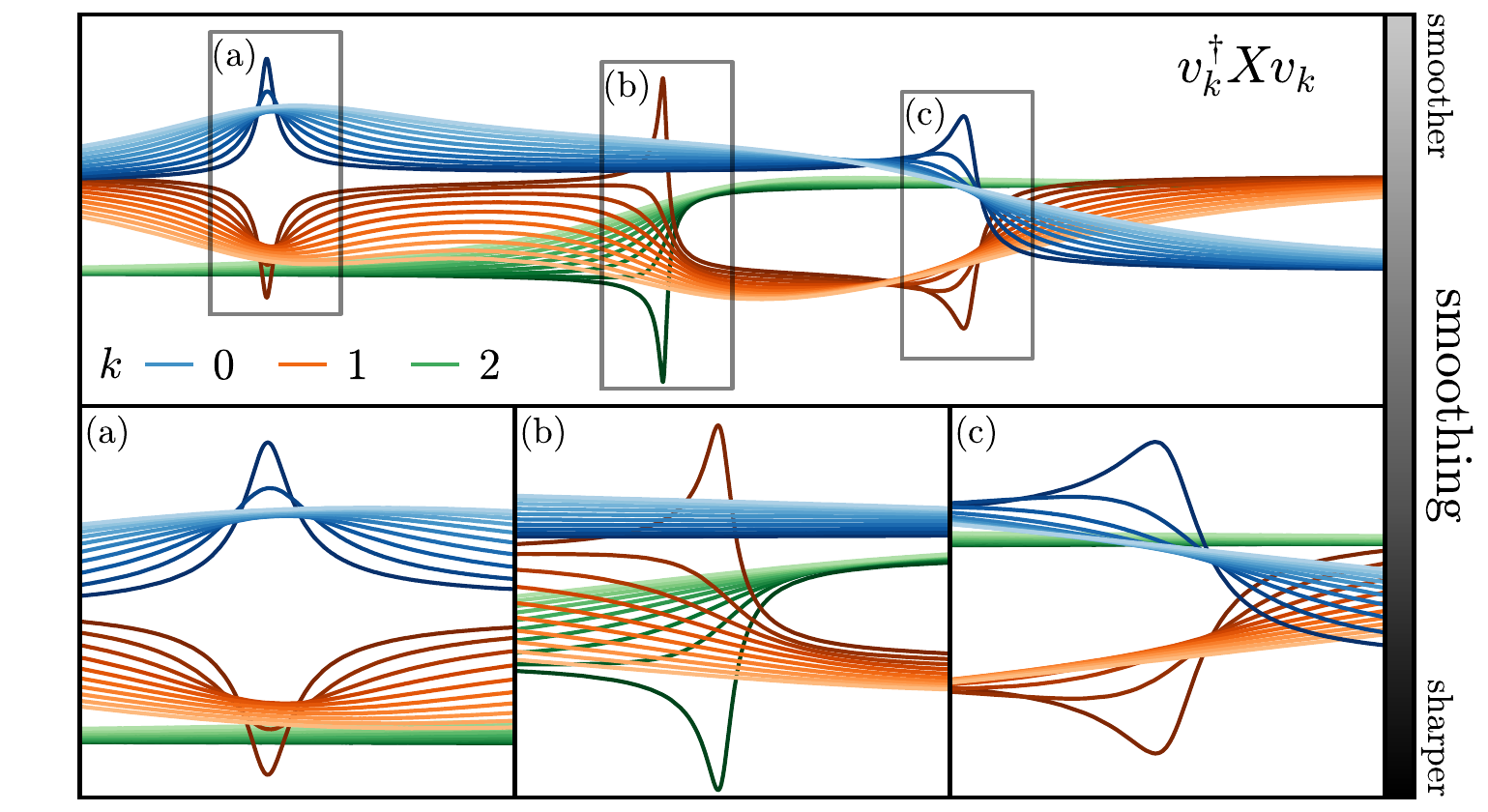}
    \caption[Demonstration of smoothing a Hermitian form of a parametric affine Hermitian eigenproblem.]{\label{fig:pmm_smoothing_o}Demonstration of the smoothing term in Eq.~\eqref{eq:affine_smoothed} on a Hermitian form of the eigenvectors of the same parametric affine Hermitian eigenproblem as in Fig.~\ref{fig:pmm_smoothing}. The top panel ($v_k^\dagger X v_k$) shows the value of a Hermitian form for each of the eigenvectors $v_k$ as a function of the control parameter for values of the smoothing parameter starting from no smoothing (darkest) to strong smoothing (lightest). The insets provide a zoomed-in view of the location of each avoided crossing.}
\end{figure}

Figures~\ref{fig:pmm_smoothing}~and~\ref{fig:pmm_smoothing_o} illustrate the sharp behavior that can occur in a parametric affine eigensystem and demonstrate the effect this smoothing term has on the eigensystem for increasing values of the smoothing parameter $s$. The figures use a one-parameter $3\times 3$ parametric affine Hermitian eigenproblem with random operators where $H_0$ and $H_1$ nearly commute. Figure~\ref{fig:pmm_smoothing} shows the behavior of the three eigenvalues as a function of $c$ and $s$. The insets in Fig.~\ref{fig:pmm_smoothing} make clear how sharp the level repulsion can be when no smoothing is present. Figure~\ref{fig:pmm_smoothing_o} shows the value of a Hermitian form of the eigenvectors associated with a Hermitian matrix $X$, equivalent to the expectation value of $X$ in each of the three eigenstates. We see that despite having no knowledge of $X$ the additional term in Eq.~\eqref{eq:affine_smoothed} is able to smooth not only the eigenvalues in Fig.~\ref{fig:pmm_smoothing} but also the Hermitian form defined by $X$ in Fig.~\ref{fig:pmm_smoothing_o}. Both figures additionally show the fine-grain and continuous control the smoothing parameter $s$ has over the induced smoothing.


\section{Data-Driven Regression}\label{sec:pmm_reg}

Despite its foundations as an \ac{rbm}, the \ac{pmm} method can be used as a purely data-driven machine learning model simply by choosing a suitable form. In Ref.~\cite{Cook25} it is shown that the eigenvalues of a parametric affine Hermitian matrix can approximate any other function. However, more efficient forms involving sesquilinear forms of the eigenvectors of the same parametric matrix are used in practice. Here we cover the most recent architecture used for data-driven \acp{pmm} which is a refinement of the forms used in Ref.~\cite{Cook25}.

To form a data-driven model for $p$ real-valued inputs $c\in \Rbb^p$ that produces $q$ real-valued outputs $z\in \Rbb^q$, we start by choosing the \ac{pmm} dimension $n$ and form the parametric affine Hermitian matrix $\ul{H(c)}$ with an optional smoothing term controlled by the smoothing parameter $s$ as in Eq.~\eqref{eq:affine_smoothed}. Denote the eigenvalues and associated eigenvectors of this matrix by $\lambda_k(c)$ and $v_k(c)$. We choose $r<n$ of these eigenvectors and for each output compute the sum of the squared magnitudes of a series of sesquilinear forms between every pair of eigenvectors defined by $l$ Hermitian matrices. Each output has an independent set of $l$ Hermitian matrices, so we have
\begin{equation}\label{eq:ddpmm_intermediate}
    y_j = \sum_{\omega=1}^l \sum_{\mu,\nu=1}^r \abs{v_\mu^\dagger \ul{D_{j\omega}} v_\nu}^2,~\text{for}~j=1,\ldots,q,
\end{equation}
where $\ul{D_{j\omega}}$ is the Hermitian matrix defining the \ord{\omega} sesquilinear form for the \ord{j} output. Altogether, there are $q \times l$ independent sesquilinear forms and $q \times l \times r^2$ terms in the sum. The specific subset of $r$ eigenvectors is usually the $r$ eigenvectors with algebraically smallest or largest associated eigenvalue. Note that choosing the subset based on the magnitude of the associated eigenvalue can result in extremely sharp and unstable output since the order of the magnitudes of the eigenvalues is not stable as a function of $c$ and smoothing via level repulsion does not alleviate this. The intermediate result of $y_j$ in Eq.~\eqref{eq:ddpmm_intermediate} is a non-negative real number. To allow for negative outputs, we introduce a centering term and trainable biases to give the final form of the data-driven \ac{pmm},
\begin{equation}\label{eq:ddpmm_final}
    z_j = \ul{b_j} + y_j - \frac{1}{2}\sum_{\omega=1}^l \left\Vert\ul{D_{j\omega}}\right\Vert_2^2,~\text{for}~j=1,\ldots,q,
\end{equation}
where $\ul{b_j}\in\Rbb$ is the trainable bias for the \ord{j} output and $\left\Vert\ul{D_{j\omega}}\right\Vert_2^2$ is the squared spectral norm of the matrix $\ul{D_{j\omega}}$. This centering term is roughly motivated by the average contribution of each sesquilinear form to the output\footnote{The actual mean value of the terms in Eq.~\eqref{eq:ddpmm_intermediate} is not known since the distribution of the eigenvectors is difficult to characterize for a generic parametric affine Hermitian matrix. This centering term is a partially heuristic choice.} and has been found to improve training performance.

The number of trainable real values for this form of data-driven \ac{pmm} is
\begin{equation}
    \underbrace{(p+1)n^2}_{\ul{H_0},\, \ul{H_i}} + \underbrace{q l n^2}_{\ul{D_{j\omega}}} + \underbrace{q}_{\ul{b_j}},
\end{equation}
and inference can be performed with a time complexity of $\bigO{qr^2n^2}$ when using an optimized partial eigensolver to compute the $r$ eigenpairs of $\ul{H(c)}$ or $\bigO{qr^2n^2 + n^3}$ when using a full eigensolver. In Ref.~\cite{Cook25} a less efficient variant of this form is shown to perform well on a variety of regression tasks and consistently outperforms existing standard machine learning regression models, including feedforward \acp{ann}, random forests, and support vector regression.

Data-driven \acp{pmm} in emulation are particularly useful for identifying hidden features that become the control parameters for physically grounded \ac{pmm} emulators. This application is central to solidifying the method as an adaptive emulation method and enabling the emulation of significantly more complex systems. Section~\ref{sec:pmm_eos} provides an example of this application in the context of the nuclear equation of state.

\subsection{Data-Driven Classification}\label{sec:pmm_class}

A model for classification can be formed by taking the data-driven regression form of Eq.~\eqref{eq:ddpmm_final} and applying the \textit{softmax} function, defined as
\begin{equation}
    \text{softmax}(z)_j = \frac{e^{z_j}}{\sum_{k=1}^q e^{z_k}},~\text{for}~j=1,\ldots,q,
\end{equation}
to the output to give a probability distribution over the classes. The resulting model is a data-driven \ac{pmm} classifier. The number of trainable parameters and inference time complexity are the same as for the regression form, with the additional cost of the softmax function, which has a time complexity of $\bigO{q}$~\cite{Goodfellow2016}. The loss functions used for training, validation, and testing of classification models are much different than those for regression and involve applications of information theory---see Refs.~\cite{Zhang2023,Geron2022} for an introduction to training classification models. Reference~\cite{Cook25} contains additional details on the architecture modifications necessary to accept images as input and demonstrates strong classification performance on a variety of standard small image classification datasets.

\section{Differential Equations}
Emulating systems described by differential equations (usually in time) with \acp{pmm} can be done in one of two complimentary ways. The first is by direct fitting of samples of the state vector and the second is by fitting to the derivative vector(s). To make the distinction clear, consider a system described by the state vector $v$ and an arbitrary explicit first-order ordinary differential equation in time,
\begin{equation}\label{eq:1eode}
    \frac{dv}{dt} = F(t, v),
\end{equation}
with initial state $v_0$. Given snapshots of the state vector $\mathset{t_i, v(t_i)}$, in the first approach the \ac{pmm} encapsulates the entirety of Eq.~\eqref{eq:1eode} and the initial state and produces as an output the emulated state vector at time $t$, $\tilde{v}(t)$. In this case, computing the loss for a single snapshot $\mathcal{L}\left(v(t),\tilde{v}(t)\right)$ requires integrating the \ac{pmm} in time. Training the \ac{pmm} in this approach can become both inefficient due to the full integration for each snapshot and difficult since snapshots at large $t$ provide very little meaningful information for gradient-descent based algorithms. This is akin to the problem of vanishing gradients in deep learning. The alternative \ac{pmm} approach avoids this issue by training directly on snapshots for the derivative of the state vector, usually approximated from finite differences of the state vector snapshots. This way, the \ac{pmm} aims only to implement a single evaluation of the right hand side of Eq.~\eqref{eq:1eode}---that is, the emulated derivative of the state vector, $\widetilde{\frac{dv}{dt}}(t)$. In this case, the loss for a single sample $\mathcal{L}\left(\frac{dv}{dt}(t), \widetilde{\frac{dv}{dt}}(t)\right)$ is efficient to compute and contributes equally for all $t$ during training. These two approaches are complimentary in that the second can be used to train a \ac{pmm} which, now producing reasonable predictions at later $t$, can then be fine-tuned by training with the first approach.

\section{Nonlinear Eigenproblems}\label{sec:dft}

\Ac{dft} is a powerful method for computing various properties of nuclear systems and fundamentally relies on a nonlinear eigenproblem~\cite{Bender2003,Dobaczewski2016,Schunck2019}. See Refs.~\cite{Bender2003,Dobaczewski2016,Schunck2019} for detailed reviews of \ac{dft} in nuclear physics, which will not be covered here\footnote{This section will greatly over-simplify some points of \ac{dft}, but the core elements relevant to emulation in the context of \acp{rbm} and specifically \acp{pmm} will remain intact.}. The central nonlinear operation in \ac{dft} comes in the form of a density-dependent interaction\footnote{Specifically, the Kohn-Sham potential or in reality an approximation of it~\cite{Schunck2019}.}, of which the lowest few eigenstates\footnote{Usually called Kohn-Sham ``orbitals'' in \ac{dft} literature~\cite{Dobaczewski2016}.} and eigenenergies are sought. The density that this Hamiltonian depends on is the sum of the individual eigenstate densities, $\rho = \sum_k \abs{\Psi^{(k)}}^2$. Computing an orthonormal set of eigenstates for such a Hamiltonian is accomplished by the self-consistent loop~\cite{Bender2003,Dobaczewski2016,Schunck2019}. A simplified description of the procedure begins with a reasonable initial guess for the eigenstates. Those eigenstates are used to compute the density-dependent parts of the Hamiltonian, yielding a now no-longer density-dependent Hamiltonian\footnote{This is not dissimilar to evaluating parameter-dependent Hamiltonians at a specific parameter realization, thereby yielding a fixed Hamiltonian without any parameters.}, which can be diagonalized to find a new set of eigenstates~\cite{Dobaczewski2016}. This new set is again used to compute the density-dependent terms of the Hamiltonian, and the iteration repeats until some convergence criteria is met.

\subsection*{Example}
{%
\newcommand{\Rmnum}[1]{\MakeUppercase{\romannumeral #1}}
\newcommand{\Pone}{\ac{pmm}--\Rmnum{1}}
\newcommand{\Ptwo}{\ac{pmm}--\Rmnum{2}}
Here, we consider a highly simplified, toy, density-dependent Hamiltonian with all the mathematical features that would need to be considered for the effective emulation of a realistic \ac{dft} Hamiltonian.
\begin{equation}\label{eq:dftham}
    H\left(\alpha,\, c;\,\mathset{\Psi^{(k)}}\right) = T + \alpha V - c \sum_k\abs{\Psi^{(k)}}^2
\end{equation}
This Hamiltonian is parametric with the parameter $\alpha$ scaling the typical non-density-dependent potential $V$ and the parameter $c$ scaling the density-dependent term which is a sum of a set of states $\mathset{\Psi^{(k)}}$. In this example, $T$ is the usual kinetic energy operator, $V$ is a harmonic oscillator potential, and we take the number of states to be $2$. To build a \ac{pmm} emulator for this system, we can treat the nonlinear term in either of the two ways described in Section~\ref{sec:pmm_linearity}.

Before we can write either form of \ac{pmm} emulator, we must disambiguate the notation in Eq.~\eqref{eq:dftham}. Specifically, at first glance the nonlinear term appears to be a vector, but the linear terms are unmistakably operators (matrices) which implies an impossible addition between these operators and the nonlinear term. The truth is that $\abs{\Psi^{(k)}}^2$ is a slight abuse of notation, instead it actually denotes either the diagonal matrix whose elements are $\abs{\Psi_k}^2$ or hides an implicit elementwise product $\odot$ after the term, so that when $H$ acts on a vector, $v$, the nonlinear term appears as
\begin{equation}
    -c\sum_k\abs{\Psi^{(k)}}^2\odot v.
\end{equation}
Both more-explicit interpretations of the nonlinear term in Eq.~\eqref{eq:dftham} are equivalent; however, since the self-consistent iteration relies on an eigenvalue decomposition of $H$ with explicit $\Psi^{(k)}$ realizations, it is almost necessary\footnote{It is possible to implement the Hamiltonian in a matrix-free way, which would enable the elementwise-product interpretation (see Appendix~\ref{app:matrixfree}) but doing so in the reduced space of the emulator is unnecessary complexity and would likely result in lesser performance.} to take the diagonal-operator interpretation to form the emulator, which we use in this example.

We adopt the usual notation where $n$ is the dimension of the reduced space and $N$ is the dimension of the full space. The first \ac{pmm}, denoted here as \Pone{}, follows the treatment of polynomially nonlinear terms in the way of the \ac{pod}-Galerkin method (see Section~\ref{sec:pod}),
\begin{equation}\label{eq:dftpmmi}
    \ul{H}^\text{\Rmnum{1}}\left(\alpha,\, c;\,\mathset{\ul{\Psi}^{(k)}}\right) = \ul{T} + \alpha \ul{V} - c \sum_k \ul{G}^{(3\ast)}\left(\ul{\Psi}^{(k)\ast}\otimes\ul{\Psi}^{(k)}\right),
\end{equation}
where $\ul{G}^{(3\ast)}\in\Cbb^{n\times n\times n\times n}$ is an order-$4$ tensor. To arrive at the nonlinear term in Eq.~\eqref{eq:dftpmmi}, we first consider each nonlinear term in the sum as a black-box function, $F\left(\Psi^{(k)}\right)$, given by
\begin{equation}
    F(\Psi^{(k)})\equiv\abs{\Psi^{(k)}}^2 \equiv \diag{\abs{\Psi^{(k)}}^2},
\end{equation}
which in the reduced space under the \ac{pod} projector $P$ becomes
\begin{equation}
    \Pdag F\left(P \ul{\Psi}^{(k)}\right) P.
\end{equation}
Now, expanding the known form of $F(\cdot)$ in Einstein summation notation yields
\begin{equation}
    \begin{aligned}
        \left[\Pdag F\left(P\ul{\Psi}^{(k)}\right)P\right]_{ij} &= \left(\Pdag\right)_{i\mu} \delta_{\mu\nu}\left(P \ul{\Psi}^{(k)}\right)^{\!\ast}_\mu\left(P\ul{\Psi}^{(k)}\right)_\nu P_{\nu j}\\
        &= \underbrace{\left(\Pdag\right)_{i\mu}P^\ast_{\mu\omega}P_{\mu\sigma}P_{\mu j}}_{\ul{G}^{(3\ast)}_{ij\omega\sigma}}\Psi^{(k)\ast}_\omega\Psi^{(k)}_\sigma\\
        \implies \abs{\Psi^{(k)}}^2 &\rightarrow \ul{G}^{(3\ast)}\left(\ul{\Psi}^{(k)\ast}\otimes\ul{\Psi}^{(k)}\right).
    \end{aligned}
\end{equation}

The second \ac{pmm}, denoted by \Ptwo{}, is
\begin{equation}
    \ul{H}^\text{\Rmnum{2}}\left(\alpha,\,c;\,\mathset{\ul{\Psi}^{(k)}}\right) = \ul{T} + \alpha \ul{V} - c \sum_k\left(\ul{G}^{(2)}\ul{\Psi}^{(k)}\right)^{\!\dagger}\left(\ul{G}^{(2)}\ul{\Psi}^{(k)}\right),
\end{equation}
where $\ul{G}^{(2)}\in\Cbb^{n\times n\times n}$ is an order-$3$ tensor. To derive this form, we first rewrite the nonlinear term in Eq.~\eqref{eq:dftham} using the canonical matrix-valued magnitude-squared operation $F_S(A) \equiv A^\dagger A = V_A \diag{\abs{\lambda_A}^2} V_A^{-1}$,
\begin{equation}
    \sum_k \abs{\Psi^{(k)}}^2 \equiv \sum_k F_S\left(\diag{\Psi^{(k)}}\right) \equiv \sum_k \diag{\Psi^{(k)}}^\dagger \diag{\Psi^{(k)}}.
\end{equation}
This equivalent form involves no nonlinear operations and so can be projected straightforwardly. Let $D^{(k)} \equiv \diag{\Psi^{(k)}}$, then
\begin{equation}
    D^{(k)} \equiv \diag{\Psi^{(k)}} \rightarrow \Pdag \diag{P\ul{\Psi}^{(k)}} P \equiv \ul{D}^{(k)},
\end{equation}
which expanded in Einstein summation notation is
\begin{equation}
    \begin{aligned}
        \ul{D}^{(k)}_{ij} = \left[\Pdag \diag{P\ul{\Psi}^{(k)}} P\right]_{ij} &= \left(\Pdag\right)_{i\mu} \delta_{\mu\nu} P_{\nu\omega}\ul{\Psi}^{(k)}_\omega P_{\nu j}\\
        &= \underbrace{\left(\Pdag\right)_{i\mu}P_{\mu\omega}P_{\mu j}}_{\ul{G}^{(2)}_{ij\omega}}\ul{\Psi}^{(k)}_\omega\\
        \implies \ul{D}^{(k)} &= \ul{G}^{(2)}\ul{\Psi}^{(k)}.
    \end{aligned}
\end{equation}
And so,
\begin{equation}
    \abs{\Psi^{(k)}}^2 \rightarrow \ul{D}^{(k)\dagger} \ul{D}^{(k)} \equiv \left(\ul{G}^{(2)}\ul{\Psi}^{(k)}\right)^{\!\dagger}\left(\ul{G}^{(2)}\ul{\Psi}^{(k)}\right).
\end{equation}

Both \acp{pmm} here represent the linear terms, $T+\alpha V$, as $\ul{T}+\alpha\ul{V}$ where $\ul{T}$ and $\ul{V}$ are trainable $n\times n$ Hermitian matrices with no further constraints. \Pone{} and \Ptwo{} approach equivalence with one another as the size of the reduced space approaches the size of the full space, and both are valid dimensionality-reduced forms of the full Hamiltonian in Eq.~\eqref{eq:dftham}. However, following the naive parameterization of each, $\ul{G}^{(3\ast)}$ in \Pone{} would require $n^4$ trainable complex parameters compared to the $n^3$ trainable complex parameters required by $\ul{G}^{(2)}$ in \Ptwo{}. Following the discussion in Section~\ref{sec:hypernetwork_param}, both of these tensors can be approximated in the reduced space by parameterizing their projector-based construction directly. Consider the definition of each tensor with the projector $P\in\mathbb{U}(N, n)$,
\begin{equation}
    \begin{aligned}
        \ul{G}^{(2)}_{ijk} &\equiv \left(\Pdag\right)_{i\mu}P_{\mu j}P_{\mu k},\\
        \ul{G}^{(3\ast)}_{ijkl} &\equiv \left(\Pdag\right)_{i\mu}P_{\mu j}P^\ast_{\mu k} P_{\mu l}.
    \end{aligned}
\end{equation}
Instead, we truncate the larger dimension of $P$ to $m$ with $n\leq m \ll N$ and call this truncated approximation of the projector\footnote{This is only an approximation of the projector in the context of computing $\ul{G}^{(2)}$ and $\ul{G}^{(3\ast)}$. It is not an approximation of the projector in the context of projecting the high-fidelity equations to create an \ac{erbm}.} $\ul{Q}\in\mathbb{U}(m,n)$ which we can use to approximate $\ul{G}^{(2)}$ and $\ul{G}^{(3\ast)}$ by
\begin{equation}\label{eq:GDFTapprox}
    \begin{aligned}
        \ul{G}^{(2)}_{ijk} &\approx \left(\ul{Q}^\dagger\right)_{i\mu}\ul{Q}_{\mu j}\ul{Q}_{\mu k},\\
        \ul{G}^{(3\ast)}_{ijkl} &\approx \left(\ul{Q}^\dagger\right)_{i\mu}\ul{Q}_{\mu j}\ul{Q}^\ast_{\mu k} \ul{Q}_{\mu l}.
    \end{aligned}
\end{equation}
This allows the number of parameters in each of the two \acp{pmm} to be equal, as now the $nm$ complex elements of $\ul{Q}$ are the only trainable parameters for the nonlinear term. However, the cost of computing the approximations in Eq.~\eqref{eq:GDFTapprox} is still more expensive in \Pone{} by a factor of $n$, as is the cost of using the tensor to compute the reduced nonlinear term. This example highlights the efficiency improvements that occur when replacing nonlinear terms by their canonical matrix-valued counterparts.

Both \acp{pmm} solve for the eigenstates and eigenenergies of their respective reduced-space state-dependent Hamiltonians via the same self-consistent iteration that would be used in the full space. Since all operations in the self-consistent loop are linear---only matrix-vector products and vector addition---nothing needs to be modified to accurately emulate this method in the reduced space.

In this example, we take $N=25$ and randomly sample $100$ training points uniformly in the domain $\left(\alpha, c\right)\in[0, 2]\times[0, 1]$. For a given $\alpha$ and $c$, the high-fidelity solution is given by the self-consistent iteration that results in $\mathlist{\Psi^{(0)},\Psi^{(1)}}$ and $\mathlist{E_0,E_1}$ such that the $\Psi^{(k)}$ are orthonormal and
\begin{equation}
    \begin{aligned}
        H\left(\alpha,\,c;\,\mathset{\Psi^{(0)},\Psi^{(1)}}\right)\Psi^{(0)} &= E_0 \Psi^{(0)},\\
        H\left(\alpha,\,c;\,\mathset{\Psi^{(0)},\Psi^{(1)}}\right)\Psi^{(1)} &= E_1 \Psi^{(1)}.
    \end{aligned}
\end{equation}
We seek an emulator for the energies and the states, so we follow the steps for vector-valued \acp{pmm} in Section~\ref{sec:vec_pmm}. The states from the training data are used to form the \ac{pod} projector, $P\in\mathbb{U}(N,n)$, for a reduced dimension of size $n=5$. The states in the training data are projected to the reduced space by $\hat{\Psi}^{(k)} \equiv \Pdag \Psi^{(k)}$, and the predictions from the emulator are projected from the reduced space to the full space by $\tilde{\Psi}^{(k)} = P \ul{\Psi}^{(k)}$. The loss between the reduced-space predictions from the emulator, $\ul{\Psi}^{(k)}$ and $\ul{E_k}$, and the high-fidelity results projected to the reduced space, $\hat{\Psi}^{(k)}$ and $E_k$, is taken to be
\begin{equation}
    \mathcal{L}\left(\mathset{\left(\hat{\Psi}^{(k)}, E_k\right)},\mathset{\left(\ul{\Psi}^{(k)}, \ul{E}_k\right)}\right) = 1 - \frac{1}{2}\sum_{k=0,1}\left[\abs{\hat{\Psi}^{(k)\dagger}\ul{\Psi}^{(k)}}^2 - \abs{E_k - \ul{E}_k}^2\right],
\end{equation}
which is the sum of the mean squared error of the energies and the mean ``overlap error'' of the states. The overlap error between two unit vectors $u$ and $v$ is simply $1-\abs{u^\dagger v}^2$ which is bounded between $0$ and $1$. Note that since the loss involves only inner products of the states, it is equivalent to computing the same loss in the full space due to the properties of the \ac{pod} projector (see Section~\ref{sec:pod}). In this simple example we are unconcerned with overfitting and so do not withhold any data for validation. After training both \acp{pmm}, we evaluate their performance on test data taken uniformly on a $50\times 50$ grid over the same domain that the training data was sampled from. The results are shown in Figs.~\ref{fig:dftenergy}~and~\ref{fig:dftoverlap}. Figure~\ref{fig:dftenergy} shows the true $E_0$ and $E_1$ values from the $N=25$ system and the percent error between both emulators and the true values, demonstrating that both are able to achieve sub-$1\%$ errors for both energies across nearly the entire parameter domain. Figure~\ref{fig:dftoverlap} shows the overlap error for each emulator's prediction of each state in the full space---that is, $1-\abs{\Psi^{(k)\dagger}\tilde{\Psi}^{(k)}}^2$, as opposed to the overlap error in the reduced space $1-\abs{\hat{\Psi}^{(k)\dagger}\ul{\Psi}^{(k)}}^2$ although these should be equal due to the properties of $P$---demonstrating that both are capable of highly accurate predictions except near $\alpha = 0$, where the system develops a continuous degeneracy.

\begin{figure}
    \centering
    \includegraphics[height=0.4\textheight]{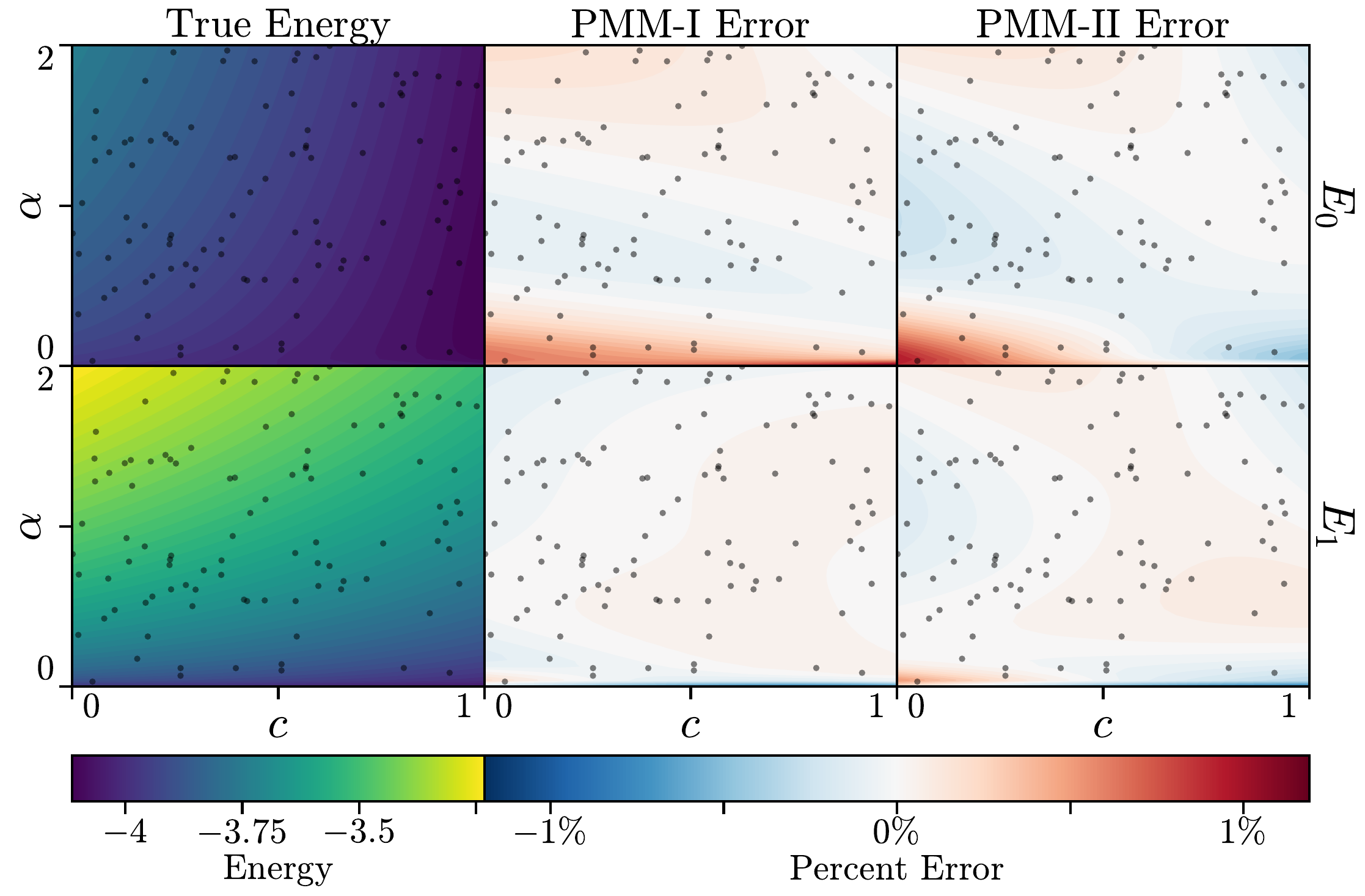}
    \caption[Results for the emulation of the energy in the nonlinear eigenproblem example.]{\label{fig:dftenergy}(Left column) high-fidelity calculations of the ground state ($E_0$, top row) and first excited state ($E_1$, bottom row) energies for the example nonlinear Hamiltonian in Eq.~\eqref{eq:dftham} throughout the phase space $(\alpha,c)\in[0,2]\times[0,1]$. (Middle column and right column) percent errors between the high-fidelity calculations and \Pone{} (middle column) and \Ptwo{} (right column) for the energies. The locations of the training data are shown as black points.}
\end{figure}

\begin{figure}
    \centering
    \includegraphics[height=0.4\textheight]{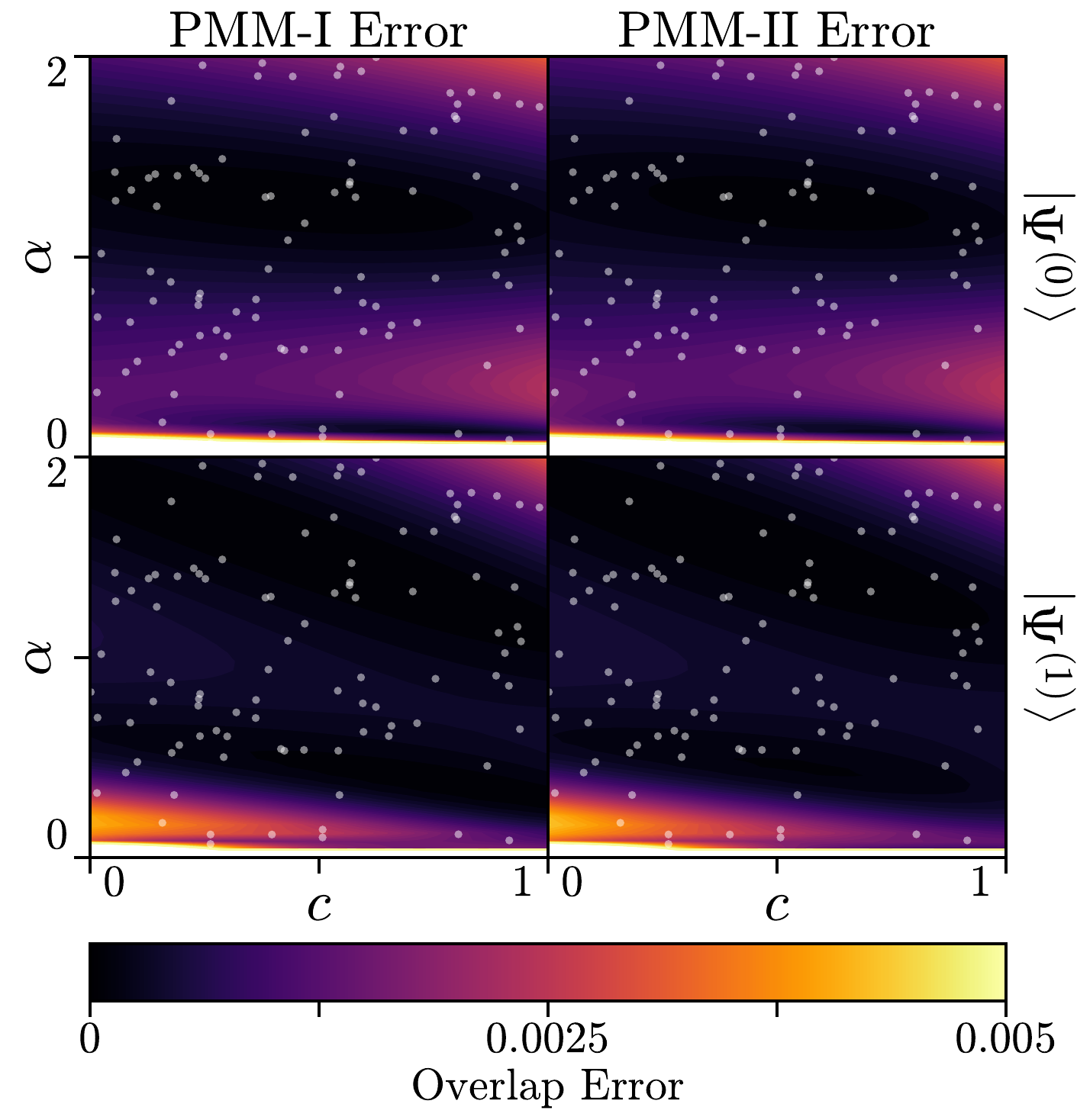}
    \caption[Results for the emulation of the eigenvectors in the nonlinear eigenproblem example.]{\label{fig:dftoverlap}Results for the overlap error, $1-\abs{\Psi^{(k)\dagger}\tilde{\Psi}^{(k)}}^2$, between the high-fidelity calculations of the ground state (top row) and first excited state (bottom row) eigenvectors of Eq.~\eqref{eq:dftham} and the corresponding predictions from \Pone{} (left column) and \Ptwo{} (right column) across the phase space $(\alpha,c)\in[0,2]\times[0,1]$. The locations of the training data are shown as white points.}
\end{figure}}

\section{Nonlinear Dynamics and Macrostate Data}\label{sec:tddft}
Nonlinear dynamics combines differential equations and nonlinear terms. This appears in \ac{tddft}, in which the time-dependent Schr\"odinger equation involves a (local) density-dependent parametric interaction~\cite{Bulgac2019}\footnote{In principle, the interaction can be non-local and depend on the full density matrices, $\Psi^{(k)}(c, t)\Psi^{(k)\dagger}(c,t)$, but this greatly complicates the problem, potentially to the point of being prohibitively computationally expensive~\cite{Bulgac2019}.}---that is, in the limit of no pairing (\ac{hf} theory),
\begin{equation}\label{eq:tddft}
    \begin{aligned}
        i\frac{\partial}{\partial t}\Psi^{(k)}(c, t) &= H\left(c, \mathset{\Psi^{(k)}(c,t)}\right)\Psi^{(k)}(c, t),\\ \text{where}~H\left(c, \mathset{\Psi^{(k)}(c,t)}\right) &= T + V(c) + U\left(c, \mathset{\abs{\Psi^{(k)}(c,t)}^2}\right).
    \end{aligned}
\end{equation}
See Ref.~\cite{Bulgac2019} for a review of \ac{tddft}\footnote{Similar to the previous section, this is a greatly oversimplified description of \ac{tddft} that focuses only on the core mathematical elements which are relevant to emulation in the context of \acp{rbm} and \acp{pmm}. Application-specific complications such as devising the nuclear energy density functional, bases with good quantum numbers, and pairing correlations do not change the core principles discussed here.}. Direct integration of this equation is exceedingly expensive for sufficiently high-fidelity systems of heavy nuclei, warranting emulation. However, a barrier in the way of emulation is the lack of available data for the wavefunctions. Snapshots of the wavefunction, not only the density, are seemingly required since the dynamics of the system are determined by the wavefunction and not by the density\footnote{It is not possible in general to determine the dynamics of any quantum system purely from the instantaneous local density. Consider the simple 1D harmonic oscillator, where---in the position basis---the momentum of the state is encoded in the complex phase of the wavefunction which is lost in the local density.}. \ac{tddft} simulations in general do not return snapshots of the single-particle wavefunctions---denoted here by $\Psi^{(k)}$---as both the number and dimension of these wavefunctions are prohibitively large~\cite{Jin2021}\footnote{In Ref.~\cite[Section 3.5.3]{Jin2021}, the authors note that reasonably sized fission studies with \ac{tddft} require $\sim 10^{11}$ double-precision floating point values per time step and $\gtrsim 10^5$ time steps. This would require more than $\si{10~\peta\byte}$ to store all the snapshots.}. Instead of taking snapshots of all the single-particle states, one might propose using snapshots of the \ac{hf} many-body wavefunction. However, the many-body wavefunction is a Slater determinant of the single-particle wavefunctions, which would require the computation of $A!{}$ terms for an $A$-nucleon system\footnote{For example, a \ac{tddft} simulation for ${}^{240}\text{Pu}$ fission, which appears in Refs.~\cite{Bulgac2016, Bjelcic2026}, would require $240!>10^{468}$ terms in the Slater determinant.}---a completely infeasible calculation. In contrast, the one-body density is comparatively cheap in terms of both computation time and storage size.

{%
    \newcommand{\PPdag}{\ensuremath{P^{(\Psi)\dagger}}}%
\newcommand{\PP}{\ensuremath{P^{(\Psi)}}}%
\newcommand{\PRdag}{\ensuremath{P^{(\rho)\dagger}}}%
\newcommand{\PR}{\ensuremath{P^{(\rho)}}}%
\newcommand{\Psik}{\ensuremath{\Psi^{(k)}}}%
\newcommand{\ulPsik}{\ensuremath{\ul{\Psi}^{(k)}}}%
In traditional \ac{erbm}-based emulation, the lack of wavefunction snapshots would prohibit the construction of a proper and effective reduced form of Eq.~\eqref{eq:tddft}. However, given the relation between the wavefunctions and the one-body density,
\begin{equation}\label{eq:sqprho}
    \rho = \sum_k \abs{\Psi^{(k)}}^2,
\end{equation}
we can still form the \ac{pmm} for this system as if we had snapshots of the wavefunctions. Denote the unknown projector for the wavefunctions by $\PP$ and the projector for the one-body density by $\PR$; then the \ac{erbm}-reduced form for Eq.~\eqref{eq:sqprho} is
\begin{equation}\label{eq:sqprho_rbm}
    \begin{alignedat}{3}
        \underbrace{\PRdag \rho}_{\ul{\rho}} &= \PRdag \sum_k \Psi^{(k)\ast}\odot\Psik &\qquad&(\text{project into}~\rho~\text{subspace})\\
        \rightarrow \ul{\rho} &= \PRdag \sum_k \left(\PP \ulPsik\right)^\ast \odot \left(\PP \ulPsik\right)&\qquad&(\text{handle nonlinearity})\\
        \iff \ul{\rho}_i &= \left(\PRdag\right)_{i\mu}\sum_k P^{(\Psi)\ast}_{\mu\nu}\ul{\Psi}^{(k)\ast}_\nu\PP_{\mu\omega}\ulPsik_\omega&\qquad&(\text{Einstein notation})\\
        &= \sum_k\underbrace{\left(\PRdag\right)_{i\mu} P^{(\Psi)\ast}_{\mu\nu}\PP_{\mu\omega}}_{\ul{W}^{(\Psi\rightarrow\rho)}_{i\nu\omega}}\ul{\Psi}^{(k)\ast}_\nu\ulPsik_\omega&\qquad&\\
        \iff \ul{\rho} &= \sum_k\ul{W}^{(\Psi\rightarrow\rho)}\left(\ul{\Psi}^{(k)\ast}\otimes\ulPsik\right),&\qquad&
    \end{alignedat}
\end{equation}
where we introduce $\ul{W}^{(\Psi\rightarrow\rho)}$, the order-$3$ tensor which encodes the nonlinear operation from $\Psik$ to $\abs{\Psik}^2$ in the reduced basis of $\rho$. This completes the reduced-basis form only of Eq.~\eqref{eq:sqprho}. The \ac{erbm} derivation of Eq.~\eqref{eq:tddft} uses only $\PP$ since the operators only act on $\Psik$,
\begin{equation}\label{eq:rbm_tddft}
    \begin{aligned}
        i\frac{\partial}{\partial t} \underbrace{\left(\PPdag \Psik \right)}_{\ul{\Psik}} &= \left[\begin{aligned}
        & \underbrace{\PPdag T \PP}_{\ul{T}}+ \underbrace{\PPdag V \PP}_{\ul{V}}\\ &+ \PPdag U\left(\PP \ul{G}^{(\Psi,\,2\ast)}\left[\left(\PPdag\Psi^{(k)\ast}\right)\otimes\left(\PPdag\Psik\right)\right]\right)\PP\end{aligned}\right]\ulPsik\\
        \iff i\frac{\partial}{\partial t} \ulPsik &= \left[\ul{T} + \ul{V} + \PPdag U\left(\PP \ul{G}^{(\Psi,\,2\ast)}\left(\ul{\Psi}^{(k)\ast}\otimes\ulPsik\right)\right)\PP\right]\ulPsik,
    \end{aligned}
\end{equation}
where the parameter and time dependences, $(c,t)$, have been left implicit for brevity. To complete the reduced form of the model, leaving all objects in the reduced space, the precise form of the functional $U$ would need to be specified. So long as $U$ is either linear in the density or can be handled by the methods for nonlinear terms discussed in Sections~\ref{sec:pod}~and~\ref{sec:pmm_linearity}, then the final $\PPdag$ and $\PP$ in Eq.~\eqref{eq:rbm_tddft} will cancel or otherwise be able to be condensed into a single tensor in the reduced space. Along with Eq.~\eqref{eq:sqprho_rbm}, this yields a complete \ac{rbm} form for an \ac{hf} \ac{tddft} calculation\footnote{The methods discussed here can be generalized to \ac{hfb} \ac{tddft} calculations with pairing interactions, as the fundamental nonlinear operations do not change.}. Computing the reduced-space objects explicitly is not possible due to the aforementioned information gap regarding snapshots of the wavefunctions; however, it is possible to take the implicit approach and train parameterized versions of these operators such that $\PR\ul{\rho}$ fits snapshots of the density.

This concept of relating the relevant state vector of the high-fidelity model to snapshots of some properties of that state vector is the generalization of the typical \ac{pmm} problem context where only data for a small set of scalar observables is available, prohibiting the application of \acp{erbm}. This more general data, that lies between scalar values and state vectors, is called \textit{macrostate data}.

\subsection*{Example}

As an example, we consider the problem of emulating the dynamics of the time-dependent Gross--Pitaevskii equation using only snapshots of the density, $\abs{\Psi}^2$. The equation describes the dynamics of the single-particle wavefunction of a Bose-Einstein condensate and is
\begin{equation}
    i\frac{\partial}{\partial t} \Psi(r, t) = \left[-\frac{1}{2m}\nabla^2 + V(r) + g\abs{\Psi(r,t)}^2\right]\Psi(r,t),
\end{equation}
where $V(r)$ is some external trapping potential and we have taken $\hbar=1$~\cite{Gross1961,Pitaevskii1961}. Given only snapshots of the density, we can see that this system has all of the components that made \ac{tddft} emulation challenging. For this example we will restrict the problem to one spatial dimension and take the external potential to be the harmonic oscillator potential with scale parameter $a$,
\begin{equation}\label{eq:gpe}
    i\frac{\partial}{\partial t} \Psi(x, t) = \left[ -\frac{1}{2m}\frac{\partial^2}{\partial x^2} + ax^2 + g\abs{\Psi(x, t)}^2\right]\Psi(x,t),
\end{equation}
and take the initial state to be a normalized off-center Gaussian
\begin{equation}
    \Psi(x, 0)\sim \exp\left[-\frac{\left(x-x_0\right)^2}{2\sigma^2}\right].
\end{equation}
For the high-fidelity simulation, we use periodic boundary conditions with $x\in[-10,10]$ discretized into $N = 1000$ points. Using explicit time integration, we compute the dynamics for $5$ values of $m$ taken on a uniform grid in the range $[1, 2]$, $10$ values of $a$ taken on a uniform grid in the range $[1/2,2]$, and $10$ values of $g$ taken on a uniform grid in the range $[-40, 0]$. The system is evolved for $500$ time steps to $t=50$. The first $100$ snapshots of the density evolution ($0\leq t < 10$) become the training snapshots, the next $100$ ($10\leq t < 20$) are reserved for validation, and the remaining $300$ ($20\leq t < 50$) are withheld for testing the emulator.

Following the discussion for emulating \ac{tddft}, we derive the form of the \ac{pmm} emulator for Eq.~\eqref{eq:gpe} similar to the example in Section~\ref{sec:dft} but instead taking the elementwise-product interpretation of the nonlinear term ($\abs{\Psi}^2\Psi\equiv \Psi^\ast\odot\Psi\odot\Psi$) rather than the diagonal-operator one.
\begin{equation}
    \begin{aligned}
        i\frac{\partial}{\partial t} \underbrace{\left(\PPdag \Psi\right)}_{\ul{\Psi}} &= \begin{aligned}[t]
            &-\frac{1}{m}\underbrace{\PPdag\left(\frac{1}{2}\frac{\partial^2}{\partial x^2}\right)\PP}_{\ul{T}}\ul{\Psi} + a \underbrace{\PPdag\left(x^2\right)\PP}_{\ul{V}}\ul{\Psi}\\ &+ g \PPdag \left[\left(\PP \ul{\Psi}\right)^\ast\odot\left(\PP\ul{\Psi}\right)\odot\left(\PP\ul{\Psi}\right)\right]
        \end{aligned}\\
        \iff i\frac{\partial}{\partial t} \ul{\Psi}_i &= -\frac{1}{m} \ul{T}_{i\mu}\ul{\Psi}_\mu + a \ul{V}_{i\mu}\ul{\Psi}_\mu + g \underbrace{\left(\PPdag\right)_{i\mu}P^{(\Psi)\ast}_{\mu\nu}\PP_{\mu\sigma}\PP_{\mu\omega}}_{\ul{G}^{(\Psi,3\ast)}_{i\nu\sigma\omega}}\underbrace{\ul{\Psi}^\ast_\nu\ul{\Psi}_\sigma\ul{\Psi}_\omega}_{\left(\ul{\Psi}^\ast\otimes\ul{\Psi}\otimes\ul{\Psi}\right)_{\nu\sigma\omega}}\\
        \iff i\frac{\partial}{\partial t} \ul{\Psi} &= -\frac{1}{m}\ul{T}\ul{\Psi} + a\ul{V}\ul{\Psi} +g\ul{G}^{(\Psi,3\ast)}\left(\ul{\Psi}^\ast\otimes\ul{\Psi}\otimes\ul{\Psi}\right),
    \end{aligned}
\end{equation}
where the density in the reduced space is given by
\begin{equation}
    \ul{\rho}=\ul{W}^{(\Psi\rightarrow\rho)}\left(\ul{\Psi}^\ast\otimes\ul{\Psi}\right),
\end{equation}
and predictions for $\rho(t)$ in the full space, denoted by $\tilde{\rho}(t)$, are given by
\begin{equation}
    \rho(t)\approx\tilde{\rho}(t)=P^{(\rho)}\ul{\rho},
\end{equation}
where $P^{(\rho)}$ is the \ac{pod} projector computed from the training snapshots of the density. We choose a reduced-space dimension of $n=8$, less than $1\%$ the size of the full space. We parameterize the reduced-space operators following Section~\ref{sec:parameterization}, and---since we are only concerned with a single initial state\footnote{In general, different initial states that, for example, come from the ground state of a separate parametric Hamiltonian could be emulated by including the reduced-space representation of their generating process in the emulator.}---we directly parameterize the reduced-space initial state $\ul{\Psi}_0$ as a trainable unit vector. We train this \ac{pmm} by minimizing the mean squared error between the reduced-space projection of the target density, $\PRdag \rho(t)$, and the reduced-space prediction $\ul{\rho}(t)$.

Figure~\ref{fig:gpe} shows the results of using this trained emulator to predict the dynamics for a parameter set included in the training data, $\mathlist{m,a,g}=\mathlist{1.50, 1.33, -17.78}$, and for a parameter set not present in either the training or validation data, $\mathlist{m,a,g}=\mathlist{1.33, 1.75, -20.00}$. Both cases show good agreement with the high-fidelity result---and importantly stability at late times---well beyond the training and validation horizons. Additionally, the aggressive dimensionality reduction provided by the emulator reduces the real-time cost of the dynamics by nearly a factor of $50$\footnote{On a consumer-grade laptop, a single high-fidelity simulation of the dynamics required $7$ minutes $22$ seconds, compared to $9$ seconds for the emulator. Since these times include data- and model-loading overhead---which would impact the emulator proportionally much more---the $50\times$ figure is a lower bound.}.

\begin{figure}
    \centering
    \includegraphics[width=0.9\linewidth]{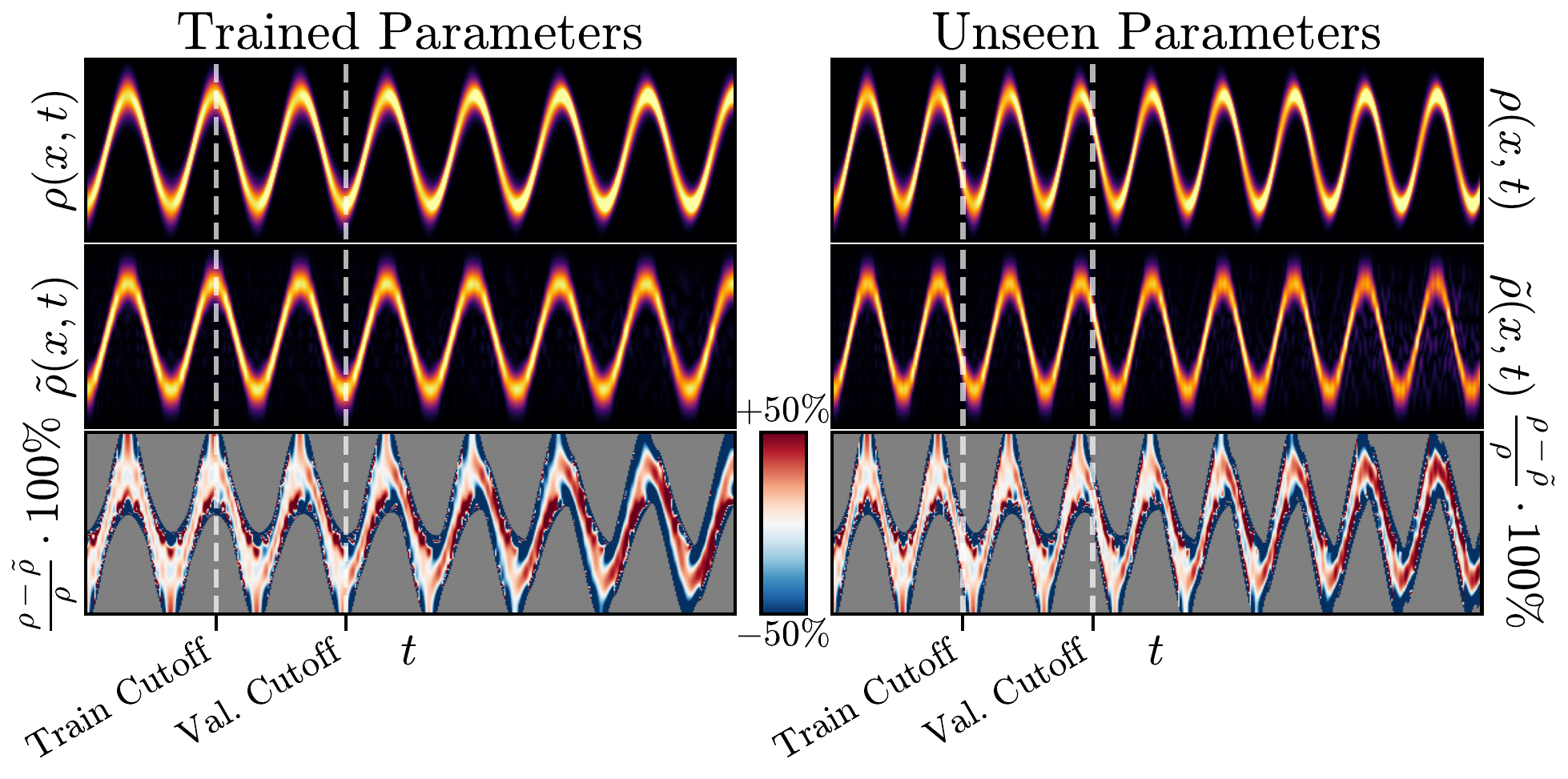}
    \caption[Results for the nonlinear dynamics and macrostate data example.]{\label{fig:gpe}High-fidelity and emulator-predicted results for the dynamics of the system in Eq.~\eqref{eq:gpe}. The left column shows the dynamics for a system with parameters that were included in the training data, $\mathlist{m,a,g}=\mathlist{1.50,1.33,-17.78}$. The right column, conversely, shows results for parameters not seen by the emulator during training or validation, $\mathlist{m,a,g}=\mathlist{1.33,1.75,-20.00}$. The top row shows the local density dynamics, $\rho(x,t)$, from the high-fidelity solution with $x$ on the vertical axis and $t$ on the horizontal axis. The middle row shows the emulator-predicted local density dynamics, $\tilde{\rho}(x,t)$. Finally, the bottom row shows the percent error between the high-fidelity dynamics and the emulated dynamics---regions where the high-fidelity density is too small are shown in grey since a relative error metric is not meaningful there. The training cutoff (initial $1/5$ of the dynamics) and validation cutoff (initial $2/5$ of the dynamics) are indicated by dashed lines. The visible range of $x$ values is $[-5,5]$ to better show the dynamics of the system.}
\end{figure}
}
\section{Incomplete Underlying Equations}\label{sec:pmm_pdd}
The applications discussed so far have operated with full, explicit knowledge of the form of the equations governing the high-fidelity model or, as in Section~\ref{sec:pmm_reg}, no knowledge of any underlying equations whatsoever, in which case a universal form was prescribed. Between these two extremes lies problems where much or some, but not all, of the form of the high-fidelity equations are known. In these cases, missing terms can be replaced with physically constrained---or entirely unconstrained---data-driven machine learning models while leaving the known terms unchanged. This essentially amounts to devising an informed guess for a full, explicit form of the high-fidelity model from which the \ac{erbm} derivation for the associated \ac{pmm} emulator can begin.

It is usually the case that significant physical insight is known about the missing terms in the underlying equations. The ability to incorporate this information is what makes the \ac{pmm} method so powerful and adaptable. Consider the case of emulating a system-size-dependent, parametric, quantum many-body interaction (parameterized by $c$) which preserves particle number. We know that the Hamiltonian for $L$ bodies is a projection of some infinitely large Hamiltonian---which describes the interaction for any possible system size---onto the states which describe $L$ particles. We can write this as
\begin{equation}\label{eq:sys_size_affine}
    H_L(c) = U^\dagger(L) H_\infty(c) U(L)
\end{equation}
where $H_\infty(c)$ represents the particle-number-independent representation of the interaction parameterized by $c$ and $U(L)$ is the projection onto the space which spans the states of $L$ particles. We may well know the form of $H_\infty(c)$ (for example, affine) but we may not know how the basis projection $U(L)$ depends on $L$. For now however, we continue with the \ac{erbm} derivation, introducing the implicit \ac{pod} projector $P_1$ which represents a suitable $n_1$-dimensional reduced space for $H_L(c)$,
\begin{equation}
    \ul{H}_L(c) \equiv P_1^\dagger H_L(c) P_1,
\end{equation}
where $n_1\leq\text{dim}\left(H_L(c)\right)$. Substituting the definition of $H_L(c)$ and introducing the implicit \ac{pod} projector $P_2$ which represents a suitable $n_2$-dimensional reduced space for $H_\infty(c)$---with $n_1 < n_2 \leq \text{dim}\left(H_\infty(c)\right)$---we have
\begin{equation}
    \begin{aligned}
        \ul{H}_L &= \underbrace{P_1^\dagger U^\dagger(L) P_2}_{\ul{U}^\dagger(L)} \underbrace{P_2^\dagger H_\infty(c) P_2}_{\ul{H}_\infty(c)} \underbrace{P_2^\dagger U(L) P_1}_{\ul{U}(L)}\\
            &= \ul{U}^\dagger(L) \ul{H}_\infty(c) \ul{U}(L),
    \end{aligned}
\end{equation}
where $\ul{U}(L)\equiv P_2^\dagger U(L) P_1$ is the reduced-space representation of the finite-particle-number basis projection and $\ul{H}_\infty(c)\equiv P_2^\dagger H_\infty(c) P_2$ is the reduced-space representation of the particle-number-independent representation for the interaction. Note that this would be systematically improvable if the form of $U(L)$ was known, and note that the exact form of $\ul{H}_\infty(c)$ would be known in this example and thus so would the form of $\ul{H}_\infty(c)$. Even without the exact form of $U(L)$, we still know one very important physical constraint on it: it is semi-unitary. Here, we introduce a data-driven, universal form able to approximate any function which maps any number of real inputs $x\in\Rbb^p$ to a semi-unitary operator that represents a basis compression from a space of size $N$ to one of size $n\leq N$,
\begin{equation}
    U_\text{univ}(x)\equiv\exp\left[i\sum_{j=1}^l f_j(x) M_j\right](:,1\mathbin{:}n),
\end{equation}
where $M_j$ are $N\times N$ Hermitian operators, $f_j(x)$ are universal functions (for example, the output of an \ac{ann}), and $A(:,1\mathbin{:}n)$ represents taking the first $n$ columns of $A$. For sufficiently large $l$---a hyperparameter akin to a ``hidden model'' size---and expressive $f_j$, this expression can represent any parameterized semi-unitary operator to arbitrary accuracy. In analogy with traditional data-driven machine learning, we refer to the $f_j(x)$ as the identified \textit{hidden features} of the emulator. We can now approximate $U(L)$ by this universal form, $U(L)\approx U_\text{univ}(L)$. From the properties of orthogonal projections, we then know that $\ul{U}(L)$ is approximated by
\begin{equation}
    \ul{U}(L)\approx\ul{U}_\text{univ}(L)\equiv\exp\left[i\sum_{j=1}^l f_j(x) \ul{M}_j\right](: 1\mathbin{:}n_1),
\end{equation}
where $\ul{M}_j$ are $n_2\times n_2$ Hermitian matrices. By substituting this into our expression for $\ul{H}_L(c)$, we see that we now have a complete, partially data-driven, \ac{pmm} for $H_L(c)$,
\begin{equation}
    \begin{aligned}
        \ul{H}_L(c) &= \ul{U}^\dagger(L)\ul{H}_\infty(c)\ul{U}(L),\\
        \text{where}~\ul{U}(L) &= \exp\left[i\sum_{j=1}^l f_j(L) \ul{M}_j\right](:,1\mathbin{:}n_1).
    \end{aligned}
\end{equation}
To use this \ac{pmm} emulator, one need only to substitute the known form of $\ul{H}_\infty(c)$, choose suitable values of $n_1$, $n_2$, and $l$, and select the type of universal functions to use for $f_j$.

\begin{figure}
    \centering
    \includegraphics[width=0.7\linewidth]{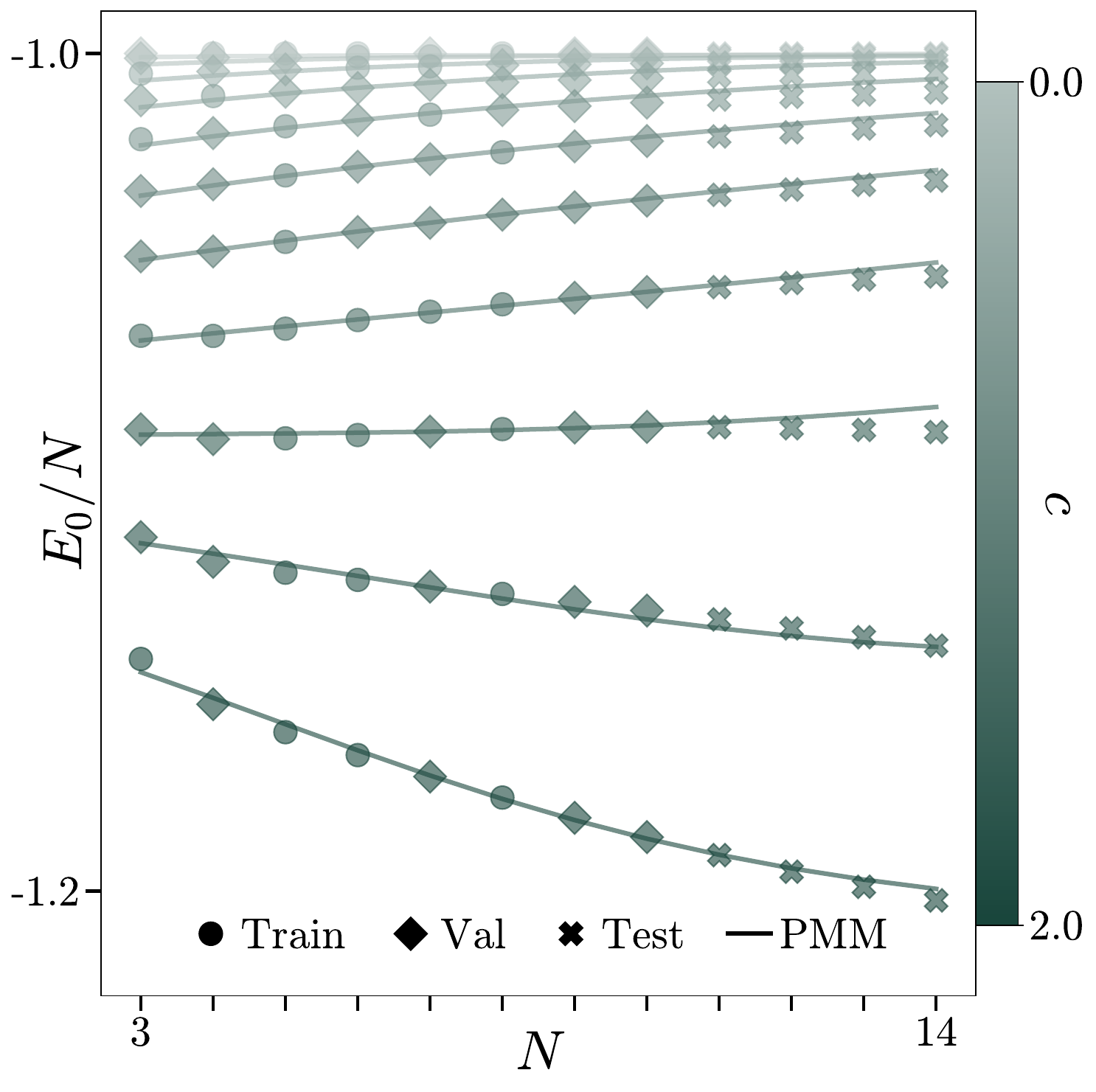}
    \caption[Results of applying a \ac{pmm} emulator to the problem of system-size extrapolation in a quantum many-body system.]{\label{fig:basispmm}High-fidelity (markers) and \ac{pmm}-emulated (lines) ground state energy per particle of the parametric Hamiltonian in Eq.~\eqref{eq:longrange_syssize} as a function of particle number ($L$, horizontal axis) and $x$-coupling ($\gamma_x$, color). Training data is denoted by circles, validation data is denoted by square diamonds, and withheld test data is denoted by crosses. Darker shades indicate a stronger $x$-coupling via larger $\gamma_x$.}
\end{figure}

To concretize this example, we again consider the Hamiltonian described in Appendix~\ref{app:spin}, the same Hamiltonian from the examples in Sections~\ref{sec:ec}~and~\ref{sec:affine}, which describes a 1D chain of $L$ spins with couplings $J_{ij}$ in the influence of a transverse field with strength $B$,
\begin{equation}\label{eq:longrange_syssize}
    H = -\frac{1}{\mathcal{J}} \sum_{i<j}^L J_{ij} \left(\gamma_x \sigma_i^x \sigma_j^x + \gamma_y \sigma_i^y \sigma_j^y\right) - B \sum_i^L \sigma_{i}^z,
\end{equation}
where $\gamma_{x/y}$ are the relative strengths of the $x$ and $y$ couplings, $\sigma_i^u$ is the Pauli spin operator acting on the \ord{i} spin in the $u$ axis, and $\mathcal{J}$ is the Kac normalization factor
\begin{equation}
    \mathcal{J}\equiv \frac{1}{L-1}\sum_{i\neq j}^L J_{ij}.
\end{equation}
Here we take $J_{ij}$ to be the power-law interaction, $J_{ij}=1/\abs{i-j}^{0.5}$, and $\gamma_y$ to be $0$. We take $n_1=5$, $n_2=7$, $l=2$, and $f_1$ and $f_2$ to be the outputs of a simple \ac{mlp} with $4$ hidden layers with $2$, $4$, $4$, and $2$ nodes and softplus\footnote{$\text{softplus}(x)\equiv\ln\left(1+\exp(x)\right)$.} activation functions. This makes the form of the \ac{pmm}
\begin{equation}
    \begin{aligned}
        H\left(\gamma_x;\,L\right) &= \ul{U}^\dagger(L) \ul{H}_\infty\left(\gamma_x\right) \ul{U}(L),\\
        \text{where}~\ul{H}_\infty\left(\gamma_x\right) &= \ul{H}_0 + \gamma_x\ul{H}_1,\\
        \text{and}~\ul{U}(L)&=\exp\left[if_1^\text{(MLP)}(L)\ul{M}_1 + if_2^\text{(MLP)}(L)\ul{M}_2\right](:, 1\mathbin{:}n_1).
    \end{aligned}
\end{equation}
To emulate the ground state energy as a function of both system size $L$ and $x$ coupling strength $\gamma_x$, we train and validate this \ac{pmm} on high-fidelity results for $E_0\left(\gamma_x;\,L\right)$ for $(\gamma_x,L)\in[0,2]\times[3,10]$. This represents high-fidelity Hamiltonian dimensions of $2^3=8$ to $2^{10} = 1024$. We test the emulator by extrapolating to $L=14$, representing a full-space dimensionality $16$ times greater than that of any of the data the emulator was trained on. The results for this example are shown in Fig.~\ref{fig:basispmm}. Note that despite the purely data-driven ``hidden'' model within the emulator depending only on $L$, the emulator is able to accurately capture the correlation between $L$ and $\gamma_x$ in the energy.

Alternatively, it may instead be the case that a term is known but is too complicated or expensive to include in the emulator. Consider the case of a differential equation which is known to be linear and parametrically affine to first order, but the form of the higher-order corrections are unknown
\begin{equation}
    \frac{d v(\mu)}{d t} = (A+\mu B)v(\mu) + F(\mu; v(\mu)),
\end{equation}
where $F(\mu; v(\mu))$ denotes the parametric higher-order (potentially nonlinear) terms whose form is unknown. The form of the \ac{rbm} would then be
\begin{equation}
    \frac{d \ul{v(\mu)}}{dt} = \left(\ul{A} + \mu \ul{B}\right)\ul{v(\mu)} + f\left(\mu; \ul{v(\mu)}\right)
\end{equation}
where $f\left(\mu;\ul{v(\mu)}\right)$ is some unknown parametric function of $\mu$ and $\ul{v(\mu)}$. We again can substitute a numerically parametric universal approximator for $f$ and train its parameters along with the reduced operators. Often, even without knowledge of the form of $F$ (and therefore $f$), properties are known that greatly constrain the form of the universal function. For example, a common case is that the dependence on $\mu$ and $v(\mu)$ is multiplicatively separable,
\begin{equation}
    f\left(\mu, \ul{v(\mu)}\right) = \sum_i g_i(\mu)h_i\left(\ul{v(\mu)}\right),
\end{equation}
for independent universal functions $g_i$ and $h_i$.

Incorporating as many constraints as possible into the form of any data-driven parts of a partially data-driven \ac{pmm} vastly improves the performance of the resulting emulator---especially in low-data regimes and extrapolation. Common choices for universal functions used as parts of a partially data-driven \ac{pmm} are polynomials, \acp{ann}, and purely data-driven \acp{pmm} of the form discussed in Section~\ref{sec:pmm_reg}.

\subsection{Nuclear Equation of State}\label{sec:pmm_eos}

This section details the, at the time of writing, most comprehensive and in-depth application of the \ac{pmm} method to the emulation of a real, non-toy, high-fidelity model of a physical system. The contents of this section largely come from a manuscript written by myself, Kang Yu, Dr.\ Christian Drischler, and Dr.\ Scott Bogner that is available as a preprint at the time of writing in Ref.~\cite{Cook2026}.

The (zero-temperature) nuclear \ac{eos} relates the energy per nucleon of nuclear matter---an infinite, homogeneous, isotropic system of protons and neutrons---to the densities of protons and neutrons in that matter. This in turn describes many thermodynamic properties of nuclear matter, such as the pressure and compressibility. The \ac{eos} is fundamental to understanding the structure of dense astrophysical objects such as neutron stars and connects nuclear physics at the microscopic scale to observables at the stellar scale~\cite{Drischler2021,Sorensen2023,Kumar2024,Chatziioannou2024,Agarwal2026}. Therefore, constraining the \ac{eos} using microscopic, \textit{ab initio} methods has drawn considerable attention and progress using \ac{ceft} paired with a wide variety of many-body methods has continued. One specific many-body method which we will use here, \ac{imsrg}, was recently extended to nuclear matter and the nuclear \ac{eos}~\cite{Udiani2025}. Constraining the nuclear \ac{eos} to match experimental observations requires repeated calculations, most often by sampling the various interaction strengths, for rigorous \ac{uq}. For sophisticated and accurate many-body methods, like \ac{imsrg}, this is computationally intractable as a single evaluation can take significantly more than $24$ hours even in a high-performance computing environment and many tens or hundreds of thousands of evaluations may be necessary. Emulators currently provide the only way of surmounting this computational resource limitation.

In this section we construct and train a comprehensive \ac{pmm} emulator for \ac{imsrg} calculations of the energy of nuclear matter both in the \ac{pnm} ($Z=0$) and \ac{snm} ($N=Z$) regimes as a function of the nuclear density with fully quantified emulator uncertainties. Additional nuclear matter properties such as the pressure, symmetry energy, slope parameter, and incompressibility can be calculated from this. This section will focus only on the creation, calibration, and evaluation of the emulator. For the application of this emulator to the problem of constraining the coupling constants of the quark-mass-dependent three-nucleon interactions in the \ac{emn} model, see Ref.~\cite{Cook2026}.

\subsubsection{The \acl{emn} Interaction}
We first consider the details of the high-fidelity interaction that we aim to emulate. As in Ref.~\cite{Drischler2019}, we use the \ac{emn} interaction developed in Ref.~\cite{Entem2017} with a momentum cutoff $\Lambda=450~\si{\mega\eV}$. The free-space Hamiltonian is affine in parameters known as \acp{lec} and is a chiral expansion with the chiral order denoted by $\nu=\mathset{0,2,3,4}$ corresponding to leading order (\nlo{0}), next-to-leading order (\nlo{1}), next-to-next-to-leading order (\nlo{2}), and next-to-next-to-next-to-leading order (\nlo{3}) respectively\footnote{There is no limit to the order of the chiral expansion in general, but here we only consider up to \nlo{3}.},
\begin{equation}\label{eq:emn}
        H_\text{free}(c, \nu) = \sum_{j=0}^\nu \left[V_0^{(j)} + \sum_{k=1} c_k V_k^{(j)}\right],
\end{equation}
where $c$ is the vector of \acp{lec}, $V_k^{(j)}$ are Hermitian operators, and $V_k^{(1)}\equiv0$ in Weinberg power counting~\cite{Yang2021}. Here we consider emulating only the \nlo{2} three-nucleon \acp{lec} $c=\mathlist{c_D, c_E, D_2, F_2}$, and so for our purposes here the free-space Hamiltonian simplifies to
\begin{equation}\label{eq:emn_3n}
    \begin{alignedat}{3}
        H_\text{free}(c, \nu) =~&V_0^{(0)}&\quad&\text{(\nlo{0})}\\
        &+ \left\{V_0^{(2)}~\text{if}~\nu\geq 2\right\}&\quad&\text{(\nlo{1})}\\
        &+ \left\{V_0^{(3)}+\sum_{k=1}^4 c_k V_k^{(3)}~\text{if}~\nu\geq 3\right\}&\quad&\text{(\nlo{2})}\\
        &+ \left\{V_0^{(4)}+\sum_{k=1}^4 c_k V_k^{(4)}~\text{if}~\nu\geq 4\right\}&\quad&\text{(\nlo{3})}.
    \end{alignedat}
\end{equation}
Additionally, the interaction does not depend on $c_D$ or $c_E$ for \ac{pnm}, in which case $V_{c_D}^{(j)}\equiv 0$ and $V_{c_E}^{(j)}\equiv 0$~\cite{Hebeler2010}.

\subsubsection{\Acl{imsrg}}
We now consider the details of the high-fidelity method used to find the energy of nuclear matter given this interaction and a realization of $(c,\nu)$. For a complete review of \ac{imsrg} and its application to nuclear matter, see Refs.~\cite{Hergert2016b,Hergert2017,Hergert2016,Stroberg2019, Udiani2025, Zhen2025}. For our purposes here, the important aspects are the ``in-medium'' representation of the Hamiltonian and the ``flow''. \ac{imsrg} works directly with the in-medium Hamiltonian as opposed to the free-space Hamiltonian. This is achieved by ``normal-ordering'' the Hamiltonian with respect to a reference state that is specific to the system under consideration. This can be re-interpreted as a projection of the free-space interaction onto the specific system. In \ac{imsrg} calculations of nuclear matter, the basis for the in-medium system is described by the density, $n$, the maximum allowed number of single-particle states, $N_s$, or equivalently the number of closed shells, and the proton fraction, $Y_p$. The density $n$ describes the number of nucleons per unit volume\footnote{We use $n$ for the density here to be consistent with nuclear matter literature. This is not to be confused with the reduced-space dimension notation used in other sections.}. The number of allowed single-particle states $N_s$ is a numerical parameter describing the size of the model space, which in practice cannot be infinite unlike in the physical system. Calculations with larger model spaces are in general more accurate but also more computationally expensive. Finally, $Y_p$ describes the ratio of the mixture of protons and neutrons in the system with $Y_p=0$ denoting \ac{pnm} and $Y_p=0.5$ denoting \ac{snm}. In the basis-projection interpretation, we can write the in-medium Hamiltonian as a parameterized projection of Eq.~\eqref{eq:emn_3n},
\begin{equation}
    H_\text{IM}\left(c,\nu;\,N_s,Y_p,n\right) = U^\dagger\left(N_s,Y_p,n\right)H_\text{free}(c,\nu)U\left(N_s,Y_p,n\right),
\end{equation}
where $U(\cdot)$ is a partial isometry---the semi-unitary operator that induces the projection $U^\dagger(\cdot)U(\cdot)$---parameterized by $N_s$, $Y_p$, and $n$. The \ac{imsrg} method aims to construct a continuous unitary transformation $V(s)$ that renders this Hamiltonian block-diagonal as $s\rightarrow\infty$,
\begin{equation}
    H_\text{IMSRG}(s) = V^\dagger(s)H_\text{IM} V(s) \equiv H_\text{d}(s) + H_\text{od}(s),
\end{equation}
where
\begin{equation}
    \lim_{s\rightarrow\infty} H_\text{od}(s) = 0,
\end{equation}
$H_\text{d}(s)$ is the block-diagonal part of the Hamiltonian, and $H_\text{od}(s)$ is the off-diagonal part. The continuous parameter $s$ is known as the flow parameter and this process of computing $H(s)$ as $s\rightarrow\infty$ is referred to as ``flowing'' the Hamiltonian. In principle, the \ac{imsrg} result at finite $s$ is the exact ground state energy of the in-medium Hamiltonian plus errors which depend on the $s=0$ Hamiltonian (and therefore the control parameters of the Hamiltonian) and decay exponentially in $s$. Denoting the set of Hamiltonian parameters as $X_H\equiv\mathlist{c,\nu;\,N_s,Y_p,n}$, then
\begin{equation}\label{eq:imsrg_e}
    \begin{alignedat}{3}
        E_0^\text{IMSRG}\left(X_H;\, s\right) =~&E_0\left(X_H\right)&\quad&\text{(true ground state energy)}\\
        &+\sum_i a_i\left(X_H\right)\exp\left[-b_i\left(X_H\right)s\right]&\quad&\text{(finite-$s$ errors)},
    \end{alignedat}
\end{equation}
where $a_i(X_H)$ and $b_i(X_H)$ are smooth, non-negative functions which depend only on the Hamiltonian parameters, and $E_0\left(X_H\right)$ is the algebraically lowest eigenvalue of $H_\text{IM}(X_H)$. Crucially, the number and exact form of $a_i(X_H)$ and $b_i(X_H)$ are not known.

\subsubsection{The \acs{pmm}--\acs{imsrg} Emulator}

\begin{figure}
    \centering
    \includegraphics[width=0.9\linewidth]{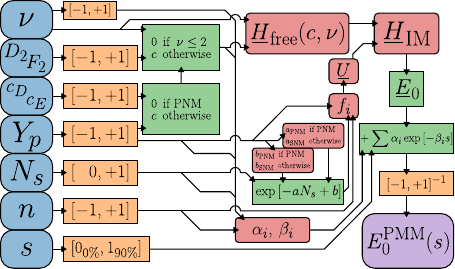}
    \caption[Comprehensive diagram of the \ac{pmm}--\ac{imsrg} emulator.]{\label{fig:pmmimsrg}Comprehensive diagram of the \ac{pmm}--\ac{imsrg} emulator. Inputs are shown on the left (blue boxes) and the predicted energy on the right (purple box). The rescaled chiral order and \acp{lec} feed into the construction of the reduced-space free-space Hamiltonian $\ul{H}_\text{free}$, while the learned projection encodes the dependence on the basis size, proton fraction, and density to form the reduced-space in-medium Hamiltonian $\ul{H}_\text{IM}$. Red-shaded blocks indicate trainable components, including data-driven PMMs that capture exponential convergence with $N_s$ and finite-$s$ corrections. Diagonalization of $\ul{H}_\text{IM}$ produces the $s\rightarrow\infty$ ground-state energy, to which finite-$s$ corrections are added and then rescaled to yield the emulator prediction $\ul{E}_0(s)\equiv\ul{E}_0\left(X_H;\,s\right)\equiv\ul{E}_0\left(c,\nu;\,N_s,Y_p,n;\,s\right)$.}
\end{figure}

From Eq.~\eqref{eq:emn_3n}, Eq.~\eqref{eq:imsrg_e}, and the discussion at the beginning of this section (Section~\ref{sec:pmm_pdd}), we can write the form of the \ac{pmm} emulator for \ac{imsrg} predictions of the energy of nuclear matter as
\begin{equation}\label{eq:pmmimsrg_flow}
    \begin{aligned}
        E_0^\text{PMM}(X_H;s) =~&\ul{E}_0(X_H)\\
                              &+\sum_i^{l_1} \alpha_i(X_H)\exp\left[-\beta_i(X_H)s\right],
    \end{aligned}
\end{equation}
where $\alpha_i(X_H)$ and $\beta_i(X_H)$ are non-negative universal functions which will be trained to approximate the exponentially decreasing error terms and $l_1$ is a hyperparameter determining the number of these universal functions to use, and
\begin{equation}\label{eq:pmmimsrg_im_e}
    \begin{aligned}
        \ul{H}_\text{IM}(X_H)\ul{\Psi}_0(X_H)&=\ul{E}_0(X_H)\ul{\Psi}_0(X_H)\\
        \text{with}~\ul{H}_\text{IM}(X_H)&=\ul{U}^\dagger\left(N_s,Y_p,n\right)\ul{H}_\text{free}(c,\nu)\ul{U}\left(N_s,Y_p,n\right),
    \end{aligned}
\end{equation}
\begin{equation}\label{eq:pmmimsrg_U}
    \ul{U}\left(N_s, Y_p, n\right)=\exp\left[i\sum_{j=1}^{l_2} f_j\left(N_s, Y_p, n\right)\ul{M}_j\right](:,1\mathbin{:}n_1),
\end{equation}
with $\ul{H}_\text{free}\in\Hbb(n_2)$ and $\ul{H}_\text{IM}\in\Hbb(n_1)$. $f_j\left(N_s, Y_p, n\right)$ are again trainable universal functions of which there are $l_2$, and
\begin{equation}\label{eq:pmmimsrg_free}
    \begin{aligned}
        \ul{H}_\text{free}(c,\nu)=~&\ul{V}_0^{(0)}\\
            &+ \left\{\ul{V}_0^{(2)}~\text{if}~\nu\geq 2\right\}\\
            &+ \left\{\ul{V}_0^{(3)}+\sum_{k=1}^4 c_k \ul{V}_k^{(3)}~\text{if}~\nu\geq 3\right\}\\
        &+ \left\{\ul{V}_0^{(4)}+\sum_{k=1}^4 c_k \ul{V}_k^{(4)}~\text{if}~\nu\geq 4\right\}.\end{aligned}
\end{equation}
An additional preprocessing step is required to enforce the physical constraints of the model and its lack of dependence on \acp{lec} at specific $Y_p$ and $\nu$,
\begin{equation}\label{eq:pmmimsrg_preprocess}
    \begin{alignedat}{3}
        c_D&\leftarrow\begin{cases}
            0 &\text{if}~\nu\leq 2~\text{or \acs{pnm}}\\
        c_D &\text{otherwise}\end{cases}&\qquad
        c_E&\leftarrow\begin{cases}
            0 &\text{if}~\nu\leq 2~\text{or \acs{pnm}}\\
        c_E &\text{otherwise}\end{cases}\\
        D_2&\leftarrow\begin{cases}
            0 &\text{if}~\nu\leq 2\\
        D_2 &\text{otherwise}\end{cases}&\qquad
        F_2&\leftarrow\begin{cases}
            0 &\text{if}~\nu\leq 2\\
        F_2 &\text{otherwise}\end{cases}.
    \end{alignedat}
\end{equation}
The complete \ac{pmm}--\ac{imsrg} emulator as described here, with additional data rescaling steps, is summarized diagrammatically in Fig.~\ref{fig:pmmimsrg}. All input features, except for $N_s$ and $s$, are rescaled to the range $[-1,+1]$ via min-max scaling (see Section~\ref{sec:preprocess}). $N_s$ is rescaled to the range $[0,+1]$ using min-max scaling. The flow parameter, $s$, is rescaled so that the minimum value in the training data becomes $0$ and the \ord{90} percentile value becomes $1$. This is similar to the robust scaling method in Section~\ref{sec:preprocess} but with the \ord{0} and \ord{90} percentiles. To help the model identify the expected exponential convergence with $N_s$ through the data-driven functions $f_j$, the value of $N_s$ is rescaled in the model by a trainable exponential decay whose coefficients vary between \ac{pnm} and \ac{snm}. The final step of the model is inverse scaling since the model will be trained on data rescaled to $[-1,+1]$ using min-max scaling. All data-driven functions in the emulator were chosen to be the output of smoothed, data-driven \acp{pmm} as detailed in Section~\ref{sec:pmm_reg} since these were found to perform the best during model selection.

\subsubsection{Data Collection, Cleaning, and Stratified Partitioning}

We collect data for the training, validation, calibration, and testing of the \ac{pmm}--\ac{imsrg} emulator from hundreds of \ac{imsrg} calculations. Each calculation specified by the parameters $X_H\equiv\mathlist{c,\nu;\,N_s,Y_p,n}$ produces an \ac{imsrg} prediction for the energy at several values of $s$: $\mathset{\left(X_H, s_i;\, E_0^\text{IMSRG}(s_i)\right)}$. The Hamiltonian parameters for the calculations uniformly span a broad range of \ac{lec} values and densities and cover a wide range of model space sizes, $N_s$, at all chiral orders from \nlo{0} to \nlo{3}, for both \ac{pnm} and \ac{snm}. For each \ac{imsrg} calculation, the flow continued until either convergence---determined by the criteria $\Delta(s)\equiv\abs{(\Delta E_2(s) + \Delta E_3(s))/E(s)}$ where $E(s)$ is the current \ac{imsrg} prediction for the energy at the given $s$ value, $\Delta E_2(s)$ and $\Delta E_3(s)$ are the second- and third-order many-body perturbation theory energy corrections of the \ac{imsrg}-evolved Hamiltonian, respectively, and $\varepsilon$ is some tolerance---or divergence, determined by the criteria $\norm{H_\text{od}(s+ds)}\geq\norm{H_\text{od}(s)}$. Any \ac{imsrg} results which diverged before $s=5$ are discarded entirely; otherwise only values after the flow diverged are discarded. Any results with $\Delta(s)\geq0.1~\si{\mega\eV}$ were discarded as they are considered to not be converged enough to be informative. Finally, de-duplication is performed---retaining only a single \ac{imsrg} flow for each unique set of $(N_s, Y_p, n)$ for \nlo{0} and \nlo{1}, since the \acp{lec} considered only enter at \nlo{2} and above.

\begin{figure}
    \centering
    \includegraphics[width=0.8\linewidth,trim={0 0 0 1.25cm},clip]{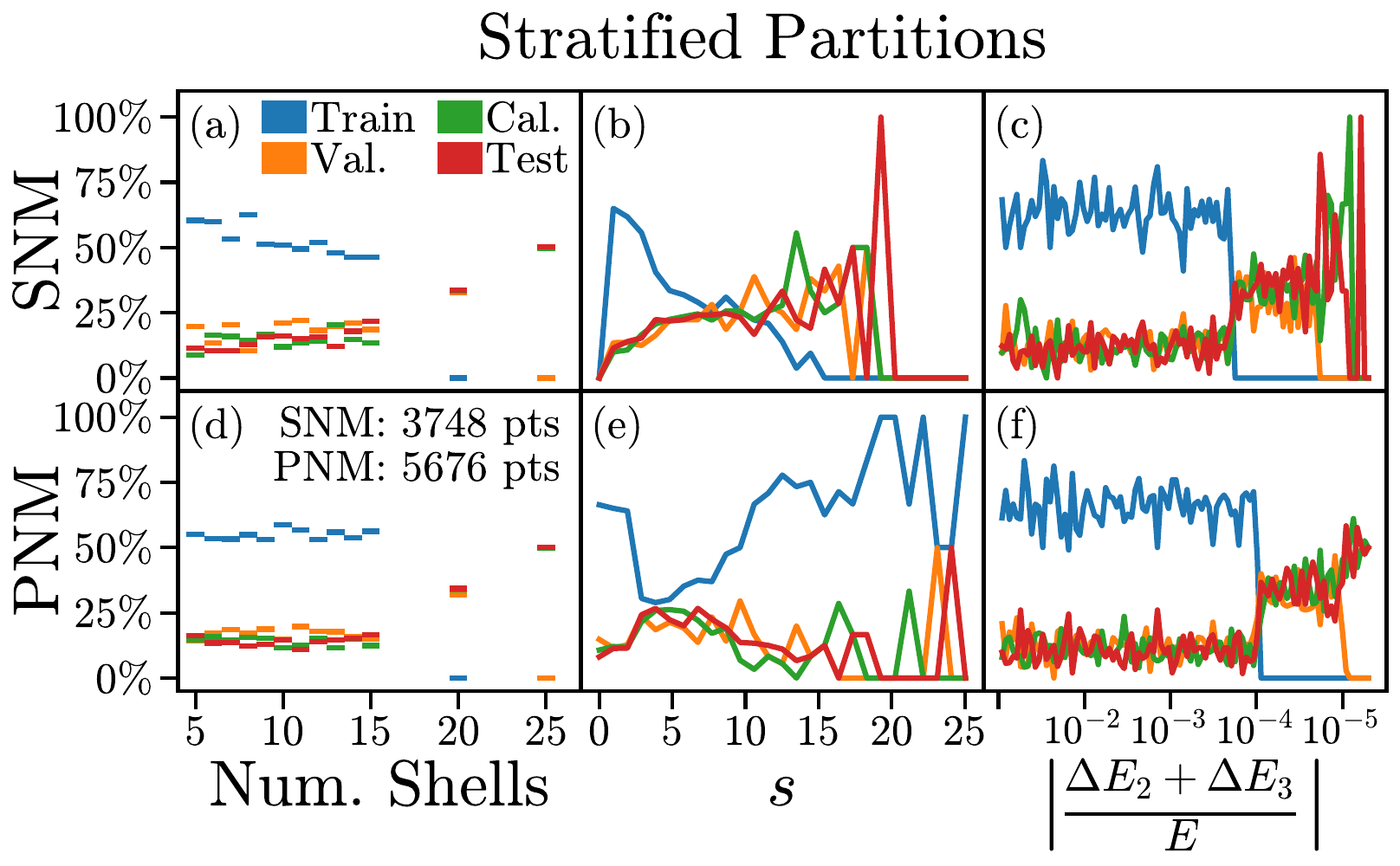}
    \caption[Distribution of training, validation, calibration, and testing data as a function of strata-determining features for the \ac{pmm}--\ac{imsrg} emulator.]{\label{fig:imsrg_split}The stratified partitioning of the \ac{imsrg} samples used for training, validating, calibrating, and testing the \ac{pmm}--\ac{imsrg} emulator. The training data (blue) is taken only from relatively unconverged samples with $15$ or fewer closed shells. Validation data (orange) are drawn from up to $20$ closed shells and include nearly converged samples. No fully converged samples or data from the largest number of closed shells, $25$, are present in the training or validation sets. The calibration (green) and testing (red) datasets span the full range of available \ac{imsrg} samples equally. All four datasets are entirely disjoint. Due to the comparative difficulty in obtaining \ac{imsrg} results for \ac{snm}, the number of \ac{snm} samples ($3748$) is roughly two-thirds that of \ac{pnm} samples ($5676$). The number of closed shells is determined by the number of single-particle states and is shown here instead of $N_s$ to allow better visualization between \ac{snm} and \ac{pnm}~\cite{Lietz2017}. Note that $s$ was not used to partition the data as it does not perfectly correlate with convergence.}
\end{figure}

The goal is for the emulator to not only reproduce \ac{imsrg} calculations but be able to accurately extrapolate in model size $N_s$ and flow $s$ with fully calibrated uncertainty estimates. To accurately assess real-world model performance, we partition the \ac{imsrg} data such that the test set is disproportionately from more-converged (generally larger $s$) results with larger $N_s$. Similarly, to encourage the training process to prefer models with more robust extrapolation, we partition the remaining data such that the validation set favors more-converged results with larger $N_s$. To satisfy the exchangeability requirement of our chosen \ac{uq} method (see Section~\ref{sec:conformal}), we ensure that the calibration set is partitioned equivalently to the test set. All data not yet allocated to the test, validation, or calibration sets are allocated to the training set. The final partitioning of the data is shown in Fig.~\ref{fig:imsrg_split}. The top row of Fig.~\ref{fig:imsrg_split} shows the percentage of the data in each partition for \ac{snm} calculations while the bottom row shows the same for \ac{pnm}. The three columns correspond, from left to right, the number of closed shells determined by the number of single particle states~\cite{Lietz2017}, the flow parameter, and the perturbative convergence estimate. At each point on the $x$-axis of Fig.~\ref{fig:imsrg_split}, the percentages across the four data partitions sum to $100\%$.

\subsubsection{Training, Model Selection, and Calibration}

The implementation, training, calibration, and deployment of the emulator is accomplished through the open-source \pypackage{pyPMM} package\footnote{I am currently both the lead and sole developer of \pypackage{pyPMM}.}~\cite{pyPMM2026}. We train the emulator via gradient descent as discussed in Section~\ref{sec:training}. The loss function is an annealed average of the \ac{wmse} and the \ac{wvar} where the weights were inverse class frequencies. Let
\begin{equation}
\Delta\mleft(X_H;\,s\mright)\equiv E^\text{IMSRG}_0\mleft(X_H;\,s\mright) - E^\text{PMM}_0\mleft(X_H;\,s\mright)
\end{equation}
denote the prediction error. Then, the \ac{wmse} over the training set $T$ is given by
\begin{equation}
    \text{wMSE}(T)\equiv\mean\limits_{\left(X_H;\,s\right)\in T} w(X_H)\Delta\mleft(X_H;\,s\mright)^2,
\end{equation}
where $w(X_H)$ is the inverse of the proportion of the training set which is represented by the specific $\nu$, $Y_p$, and $N_s$ values in $X_H$,
\begin{equation}
\hspace{-0.5em}w(X_H) \equiv \frac{\abs{T}}{\abs{\left\{X_H^\prime\in T \,\middle| \,\left(\nu,Y_p,N_s\right)=\left(\nu^\prime,Y_p^\prime,N_s^\prime\right)\right\}}}.
\end{equation}
This weighting ensures that data taken at a certain chiral order, proton fraction, or number of allowed single particle states do not dominate the model training simply due to over- or under-representation in the training set. Similarly, the \ac{wvar} is given by
\begin{equation}
    \text{wVAR}(T)\equiv\mean\limits_{\left(X_H;s\right)\in T}w(X_H)\left[\Delta\mleft(X_H;\,s\mright) - \overline{\Delta}_w\right]^2,
\end{equation}
where $\overline{\Delta}_w$ is the weighted mean of the prediction error given by
\begin{equation}
\overline{\Delta}_w\equiv\frac{\sum\limits_{\left(X_H;s\right)\in T} w(X_H) \Delta\mleft(X_H;\,s\mright)}{\sum\limits_{\left(X_H;s\right)\in T} w(X_H)}.
\end{equation}
Note that only $\overline{\Delta}_w$ is normalized by the sum of the weights. This ensures that, during mini-batch gradient descent, particular mini-batches with larger---or smaller---than average weights do not disproportionately influence the emulator's parameters. During mini-batch training, the training set in all terms in the loss function---but not in the definition of the weights---is replaced by the mini-batch. The complete loss function used during training at Adam epoch $t$ is then
\begin{equation}
    \frac{\xi\text{wMSE}(T) + \gamma(t) \text{wVAR}(T)}{\xi + \gamma(t)},
\end{equation}
where $\xi$ is a hyperparameter weighting the \ac{wmse}, and $\gamma(t)\in[0, 1]$ is the annealing given by
\begin{equation}
\gamma(t) = \clipfunc\mleft(\frac{t - t_\text{wMSE}}{t_\text{transition}}, 0, 1\mright),
\end{equation}
where $t_\text{wMSE}$ is a hyperparameter that determines how many epochs of pure \ac{wmse} loss ($\gamma(t) = 0$) to train the model with before gradually, at a rate determined by the hyperparameter $t_\text{transition}$, transitioning to a loss composed of $\xi$ parts \ac{wmse} and one part \ac{wvar} ($\gamma(t) = 1$). This choice of loss function allows training to quickly converge to an accurate but imprecise model before gradually encouraging an accurate and precise model. The validation loss, used both for quantifying training progress and comparing candidate models, is quantified by the top \ord{95} percentile of the absolute error on the validation set. This validation loss favors models with better worst-case performance over models that have good average performance but exhibit long tails in their error distributions.

The architecture (such as the choice for the universal functions), hyperparameters (such as the reduced-space dimensions $n_1$ and $n_2$), and loss function were chosen by a standard partial grid search. Several candidate models with different hyperparameters and data-driven function choices were trained and compared using the validation loss. The model with the best validation loss across all trials was selected as the final emulator. A small amount of additional validation data was then created to manually verify the performance of the selected model. This additional data is not included in any of the aforementioned datasets. This model selection procedure led to $n_1=7$, $n_2=17$, and data-driven \acp{pmm} with mild smoothing (see Section~\ref{sec:pmm_reg}) for $f_j$, $\alpha_i$, and $\beta_i$. This yields a \ac{pmm} emulator with fewer than $\num{30000}$ single-precision floating-point numbers.

We use the calibration dataset for uncertainty quantification and stratified bias corrections. Utilizing split-conformal predictions (Section~\ref{sec:conformal}) and the uncertainty heuristic described in Section~\ref{sec:pmm_uh}, we are able to turn the emulator's point-wise predictions into rigorous confidence intervals for any desired significance level. The stratified bias corrections are fixed additive corrections to the model independent of the trainable parameters. These corrections ensure that the median error over the calibration set is exactly $0$ for each of the four strata specified by the distinct values of $\nu$ and $Y_p$. While these corrections are beneficial, they are likely unnecessary as the largest relative correction---which occurred for \ac{snm} at \nlo{3}---was less than $0.037\%$ of the range of the training data\footnote{In the rescaled units where the training data energies belong to the range $[-1,1]$, the largest correction was less than $7.39\times 10^{-4}$.}\footnote{Once calculated, choosing whether or not to keep these corrections based on their values would constitute data leakage and introduce researcher bias---see Section~\ref{sec:gen_cal_eval}---and thus these corrections remain part of the model despite their negligible contributions.}.

\subsubsection{Results}

\begin{figure}
    \centering
    \includegraphics[width=0.8\linewidth]{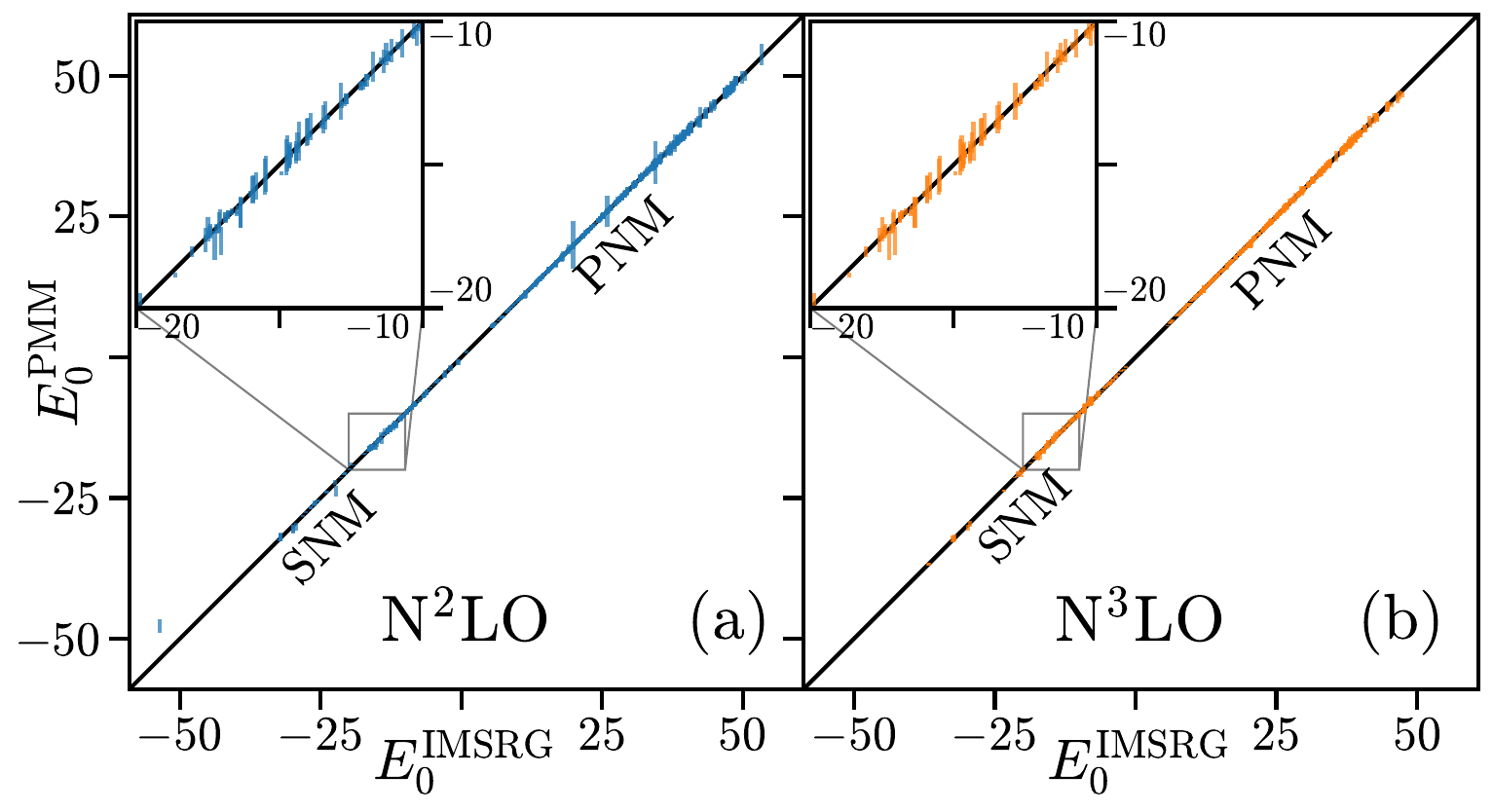}
    \caption[Comparison of the \ac{pmm}-predicted energies and \ac{imsrg} calculations on withheld test data.]{\label{fig:imsrgpmm_acc}Comparison of \ac{pmm}-predicted energies at the $95\%$ confidence level and \ac{imsrg} calculations for the withheld test set for both \ac{pnm} and \ac{snm} as all degrees of freedom, including \acp{lec} and density, are varied for both \nlo{2} (panel a) and \nlo{3} (panel b). Insets show additional detail around the empirical saturation energy. All values are in units of $\si{\mega\eV}$.}
\end{figure}

In this section we focus on the performance of the emulator in terms of accuracy and computational speedup. For further discussions of the physics applications of this emulator, see Ref.~\cite{Cook2026}. Figure~\ref{fig:imsrgpmm_acc} shows the accuracy of the emulator on the withheld test data at \nlo{2} and \nlo{3}. The test data includes both \ac{snm} and \ac{pnm}, varies all considered \acp{lec} ($c_D$, $c_E$, $D_2$, $F_2$) and the nuclear density, and includes data of the most converged \ac{imsrg} calculations for the largest model sizes---neither of the final two were represented in the training or validation data. Each point in Fig.~\ref{fig:imsrgpmm_acc} shows the $95\%$ confidence interval of the emulator-predicted energy versus the \ac{imsrg} calculation, demonstrating both high accuracy and precision.

\begin{figure}
    \centering
    \includegraphics[width=0.8\linewidth]{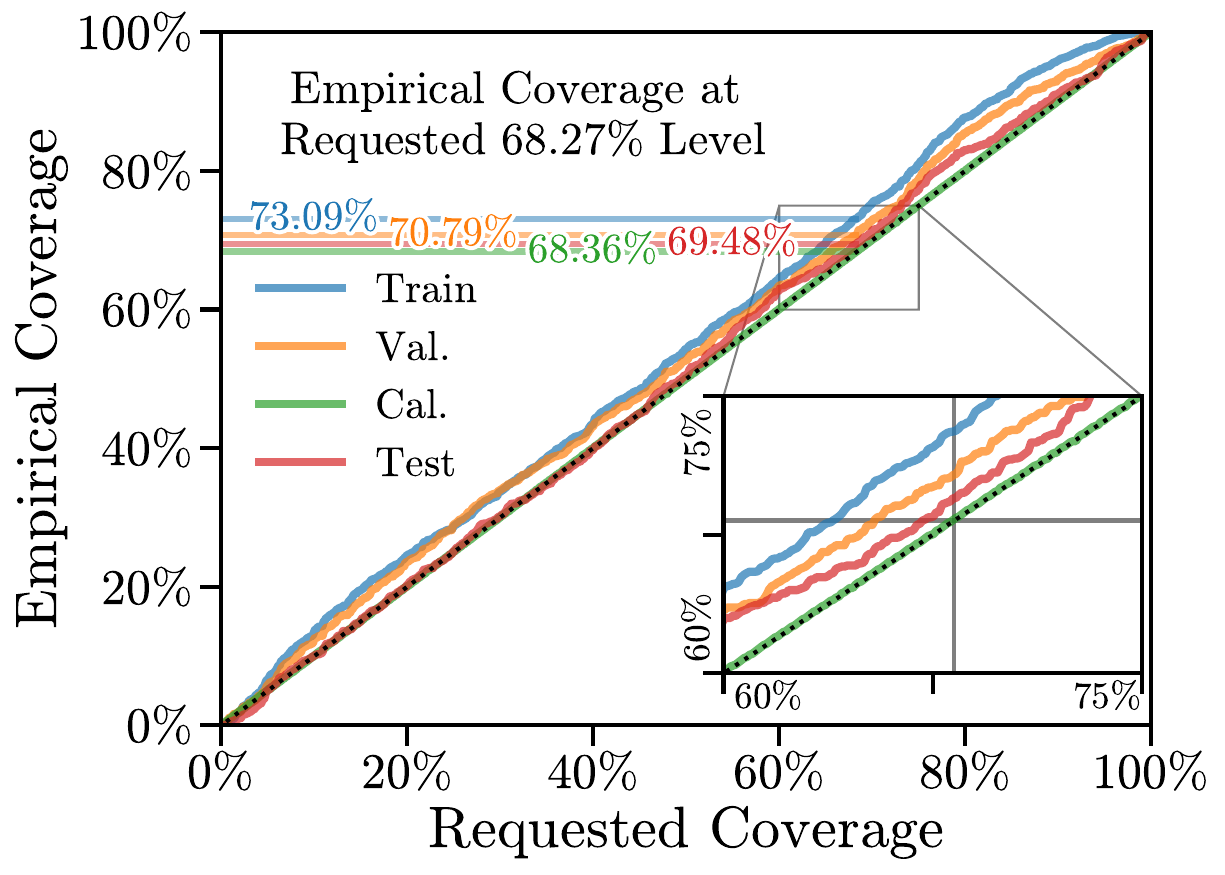}
    \caption[Empirical versus requested coverage for the confidence intervals produced by the \ac{uq} of the \ac{pmm}--\ac{imsrg} emulator.]{\label{fig:imsrgpmm_coverage}Actual coverage versus requested confidence level for the \ac{pmm} emulator on each of the four data partitions. The observed coverages at the requested $68.27\%$ level are shown as horizontal lines, and the coverage range from $60$ to $75\%$ is magnified in the inset to illustrate the small deviations from the ideal coverage.}
\end{figure}

We can quantify the trustworthiness of the \ac{uq} for this emulator by comparing the requested confidence level to the empirically observed coverage over the withheld test data. This is shown for all requested levels in Fig.~\ref{fig:imsrgpmm_coverage}. These results show that the emulator's \ac{uq} produces empirical coverages that track closely with the requested coverage. Importantly, for the withheld test data at requested levels above $10\%$, the maximum under-coverage is $0.55\%$ and the maximum over-coverage is $3.45\%$. The \ac{uq} is biased slightly toward over-coverage, yielding slightly overly cautious uncertainty estimates---this is preferable to an overly confident model.

This emulator provides a computational speedup of about five orders of magnitude\footnote{\acs{imsrg}: up to $2~\text{days}$ vs.\ \acs{pmm}: $<1~\si{\sec}$} compared to direct \ac{imsrg} calculations, enabling studies of the \ac{eos} such as Bayesian model calibration that otherwise would be intractable. Beyond raw speed, the emulator requires about six orders of magnitude less energy to perform a single prediction\footnote{\acs{imsrg}: $\sim1~\si{\kWh}$ vs.\ \acs{pmm}: $<10^{-6}~\si{\kWh}$} and $300$ times less memory to perform predictions\footnote{\acs{imsrg}: up to $600~\si{\giga\byte}$ vs.\ \acs{pmm}: $<2~\si{\giga\byte}$}. Predictions can be made with the emulator on a low-power consumer laptop, compared to the expensive high-performance computing server required for \ac{imsrg} calculations. Even the offline costs---including those of time, electricity, and hardware---associated with training the model are reasonable and less than what was required to generate the \ac{imsrg} data\footnote{\acs{imsrg}: more than $1$ week vs.\ \acs{pmm}: $1$ day}\footnote{\acs{imsrg}: $\sim 1~\si{\MWh}$ vs.\ \acs{pmm}: $<6~\si{\kWh}$}\footnote{\acs{imsrg}: high-performance computing server vs.\ \acs{pmm}: desktop computer with discrete \ac{gpu}}.

\begin{figure}
    \centering
\includegraphics[width=0.7\linewidth]{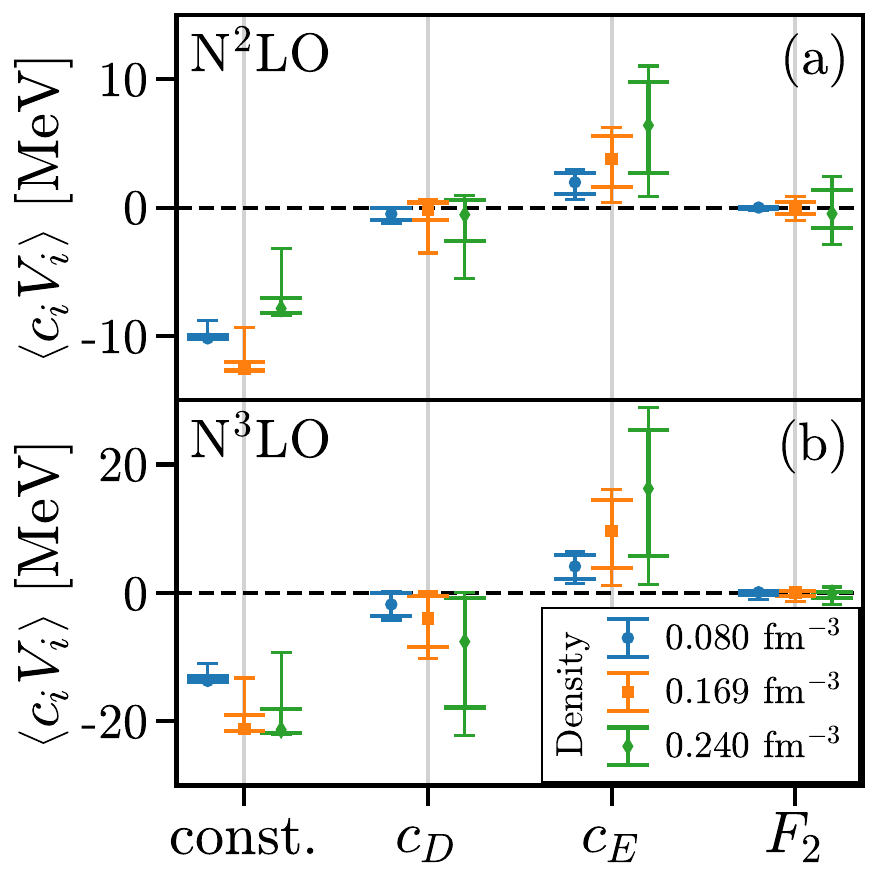}
    \caption[\ac{pmm}--\ac{imsrg} emulated expectation values for individual terms in the emulated Hamiltonian.]{Median expectation values of each term in the emulated Hamiltonian with $68\%$ and $95\%$ credible intervals from posterior \ac{lec} samples (see Ref.~\cite{Cook2026}) at \nlo{2} (panel a) and \nlo{3} (panel b) at three different densities for \ac{snm}. The labels on the horizontal axis indicate the individual $c_iV_i$ term where ``$\text{const.}$'' denotes the fixed $V_0$ term in the emulator. It is important to emphasize again that the emulator was never provided with any high-fidelity data for the individual contributions of each term.}
\label{fig:imsrgpmm_exp}
\end{figure}

Finally, the grounding of the \ac{pmm} method to the \ac{erbm} of the original system allows us to probe and extract individual operator contributions to the energy of the system from Eqs.~\eqref{eq:emn}~and~\eqref{eq:pmmimsrg_free}. Despite the training data not containing any information about the individual contributions, the emulator is still able to correctly identify effective reduced-space representations of each individual term. This is not something that would be natively possible with purely data-driven or even many hybrid methods. To demonstrate this capability, after posterior distributions of the \acp{lec} are fit to empirical properties of nuclear matter as detailed in Ref.~\cite{Cook2026}, we compute the ground-state energy contributions of the $c_D$, $c_E$, and $F_2$ terms as well as the remaining part of the Hamiltonian, denoted $V_0$, which does not depend on any of the emulated \acp{lec}\footnote{Here, $D_2$ is held constant at $0$ and so the associated terms have no contribution.}. The contributions are calculated as the expectation value of the respective emulated operators in the emulated ground state. Figure~\ref{fig:imsrgpmm_exp} shows these expectation values at both \nlo{2} and \nlo{3} at the minimum ($0.08~\si{\per\cubic\femto\meter}$) and maximum ($0.24~\si{\per\cubic\femto\meter}$) densities present in the training data as well as the emulator-predicted saturation density\footnote{This is the median of the predicted saturation density when sampling from \ac{lec} posterior distributions resulting from the aforementioned fitting to empirical properties. See Ref.~\cite{Cook2026} for details.} ($0.169~\si{\per\cubic\femto\meter}$). These results are consistent with those in Ref.~\cite{Vernik2025}, in which the contributions from the $F_2$ term are significantly smaller than those from the $c_D$, $c_E$, and $V_0$ terms.

\section{Community Applications}

Since the posting of the original preprint manuscript for \acp{pmm}, several research groups have successfully applied \ac{pmm}-based emulators to a wide range of problems~\cite{Jin2025,Armstrong2026,Munoz2026,Yu2026,Curry2025,Curry2026,Armstrong2025,Somasundaram2025Scarce,Somasundaram2025,Reed2024}. In particular, just like the work associated with Section~\ref{sec:pmm_eos} (Ref.~\cite{Cook2026}), many groups have applied \ac{pmm}-based emulators to the problem of constraining microscopic nuclear forces from various observables such as (in no particular order)
\begin{itemize}
    \item quantum Monte Carlo calculations of the binding energy and Gamow-Tellow matrix element of ${}^3\text{He}$ and charge radius of ${}^4\text{He}$ (Ref.~\cite{Curry2025}),
    \item valence-space \ac{imsrg} calculations of the binding energy, charge radius, magnetic dipole moment, and electric quadrupole moment of light- to medium-mass nuclei (Ref.~\cite{Munoz2026}),
    \item and auxiliary-field diffusion Monte Carlo calculations of the nuclear matter \ac{eos} and neutron star mass, radius, and tidal deformability from the Tolman-Oppenheimer-Volkoff equations\footnote{The emulators for the Tolman-Oppenheimer-Volkoff equations in Refs.~\cite{Somasundaram2025, Armstrong2026} used \acp{ann}.} (Refs.~\cite{Somasundaram2025, Armstrong2026}).
\end{itemize}

\section{Summary}

    This chapter introduced \acfp{pmm} as an \acf{irbm} for emulation with additional applications in general purely data-driven machine learning. A comprehensive framework for \acp{pmm} was established, built on the central idea of allowing both the subspace and the basis of \acfp{erbm} to be implicit and instead fitting parameterized versions of reduced operators directly to available data. Details for training \acp{pmm} as well as a guide to effective parameterization were provided. Applications of \acp{pmm} to the emulation of a variety of problems including standard parametric eigenproblems, linear partial differential equations, nonlinear eigenproblems, and nonlinear partial differential equations were presented. A final application demonstrating the partially data-driven-\ac{pmm} emulation of \acf{imsrg} calculations of the nuclear \acf{eos} showed the efficacy of \acp{pmm} in a real-world large-scale scientific application.

\begin{figure}
    \centering
    \setlength\tabcolsep{5pt}
    \begin{tabularx}{\linewidth}{Xccccccccccc}
        \allemsummary
        \specialrule{\lightrulewidth}{\aboverulesep}{0pt}
        \rowcolor{gray!20} \colorstrut%
        \hyperref[ch:pmm]{\textbf{\Aclp{pmm}}} & \cmark & \cmark & \cmark & \cmark & \cmark & \cmark & \cmark & \cmark & \cmark & \textbf{A} & \sumE\\
        \specialrule{\heavyrulewidth}{0pt}{\belowrulesep}
    \end{tabularx}
    \captionof{table}[Summary of the properties of emulator methods covered in Chapter~\ref{ch:emu} as well as \acp{pmm}.]{\label{tbl:emupmmsummary}Same as Table~\ref{tbl:emusummary} from the end of Chapter~\ref{ch:emu}, but with the properties of \aclp{pmm} added.}%
\end{figure}%

\chapter{Uncertainty Quantification}\label{ch:uq}
\acresetall
A central motivation for emulation is quantifying the uncertainties of high-fidelity models, a process that without emulation would often be impractical or impossible due to the large number of model evaluations. For accurate and effective \ac{uq} of high-fidelity models via emulation, the emulator itself must be not only accurate but also have well-quantified uncertainties. Beyond this application, the trustworthiness of any machine learning method is a key---though sometimes overlooked---aspect that incorporates both accuracy and (un)certainty. Well-calibrated \ac{uq} can be the difference between an impressive method and a useful method.
In this chapter, we briefly discuss two approaches to \ac{uq} in machine learning and emulation. The first method is often the ``go-to'' due to its relative simplicity. The second is particularly well-suited for \ac{uq} in \ac{pmm}-based emulators.

\section{Bootstrap Aggregation}

Bootstrap aggregation, or bagging\footnote{From ``\underline{\textbf{b}}ootstrap \underline{\textbf{agg}}regat\underline{\textbf{ing}}.''~\cite{Breiman1996}}, is a specific case of the more general bootstrap and ensemble methods~\cite{Breiman1996,Efron1979}. This method is used to create an ensemble of models whose average predictions are more accurate than any individual constituent model while additionally providing uncertainty estimates via statistics over the ensemble predictions. In bootstrap aggregation, each model in the ensemble has the same architecture, training procedure, and loss function but is trained on a different random subset of the training data~\cite{Goodfellow2016}.

The core motivation for bagging is that if the error of an individual model is a random distribution centered near zero, and the errors of the individual models are uncorrelated, then on average these errors will cancel. In the worst possible case, where the error of each model is perfectly correlated, the ensemble will have the same error as any individual model, and so the bagging procedure will neither improve nor worsen the quality of the predictions. However, correlated errors will lead to overconfident uncertainty estimates, which is a common failure mode of bagging. In practice, the errors of individual models are often correlated to some degree. Beyond \ac{uq}, bagging is often used as a form of regularization to reduce overfitting and improve generalization.

The process of bootstrap aggregation is as follows~\cite{Breiman1996,Goodfellow2016}:

\begin{enumerate}
    \item Given $m$ training samples, choose a fixed model architecture, training procedure, loss function, and number of models $K$ to include in the ensemble.
    \item For each model $k \in \{1, 2, \ldots, K\}$:
    \begin{enumerate}
        \item Randomly sample $m$ training samples with replacement from the original training set to create a new training set for model $k$. This set will likely contain duplicates and also omit some of the original training samples.
        \item Train model $k$ on this new training set using an independent random initialization of the model parameters (if applicable).
    \end{enumerate}
    \item After training all $K$ models, the ensemble prediction for a new input is some aggregation of the predictions of the individual models, such as the mean or median\footnote{The aggregation method used is dependent on the type of prediction. For example the median or mean is common in regression tasks while the mode is common in classification tasks.} of the individual predictions.
    \item The uncertainty of the ensemble prediction can be estimated by computing the variance or standard deviation of the individual model predictions.
\end{enumerate}

While both the training and prediction steps of bagging can be parallelized, the increased computational cost of training many models can be prohibitive. In the context of emulation, bagging also diminishes the interpretability of the emulator, as the ensemble predictions are no longer directly tied to the underlying model architecture.

\section{Conformal Prediction}\label{sec:conformal}

Conformal prediction\footnote{Also called conformal inference.} is a framework that turns a pointwise prediction model into a set or interval prediction model with distribution-free finite-sample coverage guarantees that rely only on the exchangeability of data. Here, we consider only the core principle and process of split conformal prediction for regression. See Refs.~\cite{Angelopoulos2023,Hulsman2022} for comprehensive introduction of conformal prediction.

The first core idea of split conformal prediction is that statistics of the predictions on a calibration set that is representative of the data the model will see in deployment should be nearly equal to the statistics of the predictions on the test set. A simplified example of this idea is that if the model achieves sub-$1\%$ error on the calibration set, then it should also achieve sub-$1\%$ error on the test set if the calibration set is representative of the test set. This division of calibration and test sets is where the ``split'' in split conformal prediction comes from. This relies on the assumption that the calibration and test sets are drawn from the same distribution and samples in each are therefore exchangeable.

The second, and more important, core principle is that of ``conformalization.'' This is the process of transforming any heuristic uncertainty measure into a rigorous uncertainty measure~\cite{Angelopoulos2023}. The heuristic uncertainty measure, called the uncertainty heuristic $\mathcal{U}(\cdot)$, is a scalar-valued function that need only be monotonically related to the true uncertainty of the model. That is, the value of $\mathcal{U}(\cdot)$ should be larger for inputs where the model is less certain. The calibration set is used to calibrate---or conformalize---this heuristic into rigorous prediction intervals for any miscoverage rate $\alpha \in (0, 1)$\footnote{The miscoverage rate $\alpha$ is the probability that the true value is not contained in the prediction interval. The confidence level is $1 - \alpha$.}. Importantly, the conformalization process guarantees that the prediction intervals will have at least $1 - \alpha$ coverage on the calibration set (and thus the test set if exchangeability holds) regardless of the underlying model or uncertainty heuristic~\cite{Angelopoulos2023,Vovk2005}.

Denoting the input features as $X$, the true output as $\hat{Y}$, and the trained model prediction as $Y(X)$, the split conformal prediction process is broadly as follows~\cite{Angelopoulos2023}:
\begin{enumerate}
    \item Devise the uncertainty heuristic $\mathcal{U}(X)$ which is a scalar-valued function that encodes some notion of uncertainty of the model prediction $Y(X)$.
    \item Calculate the \textit{score}---the residual divided by the uncertainty heuristic---for each sample in the calibration set
    \begin{equation}
        \mathcal{S}(X_i) = \frac{\abs{Y(X_i) - \hat{Y}_i}}{\mathcal{U}(X_i)}.
    \end{equation}
\item For a desired miscoverage rate $\alpha$ (or confidence level $1 - \alpha$), define $\hat{q}$ as the
    \begin{equation}
        \frac{\lceil (1-\alpha)(n+1) \rceil}{n}
    \end{equation}
        quantile of the calibration set scores $\mathset{\mathcal{S}(X_i)}$ where $n$ is the number of samples in the calibration set and $\lceil \cdot \rceil$ is the ceiling function.
    \item \label{itm:conformal_interval} $\hat{q}$ is the multiplicative correction to the uncertainty heuristic which ensures that the prediction interval \begin{equation}
        \left[Y(X) - \hat{q}\,\mathcal{U}(X),\; Y(X) + \hat{q}\,\mathcal{U}(X)\right]
    \end{equation}
    has at least $1 - \alpha$ coverage for any new input $X$.
\end{enumerate}
Note that all of the above steps except for the final step can be performed as part of the offline phase. Only step~\ref{itm:conformal_interval} must be performed online for each new input $X$ and so only requires the additional cost of the uncertainty heuristic evaluation and the multiplication by $\hat{q}$. For multi-dimensional outputs, the uncertainty heuristic is typically also multi-dimensional---matching the output dimension---and this process is applied element-wise to each output.

While the coverage of the intervals is guaranteed, the tightness of the intervals is reliant on the ability of the uncertainty heuristic to accurately reflect the true uncertainty of the model. Consider the case of a completely uninformative uncertainty heuristic. In this case $\hat{q}$ would be so large that every prediction interval would encompass the entire range of possible outputs. These intervals would have perfect coverage, as guaranteed, but would be useless for \ac{uq}. It is therefore important to devise an uncertainty heuristic that accounts for all sources of uncertainty in the model prediction.

\subsection{Uncertainty Heuristic for Parametric Matrix Models}\label{sec:pmm_uh}

The analyticity and adaptive smoothness (see Section~\ref{sec:smoothing}) of \ac{erbm}-based and data-driven \acp{pmm} make the incorporation of all sources of sensitivity and uncertainty into a single uncertainty heuristic relatively straightforward. For any \ac{pmm}, we incorporate two sources of model sensitivity and one source of data uncertainty into the uncertainty heuristic. Denoting the predictions of the trained \ac{pmm} as $Y(X)$, we include
\begin{enumerate}
    \item \label{itm:unc_params} The sensitivity of the trained \ac{pmm} to perturbations in the trainable parameters,
        \begin{equation}
            S_\theta(X) = \frac{1}{\ell(\theta)} \norm{\left.\frac{\partial Y}{\partial \theta}\right|_X \odot \theta}_2^2,
        \end{equation}
        where $\odot$ is the elementwise product, $\theta$ is the vector of all trainable parameters, and $\ell(\theta)$ is the number of trainable parameters. This term accounts for the relative sensitivity of the model to each of its trainable parameters.
    \item \label{itm:unc_input} The sensitivity of the trained \ac{pmm} to perturbations in the continuous input features,
        \begin{equation}
            S_X(X) = \frac{1}{\ell(X_\text{cont})} \norm{\left.\frac{\partial Y}{\partial X_\text{cont}}\right|_X \odot \delta(X_\text{cont})}_2^2,
        \end{equation}
        where $X_\text{cont}$ represents the subset of input features that are continuous, $\ell(X_\text{cont})$ is the number of continuous input features, and $\delta(X_\text{cont})$ is a vector of the characteristic scales of each continuous input feature, which is taken to be the interquartile range of each feature in the calibration set. This term accounts for the relative sensitivity of the model to each of its continuous input features.
    \item \label{itm:unc_data} The dissimilarity of the input $X$ to the training data, $S_d(X)$. The dissimilarity is defined by the mean unweighted Gower's distance between $X$ and all points in the training set within a distance threshold equal to the median unweighted Gower's distance between all pairs of points in the training set~\cite{Gower1971}. For the characteristic range of each continuous feature in the calculation of Gower's distance, we again use the interquartile range. Gower's distance is a similarity metric that is useful for both continuous and discrete features simultaneously. The distance threshold drastically reduces the computational cost of making predictions for new inputs and simultaneously avoids the influence of outliers. Any input feature that has an infinite or undefined value is ignored in the calculation of Gower's distance. This term accounts for the uncertainty due to the model being applied to inputs that are dissimilar to the training data.
\end{enumerate}
The uncertainty heuristic is then defined as the normalized sum of these three terms. Each term is normalized by its \ac{mad} over the calibration set,
\begin{equation}
    \mad\limits_C(S_i) = \median\limits_{X \in C} \abs{S_i(X) - \median\limits_{X \in C} S_i(X)},
\end{equation}
where $C$ is the calibration set, to ensure that each term contributes equally to the uncertainty heuristic. The final uncertainty heuristic is then
\begin{equation}
    \mathcal{U}(X) = \frac{S_\theta(X)}{\mad\limits_C(S_\theta)} + \frac{S_X(X)}{\mad\limits_C(S_X)} + \frac{S_d(X)}{\mad\limits_C(S_d)}.
\end{equation}

By computing the \ac{mad} of each term over the calibration set in the offline phase, as well as building a data structure for efficient nearest neighbor search for the dissimilarity term\footnote{Such as a $k$-d tree as described in Ref.~\cite{Maneewongvatana1999} and provided in the \pypackage{SciPy} library~\cite{SciPy}.}, the online cost of evaluating the uncertainty heuristic is determined by the cost of evaluating the derivatives of the \ac{pmm} which are computed via \ac{ad}.

This uncertainty heuristic produces exceedingly well-calibrated prediction intervals for \ac{pmm}-based emulators, as demonstrated in Section~\ref{sec:pmm_eos}. Along with the native interpretability of \ac{pmm}-based emulators, this \ac{uq} solidifies the trustworthiness of \ac{pmm}-based emulators. The open-source \pypackage{pyPMM} package---written as a part of this thesis---implements this uncertainty heuristic and the split conformal prediction process~\cite{pyPMM2026}.

\chapter{Conclusion and Outlook}\label{ch:conclusion}
\acresetall
Nuclear and many-body physics today are limited by the computational resources available. The complexities of our modern models of these systems necessitate complicated and expensive algorithms to solve these models at the highest levels of accuracy. Simultaneously, these models have free parameters that must be fit to experimental observations, and uncertainties in both these free parameters and model predictions must be rigorously quantified. This all requires hundreds of thousands to millions of evaluations of these expensive models, which today is intractable. Fast and efficient, but also accurate and precise, surrogate models---or emulators---bridge this gap.

This thesis presented a slice of the current state of emulation in nuclear and many-body physics. After establishing why we need emulation and what we want from it, several popular methods were explored. Particular attention was paid to potentially the oldest technique, Galerkin's Method. It was shown that many emulation methods in use today can be interpreted as variations or special cases of Galerkin's Method---assisting in a loosely unified analysis. However, all such methods lack a desirable property: numerical parametricity. In contrast, the method of \acp{pinn} is guaranteed this property, albeit at the cost of many others and potentially practical effectiveness altogether.

We introduced \acp{pmm}, an emulation and general \ac{ml} technique predicated on the idea of introducing numerical parametricity directly into Galerkin's Method. The interpretation of \acp{pmm} as Galerkin's Method with implicit subspaces and bases allowed for a formal mathematical framework for analyzing \ac{pmm}-based emulators as well as an explicit step-by-step process to constructing such emulators. We saw, through examples from simple linear systems to intricate models with missing information, that this method is maximally adaptable, exceedingly efficient, and controllably accurate.

As part of this thesis, I developed the open-source \pypackage{pyPMM} package~\cite{pyPMM2026}. This package facilitates the construction, training, sharing, and deployment of \acp{pmm} in a modular, extendable, and \ac{gpu}-optimized framework thanks to the \pypackage{JAX} package~\cite{JAX}.

Research is currently ongoing across several groups to apply \ac{pmm}-based emulation to ever-larger and more difficult problems. In particular, extensions of the emulator for the nuclear \ac{eos} in Section~\ref{sec:pmm_eos} to natively emulate \ac{imsrg} calculations of finite nuclei, multiple \ac{pmm} emulators for various nuclear lattice \ac{ceft} calculations, and \ac{pmm} emulation of large-scale nuclear reaction networks are directions all being pursued currently or in the immediate future. Although the method is mature enough now for research to be mainly focused on applications, there are many questions to be answered and algorithmic improvements to be made. For instance, the optimal pre-training initialization of \acp{pmm} parameters is currently unknown and could drastically reduce the cost of training. The extent to which purely data-driven \acp{pmm} are competitive with modern \ac{ann} methods and what exactly their scaling laws are in these contexts is under active research. The answers to these questions may have impacts on partially data-driven \ac{pmm}-based emulators, which are perhaps the most powerful application of the method.

\Acp{pmm} provide a unique, adaptive, and powerful tool for emulation in physics, enabling scientific progress to continue in the face of computational bottlenecks.

\printbibliography

\begin{appendices}
    \chapter{Matrix-Free Operators}\label{app:matrixfree}
    \acresetall
    A key observation in many numerical methods used across scientific computing is that rarely are the individual matrix elements of linear operators necessary~\cite{Hestenes1952, Liu1989, Knoll2004, Lehoucq1998}. Instead, often only the \textit{action} of the operator is required. For an operator $A$ and an element of the vector space in which it acts on, $v$, this is denoted by the usual left multiplication, $Av$. In numerical contexts, this is often given the more practical term of \textit{matrix-vector products}. Many core software libraries for scientific computing already support the representation of operators via functions on vectors~\cite{Lehoucq1998,SciPy}. Even when representing operators as matrices element-by-element, one must have considered the matrix-vector product to derive the matrix elements. In this sense, representing operators by their action is more natural.

When representing elements of a vector space numerically, the ubiquitous choice is to represent them as vectors (one dimensional arrays). Doing so necessitates the choice of a basis, which will also determine the representation of any operators---regardless of whether they are represented by explicit matrices or by their actions. Consider the standard quantum harmonic oscillator in one dimension. Choosing the position basis, the Hamiltonian is written as\footnote{With $\hbar=1$.}
\begin{equation}\label{eq:ho_x}
    H = -\frac{1}{2m}\frac{\partial^2}{\partial x^2} + \frac{1}{2}m\omega^2 x^2,
\end{equation}
and the state as $\Psi(x)\equiv \int\Psi(x)\delta(x-u)du$. In contrast, in the energy basis the Hamiltonian is written as
\begin{equation}\label{eq:ho_E}
    H = \omega\left(a^\dagger a + \frac{1}{2}\right),
\end{equation}
and the state as $\ket{\Psi}\equiv\sum_{k=0} c_k \ket{k}$, where $k=0,1,\ldots$ and $a^\dagger a\ket{k} = k\ket{k}$. Note that both representations of the state vector are expressed as an integral (in the case of a continuous basis) or sum (in the case of a discrete basis) of some state-specific coefficients times the basis states. These coefficients are what we store programmatically to represent the states numerically, leaving the basis entirely implicit. In the first example, we would first approximate the continuous basis via discretization,
\begin{equation}\label{eq:x_vec}
    \begin{aligned}
        \relax[-\infty,\infty] &\approx \mathlist{x_1, x_2, \ldots, x_N},\qquad x_k\equiv x_\text{min} + (k-1)\Delta x,\\
        \implies \Psi(x)&\approx \sum_k \underbrace{\Psi(x_k)}_{\text{coeff.}}\underbrace{s_k(x)}_{\text{basis st.}},\\
        \text{where}~s_k(x) &\equiv \begin{cases}
            \frac{1}{\sqrt{\Delta x}} & x_k - \frac{\Delta x}{2} \leq x < x_k + \frac{\Delta x}{2}\\
            0 & \text{otherwise}
            \end{cases},
    \end{aligned}
\end{equation}
then numerically the one dimensional array representing the state would be the column vector
\begin{equation}\label{eq:x_num}
    \Psi^{(x)}\doteq \mathlist{\Psi(x_1)~\Psi(x_2)~\cdots~\Psi(x_N)}^\mathsf{T}.
\end{equation}
Similarly, for the energy basis we would first truncate the summation so as to only retain the lowest $N$ states,
\begin{equation}\label{eq:e_vec}
        \ket{\Psi}\approx \sum_{k=0}^{N-1} c_k \ket{k},
\end{equation}
and then store the vector of coefficients,
\begin{equation}\label{eq:e_num}
    \implies \Psi^{(E)}\doteq\mathlist{c_0~c_1~\cdots~c_{N-1}}^\mathsf{T}.
\end{equation}
These approximations are necessary to represent what is fundamentally an infinitely large Hilbert space with finite computational resources. Note that in both cases, the basis states are orthonormal:
\begin{equation}
    \begin{aligned}
        \int s_i(x)s_j(x)dx &= \delta_{ij},\\
        \braket{i}{j} &= \delta_{ij}.
    \end{aligned}
\end{equation}
From here, the usual process is to examine the ``matrix elements'' of the Hamiltonian in the given basis. For the position basis, we must additionally approximate the continuous derivative by a suitable finite-difference approximation,
\begin{equation}
    \frac{\partial^2}{\partial x^2} f(x) \approx \frac{f(x+\Delta x) - 2f(x) + f(x - \Delta x)}{\left(\Delta x\right)^2}.
\end{equation}
Splitting Eq.~\eqref{eq:ho_x} into a kinetic term $T^{(x)}$ and a potential term $V^{(x)}$, the matrix element of the kinetic energy term of Eq.~\eqref{eq:ho_x} is
\begin{equation}
    \begin{aligned}
        T^{(x)}_{ij} &\doteq -\frac{1}{2m}\int s_i(x) \frac{\partial^2}{\partial x^2} s_j(x) dx\\
        &\approx -\frac{1}{2m\left(\Delta x\right)^2} \int s_i(x) \left[s_j(x+\Delta x) - 2s_j(x) + s_j(x-\Delta x)\right]dx\\
        &= -\frac{1}{2m\left(\Delta x\right)^2}\left[\delta_{i,\,j-1} -2\delta_{ij} + \delta_{i,\,j+1}\right],
    \end{aligned}
\end{equation}
and the matrix element of the potential term is
\begin{equation}
    \begin{aligned}
        V^{(x)}_{ij} &\doteq\frac{1}{2}m\omega^2\int s_i(x) x^2 s_j(x) dx\\
               &=\frac{1}{2}m\omega^2\delta_{ij}\left[x_i^2 +\frac{\Delta x^2}{12}\right]\\
               &\approx\frac{1}{2}m\omega^2\delta_{ij} x_i^2.
    \end{aligned}
\end{equation}
And so the Hamiltonian in this discrete position basis can be written as the tridiagonal matrix\footnote{We have ignored boundary conditions here, and therefore implicitly assumed Dirichlet boundary conditions where $\Psi(x)=0$ for $x<x_\text{min}-\frac{1}{2}$ or $x\geq x_\text{max}+\frac{1}{2}$, which is equivalent to placing the potential within an infinite well at the boundaries.}
\begin{equation}
    H^{(x)} \doteq -\frac{1}{2m\left(\Delta x\right)^2}\left[\sqmatrix{%
        -2 & 1 & & & \\
        1 & -2 & 1 & & \\
          & \ddots & \ddots & \ddots &\\
        & &1&-2 &1\\
        & & & 1 & -2
    }\right]
        + \frac{1}{2}m\omega^2\left[\sqmatrix{%
        x_1^2 &   & & & \\
          & x_2^2 &   & & \\
        &   & x_3^2 &   & \\
        & & &  \ddots & \\
         & & & & x_N^2
        }\right].
\end{equation}
And likewise in the energy basis,
\begin{equation}
    \begin{aligned}
        H^{(E)}_{nm} &\doteq \sandwich{n}{H}{m}\\&=\omega \sandwich{n}{a^\dagger a}{m} + \frac{\omega}{2}\braket{n}{m}\\&=\omega\left(n +\frac{1}{2}\right)\delta_{nm}\\
        \implies H^{(E)}&\doteq \omega\begin{bmatrix}0 & & & \\&1&&\\&&\ddots&\\&&&N-1
        \end{bmatrix} + \frac{\omega}{2}I_{N\times N},
    \end{aligned}
\end{equation}
where $I_{N\times N}$ is the $N\times N$ identity matrix. This is the typical ``dense-matrix'' way one would go about implementing this system. These matrices multiply their respective numerical state vector representations in Eqs.~\eqref{eq:x_num}~and~\eqref{eq:e_num}---despite only being computed from their multiplication on the basis vectors.

Consider instead describing this Hamiltonian in each of these bases by its action on the numerical state vector representations, Eqs.~\eqref{eq:x_num}~and~\eqref{eq:e_num} directly. Each element of the result $H\Psi$ can only ever be a linear combination of the elements of $\Psi$. There are many ways to derive this action, one of which is to carry out the matrix-element derivation as before and reinterpret the results as an action. The more natural approach is to interpret the states and operators as one would for analytical, as opposed to numerical, expressions. For the position basis, we know the kinetic energy term is a scalar multiple of the second derivative which we approximate with finite differences,
\begin{equation}
    \Psi(x) \stackrel{T}{\longrightarrow} -\frac{1}{2m}\Psi^{\prime\prime}(x) \approx -\frac{1}{2m} \frac{\Psi(x+\Delta x) - 2\Psi(x) + \Psi(x-\Delta x)}{\left(\Delta x\right)^2},
\end{equation}
and similarly the potential energy term is just a multiplication by $m\omega^2 x^2/2$,
\begin{equation}
    \Psi(x)\stackrel{V}{\longrightarrow} \frac{1}{2}m\omega^2 x^2\Psi(x).
\end{equation}
Since the numerical representation of the state vector is simply the evaluation of this $\Psi(x)$ on a grid of positions with spacing $\Delta x$, we know element-by-element the result of applying either of these operators to an arbitrary coefficient vector,
\begin{equation}\label{eq:matfree_x}
    \begin{aligned}
        \left(T^{(x)}\Psi^{(x)}\right)_i &= -\frac{1}{2m\left(\Delta x\right)^2}\left(\Psi^{(x)}_{i+1} - 2\Psi^{(x)}_i + \Psi^{(x)}_{i-1}\right),\\
        \left(V^{(x)}\Psi^{(x)}\right)_i &= \frac{1}{2}m\omega^2 x_i^2 \Psi^{(x)}_i.
    \end{aligned}
\end{equation}
Equation~\eqref{eq:matfree_x} is known as the \textit{matrix-free} representation of the operators $T$ and $V$ in the position basis. Both of these expressions can be implemented efficiently with vectorized operations available in many libraries\footnote{For example---with periodic boundary conditions for simplicity---the action of the kinetic energy operator in the position basis can be written in \pypackage{Python}+\pypackage{NumPy} as the anonymous function \mbox{\texttt{T = lambda v: -(numpy.roll(v, +1) - 2 * v + numpy.roll(v, -1)) / (2 * m * dx**2)}}.}. Note that acting on a standard basis vector reproduces the corresponding column of the dense representation\footnote{Taking $\Psi_i=0$ for $i<0$ or $i>N$.},
\begin{equation}
    T \begin{bmatrix}
        1\\0\\0\\\vdots\\0
    \end{bmatrix} = -\frac{1}{2m\left(\Delta x\right)^2}
    \begin{bmatrix}
        -2\\1\\0\\\vdots\\0
    \end{bmatrix},\qquad\qquad\qquad T
    \begin{bmatrix}
        0\\1\\0\\0\\\vdots\\0
    \end{bmatrix} = -\frac{1}{2m\left(\Delta x\right)^2}
    \begin{bmatrix}
        1\\-2\\1\\0\\\vdots\\0
    \end{bmatrix}.
\end{equation}
Due to the energy basis being the basis in which the Hamiltonian is diagonal, expressing the action of the Hamiltonian in that basis is less involved,
\begin{equation}
    \left(H^{(E)}\Psi^{(E)}\right)_n = \omega\left(n + \frac{1}{2}\right)\Psi^{(E)}_n.
\end{equation}
However, the individual creation and annihilation operators, $a^\dagger$ and $a$, are more interesting, since
\begin{equation}
    \begin{aligned}
        a\ket{n} &= \begin{cases} 0 & n = 0\\ \sqrt{n}\,\ket{n-1} & \text{otherwise}\end{cases}\\a^\dagger \ket{n}&=\begin{cases} 0 & n = N - 1\\ \sqrt{n+1}\,\ket{n+1} & \text{otherwise}\end{cases},
    \end{aligned}
\end{equation}
for their actions on $\ket{\Psi}$ in Eq.~\eqref{eq:e_vec} we have
\begin{equation}
    \begin{alignedat}{3}
        \ket{\Psi} &\stackrel{a}{\longrightarrow} \sum_{k=1}^{N-1} c_{k}\sqrt{k}\,\ket{k-1} &&= \sum_{k=0}^{N-2} c_{k+1} \sqrt{k+1}\,\ket{k}\\
        \ket{\Psi} &\stackrel{a^\dagger}{\longrightarrow} \sum_{k=0}^{N-2} c_{k}\sqrt{k+1}\,\ket{k+1} &&= \sum_{k=1}^{N-1} c_{k-1}\sqrt{k}\,\ket{k},
    \end{alignedat}
\end{equation}
where the final expression for each makes it clear that the matrix-free representations are
\begin{equation}
    \begin{aligned}
        \left(a \Psi^{(E)}\right)_n &= \begin{cases}0 & n = N-1\\ c_{n+1} \sqrt{n+1}&\text{otherwise}\end{cases},\\
        \left(a^\dagger \Psi^{(E)}\right)_n &= \begin{cases}0 & n = 0\\ c_{n-1} \sqrt{n}&\text{otherwise}\end{cases}.
    \end{aligned}
\end{equation}
Again, these expressions can be implemented efficiently using vectorized operations.

Compared to the dense matrix representation\footnote{As well as implementations which leverage sparsity.}, matrix-free implementations can be significantly faster and less memory-intensive. For the examples here, the direct (naive) dense matrix approach would require $\bigO{N^2}$ multiplications per matrix-vector product and require the storage of $\bigO{N^2}$ elements for the operator\footnote{Non-naive sparse approaches would require $\bigO{(1-s)N^2}$ space where $s$ is the \textit{sparsity} of the operator.}. Compared to the matrix-free implementations which, in these examples, all require $\bigO{N}$ multiplications for matrix-vector products and $\bigO{N}$ space for the operator\footnote{The matrix-free representation of $T^{(x)}$ requires only $\bigO{1}$ space.}. \Ac{ad} is compatible with matrix-free implementations~\cite{Kraemer2024} and has been implemented for \pypackage{JAX} in Ref.~\cite{matfree}.

    \chapter{Long-Range Spin Model}\label{app:spin}
    \acresetall
    The \ac{lmg} model\footnote{Also commonly referred to simply as the Lipkin model.}, named after the authors of the paper which introduced it~\cite{Lipkin1965}, is a standard ``toy model'' in nuclear and many-body theory. Following the original description of the model, we consider $L$ particles distributed between two $L$-fold degenerate energy levels. The separation between the energy levels is $\varepsilon$. The state of each particle can be described by the quantum numbers $\sigma\in\mathset{-1,1}$ and $p\in\mathset{1,2,\ldots,L}$. $\sigma$ describes which of the two levels the particle occupies, $+1$ for the upper level and $-1$ for the lower level. $p$ labels the individual particle. These particles are subject to an infinite-range two-body interaction which scatters pairs of particles between the two levels without changing $p$. This system can be described by the Hamiltonian
\begin{equation}
\begin{aligned}
    H = \frac{1}{2}\varepsilon\sum_{p\sigma} \sigma a^\dagger_{p\sigma} a_{p\sigma} &+ \frac{1}{2}V\sum_{p p^\prime \sigma} a^\dagger_{p\sigma} a^\dagger_{p^\prime \sigma} a_{p^\prime (-\sigma)} a_{p (-\sigma)}\\
    &+ \frac{1}{2}W\sum_{p p^\prime \sigma} a_{p\sigma}^\dagger a_{p^\prime (-\sigma)}^\dagger a_{p^\prime \sigma} a_{p (-\sigma)},
\end{aligned}
\end{equation}
where $a^\dagger_{p\sigma}$ and $a_{p \sigma}$ are the usual creation and annihilation operators for a single particle in the $(p, \sigma)$ state, and $V$ and $W$ parameterize the strength of each type of interaction~\cite{Lipkin1965}. The first term is the energy contribution of the single particle states, here $-\varepsilon/2$ or $\varepsilon/2$. The second term represents the interaction that sends two different particles from the same level to the opposing level. The third term exchanges particles on different levels, sending one particle to the lower level from the upper level and the other to the upper level from the lower level. Up to an additive constant---which merely shifts the single particle energy levels---this is equivalent to describing the system as $L$ distinguishable spin-$1/2$ particles under the re-parameterized Hamiltonian with infinite-range interactions,
\begin{equation}\label{eq:lmg_spin}
    H = -\frac{J}{L}\sum_{i<j}^L\left(\gamma_x \sigma_i^x \sigma_j^x + \gamma_y \sigma_i^y\sigma_j^y\right) - B \sum_i^L \sigma_i^z,
\end{equation}
where $\sigma_k^u$ is the usual Pauli $u$ operator for particle $k$, $B$ parameterizes the single-particle energy levels, and $J$, $\gamma_x$, and $\gamma_y$ parameterize the interaction\footnote{This is equivalent to the original representation of the \ac{lmg} model with $\varepsilon=-2B$, $V=-J\left(\gamma_x-\gamma_y\right)/L$, and $W=-J\left(\gamma_x+\gamma_y\right)/L$---with an additional additive constant $J\left(\gamma_x+\gamma_y\right)/2$.}. To completely disambiguate the notation,
\begin{equation}
    \begin{aligned}
        \ket{\sigma_{L-1},\ldots,\sigma_1,\sigma_0}&\equiv\ket{\sigma_{L-1}}\otimes\cdots\otimes\ket{\sigma_1}\otimes\ket{\sigma_0},\\
        \sigma_k^u &\equiv \underbrace{I}_{L-1}\otimes \underbrace{I}_{L-2}\otimes\cdots\otimes\underbrace{\sigma^u}_k \otimes\cdots\otimes \underbrace{I}_1 \otimes \underbrace{I}_0,
    \end{aligned}
\end{equation}
which means
\begin{equation}
    \sigma_k^u\ket{\sigma_{L-1},\ldots,\sigma_1,\sigma_0}=\left(I\ket{\sigma_{L-1}}\right)\otimes\cdots\otimes\left(\sigma^u\ket{\sigma_k}\right)\otimes\cdots\otimes\left(I\ket{\sigma_1}\right)\otimes\left(I\ket{\sigma_0}\right).
\end{equation}
This formulation bears striking resemblance to the Ising model, and in practice the \ac{lmg} model is sometimes referred to as the ``infinite-range Ising model''. This individual-particle representation would require basis states described by $L$ quantum numbers, $\ket{\sigma_1\sigma_2\cdots\sigma_L}$, which each could take on two possible values. Thus the Hilbert space would be of dimension $2^L$.

Alternatively, it is more common to see the \ac{lmg} model refer only to the Hamiltonian expressed in the quasi-spin basis,
\begin{equation}
    H = \varepsilon J_0 + \frac{1}{2}V\left(J_+^2 + J_-^2\right) + \frac{1}{2}W\left(J_+J_- + J_-J_+\right),
\end{equation}
where
\begin{equation}
    J_+ = \sum_p a^\dagger_{p+1} a_{p-1},\qquad J_- = \sum_p a^\dagger_{p-1}a_{p+1},\qquad J_0 = \frac{1}{2}\sum_{p\sigma}\sigma a^\dagger_{p\sigma}a_{p\sigma}.
\end{equation}
The basis states are then represented by just two quantum numbers,
\begin{equation}
    \begin{aligned}
        J&\in\left\{\frac{L}{2}, \frac{L}{2}-1, \ldots, \left(\begin{aligned}0~&\text{if}~L~\text{even}\\1/2~&\text{if}~L~\text{odd}\end{aligned}\right)\right\},\\
    J_0&\in\left\{-\frac{J}{2}, -\frac{J}{2}+1, \ldots, \frac{J}{2}\right\}.
    \end{aligned}
\end{equation}
The Hamiltonian does not change $J$ since the operator $J^2=\frac{1}{2}\left(J_+J_-+J_-J_+\right)+J_0^2$ commutes with the Hamiltonian. Because the ground state of the non-interacting ($V=0,\,W=0$) Hamiltonian has $J=L/2$---as the non-interacting ground state must have $J_0=-L/2$---it is common to consider only the \textit{sector} with $J=L/2$. This sector is completely described by only $L+1$ basis states, significantly fewer than the full Hilbert space.

While the \ac{lmg} model exhibits interesting phase-transition physics, it is fundamentally a mean-field model~\cite{Titum2020}. However, a beyond-mean-field generalization of Eq.~\eqref{eq:lmg_spin} exists that exhibits much of the same interesting physics and is physically realizable on a quantum simulator~\cite{De25}. By allowing the pairwise interactions to vary---therefore no longer necessarily being infinite range---one arrives at
\begin{equation}\label{eq:longrange_spin}
    H = -\frac{1}{\mathcal{J}}\sum_{i<j}^L J_{ij}\left(\gamma_x \sigma_i^x \sigma_j^x + \gamma_y \sigma_i^y\sigma_j^y\right) - B \sum_i^L \sigma_i^z,
\end{equation}
where $\mathcal{J}$ is a normalization factor introduced to ensure a well-defined thermodynamic limit---called the Kac normalization factor---given by
\begin{equation}
    \mathcal{J}=\frac{1}{L-1}\sum_{i\neq j}^LJ_{ij}.
\end{equation}
For a general interaction $J_{ij}$, the only reasonable choice of basis is the individual particle basis, usually specified by the $z$-projection of each spin, $\ket{\sigma^z_1\cdots\sigma^z_N}$.

Reference~\cite{De25} includes details of the quantum-simulator implementation of this model. My contributions to Ref.~\cite{De25} included predicting critical exponents for non-equilibrium critical phenomena and implementing numerical simulations of the experimental system.

Here we focus on the computational aspects of this model, specifically the matrix-free implementation of the Hamiltonian and relevant operators (see Appendix~\ref{app:matrixfree}). Since the dimensionality of the Hilbert space is $2^L$, intractable for even modest $L$, either approximate methods\footnote{Such as Density Matrix Renormalization Group (DMRG).} must be used or great care must be taken to implement the matrix-vector products as efficiently as possible.

A dense-matrix implementation of the Hamiltonian would require storing $4^L$ matrix elements, cost $\bigO{4^L}$ per matrix-vector product, and cost $\bigO{8^L}$ to compute the ground state or time dynamics. Sparse methods are not easily applicable for an arbitrary $J_{ij}$. However, with a matrix-free implementation of the Hamiltonian combined with a suitable eigensolver or time evolution method (see Section~\ref{sec:apdr}) the memory cost can be reduced to $\bigO{L^2}$, and the matrix-vector product, ground state computation, and time dynamics costs can all be reduced to $\bigO{L^2 2^L}$. We start with a change in notation: Let $\sigma_i\in\mathset{0,1}$, where $0$ labels the $\expect{\sigma_i^z}=-1$ state and $1$ labels the $\expect{\sigma_i^z}=+1$ state. Additionally, begin the indexing of the particles at $0$. Now, consider the basis expansion of a general state,
\begin{equation}
    \begin{aligned}
        \ket{\Psi} &= \sum_{\sigma_{L-1},\ldots,\sigma_1,\sigma_0\in\mathset{0,1}} c_{\sigma_{L-1},\ldots,\sigma_1,\sigma_0} \ket{\sigma_{L-1},\ldots,\sigma_1,\sigma_0}\\
        &= c_{00\cdots00}\ket{00\cdots00} \begin{aligned}[t]&+ c_{00\cdots01}\ket{00\cdots01}\\ &+ c_{00\cdots10}\ket{00\cdots10}\\ &+ \cdots\\ &+ c_{11\cdots11}\ket{11\cdots11},\end{aligned}
    \end{aligned}
\end{equation}
which strongly suggests the re-labeling of the basis by the base-$10$ number represented by the binary string of $\sigma^z_i$. For example, $\ket{0011}\rightarrow\ket{\bin{0011}}=\ket{\dec{3}}$. This leads us to mapping non-negative integers to the quantum numbers by $k=\bitstr{\sigma_{L-1}\cdots\sigma_1\sigma_0}$. Writing the general state in this way yields
\begin{equation}
    \ket{\Psi} = \sum_k^{2^L} c_k \ket{k},\qquad\text{where}\qquad\ket{k}=\ket{\bitstr{\sigma_{L-1}\cdots\sigma_1\sigma_0}}.
\end{equation}
Not only is this a natural way to order the coefficients of the numerical representation of the state ($c_k$), but it is simultaneously a convenient basis for describing the action of the $\sigma_k^u$ operators. Examine the action of each of the single-particle $\sigma^u$ on the single-particle basis states $\ket{0}$ and $\ket{1}$,
\begin{equation}
    \begin{alignedat}{2}
        \sigma^x\ket{0} &= \ket{1}, &\qquad \sigma^x\ket{1} &= \ket{0},\\
        \sigma^y\ket{0} &= -i\ket{1}, &\qquad \sigma^y\ket{1} &= i\ket{0},\\
        \sigma^z\ket{0} &= -\ket{0}, &\qquad \sigma^z\ket{1} &= \ket{1}.
    \end{alignedat}
\end{equation}
That is, $\sigma^x$ ``flips'' the state, $\sigma^y$ flips the state and multiplies by $\mp i$, and $\sigma^z$ multiplies by $\mp 1$. For the full $L$-particle state then, the actions of $\sigma_k^u$ on the state $\ket{n}$ are
\begin{equation}
    \begin{aligned}
        \sigma_k^x\ket{n} &= \sigma_k^x\ket{\bitstr{\sigma_{L-1}\cdots\sigma_1\sigma_0}}\\
        &= \ket{\bitstr{\sigma_{L-1}\cdots\overline{\sigma_k}\cdots\sigma_1\sigma_0}}\\
        &= \ket{n\oplus\left(1\ll k\right)},\\
        \sigma_k^y\ket{n} &= \sigma_k^y\ket{\bitstr{\sigma_{L-1}\cdots\sigma_1\sigma_0}}\\
        &= (-1)^{\sigma_k+1}i\ket{\bitstr{\sigma_{L-1}\cdots\overline{\sigma_k}\cdots\sigma_1\sigma_0}}\\
        &= (-1)^{\sigma_k+1}i\ket{n\oplus\left(1\ll k\right)},\\
        \sigma_k^z\ket{n} &= \sigma_k^z\ket{\bitstr{\sigma_{L-1}\cdots\sigma_1\sigma_0}}\\
        &= (-1)^{\sigma_k+1}\ket{\bitstr{\sigma_{L-1}\cdots\sigma_k\cdots\sigma_1\sigma_0}}\\
        &= (-1)^{\sigma_k+1}\ket{n},
    \end{aligned}
\end{equation}
where $\overline{b}$ is the negation of $b$\footnote{$\overline{0}=1$ and $\overline{1}=0$.}, $\oplus$ is the bitwise XOR operation\footnote{For example, $3\oplus 2=\bin{11}\oplus\bin{01}=\bitstr{(1\oplus1)(1\oplus0)}=\bin{01}=1$.}, and $a\ll b$ denotes the logical left-shift of $a$ by $b$ bits\footnote{For example $1\ll 3=\bin{0001}\ll 3=\bin{1000}=8$.}. Bitwise operations like these are among the fastest operations for computers. In practice, $(-1)^{\sigma_k+1}$ is replaced with a much faster boolean statement,
\begin{equation}
    (-1)^{\sigma_k+1} \equiv \begin{cases} +1 & \text{if}~\sigma_k\\
        -1 & \text{otherwise} \end{cases}.
\end{equation}
Similar to the single-particle operators, we can interpret $\sigma_k^x$ as flipping the \ord{k} particle, $\sigma_k^y$ as flipping the \ord{k} particle and multiplying by $\mp i$, and $\sigma_k^z$ as multiplying by $\mp 1$. Extending this to $\sigma_i^u\sigma_j^u$ with $i\neq j$ and $u\in\mathset{x,y}$, which appear in Eq.~\eqref{eq:longrange_spin}, yields
\begin{equation}
    \begin{aligned}
        \sigma_i^x\sigma_j^x\ket{n} &= \ket{n \oplus (1 \ll i) \oplus (1 \ll j)},\\
        \sigma_i^y\sigma_j^y\ket{n} &= -(-1)^{\sigma_i + \sigma_j}\ket{n \oplus (1 \ll i) \oplus (1 \ll j)},
    \end{aligned}
\end{equation}
where $-(-1)^{\sigma_i+\sigma_j}$ is $+1$ if the bits at index $i$ and $j$ in the binary representation of $n$ are different and $-1$ if they are the same\footnote{A more efficient implementation to check if the bits at $i$ and $j$ are equal is checking the truth value of the expression $(n\gg i)\mathbin{\&}1=(n\gg j)\mathbin{\&}1$ where $\&$ is the bitwise AND operation and $a\gg b$ denotes the logical right-shift of $a$ by $b$ bits. So $-(-1)^{\sigma_i + \sigma_j}$ would be written as $\begin{cases}-1&\text{if}~(n\gg i)\mathbin{\&}1=(n\gg j)\mathbin{\&}1\\+1&\text{otherwise}\end{cases}$.}.
Additionally for $\sum_i\sigma_i^z$,
\begin{equation}
    \begin{aligned}
        \sum_i^L\sigma_i^z \ket{n} &= \sum^L_i\sigma_i^z \ket{\bitstr{\sigma_{L-1}\cdots\sigma_1\sigma_0}}\\
        &= -\left(\sum^L_i (-1)^{\sigma_i}\right)\ket{n}\\
        &= \left(2 w_H(n) - L\right)\ket{n},
    \end{aligned}
\end{equation}
where $w_H(n)$ is the number of ones in the binary representation of $n$, known as the Hamming weight of $n$.

We now have the action of all operators in the Hamiltonian on the basis states. Applying these to the general state vector, $\ket{\Psi}=\sum_k^{2^L}c_k\ket{k}$,
\begin{equation}
    \begin{aligned}
        \sigma_i^x\sigma^x_j\ket{\Psi}&=\sum_k^{2^L}c_k\ket{k\oplus\left(1\ll i\right)\oplus\left(1\ll j\right)},\\
        \sigma_i^y\sigma_j^y\ket{\Psi}&=\sum_k^{2^L}-(-1)^{\sigma_i + \sigma _j}c_k\ket{k\oplus\left(1\ll i\right)\oplus\left(1\ll j\right)},\\
        \sum_i^L\sigma_i^z\ket{\Psi} &= \sum_k^{2^L}\left(2 w_H(k) - L\right)c_k\ket{k}.
    \end{aligned}
\end{equation}
We can then ``reverse'' these relations to get the action of the operators on the numerical representation of the state vector, $\Psi\doteq\left[c_0\,c_1\,\cdots\,c_{L-1}\right]^\mathsf{T}$,
\begin{equation}
    \begin{aligned}
        \left(\sigma_i^x\sigma_j^x \Psi\right)_k &= c_{k\oplus\left(1\ll i\right)\oplus\left(1\ll j\right)},\\
        \left(\sigma_i^y\sigma_j^y\Psi \right)_k &= -(-1)^{\sigma_i + \sigma_j} c_{k\oplus\left(1\ll i\right)\oplus\left(1\ll j\right)},\\
        \left(\sum_i^L\sigma_i^z \Psi\right)_k &= \left(2 w_H(k) - L\right)c_k.
    \end{aligned}
\end{equation}
Finally, combining these actions to form the action of the full Hamiltonian in Eq.~\eqref{eq:longrange_spin} on the numerical state vector yields
\begin{equation}
    \begin{aligned}
        (H\Psi)_k = &-\frac{1}{\mathcal{J}}\sum_{i<j}^L J_{ij} \left(\gamma_x - (-1)^{\sigma_i+\sigma_j}\gamma_y\right) c_{k\oplus\left(1\ll i\right)\oplus\left(1\ll j\right)}\\
                    &-B\left(2w_H(k)-L\right)c_k,
    \end{aligned}
\end{equation}
or by using the more performant expression for the sign of the $\gamma_y$ term\footnote{Note that, in either expression for the sign of the $\gamma_y$ term, we are really checking to see if the \ord{i} and \ord{j} bits in $k\oplus\left(1\ll i\right)\oplus\left(1\ll j\right)$ are equal, but this is always equivalent to checking the same bits in $k$.},
\begin{equation}
    \begin{aligned}
        (H\Psi)_k = &-\frac{1}{\mathcal{J}}\sum_{i<j}^L J_{ij} \left[\gamma_x + \left(\begin{aligned}-1~&\text{if}~(k\gg i)\mathbin{\&}1=(k\gg j)\mathbin{\&}1\\+1~&\text{otherwise}\end{aligned}\right)\gamma_y\right] c_{k\oplus\left(1\ll i\right)\oplus\left(1\ll j\right)}\\
                    &-B\left(2w_H(k)-L\right)c_k.
    \end{aligned}
\end{equation}
Each term in the sum in this implementation for the action of the Hamiltonian requires only a handful of bitwise operations and array accesses which are all $\bigO{1}$. Since there are $\bigO{L^2}$ terms in the sum and $\bigO{2^L}$ elements in the numerical state vector, the complete action can be calculated in $\bigO{L^2 2^L}$ time, a significant improvement over the $\bigO{4^L}$ complexity of the dense-matrix implementation. Additionally, the Hamiltonian itself no longer requires the storage of any data besides the handful of parameters describing the interaction. Altogether, these efficiency improvements nearly double the achievable system size compared to the dense-matrix approach and allowed for the simulations of the experimental system in Ref.~\cite{De25} up to $L=25$ to be carried out on a low-power consumer-grade laptop.

\end{appendices}
\end{document}